\RequirePackage{lineno}   % This is provided with the package

\documentclass[12pt]{article}
\usepackage{epsfig}
\usepackage{amsmath}
\usepackage{hhline}
\usepackage{amssymb}
\usepackage{times}
\usepackage{cite}
\usepackage{rotate}
\usepackage{rotating}  %% to allow sidewaystable 
\usepackage{multirow}
\usepackage{subfigure}
\usepackage{hyperref} %allow html
\hypersetup{colorlinks=true,dvips,backref=true}
\usepackage[all]{hypcap} %allows hyperref to point to fig/tab not at
\newlength{\dinwidth}
\newlength{\dinmargin}
\begin{document}  

%\linenumbers   % turn on line numbering

% The rest
\newcommand\brabar{\raisebox{-3.0pt}{\scalebox{.4}{
\textbf{(}}}\raisebox{-2.0pt}{{\_}}\raisebox{-3.0pt}{\scalebox{.4}{\textbf{)}}}}

\newcommand{\pom}{{I\!\!P}}
\newcommand{\reg}{{I\!\!R}}
\newcommand{\slowpi}{\pi_{\mathit{slow}}}
\newcommand{\fiidiii}{F_2^{D(3)}}
\newcommand{\fiidiiiarg}{\fiidiii\,(\beta,\,Q^2,\,x)}
\newcommand{\n}{1.19\pm 0.06 (stat.) \pm0.07 (syst.)}
\newcommand{\nz}{1.30\pm 0.08 (stat.)^{+0.08}_{-0.14} (syst.)}
\newcommand{\fiidiiiful}{F_2^{D(4)}\,(\beta,\,Q^2,\,x,\,t)}
\newcommand{\fiipom}{\tilde F_2^D}
\newcommand{\ALPHA}{1.10\pm0.03 (stat.) \pm0.04 (syst.)}
\newcommand{\ALPHAZ}{1.15\pm0.04 (stat.)^{+0.04}_{-0.07} (syst.)}
\newcommand{\fiipomarg}{\fiipom\,(\beta,\,Q^2)}
\newcommand{\pomflux}{f_{\pom / p}}
\newcommand{\nxpom}{1.19\pm 0.06 (stat.) \pm0.07 (syst.)}
\newcommand {\gapprox}
   {\raisebox{-0.7ex}{$\stackrel {\textstyle>}{\sim}$}}
\newcommand {\lapprox}
   {\raisebox{-0.7ex}{$\stackrel {\textstyle<}{\sim}$}}
\def\gsim{\,\lower.25ex\hbox{$\scriptstyle\sim$}\kern-1.30ex%
\raise 0.55ex\hbox{$\scriptstyle >$}\,}
\def\lsim{\,\lower.25ex\hbox{$\scriptstyle\sim$}\kern-1.30ex%
\raise 0.55ex\hbox{$\scriptstyle <$}\,}
\newcommand{\pomfluxarg}{f_{\pom / p}\,(x_\pom)}
\newcommand{\dsf}{\mbox{$F_2^{D(3)}$}}
\newcommand{\dsfva}{\mbox{$F_2^{D(3)}(\beta,Q^2,x_{I\!\!P})$}}
\newcommand{\dsfvb}{\mbox{$F_2^{D(3)}(\beta,Q^2,x)$}}
\newcommand{\dsfpom}{$F_2^{I\!\!P}$}
\newcommand{\gap}{\stackrel{>}{\sim}}
\newcommand{\lap}{\stackrel{<}{\sim}}
\newcommand{\fem}{$F_2^{em}$}
\newcommand{\tsnmp}{$\tilde{\sigma}_{NC}(e^{\mp})$}
\newcommand{\tsnm}{$\tilde{\sigma}_{NC}(e^-)$}
\newcommand{\tsnp}{$\tilde{\sigma}_{NC}(e^+)$}
\newcommand{\st}{$\star$}
\newcommand{\sst}{$\star \star$}
\newcommand{\ssst}{$\star \star \star$}
\newcommand{\sssst}{$\star \star \star \star$}
\newcommand{\tw}{\theta_W}
\newcommand{\sw}{\sin{\theta_W}}
\newcommand{\cw}{\cos{\theta_W}}
\newcommand{\sww}{\sin^2{\theta_W}}
\newcommand{\cww}{\cos^2{\theta_W}}
\newcommand{\trm}{m_{\perp}}
\newcommand{\trp}{p_{\perp}}
\newcommand{\trmm}{m_{\perp}^2}
\newcommand{\trpp}{p_{\perp}^2}
\newcommand{\alp}{\alpha_s}

\newcommand{\alps}{\alpha_s}
\newcommand{\sqrts}{$\sqrt{s}$}
\newcommand{\LO}{$O(\alpha_s^0)$}
\newcommand{\Oa}{$O(\alpha_s)$}
\newcommand{\Oaa}{$O(\alpha_s^2)$}
\newcommand{\PT}{p_{\perp}}
\newcommand{\JPSI}{J/\psi}
\newcommand{\sh}{\hat{s}}
\newcommand{\uh}{\hat{u}}
\newcommand{\MP}{m_{J/\psi}}
\newcommand{\PO}{I\!\!P}
\newcommand{\xbj}{x}
\newcommand{\xpom}{x_{\PO}}
\newcommand{\ttbs}{\char'134}
\newcommand{\xpomlo}{3\times10^{-4}}  
\newcommand{\xpomup}{0.05}  
\newcommand{\dgr}{^\circ}
\newcommand{\pbarnt}{\,\mbox{{\rm pb$^{-1}$}}}
\newcommand{\gev}{\,\mbox{GeV}}
\newcommand{\WBoson}{\mbox{$W$}}
\newcommand{\fbarn}{\,\mbox{{\rm fb}}}
\newcommand{\fbarnt}{\,\mbox{{\rm fb$^{-1}$}}}

\newcommand{\vtab}{\rule[-1mm]{0mm}{4mm}}
\newcommand{\htab}{\rule[-1mm]{0mm}{6mm}}
\newcommand{\photoproduction}{$\gamma p$}
\newcommand{\ptmiss}{$P_{T}^{\rm miss}$}
\newcommand{\epz} {$E{\rm-}p_z$}
\newcommand{\vap} {  $V_{ap}/V_p$}
\newcommand{\Zero}   {\mbox{$Z^{\circ}$}}
\newcommand{\Ftwo}   {\mbox{$\tilde{F}_2$}}
\newcommand{\Ftwoz}   {\mbox{$\tilde{F}_{2,3}$}}
\newcommand{\Fz}   {\mbox{$\tilde{F}_3$}}
\newcommand{\FL}   {\mbox{$\tilde{F}_{_{L}}$}}
\newcommand{\wtwogen} {W_2}
\newcommand{\wlgen} {W_L}
\newcommand{\xwthreegen} {xW_3}
\newcommand{\Wtwo}   {\mbox{$W_2$}}
\newcommand{\Wz}   {\mbox{$W_3$}}
\newcommand{\WL}   {\mbox{$W_{_{L}}$}}
\newcommand{\Fem}  {\mbox{$F_2$}}
\newcommand{\Fgam}  {\mbox{$F_2^{\gamma}$}}
\newcommand{\Fint} {\mbox{$F_2^{\gamma Z}$}}
\newcommand{\Fwk}  {\mbox{$F_2^{Z}$}}
\newcommand{\Ftwos} {\mbox{$F_2^{\gamma Z, Z}$}}
\newcommand{\Fzz} {\mbox{$F_3^{\gamma Z, Z}$}}
\newcommand{\Fintz} {\mbox{$F_{2,3}^{\gamma Z}$}}
\newcommand{\Fwkz}  {\mbox{$F_{2,3}^{Z}$}}
\newcommand{\Fzint} {\mbox{$F_3^{\gamma Z}$}}
\newcommand{\Fzwk}  {\mbox{$F_3^{Z}$}}
\newcommand{\Gev}  {\mbox{${\rm GeV}$}}
\newcommand{\Gevv}{\mbox{${\rm GeV}^2$}}
\newcommand{\QQ}  {\mbox{${Q^2}$}}
\newcommand{\gv}{GeV$^2\,$}
\newcommand{\bs}{\bar{s}}
\newcommand{\bc}{\bar{c}}
\newcommand{\bu}{\bar{u}}
\newcommand{\bb}{\bar{b}}
\newcommand{\bU}{\bar{U}}
\newcommand{\bD}{\bar{D}}
\newcommand{\bd}{\bar{d}}
\newcommand{\bq}{\bar{q}}    
\newcommand{\FLc}{$ F_{L}\,$} 
\newcommand{\xg}{$xg(x,Q^2)\,$}
\newcommand{\xgc}{$xg\,$}
\newcommand{\ipb}{pb$^{-1}\,$}               
\newcommand{\TOSS}{x_{{i}/{\PO}}}                                              
\newcommand{\un}[1]{\mbox{\rm #1}}
\newcommand{\pdsi}{$(\partial \sigma_r / \partial \ln y)_{Q^2}\,$}
\newcommand{\pdff}{$(\partial F_{2} / \partial \ln  Q^{2})_x\,$ }
\newcommand{\Fc}{$ F_{2}~$}
\newcommand{\amz}{$\alpha_s(M_Z^2)\,$} 

%
% Some useful tex commands
%
\newcommand{\qsq}{\ensuremath{Q^2} }
\newcommand{\gevsq}{\ensuremath{\mathrm{GeV}^2} }
\newcommand{\et}{\ensuremath{E_t^*} }
\newcommand{\rap}{\ensuremath{\eta^*} }
\newcommand{\gp}{\ensuremath{\gamma^*}p }
\newcommand{\dsiget}{\ensuremath{{\rm d}\sigma_{ep}/{\rm d}E_t^*} }
\newcommand{\dsigrap}{\ensuremath{{\rm d}\sigma_{ep}/{\rm d}\eta^*} }
% Journal macro
\def\Journal#1#2#3#4{{#1} {\bf #2} (#3) #4}
\def\NCA{\em Nuovo Cimento}
\def\NIM{\em Nucl. Instrum. Methods}
\def\NIMA{{\em Nucl. Instrum. Methods} {\bf A}}
\def\NPB{{\em Nucl. Phys.}   {\bf B}}
\def\PLB{{\em Phys. Lett.}   {\bf B}}
\def\PRL{\em Phys. Rev. Lett.}
\def\PRD{{\em Phys. Rev.}    {\bf D}}
\def\ZPC{{\em Z. Phys.}      {\bf C}}
\def\EJC{{\em Eur. Phys. J.} {\bf C}}
\def\CPC{\em Comp. Phys. Commun.}

\begin{titlepage}

\noindent
\begin{tabular}{ll}
%Date:    &  \today\\
%Version: &  0.9\\
\end{tabular}

\vspace{2cm}

\begin{center}
\begin{Large}

{\bf The Quark and Gluon Structure of the Proton}
\vspace{2cm}

E. Perez$^{a}$ and E. Rizvi$^b$\\

\end{Large}
\vspace{1cm}
{\it{$^a$ CERN, PH Department, Geneva, CH}} \\
{\it{$^b$ Queen Mary University of London, School of Physics and Astronomy,\\ London, UK }}
\end{center}

\vspace{2cm}

\begin{abstract}
\noindent

In this article we present a review of the structure of the proton and
the current status of our knowledge of the parton distribution functions
(PDFs). The lepton-nucleon scattering experiments which provide the
main constraints in PDF extractions are introduced and their
measurements are discussed. Particular emphasis is given to the HERA
data which cover a wide kinematic region. Hadron-hadron scattering
measurements which provide supplementary information are also
discussed. 
The methods used by various groups to
extract the PDFs in QCD analyses of hard scattering data are presented
and their results are compared.
The use of existing measurements allows predictions
for cross sections at the LHC to be made. A comparison of these
predictions for selected processes is given. First measurements from
the LHC experiments are compared to predictions and some initial
studies of the impact of this new data on the PDFs are
presented. 

\end{abstract}

\vspace{1.5cm}

\begin{center}
Submitted to Reports on Progress in Physics.
\end{center}

\end{titlepage}

\newpage

\clearpage
%%%%%%%%%%%%%%%%%%%%%%%
\section{Introduction}
%%%%%%%%%%%%%%%%%%%%%%%
\label{sec:intro}

The birth of modern experimental particle physics in which particles
were used to probe the structure of composite objects began with the
famous alpha particle scattering experiment of Geiger and Marsden
under the direction of Rutherford. In 1911 Rutherford published an
analysis of the data providing evidence for atomic structure consisting
of a massive positively charged nucleus surrounded by
electrons~\cite{Rutherford:1911zz}. Since then the use of particle
probes to deduce structure has become standard, albeit at increasingly
higher energy and intensity which brings its own technological and
experimental challenges.

Experiments of point-like electrons scattering off extended
objects such as nuclei were expected to deviate from the predictions
of Mott scattering - relativistic electron-electron Coulomb
scattering~\cite{mott}. This deviation, the nuclear form
factor (expressed in terms of the 4-momentum transfer $Q$ between
initial and final state electrons), was shown to be related to the
Fourier transform of the nuclear charge
density~\cite{Rose:1948zz,Rosenbluth:1950yq}. In 1955 Hofstadter
measured nuclear form factors with a $100-500$~MeV electron beam and
obtained the charge density of the proton and other atomic
nuclei~\cite{Hofstadter:1955ae,Chambers:1956zz,Hahn:1956zz}. The
experiment was able to resolve the proton's charge radius to
$\simeq~0.7$~fm, at least an order of magnitude better than
Rutherford's experiment.

The idea that nucleons were composite particles was first
proposed in 1964 by Zweig~\cite{Zweig:1981pd} and
Gell-Mann~\cite{GellMann:1964nj}. Their quark model represented an
underlying schema to classify the static properties of the known
hadrons. However, this model had difficulties explaining why direct
production of quarks did not occur.

Detailed study of the structure of the proton advanced in 1967 when a
$20$~GeV linear electron accelerator commenced operation at the
Stanford Linear Accelerator Centre (SLAC) with the aim of studying
inelastic proton scattering, resonance production and the inelastic
continuum in the region of $0.7 \leq Q^2 \leq 25$~GeV$^2$.  This
opened the field of deep inelastic scattering (DIS) in which the
nucleon target was dissociated to large invariant mass states in the
interaction. First observations of elastic scattering showed a rapid
$1/Q^4$ behaviour of the cross section~\cite{Coward:1967au} as
expected from earlier low energy elastic form factor measurements at
SLAC, Cornell, DESY and
CEA~\cite{Goitein:1967zz,Berkelman:1963zz,Albrecht:1967zz,Albrecht:1965ki,Bartel:1966zz,Janssens:1965kd}.
This was found to be in stark contrast to the weaker $Q^2$ dependence of
the inelastic cross section in the same energy
range~\cite{Bloom:1969kc,Breidenbach:1969kd}.

For inelastic Coulomb scattering two form factors are required to
describe the cross section, the so-called structure functions, which
at fixed lepton beam energy can only depend on two kinematic quantities
taken to be $Q^2$ and the electron energy loss in the nucleon rest
frame, $\nu$~\cite{Drell:1963ej}. The SLAC structure
function measurements were found to exhibit scaling behaviour i.e. were
independent of $Q^2$. This behaviour had been predicted by
Bjorken~\cite{Bjorken:1968dy} in the same year.

The new SLAC data prompted Feynman to develop the parton model of deep
inelastic scattering~\cite{Feynman:1969ej} in which the scaling
behaviour is naturally explained as the point-like elastic scattering
of free partons within the protons. Bjorken and Paschos then further
developed the quark-parton model~\cite{Bjorken:1969ja}. In the same year Callan and
Gross showed that the behaviour of the
longitudinally polarised part of the virtual photon scattering cross
section required the constituents to be spin $\frac{1}{2}$
fermions~\cite{Callan:1969uq}. The association of these point-like constituents as the
quarks of Gell-Mann and Zweig was gradually made and widely accepted
by 1974.

The inability to observe free quarks (confinement) and the free quarks
of the parton model was a contradiction that was solved
through the idea of a scale dependant coupling which was
large at low energy and feeble at high energies~\cite{Gross:1973id,Politzer:1973fx}.
This lead to the rapid development of quantum-chromodynamics (QCD) which
was soon established as the correct theory of strong interactions.

A wider programme of scattering experiments followed providing
detailed insight into the structure of the proton and electroweak (EW)
interactions between the quark and lepton sectors of the Standard
Model. First observations of weak neutral currents by the Gargamelle
neutrino-nucleon experiment~\cite{Hasert:1973ff,Deden:1974uw} were
made in 1973, scaling violations of DIS cross sections were observed in
1974~\cite{Fox:1974ry}, and the discovery of the gluon was made in 1979
by the TASSO $e^+e^-$ experiment~\cite{Brandelik:1979bd} at DESY.

In 1993-2007 HERA, the only $ep$ collider, operated at DESY and opened
up a wide kinematic region for precision measurements of proton
structure and was capable of resolving structures to $10^{-3}$~fm. The
precise HERA data together with fixed target DIS measurements 
and data from hadro-production experiments
strongly constrain the proton's parton distributions.

Experiments at CERN, DESY, SLAC and
JLab (see for example~\cite{Ashman:1987hv,Adeva:1993km,Alexakhin:2005iw,Airapetian:2004tw,Abe:1997cx,Dharmawardane:2006zd})
have extensively studied polarised DIS to understand how the proton's
spin arises from the orbital and intrinsic angular momenta of the
constituent partons. In this article we omit any discussion of spin
although a review of the field can be found for example
in~\cite{Burkardt:2008jw,Myhrer:2009uq}.

The knowledge that we now have of QCD and proton structure is a vital
tool in helping disentangle and interpret potential signals of new physics at the
Large Hadron Collider (LHC) at CERN which has commenced operation colliding
protons at a centre-of-mass energy $\sqrt{s}$ of up to $8$~TeV, and is expected to reach the design
value of about $14$~TeV in the next few years.

In this report we review the current status of our understanding of
proton structure. In section~\ref{sec:theory} and the remaining
sections of this chapter the formalism for deep inelastic scattering
is given and the parton distribution functions (PDFs) are introduced. 
A more formal introduction can be found for example in~\cite{ReviewThorne}.
The experimental constraints on
the proton structure measurements are discussed in detail in
chapter~\ref{sec:expt} including data from non-DIS
experiments. Chapter~\ref{sec:qcdana} provides an overview of the
methods used to extract the PDFs from the various experimental
measurements. Finally in chapter~\ref{sec:lhc} the potential of the
LHC in constraining our knowledge of proton structure is discussed.

%%%%%%%%%%%%%%%%%%%%%%%%%%%%%%%%%%%%%%%%%%%%%%%%%%%%%%%
\subsection{Kinematic Quantities in DIS}
%%%%%%%%%%%%%%%%%%%%%%%%%%%%%%%%%%%%%%%%%%%%%%%%%%%%%%%
\label{sec:theory}

In this section we outline the general formalism for unpolarised
deeply inelastic lepton-nucleon scattering in perturbative
QCD. Inclusive neutral current (NC) scattering of a charged lepton $l$
off a nucleon $N$ proceeds via the reaction $l^{\pm}N\rightarrow
l^{\pm}X$ and the exchange of virtual neutral electroweak vector
bosons, $\gamma$ or $Z^0$. Here $X$ represents any final state. The purely weak charged current (CC) process is
$l^{\pm}N\rightarrow \overset{\brabar}{\nu}X$ and occurs via exchange of a
virtual $W^{\pm}$ boson.

\begin{figure}[htb]
\begin{center}
\includegraphics[width=0.40\columnwidth]{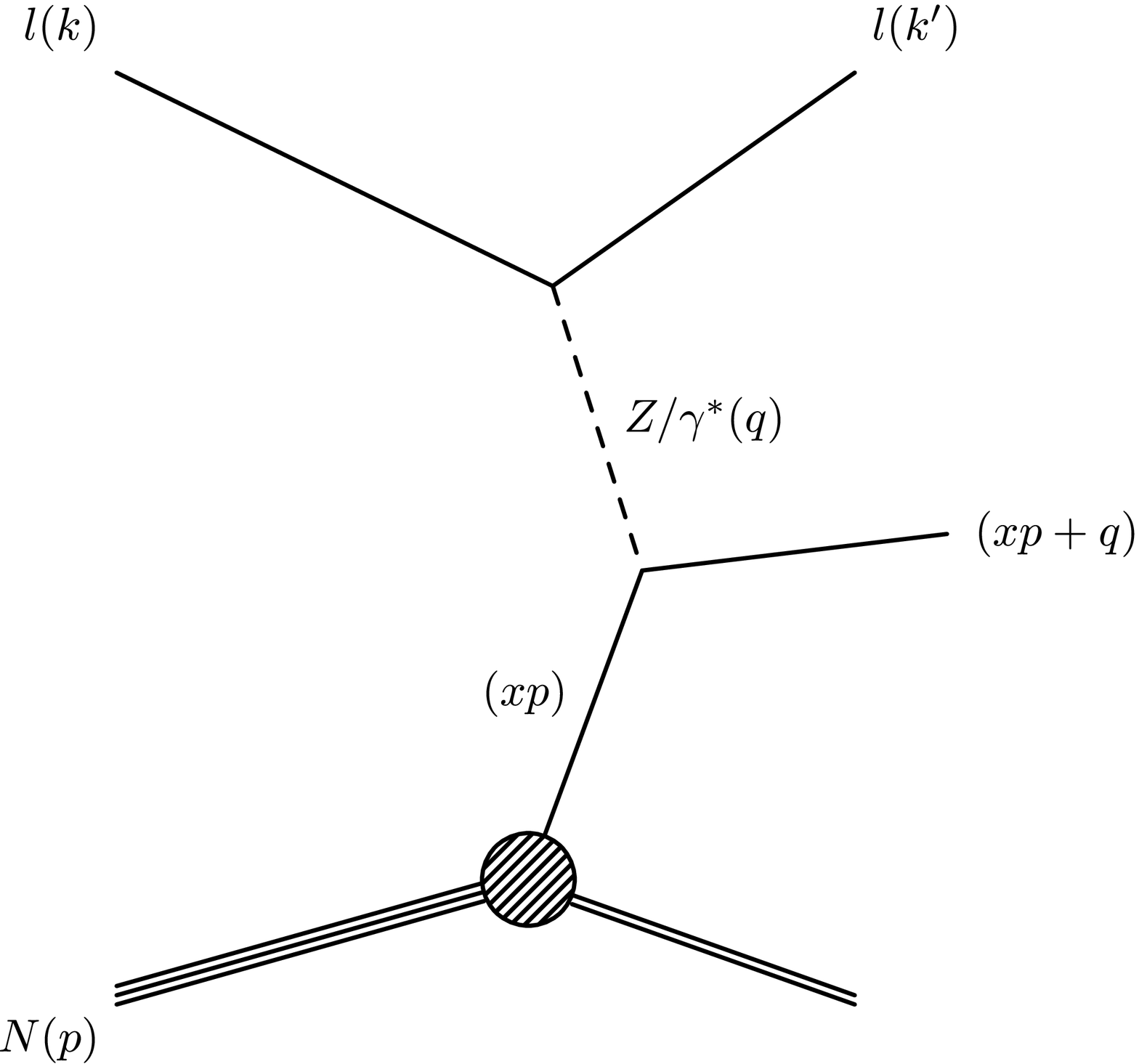}
\includegraphics[width=0.40\columnwidth]{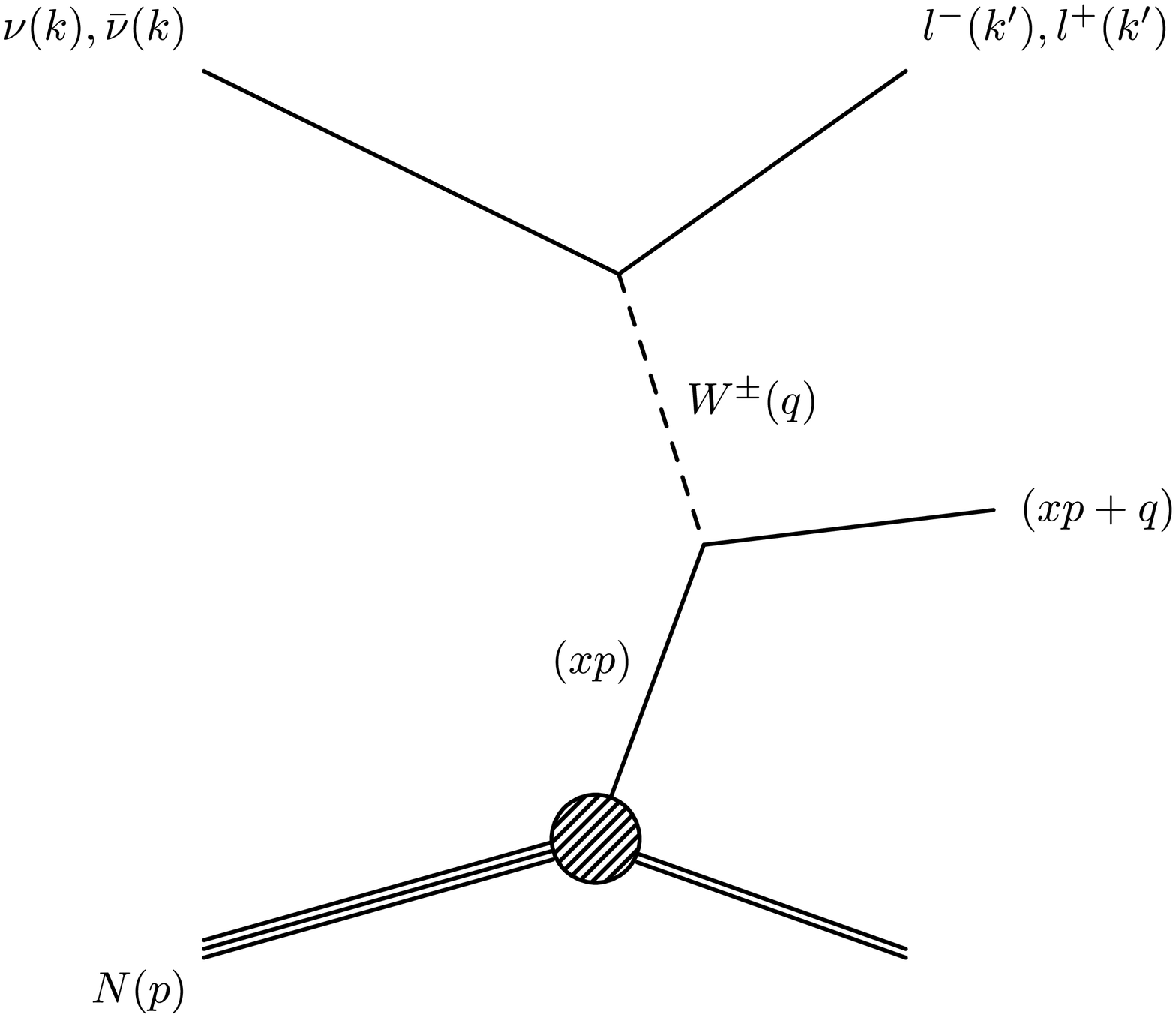}
\end{center}
\caption{\sl Left: Schematic diagram of NC DIS of a charged lepton $l$ with incoming
and outgoing momenta $k$ and $k^{\prime}$ interacting with nucleon $N$
with momentum $p$ via the exchange of a virtual $Z/\gamma^*$ boson ($q$)
between the lepton and a parton carrying fractional momentum $xp$ (see
text). Right: Schematic of CC neutrino induced DIS via the exchange of
a virtual $W^{\pm}$ boson.}
\label{fig:DISfeyn}
\end{figure}

The measured cross sections are usually expressed in terms of three variables $x$,
$y$ and $Q^2$ defined as
\begin{eqnarray}
x = \frac{Q^2}{2p\cdot q} \hspace{2cm}
y=\frac{p \cdot q}{p \cdot k} \hspace{2cm}
Q^2 = -q^2 \equiv -(k-k^{\prime})^2 \,\,,
\end{eqnarray}
where $k$ and $k^{\prime}$ are the momenta of the initial and
scattered lepton, $p$ is the nucleon momentum, and $q$ the momentum of the
exchanged boson (see
Fig.~\ref{fig:DISfeyn}). The first of these, also known as
$x_{Bjorken}$, is the fraction of the target nucleon's momentum taken
by the parton in the infinite momentum frame where the partons have
zero transverse momenta. The inelasticity is measured by the quantity
$y$ and is the fractional energy loss of the lepton in the rest frame
of the target nucleon. It also quantifies the lepton-parton scattering
angle $\theta^*$ measured with respect to the lepton direction in
the centre-of-mass frame since $y=\frac{1}{2}(1-\cos\theta^*)$. The
4-momentum transfer squared from the lepton $Q^2$, quantifies
the virtuality of the exchange boson. The three quantities are related
to the lepton-nucleon centre-of-mass energy, $\sqrt{s}$, via $Q^2=sxy$
which holds in the massless approximation since $s=(k+p)^2$. Thus for
fixed centre-of-mass energy the cross sections are dependant on only
two quantities. Modern experiments typically publish measured cross
sections differentially in two variables, usually $x$ and $Q^2$ or $x$
and $y$.

Two other related kinematic quantities are also sometimes used and are given
here for completeness. They are
\begin{eqnarray}
W^2 = (q+p)^2 = Q^2 \frac{1-x}{x} + m_N^2 \hspace{3cm} \nu = \frac{p \cdot q}{m_N} \,\,\,. 
\end{eqnarray}
$W^2$ is the invariant mass squared of the final hadronic
system, and $\nu$ is the lepton energy loss in the rest frame of the
nucleon, with $m_N$ the nucleon mass.

%%%%%%%%%%%%%%%%%%%%%%%%%%%%%%%%%%%%%%%%%%%%%%%%%%%%%%%
\subsection{The Quark Parton Model}
%%%%%%%%%%%%%%%%%%%%%%%%%%%%%%%%%%%%%%%%%%%%%%%%%%%%%%%

The Quark Parton Model (QPM) ~\cite{Feynman:1969ej,Bjorken:1969ja}
describes nucleons as consisting of massless point-like spin
$\frac{1}{2}$ quarks which are free within the nucleon. The gluon is
completely neglected, but despite this failing it is nevertheless a
useful conceptual model with which to illustrate a discussion of
proton structure. Nucleon and quark masses are also neglected, an
approximation that is valid provided the momentum scale of the
scattering process $Q$ is large enough. The parton distribution
functions $f_i(x)$ in the QPM are number densities of parton
flavour $i$ with fraction $x$ of the parent nucleon's energy and
longitudinal momentum. Often the momentum weighted distributions
$xf_i(x)$ are used. In standard notation the anti-quark PDFs are
denoted $x{\overline f_i}(x)$, and PDFs for each quark flavour are
written as $u,d,s,c,b$ for the up, down, strange, charm and bottom
respectively. The PDFs obey counting sum rules which for
the proton are written as
\begin{eqnarray}
\int_0^1 \left[u(x)-\bar{u}(x) \right] \, {\rm d}x = 2   \hspace{2cm}  
\int_0^1 \left[d(x)-\bar{d}(x) \right] \, {\rm d}x = 1 \,\,\,   \label{eq:countingrule} \\
{\rm{and}} \hspace{1cm} \int_0^1 \left[q(x)-\bar{q}(x) \right] \, {\rm d}x = 0 \hspace{1cm} {\rm{for }} \quad q=s,c,b 
\label{eq:countingrule2}
\end{eqnarray}
and require the valence structure of the proton to correspond to
$uud$. The valence distributions are defined as $u_v=u-{\bar u}$ and 
$d_v=d-{\bar d}$. The constraint of momentum conservation in the QPM
is written as 
\begin{eqnarray}
\int_0^1 \sum_i^{n_f} x\left[ q_i(x) + \bar{q_i}(x) \right] \, {\rm d}x = 1
\label{eq:sumrule_QPM}
\end{eqnarray}
where the sum runs over all active parton flavours $n_f$. 

Deeply inelastic lepton-nucleon
scattering cross sections are calculated from incoherent sums of
elastic lepton-parton processes.
More generally, the hadronic interaction cross section for a process
$A+B\rightarrow X$ can be written as
\begin{eqnarray}
\sigma_{A,B\rightarrow X} = \sum_{i,j}\iint f_i^A(x_1)\cdot
f_j^B(x_2) \cdot \hat{\sigma}_{i,j\rightarrow X}\, 
dx_{1} dx_{2} + [x_1 \leftrightarrow x_2]
\label{eq:factor}
\end{eqnarray}
where $\hat{\sigma}_{i,j\rightarrow X}$ is the partonic cross section for
interactions of two partons with flavour $i$ and $j$.
The fact that the PDFs in Eq.~\ref{eq:factor} are universal is known as the
factorisation property:
PDFs extracted from an analysis of e.g. inclusive DIS measurements can be
used to calculate the cross sections of other processes in lepton-hadron or hadron-hadron
interactions.
A proof of the factorisation theorem in perturbative QCD can be found in~\cite{Collins:1989gx}.

The QPM represents the lowest order approximation of QCD and as such
does not take into account gluon contributions to the
scattering process. This implies a scale ($Q$) invariance to all QPM
predictions known as Bjorken scaling - the quarks are free within the
nucleon and thus do not exchange momenta. Early DIS
measurements~\cite{Bloom:1969kc,Breidenbach:1969kd} demonstrated
approximate scaling behaviour for $x\approx0.3$ indicating the
scattering of point-like constituents of the proton, but subsequent
measurements extended over a wider $x$ range showed that scaling
behaviour is violated~\cite{Fox:1974ry}. This is interpreted as
being due to gluon radiation suppressing high $x$ partons and creating
a larger density of low $x$ partons with increasing $Q^2$. Thus in the
absence of scaling, the PDFs $f_i(x)$ become $Q^2$ dependent, $f_i(x,Q^2)$.

It is usual to consider the neutron PDFs ($f_i^n$) as being related to
the proton PDFs ($f_i^p$) by invoking strong isospin symmetry, i.e.
that $u^p = d^n$, $d^p = u^n$, $\bar{u}^p = \bar{d}^n$, $\bar{d}^p = \bar{u}^n$ and that the PDFs
of other flavours remain the same for neutron and proton.
This is used in order to exploit the structure function measurements performed
in DIS off a deuterium target.
For the remainder of this article PDFs always refer to proton PDFs.

The QPM provides a good qualitative description of scattering
data. It was convincingly tested in the prediction of the $pp\rightarrow
\mu^+\mu^-+X$ process ~\cite{drell-yan}. Sum rules similar to those in
Eqs.~\ref{eq:countingrule} and~\ref{eq:sumrule_QPM} such as the Gross-Llewellyn-Smith
(GLS)~\cite{gls}, Adler~\cite{adler}, Bjorken~\cite{bjorkensumrule}
and Gottfried~\cite{gottfried} sum rules were also well
predicted by the QPM (for details
see~\cite{Roberts:1990ww,Devenish:2004pb}). 

%%%%%%%%%%%%%%%%%%%%%%%%%%%%%%%%%%%%%%%%%%%%%%%%%%%%%%%
\subsection{DIS Formalism}
%%%%%%%%%%%%%%%%%%%%%%%%%%%%%%%%%%%%%%%%%%%%%%%%%%%%%%%
\label{sec:formalism}

The scattering of a virtual boson off a nucleon can be written in
terms of a leptonic and a hadronic tensor with appropriate couplings
of the exchanged boson to the lepton and the partons (see for
example~\cite{Devenish:2004pb}). The hadronic tensor is not calculable
from first principles and is expressed in terms of three general
structure functions which for NC processes
are $\tilde{F}_2$, $x\tilde{F}_3$ and
$\tilde{F}_L$. The total virtual boson absorption cross section is
related to the $\tilde{F}_2$ and $x\tilde{F}_3$ parts in which both the longitudinal
and the transverse polarisation states of the virtual boson
contribute, whereas only the longitudinally polarised piece
contributes to $\tilde{F}_L$. These NC structure functions can be further
decomposed into pieces relating to pure photon exchange, pure $Z^0$
exchange and an interference piece.

At the Born level the NC cross section for the process\footnote{For DIS of a charged
lepton, the example of an electron or positron beam is taken here.}
$e^{\pm}p\rightarrow e^{\pm}X$ 
is given by
\begin{eqnarray}
\label{Snc1}
\frac{{\rm d}^2\sigma_{NC}^{\pm}}{{\rm d}x\;{\rm d}\QQ}
& = & \frac{2\pi \alpha^2}{xQ^4} 
\left[Y_+ \Ftwo \mp Y_{-}x\Fz -y^2 \FL \right]
\end{eqnarray}
where $\alpha \equiv \alpha(Q^2=0)$ is the fine structure constant.
The helicity dependencies of the electroweak interactions are contained
in \mbox{$Y_{\pm} \equiv 1 \pm (1-y)^2$}.  

The generalised proton structure functions~\cite{klein},
$\tilde{F}_{L,2,3}$, may be written as linear combinations of the
hadronic structure functions $F_{L,2}$, $F_{L,2,3}^{\gamma Z}$, and
$F_{L,2,3}^{Z}$ containing information on QCD parton dynamics as well
as the EW couplings of the quarks to the neutral vector bosons. The
function $F_{2}$ is associated to pure photon exchange terms,
$F_{L,2,3}^{\gamma Z}$ correspond to photon-$Z^0$ interference and
$F_{L,2,3}^{Z}$ correspond to the pure $Z^0$ exchange terms.
Neglecting $\tilde{F}_{L}$, the linear combinations for arbitrary
longitudinal lepton polarised $e^{\pm}p$ scattering are given by
\begin{eqnarray}
 \tilde{F}^{\pm}_2 = F_2 - (v_e \pm P a_e) \kappa  \frac{Q^2}{Q^2+M_Z^2}    F_2^{\gamma Z} 
            + (v_e^2+a_e^2 \pm 2Pv_e a_e) \kappa^2 \left[\frac{Q^2}{Q^2+M_Z^2}\right]^2 F_2^Z \\
 x\tilde{F}^{\pm}_3 = -(a_e \pm P v_e)   \kappa    \frac{Q^2}{Q^2+M_Z^2}    xF_3^{\gamma Z} 
       + (2a_ev_e \pm P [v_e^2 + a_e^2] ) \kappa^2 \left[\frac{Q^2}{Q^2+M_Z^2}\right]^2 xF_3^Z 
\label{SF1}
\end{eqnarray}

where $P$ is the degree of lepton polarisation, $a_e$ and
$v_e$ are the usual leptonic electroweak axial and vector couplings to the
$Z^{0}$
and $\kappa$ is defined by
$\kappa^{-1}=4\frac{M_W^2}{M_Z^2}(1-\frac{M_W^2}{M_Z^2})$,
$M_W$ and $M_Z$ being the masses of the electroweak bosons.

The structure function $\FL$ (proportional to the longitudinally
polarised virtual photon scattering cross section, $\sigma_L$) may be
decomposed in a manner similar to $\Ftwo$ (proportional to the sum of
longitudinally and transversely polarised virtual photon scattering
cross sections $\sigma_L+\sigma_T$). Its contribution is significant
only at high $y$ (see Eq.~\ref{Snc1}). The ratio $R$ defined as
$R(x,Q^2)=\sigma_L(x,Q^2)/\sigma_T(x,Q^2) = F_L/(F_2-F_L)$ is often used instead of
$F_L$ to describe the scattering cross section. The QPM predicts that
longitudinally polarised virtual photon scattering is forbidden due to
helicity conservation considerations, i.e. $F_L=0$ for spin half
partons. The fact that $F_L$ is non-zero is a consequence of the
existence of gluons and QCD as the underlying theory.

Over most of the experimentally accessible kinematic domain the
dominant contribution to the NC cross section comes from the
electromagnetic structure function $F_2$. Only at large values of
$Q^2$ do the contributions from $Z^0$ boson exchange become
important.  For longitudinally unpolarised lepton beams $\Ftwo$ is the
same for $e^-$ and for $e^+$ scattering, while the $x \Fz$
contribution changes sign as can be seen in Eq.~\ref{Snc1}.

In the QPM the structure functions $F_2$,
$F_2^{\gamma Z} $ and $F_2^Z$ are related to the sum of the quark and
anti-quark densities
\begin{equation}
\label{eq:f2}
[F_2,F_2^{\gamma Z},F_2^{Z}] = x \sum_q 
[e_q^2, 2 e_q v_q, v_q^2+a_q^2] 
\{q+\bar{q}\} 
\end{equation}
and the structure functions $xF_3^{\gamma Z} $ and $xF_3^Z$ to their
difference which determines the valence quark distributions $q_v$
\begin{equation}
\label{eq:xf3}
[ x F_3^{\gamma Z},x F_3^{Z} ] = 2x \sum_q 
[e_q a_q, v_q a_q]
\{q -\bar{q} \} = 2 x \sum_q [e_q a_q, v_q a_q] q_v\,\,\,.
\end{equation}
Here $e_q$ is the charge of quark $q$ in
units of the positron charge and $v_q$ and $a_q$ are the vector and
axial-vector weak coupling constants of the quarks to the $Z^0$. 

For CC interactions the Born cross section  may  be expressed as
\begin{eqnarray}
\label{eq:cccross}
\frac{{\rm d} ^2 \sigma_{\rm CC}^{\pm}}{{\rm d} x\; {\rm d} Q^2} & = &
(1 \pm P)\frac{G_F^2 }{4 \pi x} \left[ \frac{M_W^2}{Q^2+M_W^2} \right]^2
\;(Y_+ W_2(lp)   \mp Y_- xW_3(lp) - y^2 W_L(lp))
\end{eqnarray}
where $\sigma^+_{CC}$ ($\sigma^-_{CC}$) denotes the cross section for $e^+ p$
or $\bar{\nu} p$ ($e^- p$ or $\nu p$) interactions. In the case of (anti-)neutrino
interactions, $1 \pm P = 2$ in Eq.\ref{eq:cccross}.
The weak coupling is expressed here as the Fermi constant $G_F$.  The
CC structure functions $W_2$, $W_3$ and $W_L$ are defined in a similar
manner to the NC structure functions~\cite{hector}. In the QPM (where
$W_L \equiv 0$) they may be interpreted as sums and differences of
quark and anti-quark densities and are given by
\begin{eqnarray}
\label{ccstf}
   W_2(e^+ p) = W_2 (\nu p)  =  x (\bar{U}+D)\hspace{0.1cm}\mbox{,}\hspace{0.3cm}
 & xW_3 (e^+ p) = xW_3 (\nu p)  =  x (D-\bar{U})\hspace{0.1cm}\mbox{,}\hspace{0.3cm} \\
   W_2(e^- p) = W_2 (\bar{\nu} p)  =  x (U+\bar{D})\hspace{0.1cm}\mbox{,}\hspace{0.3cm}
 & xW_3 (e^- p) = xW_3(\bar{\nu} p)  =  x (U-\bar{D})\,\,\,  \label{ccstf2}
\end{eqnarray}

where $U$ represents the sum of up-type, and $D$ the sum of down-type
quark densities, 
\begin{eqnarray}
\label{ud}
 U &=& u + c  \nonumber\\
\bU&=& \bu + \bc \nonumber\\
 D &=& d + s + b \nonumber\\
\bD&=& \bd + \bs + \bb. 
\end{eqnarray}

Neutrino DIS experiments generally use a heavy target to compensate for the
low interaction rates. For an isoscalar target, i.e. with the same number of
protons and neutrons, the structure functions become, assuming
$q = \bar{q}$ for $q = s, c, b$ :
\begin{eqnarray}
W_2(\nu N) = W_2( \bar{\nu} N) &= &x ( U + D + \bar{U} + \bar{D}) \qquad ,\\
W_3(\nu N) &= & x ( u_v + d_v + 2s + 2b - 2 \bar{c})  \qquad  ,  \\
W_3(\bar{\nu} N) &= & x ( u_v + d_v - 2\bar{s} - 2\bar{b} + 2 c) \qquad  ,  
\end{eqnarray}
such that the difference between $\nu N$ and $\bar{\nu} N$ cross sections
is directly sensitive to the total valence distribution.

Some analyses of DIS cross sections present the data
in terms of ``reduced cross sections'' where kinematic pre-factors are
factored out to ease visualisation. Typically they are defined as:
\begin{equation}
\label{Rnc}
\tilde{\sigma}_{NC}(x,Q^2) \equiv  \frac{1}{Y_+} \ 
\frac{ Q^4 \ x  }{2 \pi \alpha^2}
\          \frac{{\rm d}^2 \sigma_{NC}}{{\rm d}x{\rm d}Q^2},\;\;\;
\tilde{\sigma}_{CC}(x,Q^2) \equiv  
\frac{2 \pi  x}{ G_F^2}
\left[ \frac {M_W^2+Q^2} {M_W^2} \right]^2
          \frac{{\rm d}^2 \sigma_{CC}}{{\rm d}x{\rm d}Q^2}.\;\;\;
\end{equation}

%%%%%%%%%%%%%%%%%%%%%%%%%%%%%%%%%%%%%%%%%%%%%%%%%%%%%%%
\subsection{QCD and the Parton Distribution Functions}
%%%%%%%%%%%%%%%%%%%%%%%%%%%%%%%%%%%%%%%%%%%%%%%%%%%%%%%

The QPM is based on an apparent contradiction that DIS scattering
cross sections may be determined from free quarks which are bound
within the nucleon. Despite this the QPM was very successful at being
able to take PDFs from one scattering process and predicting cross
sections for other scattering experiments; it nevertheless has
difficulties. The first of these is the failure of the model to
accurately describe violations of scaling and scale dependence of DIS
cross sections. The fact that partons are
strongly bound into colourless states is an experimental fact, but
why they behave as free particles when probed at high momenta is not
explained. The QPM is also unable to account for the full
momentum of the proton via measurements of the momentum
sum rule of Eq.~\ref{eq:sumrule_QPM} indicating the
existence of a new partonic constituent which does not couple to
electroweak probes, the gluon ($g$), which modifies the momentum sum
rule as below:
\begin{eqnarray}
\int_0^1 \sum_i x\left[ q_i(x) + \bar{q_i}(x) \right] +xg(x)\, {\rm d}x = 1
\label{eq:sumrule}
\end{eqnarray}
It is only by including the effects of the gluon and gluon radiation
in hard scattering processes that an accurate description of
experimental data can be given. These developments led to the
formulation of quantum-chromodynamics.

Any theory of QCD must be able to accommodate the twin concepts of
asymptotic freedom and confinement. The former applies at large scales
($Q$) where experiments are able to resolve the partonic content of
hadrons which are quasi-free in the high energy (short time-scale)
limit, and is a unique feature of non-abelian theories. The latter
explains the strong binding of partons into colourless observable
hadrons at low energy scales (or equivalently, over long
time-scales). In 1973 Gross, Wilczek and
Politzer~\cite{Gross:1973id,Politzer:1973fx} showed that perturbative
non-abelian field theories could give rise to asymptotically free
behaviour of quarks and scaling violations. By introducing a scale
dependent coupling strength $\alpha_s(Q)=g^2/4\pi$ (where $g$ is the
QCD gauge field coupling and depends on $Q$), confinement and asymptotic freedom can be
accommodated in a single theory. Perturbative QCD (pQCD) is restricted to the
region where the $\alpha_s$ coupling is small enough to allow cross sections to
be calculated as a rapidly convergent power series in
$\alpha_s$. Furthermore any realistic theory of QCD must be
renormalisable in order to avoid divergent integrals arising from
infinite momenta circulating in higher order loop diagrams. The
procedure chosen for removing these ultraviolet divergences fixes a
renormalisation scheme and introduces an arbitrary renormalisation
scale, $\mu_R$. A convenient and widely used scheme is the modified
minimal subtraction ($\overline{\rm MS}$) scheme~\cite{msbar}.
When the perturbative expansion is summed to all orders the
scale dependence of observables on $\mu_R$ vanishes, as expressed
by the renormalisation group equation. For truncated
summations the scale dependence may be absorbed into the coupling
i.e. $\alpha_s\rightarrow \alpha_s(\mu_R)$.

In addition to the problems of ultraviolet divergent integrals, infrared singularities also
appear in QCD calculations from soft collinear gluon radiation as the
gluon transverse momentum $k_T\rightarrow 0$. These singularities are
removed by absorbing the divergences into redefined PDFs within a
given choice of scheme (the factorisation scheme) and choice of a second arbitrary momentum scale
$\mu_F$ (the factorisation scale). By separating out the short
distance and long distance physics at the factorisation scale $\mu_F$
the hadronic cross sections may be separated into perturbative and non-perturbative
pieces. The non-perturbative piece is not {\em a priori} calculable,
however it may be parametrised at a given scale from experimental
data. This procedure introduces a scale dependence to the PDFs. By
requiring that the $\mu_F$ scale dependence of $F_2$ vanishes in a calculation summing
over all orders in the perturbative expansion a series of
integro-differential equations may be derived that relates the PDFs at
one scale to the PDFs at another given scale. These evolution
equations obtained by Dokshitzer, Gribov, Lipatov, Altarelli and
Parisi (DGLAP)~\cite{dglap1,dglap2,dglap3,dglap4} are given in terms of a perturbative
expansion of splitting functions ($P_{ba}$) which describe the
probability of a parent parton $a$ producing a daughter parton $b$
with momentum fraction $z$ by the emission of a parton with momentum
fraction $1-z$.  Three leading order equations are derived for the
non-singlet \mbox{($q_{ij}^{NS}=q_{i}-{\bar q_{j}}$)},
singlet \mbox{($q_i^S=q_i+{\bar q_i}$) }, and gluon distributions:
\begin{eqnarray}
\frac{\partial q^{NS}(x,\mu_F^2)}{\partial \log \mu_F^2} &=& 
\frac{\alpha_s(\mu_R^2)}{2\pi}\int_x^1 \frac{dy}{y}\left[ q^{NS}(y,\mu_F^2)P_{qq}(x/y) \right] \label{eq:dglap_ns} \\
\frac{\partial q^{S}(x,\mu_F^2)}{\partial \log \mu_F^2} &=& 
\frac{\alpha_s(\mu_R^2)}{2\pi}\int_x^1 \frac{dy}{y}\left[ q^S(y,\mu_F^2)P_{qq}(x/y)+g(y,\mu_F^2)P_{qg}(x/y) \right] \label{eq:dglap_s} \\
\frac{\partial g(x,\mu_F^2)}{\partial \log \mu_F^2}  &=& 
\frac{\alpha_s(\mu_R^2)}{2\pi}\int_x^1 \frac{dy}{y}\left[ q^S(y,\mu_F^2)P_{gq}(x/y)+g(y,\mu_F^2)P_{gg}(x/y) \right] \,\, . 
\label{eq:dglap_g}
\end{eqnarray}

The corresponding splitting functions are given, at leading order (LO), by
\begin{eqnarray}
P_{qq}&=&\frac{4}{3}\left[\frac{1+x^2}{(1-x)_+} + \frac{3}{2}\delta(1-z)\right] \nonumber\\
P_{qg}&=&\frac{1}{2}\left[ x^2+(1-x)^2\right] \nonumber\\
P_{gq}&=&\frac{4}{3}\left[\frac{1+(1-x)^2}{x}\right] \nonumber\\
P_{gg}&=& 6\left[\frac{1-x}{x}+x(1-x)+\frac{x}{(1-x)_+}\right] +\left[\frac{11}{2}-\frac{n_f}{3}\right]\delta(1-x)\,\,,\nonumber\\
\end{eqnarray}
where $[f(x)]_+=f(x)-\delta(1-x)\int_0^1f(y){\rm d}y$. The DGLAP
evolved PDFs then describe the PDFs integrated over transverse
momentum $k_T$ up to the scale $\mu_F$. The splitting functions have
also been calculated at next-to-leading order (NLO)~\cite{nlo} and more recently in
next-to-next-to-leading order (NNLO)~\cite{nnlo}.

\begin{figure}[htb]
\begin{center}
\begin{tabular}{cc}
\includegraphics[width=0.50\columnwidth]{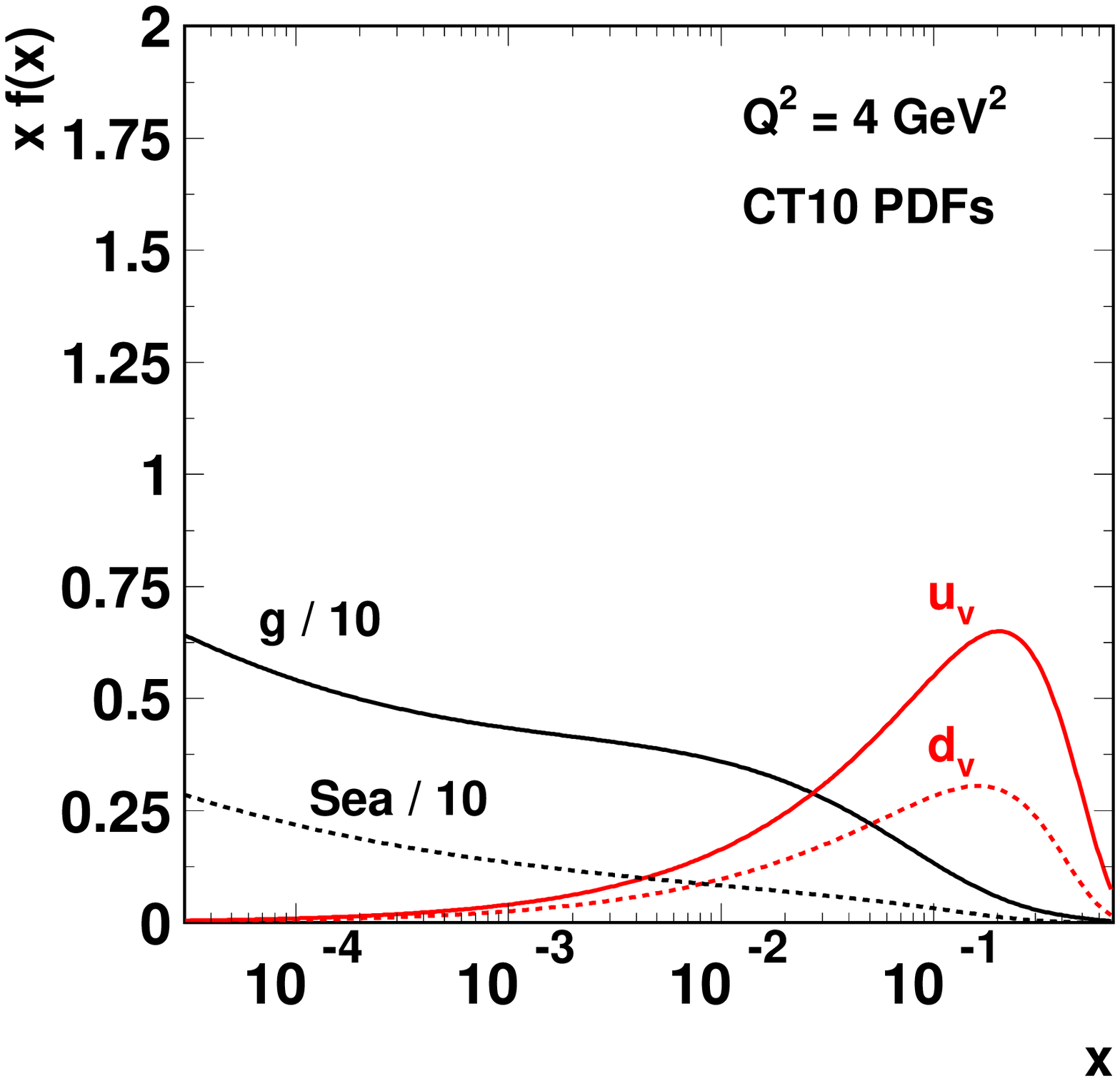} &
\includegraphics[width=0.50\columnwidth]{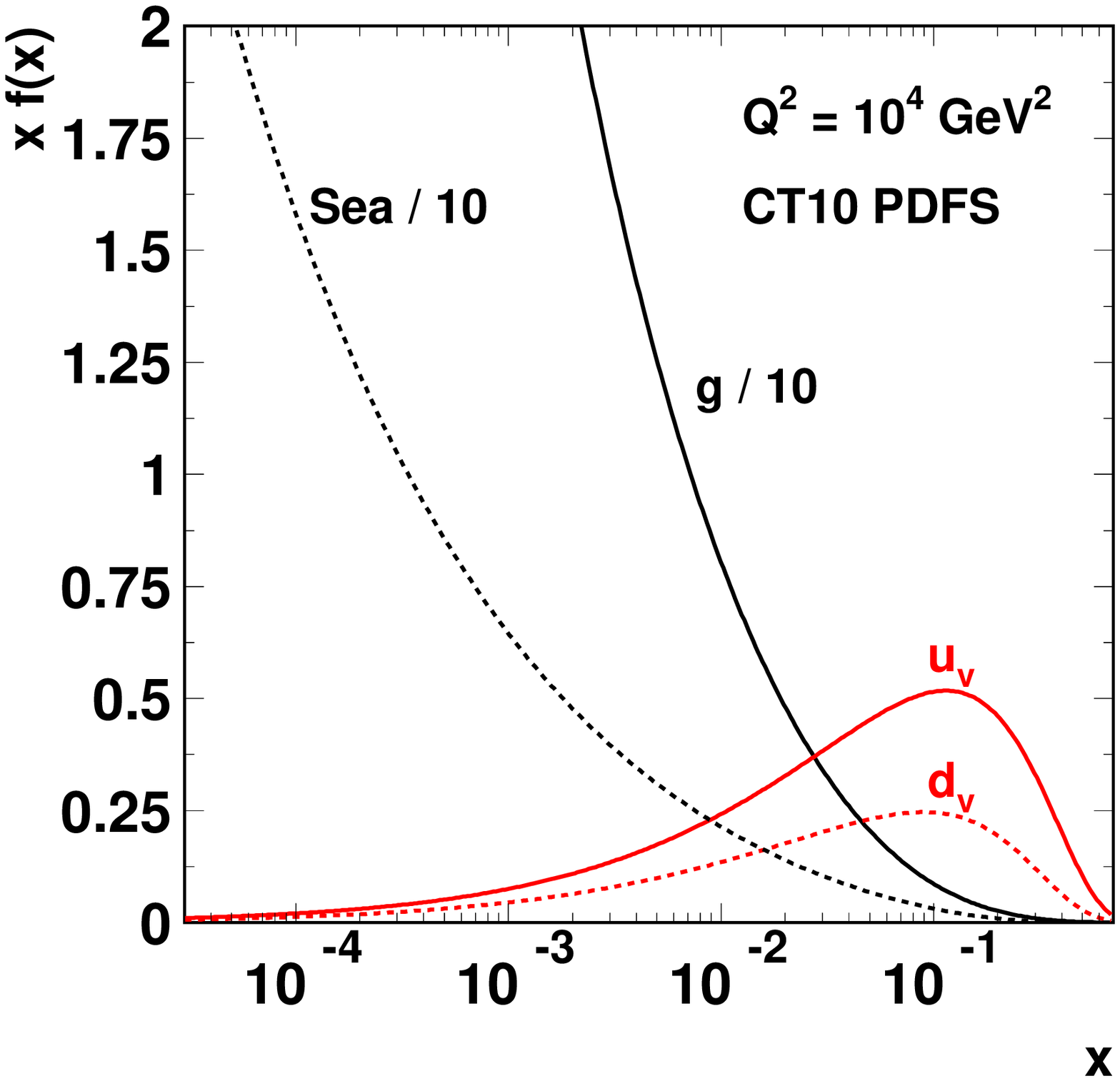}
\end{tabular}
\end{center}
\caption{\sl Example PDFs at NLO at $Q^2 = 4$~GeV$^2$ and $Q^2 = 10^4$~GeV$^2$:
the gluon density, the sea density $\sum \bar{q}$, and the valence densities
$u_v  = u - \bar{u}$ and $d_v = d - \bar{d}$.
}
\label{fig:example_pdfs}
\end{figure}
The DGLAP equations allow the PDFs to be calculated perturbatively at any scale, once they have
been measured at a given scale. Figure~\ref{fig:example_pdfs} shows example NLO PDFs for the gluon,
the valence quarks and the sea quarks, for two values of $Q^2$. While the gluon and the
sea distributions increase very quickly with $Q^2$, the non-singlet valence distributions
are much less affected by the evolution.

Higher orders should also be accounted for in the partonic
matrix element $\hat{\sigma}$ of Eq.~\ref{eq:factor}, such
that $\hat{\sigma} \rightarrow \hat{\sigma}_0
+ \hat{\sigma}_1(\mu_F,\alpha_s(\mu_R))+...$. At LO in $\alpha_s$ the $\mu_F$
dependence can be absorbed into the PDFs but beyond LO this cannot be
done in a process independent way, and hence $\hat{\sigma}$ depends on
$\mu_F$ and on the factorisation scheme. An all orders calculation
then cancels the $\mu_F$ dependence of the PDFs with the $\mu_F$
dependence of $\hat{\sigma}$. The factorisation scheme most commonly
used is also the $\overline{\rm MS}$ scheme\footnote{The DIS
scheme~\cite{Altarelli:1978id} is also useful since it is defined such
that the coefficients of higher order terms for the structure function
$F_2$ are zero and so $F_2$ retains its QPM definition in this scheme.}, and a common choice of factorisation and
renormalisation scales is $\mu_F=\mu_R = Q$. It is conventional to
estimate the influence of uncalculated higher order terms by varying
both scales by factors of two.

At NLO
the relationships between the structure functions (or any
other scattering cross section) and the PDFs are modified in a
factorisation scheme dependent way. The modifications are
characterised by Wilson coefficient functions $C$ for the hard scattering
process expressed as a perturbative series.

In the $\overline{\rm MS}$ scheme Eqs.~\ref{eq:f2} and~\ref{eq:xf3}
take additional corrections to the parton terms such that for $F_2$
the relation becomes, with the factorisation and renormalisation scales
both set to $Q^2$ :
\footnotesize
\begin{eqnarray}
F_2(x,Q^2) &=& x \sum_i e_i^2 q^S_i(x,Q^2) + \nonumber \\
&&\frac{\alpha_s(Q^2)}{2\pi} x \int_x^1 \frac{dy}{y}\left[\sum_i e_i^2
C_{2,q}\left(\frac{x}{y}\right)q^S_i(y,Q^2) + 
e^2_i C_{2,g}\left(\frac{x}{y}\right) g(y,Q^2)\right] \,\,\,.
\label{eq:f2_full_expr}
\end{eqnarray}
\normalsize
The $C_{2,q}$ and $C_{2,g}$ terms are the coefficient
functions
for $q$ induced and $g$ induced scattering contributing
to $F_2$.
Here the scaling violations are seen
explicitly in the additional term proportional to $\alpha_s$ where the
integrand is sensitive to partonic momentum fractions $y>x$.
For low and medium $x$, the integral 
is dominated by the second term and $y \sim x$. Hence, from the scale dependence of
$\alpha_s$,
the derivative $\partial F_2 / \partial \ln Q^2$, to the first
order in $\ln Q^2$, is driven by the 
product of $\alpha_s$ and the gluon density\footnote{At LO,
$\partial F_2 / \partial \ln Q^2 (x, Q^2)$ at low $x$ is proportional to
$xg(x, Q^2)$ to a good approximation. At NLO and beyond, a larger range of values 
contributes to the integral, and $\partial F_2 / \partial \ln Q^2 (x, Q^2)$ is
sensitive to $g(y, Q^2)$ for $y \geq x$.}.
By replacing the $e_i^2$ couplings with the
corresponding ones for $Z/\gamma^*$ interference and pure $Z$ exchange
as in Eq.~\ref{eq:f2} the QCD corrected formulae for $F_2^{\gamma/Z}$
and $F^{Z}_2$ are obtained.  

QCD corrections to $F_L$ and $xF_3$ must also be taken into
account. For $xF_3$ the LO QCD corrected formula is
\footnotesize
\begin{eqnarray}
xF_3(x,Q^2) &=& x\sum_i q^{NS}_i(x,Q^2)
+ \frac{\alpha_s(Q^2)}{2\pi} x \int_x^1 \frac{dy}{y} C_{3,q}\left(\frac{x}{y}\right) q^{NS}_i(y,Q^2) \,\,,
\end{eqnarray}
\normalsize
and is independent of the $g$ density resulting in much weaker scaling
violations than for $F_2$. Finally for $F_L$
\footnotesize
\begin{eqnarray}
F_L(x,Q^2) &=& \frac{\alpha_s(Q^2)}{2\pi} x \int_x^1 \frac{dy}{y}\left[\sum_i e_i^2
C_{L,q}\left(\frac{x}{y}\right)q^S_i(y,Q^2) + 
e^2_i C_{L,g}\left(\frac{x}{y}\right) g(y,Q^2)\right] \,\,.
\end{eqnarray}
\normalsize
The coefficient functions at LO in the $\overline{\rm MS}$ scheme are given below for completeness
\footnotesize
\begin{eqnarray}
C_{2,q}(x) &=& \frac{4}{3}\left[ \frac{1+x^2}{1-x}\left(\ln{\frac{1-x}{x}}-\frac{3}{4} \right) + \frac{1}{4}(9+5x)\right]_+ \nonumber \\
C_{2,g}(x) &=& n_f\left[ (x^2+(1-x)^2)\ln{\frac{1-x}{x}}-1+8x(1-x)\right] \nonumber \\
C_{L,q}(x) &=& \frac{8}{3}x \nonumber \\
C_{L,g}(x) &=& 4n_f x (1-x) \nonumber \\
C_{3,q}(x) &=& C_{2,q}(x)-\frac{4}{3}(1+x) \,\,.
\end{eqnarray}
\normalsize
In the DIS scheme, $C_{2,q}(x)$ and $C_{2,g}(x)$ are zero.

To date an enormous variety of processes have been measured at
colliders and confronted with the predictions of pQCD. These
include not only inclusive measurements of DIS cross sections, but
also semi-inclusive measurements of jet production rates, angular
distributions, and multiplicities in DIS and in hadronic colliders
as well as in $e^+e^-$ collisions. In all cases pQCD provides a
good description of the measurements.

%Chapter 2
%%%%%%%%%%%%%%%%%%%%%%%%%%%%%%%%%%%%%%%%%%%%%%%%%%%%%%%
\section{Experimental Constraints}
%%%%%%%%%%%%%%%%%%%%%%%%%%%%%%%%%%%%%%%%%%%%%%%%%%%%%%%
\label{sec:expt}

In the following sections we describe the main features and results of
experiments that put constraints on unpolarised proton PDFs. 
For reasons of space we do not provide
a complete description of all experimental data, but rather focus on
those experiments whose data is used in global approaches to
extract the PDFs. The chapter naturally
divides into common measurement techniques in DIS, measurements from
fixed target DIS experiments, results from the HERA $ep$ collider
which dominates the bulk of precision proton structure data, and results from 
hadro-production experiments.
The LHC experiments will be described in chapter~\ref{sec:lhc}.
A convenient online repository of experimental scattering data for PDF
determinations can be found at~\cite{hepdata}.
A summary of the experimental constraints described here will be given
in section~\ref{sec:exp_summary} that concludes this chapter.

%%%%%%%%%%%%%%%%%%%%%%%%%%%%%%%%%%%
\subsection{Measurement Techniques in DIS}
%%%%%%%%%%%%%%%%%%%%%%%%%%%%%%%%%%%

The following sections outline some of the general issues faced by
experimentalists in performing their measurements. These include
corrections to the measured event rates to account for detector losses
due to inefficiency and resolution effects, as well as theory-based
corrections to extract structure functions and to take into account
the sometimes large effects of QED radiation. Finally fixed target
measurements are discussed which often require corrections for nuclear targets and
kinematic effects that arise at low $Q^2$.

%%%%%%%%%%%%%%%%%%%%%%%%%%%%%%%%%%%%%%
\subsubsection{Detector efficiency and resolution corrections}
%%%%%%%%%%%%%%%%%%%%%%%%%%%%%%%%%%%%%%

Experimental corrections to account for the limited and imperfect
detector acceptance are performed using Monte Carlo simulations but
require an input PDF to be used. Thus the acceptance corrections are
weakly dependent on these input PDFs. Experimenters circumvent this
problem using an iterative approach whereby the measured structure
functions are then further used to tune the MC input which leads to a
modified measurement. The procedure is stopped when the iterations
converge, i.e. the difference in the measurements changes by a small
amount. Typically this occurs after one or two
iterations~\cite{Arneodo:1996qe}.

%%%%%%%%%%%%%%%%%%%%%%%%%%%%%%%%%%%%%%
\subsubsection{Extraction of structure functions}
%%%%%%%%%%%%%%%%%%%%%%%%%%%%%%%%%%%%%%

The measured structure functions and differential cross sections are
quoted at a point in $x$ and $Q^2$ and are derived from bin integrated
values. A correction is needed to convert the measurement to a
differential one. This is usually performed with a parameterisation of
the cross section derivatives across the bin volume. This can be done
by weighting each event by the ratio of structure functions at the bin
centre and the $x$, $Q^2$ of that
event~\cite{Adams:1996gu}. Alternatively the bin integrated
measurement can be corrected by a single factor which is the ratio of
the structure function at the bin centre to the bin integrated value
derived from an analytical calculation~\cite{Aaron:2009kv}.
The dependence of the
correction on the input parameterisation is usually small.

Early experiments often presented results as final corrected values of
the electromagnetic structure functions $F_2$ and $F_L$ (or $R$). In
order to make this decomposition of the cross section experimentalists
restricted themselves to the phase space region of low $y$ where the
contribution of $F_L$ is strongly suppressed in order to extract
$F_2$. Alternatively a value of $F_L$ may be assumed in order to extract
$F_2$. Both of these approaches have been used, but 
recent publications focus on the measurement of the differential cross
section as the primary measurement which necessarily has fewer
assumptions. Extractions of the individual structure functions are
also provided for convenience. The H1 and ZEUS experiments recommend
the use of differential cross sections only as input to further QCD
analyses of the data~\cite{:2009wt}.

By utilising different beam energies, scattering cross sections can be
measured at fixed points in $x$ and $Q^2$ but different $y$ thus
allowing direct measurements of the structure function $F_L$ to be
made\footnote{Additional techniques at fixed $\sqrt{s}$ have also been
employed to determine $F_L$ indirectly, primarily as consistency
checks of QCD.}. Since the technique relies on the measurement of the
difference between cross section measurements for two or more values
of $\sqrt{s}$ the experimental uncertainties on $F_L$ are sensitive to
systematic uncertainties in the relative normalisation of the data
sets, and are often highly correlated point-to-point.

%%%%%%%%%%%%%%%%%%%%%%%%%%%%%%%%%%%%%%
\subsubsection{Reconstruction Methods}
%%%%%%%%%%%%%%%%%%%%%%%%%%%%%%%%%%%%%%

When the centre-of-mass energy of the interaction is known, the DIS cross section
depends on two variables only, and the kinematic variables $x$, $y$
and $Q^2$ can be fully reconstructed from two independent
measurements. Fixed target experiments of charged lepton DIS generally
used the measurement of the energy and angle of the scattered lepton
to reconstruct the kinematics - the lepton method.

The use of colliding beams to measure DIS cross sections allowed new
detector designs to be employed whereby the HERA experiments could
fully reconstruct the hadronic final state in most of the accessible
kinematic domain. Hence, $x$, $y$ and $Q^2$ in NC interactions may be
determined using energy and angular measurements of the scattered
lepton alone, measurements of the inclusive hadronic final state
(HFS), or some combination of these.  This redundancy allows very
good control of the measurements and of their systematic
uncertainties. In contrast charged lepton CC interactions may only be
reconstructed using measurements of the hadronic final state since the
final state neutrino is unobserved. Each method has different
experimental resolution and precision as well as different influence
from QED radiative corrections (see below). A convenient summary of
the different methods is given in~\cite{:2009wt} and are compared
in~\cite{Bassler:1997tv}.

At HERA the main NC reconstruction methods used are the double-angle
method~\cite{damethod} (using the polar angles of the lepton and HFS),
and the $e\Sigma$ method~\cite{esigma} which combines the scattered
lepton energy and angle with the total energy and longitudinal momentum
difference, $E-P_Z$, of the HFS.

DIS experiments using a (wide band) muon neutrino beam face an
additional problem in determining the incident neutrino energy.  It
is usually reconstructed as the sum of the momentum of the scattered muon and
of the energy of the HFS measured in a calorimeter. A third independent
measurement, usually taken to be the angle of the scattered muon, is then needed 
to fully reconstruct the kinematics.

%%%%%%%%%%%%%%%%%%%%%%%%%%%%%%%%%%%%%%%%%%%%%%%%%%%%%%%
\subsubsection{QED radiative corrections}
%%%%%%%%%%%%%%%%%%%%%%%%%%%%%%%%%%%%%%%%%%%%%%%%%%%%%%%
The treatment of radiative corrections is an important aspect of DIS
scattering cross sections measurements and was first discussed
in~\cite{Mo:1968cg}. The corrections allow measured data to be
corrected back to the Born cross section in which the influence of
real photon emission and virtual QED loops are removed.  It is the
Born cross sections that are then used in QCD analyses of DIS data to
extract the proton PDFs (see section~\ref{sec:qcdana}).  This topic
has been extensively discussed for HERA data in~\cite{workshop} and
the references therein.

Corrections applied to the measured data are usually expressed as the
ratio of the Born cross section to the radiative cross
section and can have a strong kinematic
dependence since for example, the emission of a hard real photon can
significantly skew the observed lepton momentum. Thus the corrections
also depend on the detailed experimental treatment and the choice of
reconstruction method used to measure the kinematic quantities. In $ep$
scattering, hard final state QED radiation from the scattered electron
is experimentally observable only at emission angles which are of
the size of the detector spatial resolution.

Complete QED calculations at fixed order in $\alpha$ are involved and
often approximations are used, particularly for soft collinear photon
emission. These approximations are readily implemented into Monte
Carlo simulations allowing experimentalists to account for radiative
effects easily. For the HERA measurements~\cite{:2009wt} 
$\mathcal{O} (\alpha)$ diagrams are corrected for with the
exception of real photon radiation off the quark lines.
This is achieved using Monte Carlo implementations
~\cite{django,lepto} checked against analytical
calculations~\cite{hector,eprc} which agree to within ~$0.3-1\%$ in
the NC case ($2\%$ for $x>0.3$) and to within $2\%$ for the CC
case. The quarkonic radiation piece is known to be small and is
accounted for in the uncertainty given above.

The real corrections are dominated by emission from the lepton lines
and are sizable at high and low $y$~\cite{workshop}. For example, at
$\sqrt{s}=301$~GeV and $Q=22~$GeV the leptonic $ep$ corrections are
estimated to be $+40\%$ at $y=0.75$ when using the lepton
reconstruction method. This is dramatically reduced to $+15\%$ if the
$e\Sigma$ reconstruction method is used, and if an analysis cut of
$E-P_Z>35$~GeV is employed the correction is further reduced to
$+8\%$~\cite{Heinemann:1999ry}.

The vacuum polarisation effects are also
corrected for such that published cross sections correspond to
$\alpha(Q\equiv0)=1/137.04$. These photon self energy contributions
depend only on $Q$ and amount to a correction of $-6\%$ for $Q=M_Z$
and $-4\%$ for $Q=12$~GeV.

%%%%%%%%%%%%%%%%%%%%%%%%%%%%%%%%%%%%%%%%%%%%%%%%%%%%%%%
\subsubsection{Higher order weak corrections}
%%%%%%%%%%%%%%%%%%%%%%%%%%%%%%%%%%%%%%%%%%%%%%%%%%%%%%%

The weak corrections are formally part of the complete set of
$\mathcal{O}(\alpha)$ radiative corrections to DIS processes but are
often experimentally treated separately to the QED radiative
corrections discussed above. The weak parts include the self energy
corrections, weak vertex corrections and so-called box diagrams in
which two heavy gauge bosons are exchanged~\cite{workshop}. The
self energy corrections depend on internal loops including all
particles coupling to the gauge bosons e.g. the Higgs boson, the top
quark and even new particle species.  For this reason experimentalists
sometimes publish measurements in which no corrections for higher
order weak corrections are accounted for. Rather, comparisons to
theoretical predictions are made in which the weak corrections are
included in the calculations. Care must be taken to define the scheme
(i.e. the set of input electroweak parameters used) within which the
corrections are defined.

Two often used schemes are the on-mass-shell scheme~\cite{oms}, and
the $G_\mu$ scheme~\cite{gmu}. In the former the EW parameters are all
defined in terms of the on shell masses of the EW bosons. The weak
mixing angle $\theta_W$ is then related to the weak boson masses by
the relation $\sin^2{\theta_W}=1-M_W^2/M_Z^2$ to all perturbative
orders.  In the $G_\mu$ scheme, the Fermi constant which is very
precisely known through the measurements of the muon
lifetime~\cite{pdg}, is used instead of $M_W$.

The scheme dependence is of particular importance in the CC scattering
case where the influence of box diagrams is relatively small and the
corrections are dominated by the self energy terms of the $W$
propagator affecting the normalisation of the cross section. In the
$G_{\mu}$ scheme these leading contributions to the weak corrections
are already absorbed in the measured value of $G_{\mu}$ and the
remaining corrections are estimated to be at the level of $0.5\%$ at
$Q^2=10\,000$~GeV$^2$~\cite{workshop2} where experimental
uncertainties are an order of magnitude larger. For the HERA NC
structure function measurements the EW corrections are estimated to
reach the level of $\sim 3\%$ at the highest
$Q^2$~\cite{Adloff:2003uh} and should be properly accounted for in
fits to the data.

%%%%%%%%%%%%%%%%%%%%%%%%%%%%%%%%%%%%%%%%%%%%%%%%%%%%%%%
\subsubsection{Target mass corrections and higher twist corrections}
%%%%%%%%%%%%%%%%%%%%%%%%%%%%%%%%%%%%%%%%%%%%%%%%%%%%%%%
\label{sec:highertwists}

For scattering processes at low scales approaching soft hadronic
scales such as the target nucleon mass, additional hadronic effects
lead to kinematic and dynamic $1/Q^2$ power corrections to the
factorisation ansatz Eq.~\ref{eq:factor}.  Both of these corrections
are important for DIS at low to moderate $Q^2$, in particular in the
kinematic domain covered by fixed target DIS experiments.

In DIS, power corrections of kinematic origin, the target mass corrections (TMC),
arise from the finite nucleon mass.
For a mass $m_N$ of the target nucleon, the Bjorken $x$ variable is no longer
equivalent to the fraction of the nucleon's momentum carried by the interacting parton
in the infinite momentum frame. This momentum fraction is instead given
by the so-called Nachtmann variable $\xi$:
$$ \xi = \frac{2x} {1 + \gamma}  \qquad {\mbox{ with }} \qquad \gamma = \sqrt{1 + 4 x^2 m_N^2 / Q^2} $$
which differs from $x$ at large $x$ (above $\sim 0.5$) and low to moderate
$Q^2$.
Approximate formulae which relate the structure functions on a massive nucleon
$F_i^{TMC} (x, Q^2)$ to the massless limit structure functions $F_i$ can be found in~\cite{Georgi:1976ve,Schienbein:2007gr},
for example:
$$ F_2^{TMC} (x, Q^2) = \frac{x^2}{\xi^2 \gamma^3} F_2(\xi, Q^2) + \frac{m_N^2}{Q^2} \frac{6 x^3}{\gamma^4} \int_{\xi}^1 d\xi' / \xi'^2 F_2(\xi', Q^2) \qquad .$$
The ratios $F_i^{TMC} / F_i$ rise above unity at large $x$, with the rise beginning 
at larger values of $x$ as $Q^2$ increases.
The target mass correction can be quite large: for $x = 0.8$ it reaches $\sim 30 \%$
at $Q^2 = 5$~GeV$^2$. 

In addition, power corrections of dynamic origin, arising from correlations of the partons within
the nucleon, can also be important at low $Q^2$. The contribution of these
higher twist terms~\cite{Ellis:1982wd} to the experimentally measured structure functions $F_i^{exp}$
can be written as
\begin{equation}
 F_i^{exp} (x, Q^2) = F_i^{TMC} (x, Q^2) + H_i(x) / Q^2 + ... 
 \label{eq:highertwists}
\end{equation}
These terms have been studied in~\cite{Virchaux:1991jc} and more recently in~\cite{Blumlein:2006be,ABM11}
and found to be sizable at large $x$, however they remain poorly known.
QCD analyses of proton structure that make use of low $Q^2$ data may impose kinematic cuts
to exclude the measurements made at low $Q^2$ and high $x$ (i.e. at low $W^2$) that may be affected
by these higher twist corrections. Alternatively,
a model can be used for the $H_i(x)$ terms, whose parameters can be adjusted to the data.

%%%%%%%%%%%%%%%%%%%%%%%%%%%%%%%%%%%%%%%%%%%%%%%%%%%%%%%
\subsubsection{Treatment of data taken with a nuclear target} 
%%%%%%%%%%%%%%%%%%%%%%%%%%%%%%%%%%%%%%%%%%%%%%%%%%%%%%%

Neutrino DIS experiments have used high $Z,A$ targets such as iron or lead\footnote{An
exception being the WA25 experiment~\cite{WA25}, which measured $\nu d$ and $\bar{\nu}d$
cross sections using a bubble chamber exposed to the CERN SPS wide-band neutrino and
anti-neutrino beams, in
the mid-eighties.}, which
provide reasonable event rates despite the low $\nu$ interaction cross section.
The expression of the measured quantities, $F_2^A$, in terms of the
proton parton densities, has to account for the facts that:
\begin{itemize}
\item the target is not perfectly isoscalar; e.g. in iron, there is a
 $6.8 \%$ excess of neutrons over protons;
\item nuclear matter modifies the parton distribution functions; i.e. the
 parton distributions in a proton bound within a nucleus of mass number $A$,
 $f^A (x, Q^2)$, differ from the proton PDF $f(x,Q^2)$.  
 Physical mechanisms of these nuclear modifications (shadowing effect at low $x$,
 the nucleon's Fermi motion at high $x$, nuclear binding effects at medium $x$)
 are summarised e.g. in~\cite{Geesaman:1995yd, Armesto:2006ph}.
 The ratios $R^A=f^A(x,Q^2) / f(x, Q^2)$ are called nuclear corrections, and can
 differ from unity by as much as $10 - 20 \%$ in medium-size nuclei.
\end{itemize}
Nuclear corrections are obtained by dedicated groups, from fits to
data of experiments that used nuclear ($A$) and deuterium ($d$)  targets:
the structure function ratios $F_2^A / F_2^d$ measured in DIS (at SLAC by
E139, and at CERN by
the EMC and NMC experiments), and the ratios 
of Drell-Yan $q\bar{q}$ annihilation cross sections $\sigma_{DY}^A / \sigma_{DY}^d$ measured by
the E772, E866 experiments. 
Recent analyses also include the measurements of inclusive pion production obtained
by the RHIC experiments in deuterium-gold collisions at the Brookhaven National Laboratory (BNL).
The measurements of charged current DIS structure functions at experiments
using a neutrino beam (see sections~\ref{sec:CCFR-NuTev} and~\ref{sec:Chorus})
can also be included, as done in~\cite{deFlorian:2011fp} for example.

\begin{figure}[bt]
 \centerline{
  \includegraphics[width=0.8\columnwidth]{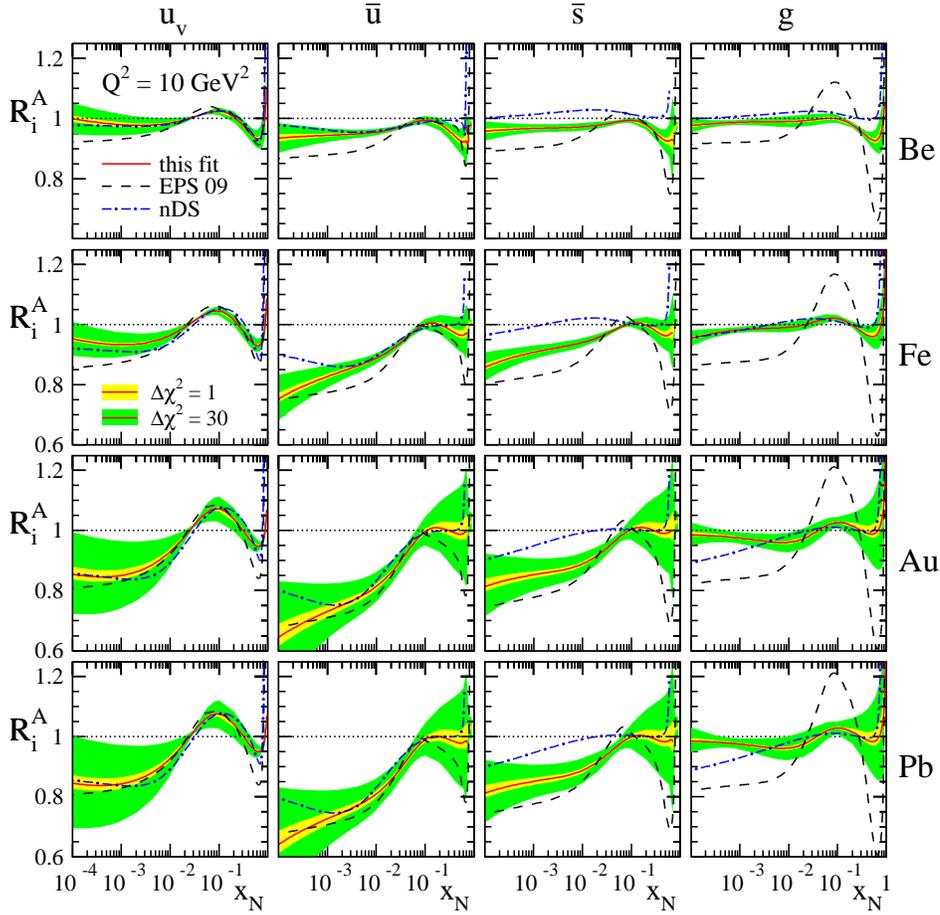}
 }
 \caption{\sl
   Example of nuclear corrections obtained from the analysis of~\cite{deFlorian:2011fp},
   at $Q^2 = 10$~GeV$^2$, for four different nuclei. From~\cite{deFlorian:2011fp}. }
 \label{fig:nuclear_corrections}
 \end{figure}

Figure~\ref{fig:nuclear_corrections} shows an example of nuclear corrections for
the $u_v$, $\bar{u}$, $\bar{s}$ and $g$ densities at a scale of $Q^2 = 10$~GeV$^2$,
as obtained from the recent analysis
described in~\cite{deFlorian:2011fp}.
They are shown for beryllium ($A = 9$), iron ($A = 56$), gold ($A = 197$) and for
lead ($A = 208$).
The size of the nuclear corrections is larger for heavier nuclei.
Nuclear effects in deuterium are usually neglected, although some QCD analyses~\cite{ABM11}
account for them explicitly.
They were studied in~\cite{Hirai:2007sx} by analysing data on
$F_2^d / F_2^p$ and were found to be small, ${\cal{O}}( 1 - 2) \%$.
In~\cite{ABM11} nuclear corrections on $F_2^d$ were found to be below $2 \%$
for $x < 0.7$, rising above $10 \%$ for $x > 0.8$.

Nuclear corrections derived from other analyses are also overlaid in Fig.~\ref{fig:nuclear_corrections}.
While the correction factors obtained by these analyses are in reasonable agreement for the up and
down quarks, sizable differences are observed for other flavours and for the
gluon\footnote{In the case of the strange density, the differences seen in
Fig.~\ref{fig:nuclear_corrections} are largely due to the fact that these analyses used
different proton PDFs, for which the strange density differ significantly.}.

Concerns have been raised in~\cite{Schienbein:2007fs, Schienbein:2009kk, Kovarik:2010uv}, regarding the possibility that
nuclear corrections may be different for NC and CC DIS. Such a breaking of
factorisation, if true, would cast serious doubts on the constraints on proton PDFs
derived from neutrino DIS experiments\footnote{As will be seen later, these experiments
set important constraints on the separation between valence and sea densities, and on the strange
PDF that, otherwise, is largely unconstrained.}.
However, this was not confirmed by the recent analysis described in~\cite{deFlorian:2011fp}.
The analysis of~\cite{Paukkunen:2010hb} also reported no tension between the NC and CC DIS data off 
heavy nuclei.

%%%%%%%%%%%%%%%%%%%%%%%%%%%%%%%%%%%%%%%%%%%%%%%%%%%%%%%%%%%%%%%%%%%%%%%%%%%%%%
\subsection{Measurements from fixed target DIS experiments}
%%%%%%%%%%%%%%%%%%%%%%%%%%%%%%%%%%%%%%%%%%%%%%%%%%%%%%%%%%%%%%%%%%%%%%%%%%%%%%
\label{sec:FixedTargetDIS}
A brief description of the main fixed target
DIS experiments is given in this section. Further details can also be
found in~\cite{CooperSarkar:1997jk}.
The $x$, $Q^2$ ranges and beam energies of
the measurements are summarised in Tab.~\ref{dataset} in section~\ref{sec:exp_summary}.
This latter section also contains a representative compilation of the measurements
described here (see Fig.~\ref{fig:f2summary} and Fig.~\ref{fig:flsummary}).

\subsubsection{SLAC}
%%%%%%%%%%%%%%%%%%%%%
The first experiments to probe the region of deep inelastic scattering
were conducted by a collaboration between the Stanford Linear
Accelerator group, the Massachusetts Institute of Technology, and the
California Institute of Technology. The experiments used an electron linac capable of
accelerating electrons up to $20$~GeV and a momentum analysing
spectrometer arm equipped with scintillator hodoscopes and multi-wire
proportional chambers (MWPCs) for electron detection, triggering and
background rejection. The experiments were performed in
the period 1970 to 1985 using one of three spectrometer arms selecting
scattered electron momenta up to $1.6$, $8$, and $20$~GeV. All three were
mounted on a common pivot around the target area and able to measure
different scattering angles. The spectrometers were designed to
decouple the measurement of scattering angle of the electron and its
momentum. This was achieved by careful design of the spectrometer
optics in which dipole and focusing quadrupole magnets were used to
deflect electrons in the vertical plane depending on momentum, and
causing horizontal dispersion of the electrons depending on the
scattering angle.

The major experiments relevant to QCD analyses of proton structure are
E49a/b, E61, E87, E89a/b, E139 and E140. In total some $6,000$ data
points were measured for $ep$ and $ed$ scattering. The latter two high
statistics experiments used the $8$~GeV spectrometer, with a
$30^{\circ}$ vertical bend to deflect scattered electrons into the
detector assembly region.

Using improved methods of applying radiative corrections, and better
knowledge of $R$~\cite{Whitlow:1990gk}, the SLAC data were
re-evaluated with a more rigorous error treatment yielding smaller
uncertainties for the relative normalisations between the individual
experiments.  A final summary dataset of all SLAC experiments combined
with precise determinations of $F_2^p$ and $F_2^d$ were
published~\cite{Whitlow:1991uw,Whitlow:1990dr}.  This combined data set achieved a
typical $3$\% statistical uncertainty, and similar systematic
uncertainties. The measurements cover a region extending to high $x$,
$0.06\leq x \leq 0.9$, and $0.6 \leq Q^2 \leq 30$~GeV$^2$.  Despite
their very good precision, the measurements at highest $x$ are usually
not included in QCD analyses because higher twist effects are
important in the domain where they were made (see
section~\ref{sec:fits_expinput}).

\subsubsection{BCDMS}
%%%%%%%%%%%%%%%%%%%%%
The BCDMS experiment~\cite{Benvenuti:1984du} was a collaboration
between the research institutes of Bologna, CERN, Dubna, Munich and
Saclay, formed in 1978 and utilised the CERN SPS M2 muon beam with
energies of $100$, $120$, $200$, $280$~GeV. The experiment was
designed to enable precise measurements of $R$ to be made with tight
control of systematic uncertainties using the high intensity muon
beam. The intense beam spills placed stringent requirements on the
experimental trigger and background rejection abilities.  The
experiment collected high statistics data on proton and deuteron
targets~\cite{Benvenuti:1989fm,Benvenuti:1989rh}.  The targets were
located serially along the common axis of eight iron toroid modules, with each
module consisting of scintillator hodoscopes and MWPCs.

The primary measurements were the inclusive double differential cross
sections corrected for radiative effects and presented as
$F_2(x,Q^2)$. Measurements of the total inclusive cross section at
different centre of mass energies allowed $R$ to be determined.  The
data cover the region $0.06 \leq x \leq 0.8$ and $7 \leq Q^2 \leq
260$~GeV$^2$. The final measured values of $F_2$ have a typical
statistical precision of $1-2\%$ and a similar systematic uncertainty,
which at high $x$ reaches up to $5\%$ arising from the spectrometer
field calibration and resolution.

The BCDMS data provide a precise measurement of the $F_2$
structure function in the valence region of high $x$. The $R$ data
show similar $x$ dependence of the two polarised pieces of the cross
section at high $x$, but indicate an increasing longitudinal component
at low $x$ consistent with the expectation of an increasing gluon
component of the proton.

%%%%%%%%%%%%%%%%%%%%%
\subsubsection{NMC}
%%%%%%%%%%%%%%%%%%%%%
\label{section:NMC}
The New Muon Collaboration (NMC) was a muon scattering DIS experiment
at CERN that collected data from $1986-1989$ using the M2 muon beam
line from the CERN SPS. It was designed to measure structure function
ratios with high precision.

The experimental apparatus~\cite{Amaudruz:1991nw} consisted of an
upstream beam momentum station and hodoscopes, a downstream beam
calibration spectrometer, a target region and a muon spectrometer.
The muon beam ran at beam energies of $90, 120, 200,$ and $280$~GeV.
The muon beams illuminated two
target cells containing liquid hydrogen and
liquid deuterium placed in series along the beam axis. Since the
spectrometer acceptance was very different for both targets they were
regularly alternated. The muon spectrometer was surrounded by several
MWPCs and drift chambers to allow a full reconstruction of the
interaction vertex and the scattered muon trajectory. Muons were
identified using drift chambers placed behind a thick iron
absorber.

The experiment published measurements of the proton and deuteron
differential cross sections $d^2 \sigma /dx dQ^2$ in the
region $0.008 < x < 0.5$ and $0.8 < Q^2 < 65$ GeV$^2$, from which the
structure functions $F_2^{p}$ and $F_2^{d}$ were
extracted~\cite{Arneodo:1996qe}. A statistical precision of $~2\%$
across a broad region of the accessible phase space was achieved, and
a systematic precision of between $2$ and $5\%$. NMC have also
published direct measurements of $R(x,Q^2)$ in the range
$0.0045<x<0.11$~\cite{Arneodo:1996qe} which provides input to the
gluon momentum distribution.

In addition the collaboration published precise measurements of the
ratio $F_2^{d}/F_2^{p}$ \cite{Arneodo:1996kd} which is sensitive to
the ratio of quark momentum densities $d/u$. By measuring the ratio of
structure functions several sources of systematic uncertainty
cancel including those arising from detector acceptance effects and
normalisation. Thus measurements in regions of small detector
acceptance could be performed and these cover the region $0.001 < x <
0.8$ and $0.1 < Q^2 < 145$ GeV$^2$ with a typical systematic
uncertainty of better than $1\%$.
The ratio $F_2^{d}/F_2^{p}$ was seen to decrease as $x \rightarrow 1$, indicating that
$d(x)$ falls more quickly than $u(x)$ at high $x$; the behaviour of
$d / u$ as $x$ approaches $1$ remains however unclear.

In 1992 NMC published the first data on the Gottfried
sum rule\cite{Amaudruz:1991at} which in the simple quark parton model states that
$\int_0^1 \frac{dx}{x} F_2^p - F_2^n = \frac{1}{3}\int_0^1 dx (u_v-d_v) 
		                      + \frac{2}{3}\int_0^1 dx (\bar{u} - \bar{d}) \,\,,$
and assuming $\bar{u}-\bar{d}=0$, should take on a value of $\frac{1}{3}$. The
initial NMC measurement indicated a violation of this assumption of a flavour symmetric
sea. This was verified by the final NMC analysis\cite{Arneodo:1994sh}
in which the Gottfried sum was determined to be $0.235 \pm 0.026$ at
$Q^2 = 4$~GeV$^2$, which implies that
$\int dx (\bar{d} - \bar{u}) \sim 0.15$,  indicating a significant excess of $\bar{d}$ over
$\bar{u}$.

%%%%%%%%%%%%%%%%%%%%%
\subsubsection{CCFR/NuTeV}
%%%%%%%%%%%%%%%%%%%%%

\label{sec:CCFR-NuTev}
The Chicago-Columbia-Fermilab-Rochester detector (CCFR) was
constructed at Fermilab to study DIS in neutrino induced lepton beams
on an almost isoscalar iron target. The detector used the wide band
mixed $\nu_{\mu}$ and $\bar{\nu}_{\mu}$ beam reaching energies of up
to $600$~GeV. The CCFR experiment collected data in 1985 (experiment
E744) and in 1987-88 (E770). 

In 1996 the NuTeV experiment (E815), using the same detector, was used
in a high statistics neutrino run with the primary aim of making a
precision measurement of $\sin^2{\theta_W}$. The major difference
between NuTeV and its predecessor CCFR was ability to select
$\nu_{\mu}$ or ${\bar{\nu}}_{\mu}$ beams which also limited the upper
energy of the wide band beam to $\simeq~500$~GeV. The neutrino beam was
alternated every minute with calibration beams of electrons and
hadrons throughout the one year data taking period. This allowed a
precise calibration of the detector energy scales and response
functions to be obtained.

The neutrino beam was produced by protons interacting with a beryllium
target. Secondary pions and kaons were sign selected and focused into
a decay volume. The detector was placed $1.4$~km downstream of
the target region and consisted of a calorimeter composed of
square steel plates interspersed with drift
chambers and liquid scintillator counters. A toroidal iron spectrometer downstream of the
calorimeter provided the muon momentum measurement using a $1.5$~T
magnetic field. In total NuTeV logged $3 \cdot 10^{18}$ protons on target.

\paragraph{Structure function measurements}

CCFR published measurements of $F_2$ and $xF_3$ ~\cite{Oltman:1992pq}
with a typical precision of $2-3$\% on $F_2$ which is largely dominated
by the statistical uncertainty on the data. The
data cover the region $0.015 \leq x \leq 0.65$ and $1.26 \leq Q^2 \leq
126$~GeV$^2$.
As discussed in~\ref{sec:formalism}, these measurements are a direct test 
of the total valence density.
NuTeV measured the double differential cross sections
$d^2 \sigma / dx dy$ from which the structure functions
$F_2$ and $xF_3$ were determined~\cite{Tzanov:2005kr,Yang:2000ju,unki}
from linear fits to the neutrino and anti-neutrino cross section
data. The data generally show good agreement between the two
experiments and the earlier low statistics CDHSW
experiment~\cite{Berge:1989hr}. However, at $x>0.4$ an increasing
systematic discrepancy between CCFR and NuTeV was observed.
A mis-calibration of the magnetic field map of the toroid in CCFR explains
a large part of this discrepancy~\cite{Tzanov:2005kr}, and the NuTeV 
measurements are now believed to be more reliable.

\paragraph{Semi-inclusive di-muon production}

\begin{figure}[bt]
 \centerline{
  \includegraphics[width=0.5\columnwidth]{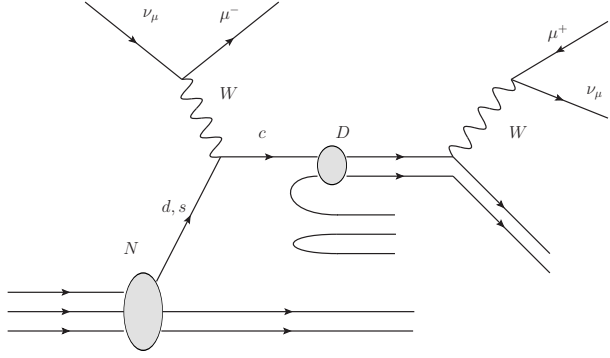}
 }
 \caption{ \sl
   Exclusive di-muon production in $\nu_{\mu}$ deep inelastic scattering.
 }
 \label{fig:dimuon_nudis}
 \end{figure}

In addition to providing inclusive cross section measurements, both
experiments also measured the semi-inclusive $\mu^+\mu^-$ production
cross section, in $\nu_{\mu}-$ and $\bar{\nu}_{\mu}-$nucleon
interactions~\cite{Goncharov:2001qe}.  Such di-muon events arise predominantly from
charged-current interactions off a strange quark, with the outgoing
charmed meson undergoing a semi-leptonic decay, as illustrated in
Fig.~\ref{fig:dimuon_nudis}.  These measurements thus provide a direct
constraint on the strange quark density in the range $0.01 < x < 0.4$.
Moreover, the separation into $\nu_{\mu}$ and $\bar{\nu}_{\mu}$
cross sections allows a separation of the $s$ and $\bar{s}$
contributions to be made, since
di-muon events are mainly produced from
$W^+ s \rightarrow c \rightarrow \mu^+ + X$ with an
incoming $\nu_{\mu}$ beam, or from
$W^- \bar{s} \rightarrow \bar{c} \rightarrow \mu^- + X$
with a $\bar{\nu}_{\mu}$ beam.
These data favour a non-vanishing asymmetry $s - \bar{s}$, as
discussed further in section~\ref{sec:StrangeSea}.

%%%%%%%%%%%%%%%%%%%%%
\subsubsection{E665}
%%%%%%%%%%%%%%%%%%%%%
This muon scattering experiment at Fermilab operated from $1987-1992$
measuring deep inelastic scattering of muons off proton and deuteron
targets in regularly alternating target cells~ \cite{Adams:1996gu}.
The data cover the range $0.0009 < x < 0.4$ and $0.2 < Q^2 < 64$ GeV$^2$.

The experiment consisted of a beam spectrometer, target region and
main spectrometer. The beam spectrometer was designed to detect and
reconstruct the beam muon momentum using trigger hodoscopes,
multi-wire proportional chambers and a dipole magnet. The target
region consisted of cells filled with liquid hydrogen
and liquid deuterium placed in a field free region and which were
alternated regularly. The main spectrometer was located immediately
downstream of the target region and consisted of two large dipole
magnets with reversed polarity. A series of drift and multi-wire
proportional chambers placed inside and downstream of both magnets
provided comprehensive tracking coverage. Further downstream a
lead-gas sampling electromagnetic calorimeter was placed in front of
iron absorbers followed by the muon detectors consisting of planes
of proportional tubes and trigger hodoscopes. 

The measurements of $F_p^p$ and $F_2^d$ typically have statistical
uncertainties of $6$\% and $5$\% respectively and systematic
uncertainties of better than $4$\%. The E665 $\mu p$ and $\mu d$ data
partially overlap with measurements from NMC at higher $Q^2$. The $x$
range of E665 data 
overlaps with that covered by the HERA experiments H1 and ZEUS (see section~\ref{hera})
though these data on $F_2^p$ lie at
higher values of $Q^2$. Comparisons between the experiments show good
agreement between NMC and E665, and the HERA data show a smooth
continuous evolution for fixed $x$ with increasing $Q^2$.

\subsubsection{CHORUS}
%%%%%%%%%%%%%%%%%%
\label{sec:Chorus}
The CERN Hybrid Oscillation Research ApparatUS (CHORUS)~\cite{Eskut:1997ar} was originally a $\nu_{\mu}
\rightarrow \nu_{\tau}$ appearance experiment in operation at CERN from
1994-1997~\cite{Eskut:2007rn}. In 1998 the run was exclusively
used for differential measurements of neutrino induced CC DIS using
the lead-scintillator calorimeter as an active
target~\cite{Onengut:2005kv} as well as studying the $Z/A$ dependence
of the total CC cross section~\cite{KayisTopaksu:2003mt}. The
experiment utilised the $450$~GeV proton beam from the SPS which was directed to a
target producing charged particles. These were sign selected and
focused into a decay volume followed by iron and earth to
filter out the neutrinos which emerged with a wide energy range
$10<E_{\nu}<200$~GeV. The detector consisted of a lead-fibre
scintillator calorimeter with nine planes of modules with alternating
orientation in the plane transverse to the beam. The muon spectrometer
was made of six toroidal iron magnets interspersed with drift chambers
scintillators and streamer tubes to reconstruct the muon momentum.

The differential cross sections in $x$, $y$ and $E_{\nu}$ are measured
in the range $0.02 \leq x \leq 0.65$ and in $0.3 \leq Q^2 \leq
82$~GeV$^2$. These are used to extract the structure functions $F_2$
and $xF_3$ in a linear fit to the $y$ dependence of the cross sections
for each $x,Q^2$ bin. The statistical uncertainty on $F_2$ is in the
region of $1\%$ and the systematic contribution to the uncertainty is
typically below $3\%$ for $x>0.1$ and increases at lower $x$. The data
for $xF_3$ are in agreement with earlier measurements from
CCFR~\cite{Seligman:1997mc} and the hydrogen target neutrino
experiment CDHSW~\cite{Berge:1989hr}. The $F_2$ measurements
are in better agreement with those from CCFR than with NuTeV.

%\clearpage
%%%%%%%%%%%%%%%%%%%%%%%%%%%%%%%%%%%%%%%%%%%%%%%%%%%%%%%
\subsection{The H1 and ZEUS experiments}
%%%%%%%%%%%%%%%%%%%%%%%%%%%%%%%%%%%%%%%%%%%%%%%%%%%%%%%
\label{hera}

The HERA collider was the first colliding beam $ep$ accelerator
operating at centre-of-mass energies of $301$ and later at $319$
GeV. At the end of the operating cycle two short low energy runs at
$\sqrt{s}=225$ and $250$ GeV were taken for a dedicated $F_L$
measurement. At the highest centre of mass energy the beams had
energies of $920$ GeV for the protons and $27.6$~GeV for the
electrons. The two experiments utilising both HERA beams were H1 and
ZEUS and they provide the bulk of the precision DIS structure function
data over a wide kinematic region. In particular, HERA opened up the 
domain of $x$ below a few $10^{-3}$ which had been mostly unexplored by
the fixed target experiments.
The fixed target experiments HERMES and HERA-B will not be discussed in this article.

The accelerator operation is divided into three periods or datasets:
HERA-I from 1992 to 2000, HERA-II from 2003 to 2007, and the dedicated Low Energy
Runs taken in 2007 after which the accelerator was
decommissioned. During the 2001-2003 upgrade of the accelerator and the
experiments, spin rotators were installed in the lepton beam line
allowing longitudinally polarised lepton beam data to be collected, with
a polarisation of up to $\pm 40$\%. In
total H1 and ZEUS together collected almost $1$~fb$^{-1}$
of data evenly split between lepton charges and polarisations.

A review of the physics results of the H1 and ZEUS experiments 
can be found in~\cite{MaxReview}, and the HERA structure function results
have been recently reviewed in~\cite{MandyReview}.

\subsubsection{Experimental Apparatus}
The two experiments were designed as general purpose detectors,
nearly $4 \pi$ hermetic, to
analyse the full range of $ep$ physics with well controlled systematic
uncertainties. The highly boosted proton
beam led to asymmetric detector designs with more hadronic
instrumentation in the forward (proton) direction which had to
withstand high rates and high occupancies. 

The most significant differences between them are the
calorimeters, which had an inner electromagnetic section and an outer hadronic part. 
ZEUS employed a compensating Uranium scintillator
calorimeter located outside the solenoidal magnet providing a
homogeneous field of $1.4$~T. H1 used a lead/steel liquid argon sampling
calorimeter located in a cryostat within the solenoid field of
$1.16$~T and a lead/scintillating fibre backward electromagnetic
calorimeter for detection of scattered leptons in neutral current
processes. 
In both H1 and ZEUS, a muon detector was surrounding the calorimeter.

Both experiments utilised drift chambers in the central regions for
charged particle detection and momentum measurements which were
enhanced by the installation of precision silicon trackers.
They allowed the momenta and polar angles $\theta$ of charged particles to be measured in
the range of $7^{\circ} < \theta < 165^{\circ}$, the backward region of large $\theta$ being
where the scattered electron was detected in low $Q^2$ NC DIS events.
In H1 an additional drift chamber gave access to larger angles of up to $\sim 172^{\circ}$.

\subsubsection{Neutral Current measurements from H1 and ZEUS}

\begin{figure}[tb]
\centerline{
  \begin{tabular}{ccc}
    \includegraphics[width=0.32\columnwidth]{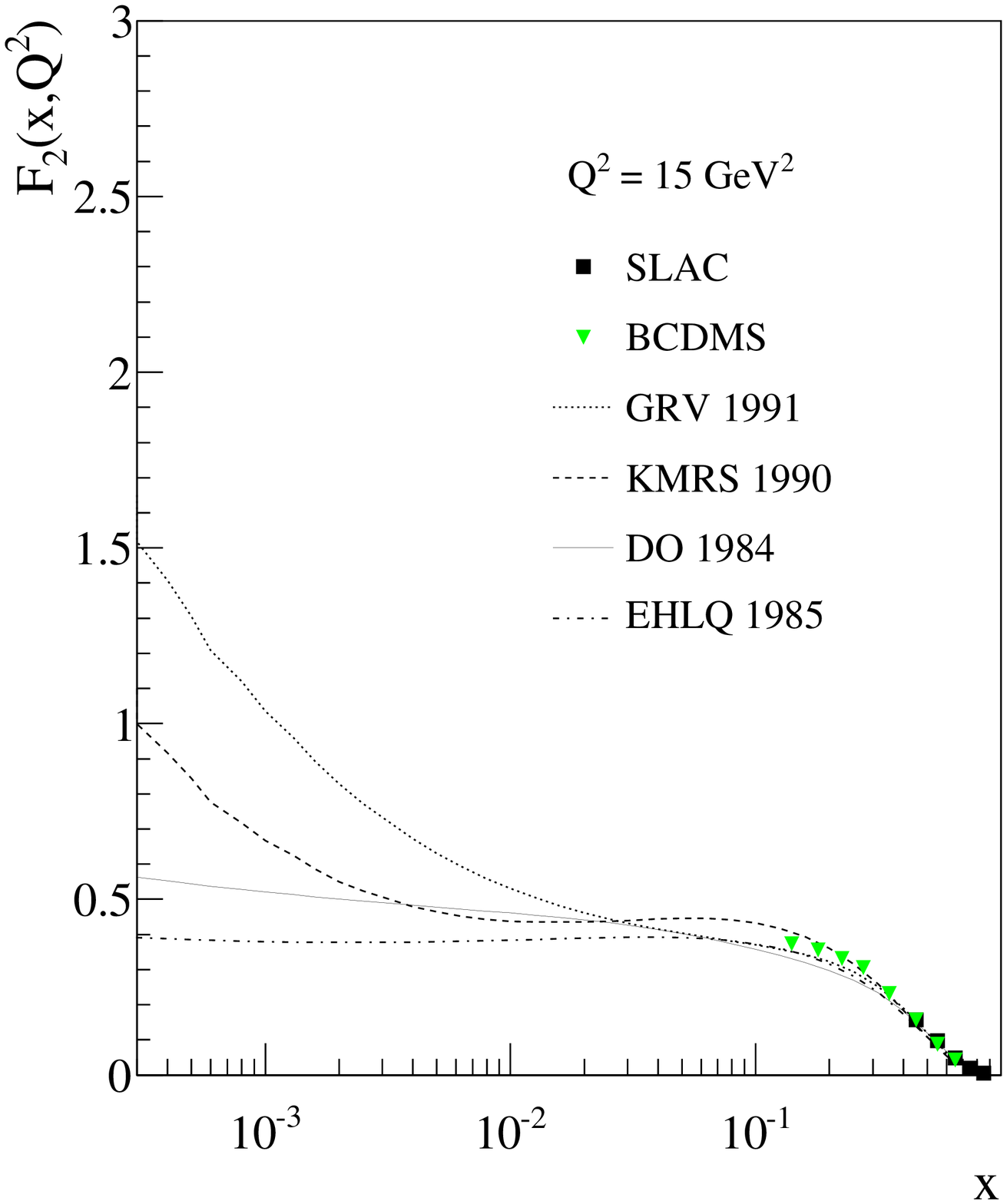}
    \includegraphics[width=0.32\columnwidth]{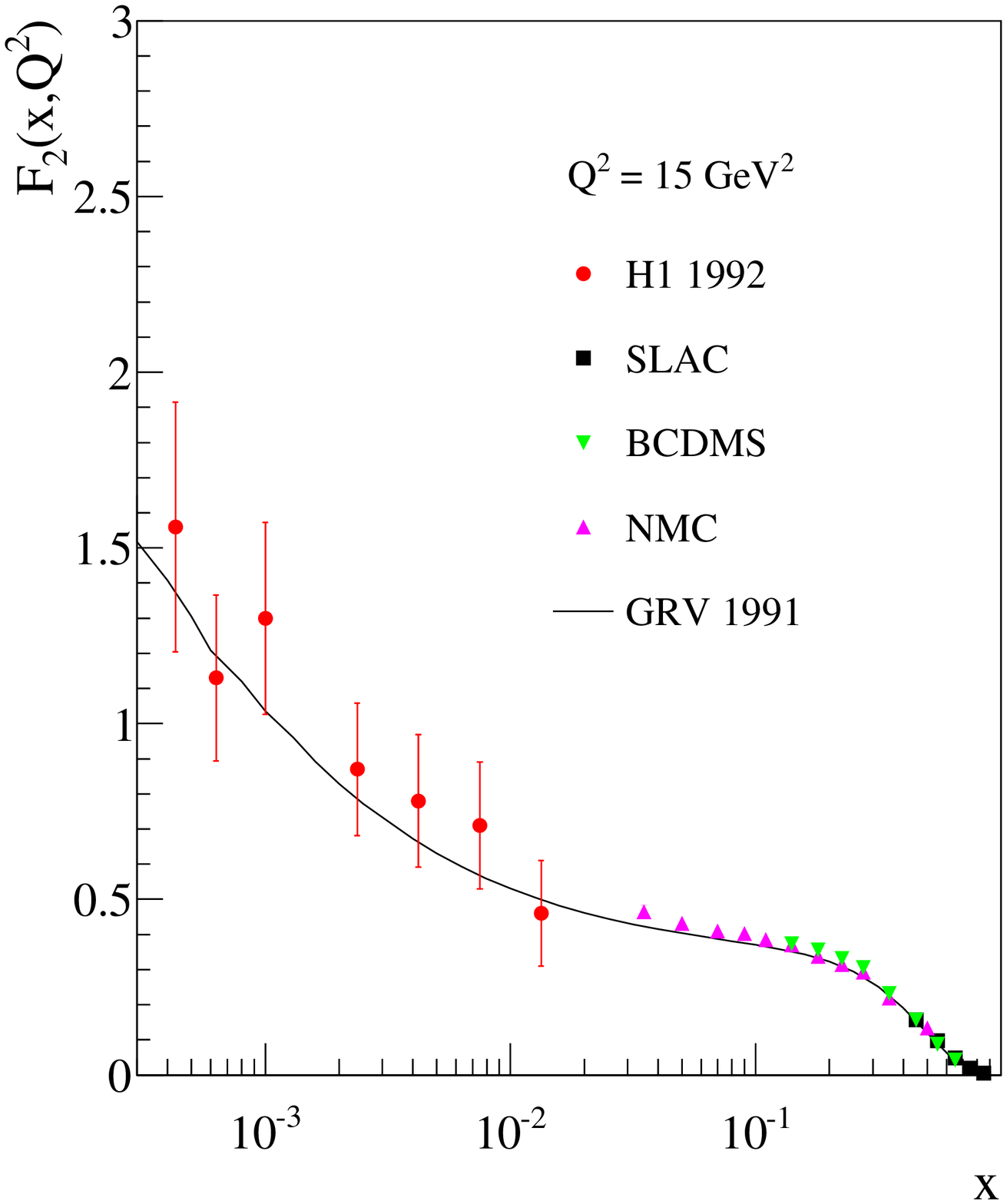}
    \includegraphics[width=0.32\columnwidth]{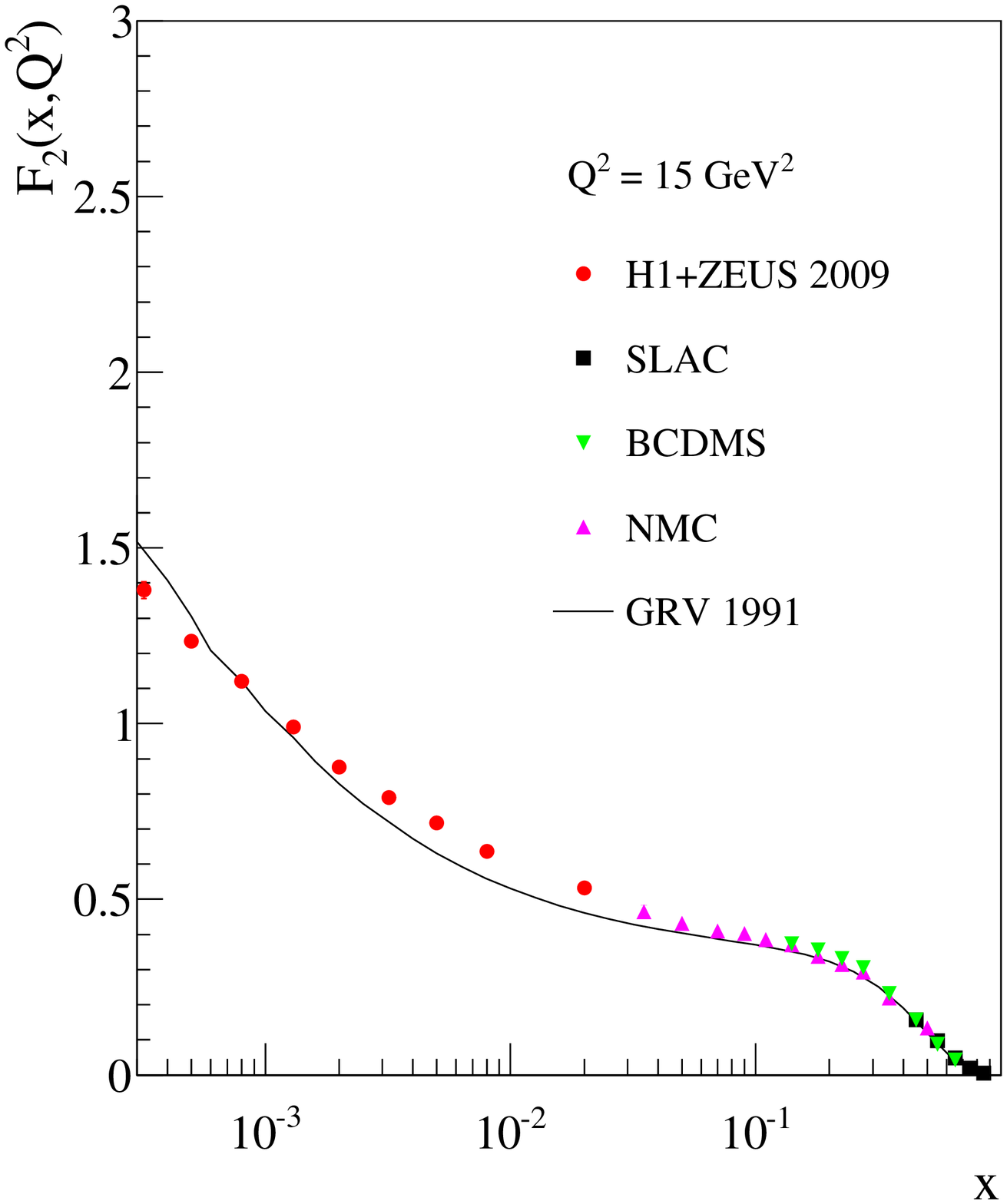}
  \end{tabular}
}
\caption{\sl
         Left: the spread of the theoretical predictions for $F_2$
         which were consistent with pre-HERA data for $Q^2=15$~GeV$^2$.
         Centre: The first $F_2$ measurements from H1 in 1992.
         Right: The complete HERA-I measurements of $F_2$.
}
\label{fig:F2_1992}
\end{figure}

Figure~\ref{fig:F2_1992} (left) shows the spread in
parameterisations of $F_2$ which existed prior to the first HERA data.
Most extrapolations from pre-HERA data indicated a ``flattish'' $F_2$ at low $x$ - which
was also expected from Regge-like arguments.
The first HERA results~\cite{Abt:1993cb,Derrick:1993fta} presented 
in $1993$ 
% and shown (for H1) in Fig.~\ref{fig:F2_1992} (centre), 
were based on $30$~nb$^{-1}$ of
data taken in $1992$ and showed a surprising, strong rise of $F_2$ towards low $x$.
An example~\cite{Abt:1993cb} of these measurements is shown in Fig.~\ref{fig:F2_1992} (centre).
With the full HERA-I dataset, the statistical uncertainty of these low
$x$ and low $Q^2$ measurements could be reduced below $1 \%$, with a systematic
error of about $2 \%$; the measurements are shown in Fig.~\ref{fig:F2_1992} (right).

With increasing luminosity, high statistics were accumulated over the
whole kinematic domain~\cite{:2009wt}.
Fig.~\ref{fig:f2summary} in the summary section~\ref{sec:exp_summary}
shows an overview of HERA $F_2$ measurements
together with data points from fixed target experiments. The very
strong scaling violations are clearly observed at low $x$. This
indicates a large gluon
density since at leading order 
$\partial F_2 / \partial \ln Q^2$ is driven at low $x$ by the product of $\alpha_s$ and
the gluon density $g(x, Q^2)$ (see Eq.~\ref{eq:f2_full_expr}).
At high $x$ the scaling violations are negative: high $x$ quarks
split into a gluon and a lower $x$ quark.
The curves overlaid are the result of QCD fits (see section~\ref{sec:qcdana}) 
based on the DGLAP evolution equations.
The data show an excellent agreement with DGLAP predictions, over five orders of magnitude in $Q^2$
and four orders of magnitude in $x$.

\begin{figure}[tb]
\centerline{
 \begin{tabular}{cc}
 \includegraphics[width=0.5\columnwidth]{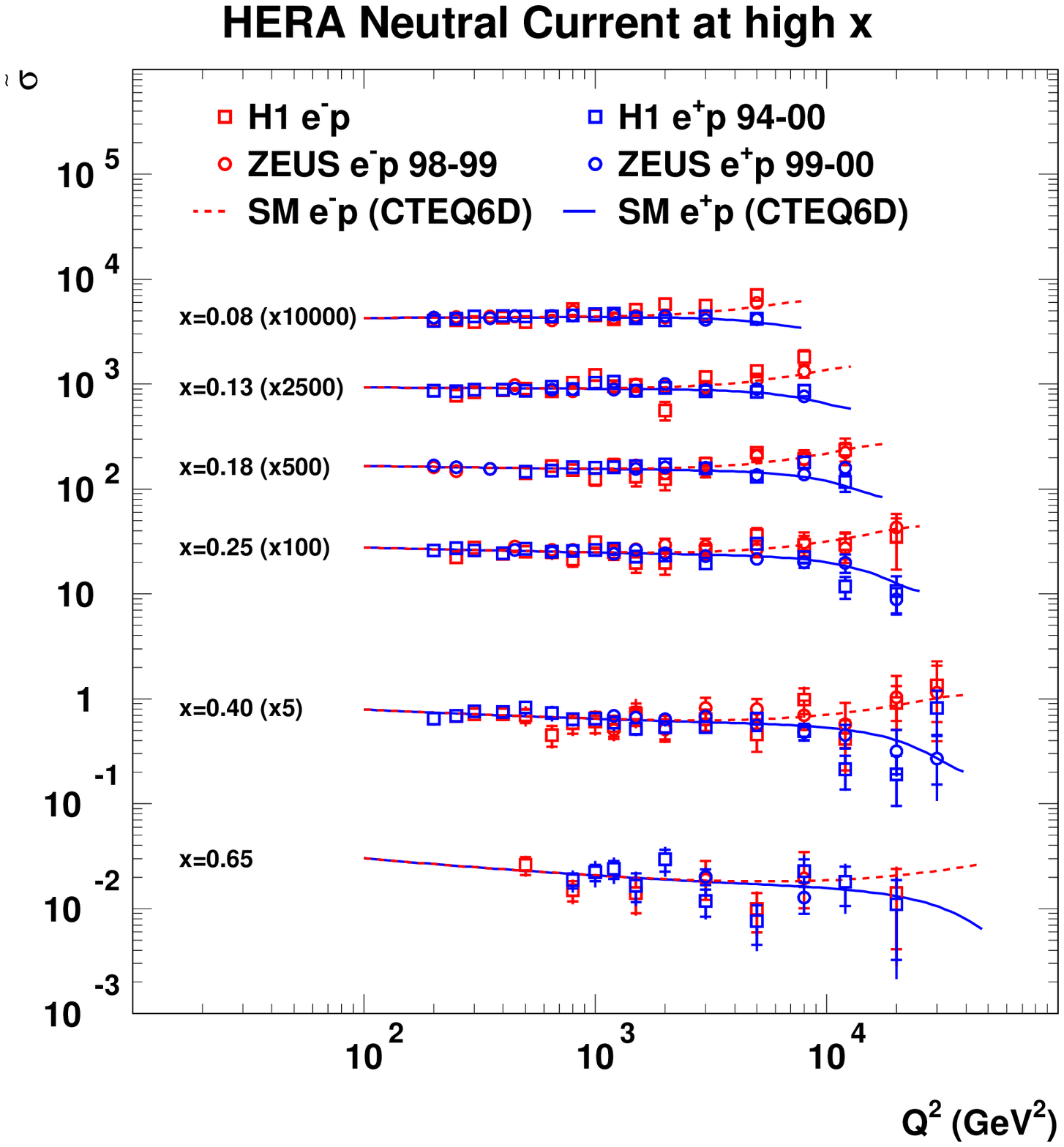}
    &
  \includegraphics[width=0.5\columnwidth]{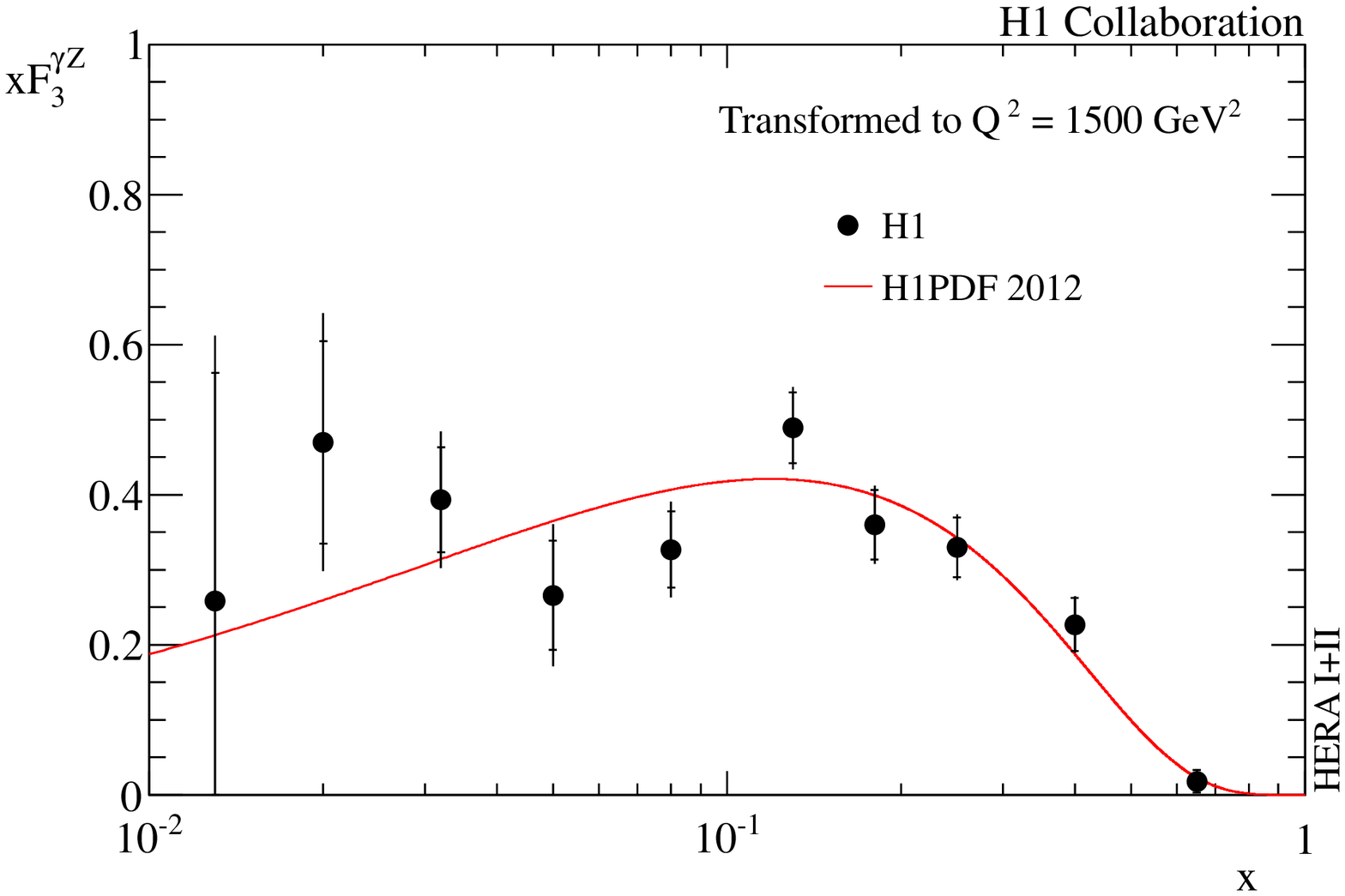}
 \end{tabular}
}
\caption{ \sl
Left: The NC DIS cross section as a function of $Q^2$ for several values
of $x$, measured in $e^+ p$ (blue symbols) and in $e^- p$ (red symbols) collisions.
The Standard Model predictions are overlaid, as the full and dashed curves, respectively.
Right: The structure function $xF^{\gamma Z}_3$ extracted from the
complete H1 HERA-I+II dataset.
From~\cite{Adloff:2003uh,h1new}.
}
\label{fig:NC_ep_em}
\end{figure}
At very high $Q^2$, the NC cross sections are sensitive to the
$Z$-exchange, resulting in $\sigma_{NC} (e^- p ) \neq \sigma_{NC} (
e^+ p)$ as was seen in Sec.~\ref{sec:formalism}.  The NC cross
sections have been measured at high $Q^2$ both in $e^+ p$ and in $e^-
p$ collisions~\cite{Adloff:1999ah, Adloff:2000qj,Adloff:2003uh,Chekanov:2002ej,Chekanov:2003yv,Chekanov:2009gm,h1new},
as shown in Fig.~\ref{fig:NC_ep_em}.  The contribution
of $Z$ exchange is clearly visible for $Q^2$ above about
$10^3$~GeV$^2$, with the $\gamma-Z$ interference being constructive
(destructive) in $e^- p$ ( $e^+ p$) collisions.  The difference
between both measurements gives access to the structure function
$x\tilde{F}_3$ which is a direct measure of the valence quark
distributions (see Eqs.~\ref{Snc1} and \ref{eq:xf3}). The H1
measurement using the full HERA-II luminosity~\cite{h1new} is shown in
Fig.~\ref{fig:NC_ep_em} (right).

HERA collider operation concluded with data taking runs at two reduced
proton beam energies in order to facilitate a direct measurement of
$F_L$. This structure function gives a larger contribution to the
cross section with increasing $y$ (see Eq.~\ref{Snc1}). It can
therefore be determined by measuring the differential cross section at different
$\sqrt{s}$, i.e. at the same $x$ and $Q^2$ but different
$y$. Measurements from H1 and ZEUS have been
published~\cite{H1FL,Chekanov:2009na} covering the low $x$ region of
$3 \times 10^{-5} - 10^{-2}$ and $Q^2$ from $1.5-120$~GeV$^2$.

\subsubsection{Charged Current DIS measurements}

\begin{figure}[tb]
\centerline{
  \includegraphics[width=0.7\columnwidth]{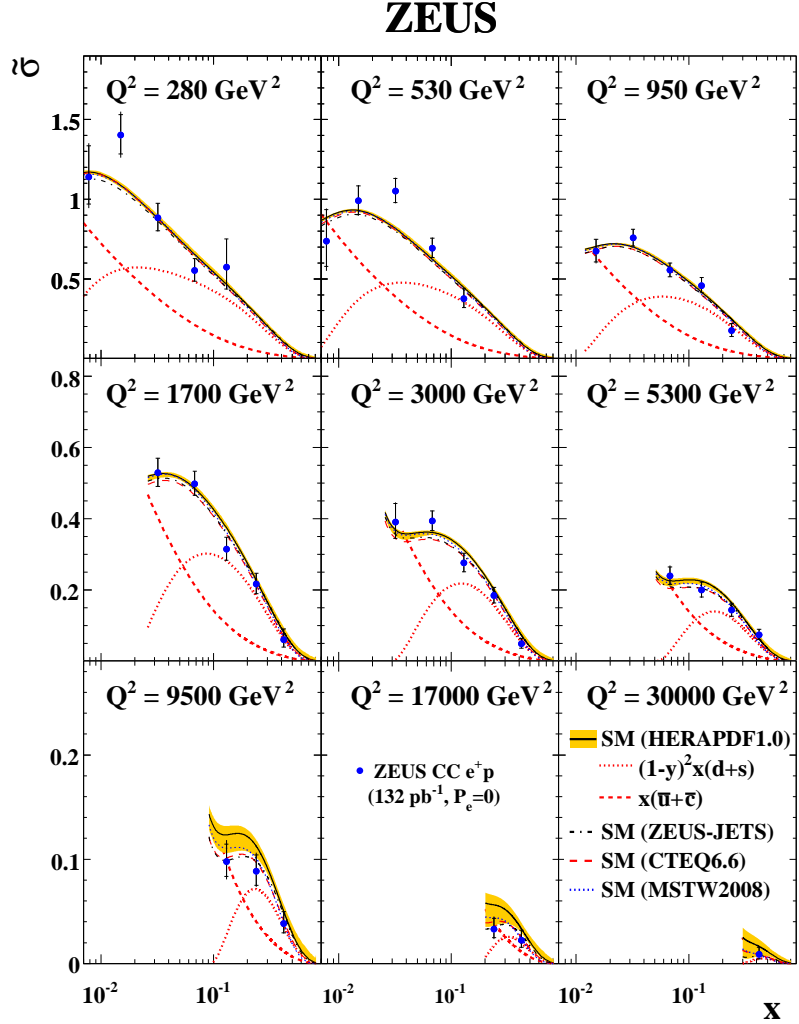}
}
\caption{ \sl
 The CC DIS cross section measured in $e^+ p$ collisions.
 The overlaid curves show how these measurements disentangle the
 contributions from up and down quarks. From~\cite{Collaboration:2010xc}.
}
\label{fig:CC_ep_em}
\end{figure}

Measurements of charged current DIS provide important constraints on the
flavour separation, which are missing from the measurement of $F_2$
alone, as the latter mostly constrains one single combination of PDFs ($4U + D$). 
Indeed (see Eq.~\ref{eq:cccross} and Eq.~\ref{ccstf}-\ref{ccstf2}), 
$\sigma^-_{CC}$ goes as $(1-y)^2 x\bar{D} + xU$ and probes mainly the $u$ density,
while $\sigma^+_{CC}$ goes as $(1-y)^2 xD + x\bar{U}$
and probes mainly the $d$ density, with some constraints being also set on
$\bar{U}$ via the high $y$ measurements.
An example of CC measurements
is shown in Fig.~\ref{fig:CC_ep_em}. 
Although the statistical precision of the HERA-II CC measurements~\cite{h1new,Chekanov:2008aa,Collaboration:2010xc}
is much better than what was achieved with HERA-I~\cite{H1PDF2K, ZEUS-CC},
these measurements remain statistically limited. For example, the precision
reaches $\sim 10\%$ for $x \sim 0.1$.
Despite this moderate precision, the constraints brought by CC DIS at HERA
are interesting since the experimental input is completely free of any correction,
in contrast to those obtained by comparing DIS measurements on a
proton and a deuterium target.

\subsubsection{The averaged H1 and ZEUS DIS dataset}
\label{sec:HERAaverage}
Recently the two collaborations have embarked on a programme of 
data combination leading to joint publications of combined data which
profit from improved uncertainties over the individual measurements. A
novel, model independent, statistical method has been employed, which
was introduced in~\cite{Glazov:2005rn} and further refined in~\cite{Aaron:2009bp}.
By taking into account the variations of the
measurements arising from different experimental sources of
uncertainty an improvement in the statistical and systematic
uncertainties is obtained. This arises from the fact that each
experiment uses different methods of measurement and each method can
act as a calibration of the other.

The unique assumption of the averaging method is that
both experiments measure the same quantity or
cross section at a given $x$ and $Q^2$. 
The averaging procedure is based on the minimisation of a $\chi^2$ function
with respect to the physical cross sections in all $(x, Q^2)$ bins of the measurement.
Each experimental systematic error source is assigned a nuisance
parameter 
with a corresponding penalty term in the $\chi^2$
function to restrict large deviations of the parameter from
zero. These parameters induce coherent shifts of the measured cross
sections according to the correlated systematic uncertainties provided
by the experiments.
The distribution of the fitted nuisance parameters in an ideal case should
be Gaussian distributed with a mean of zero and variance of one.

Several types of cross section measurement can be combined simultaneously
e.g. NC $e^+p$, NC $e^-p$, CC $e^+p$ and CC $e^-p$, yielding four
independent datasets all of which benefit from a reduction in the
uncertainty. In this case the reduction arises from correlated sources
of uncertainty common to all cross section types.  This data
combination method has been described in detail and used in several
publications~\cite{Aaron:2009bp,Aaron:2009kv,:2009wt}.

This procedure also has the advantage of producing a single set of
combined data for each cross section type which makes analysis of the
data in QCD fits practically much easier to handle. The first such
combination of H1 and ZEUS inclusive neutral and charged current cross
sections has been published using HERA-I data~\cite{:2009wt}. Further
combination updates are expected to follow as final cross sections
using HERA-II data are published by the individual experiments.

\begin{figure}[tb]
\centerline{
  \includegraphics[width=0.6\columnwidth]{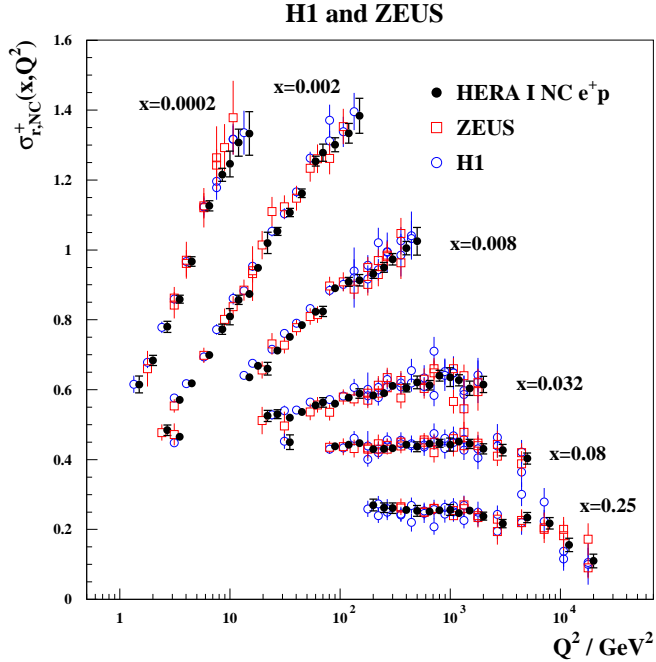}
}
\caption{ \sl
HERA combined NC $e^+p$ reduced cross section as a function of $Q^2$ for six $x-$bins,
compared to the separate H1 and ZEUS data input to the averaging procedure. The
individual measurements are displaced horizontally for better visibility. From~\cite{:2009wt}.
}
\label{fig:HERA_Combined}
\end{figure}

As an example Fig.~\ref{fig:HERA_Combined} shows the neutral current
cross section for unpolarised $e^+p$ scattering. The combined data are
shown compared to the individual H1 and ZEUS measurements. The overall
measurement uncertainties are reduced at high $x$ mainly from improved
statistical uncertainties. However at low $x$ where the data precision
is largely limited by systematic uncertainties, a clear improvement is
also visible. In the region of $Q^2 \sim 30$~GeV$^2$ the overall
precision on the combined NC cross sections has reached $1.1$\%~\cite{:2009wt}.
In the CC $e^+p$ channel the measurement accuracy is limited by the
statistical sample sizes and the combined data reduces the uncertainty
to about $10\%$ for $x\sim 0.1$. A further significant reduction in
uncertainty is expected once the combination of H1 and ZEUS data
including the complete HERA-II datasets is available.

The combined HERA datasets have been used in QCD analyses~\cite{:2009wt} to
determine proton PDFs with HERA data alone. This is described in more
detail in section~\ref{sec:HERAPDF1.0}.

\subsubsection{Heavy flavour measurements: $\boldsymbol{F_2^{c \bar{c}}}$ and $\boldsymbol{F_2^{b \bar{b}}}$ }

The charm and beauty contents of the proton have been measured at
HERA via exclusive measurements (exploiting
\begin{figure}[tb]
\centerline{
 \begin{tabular}{cc}
  \includegraphics[width=0.5\columnwidth]{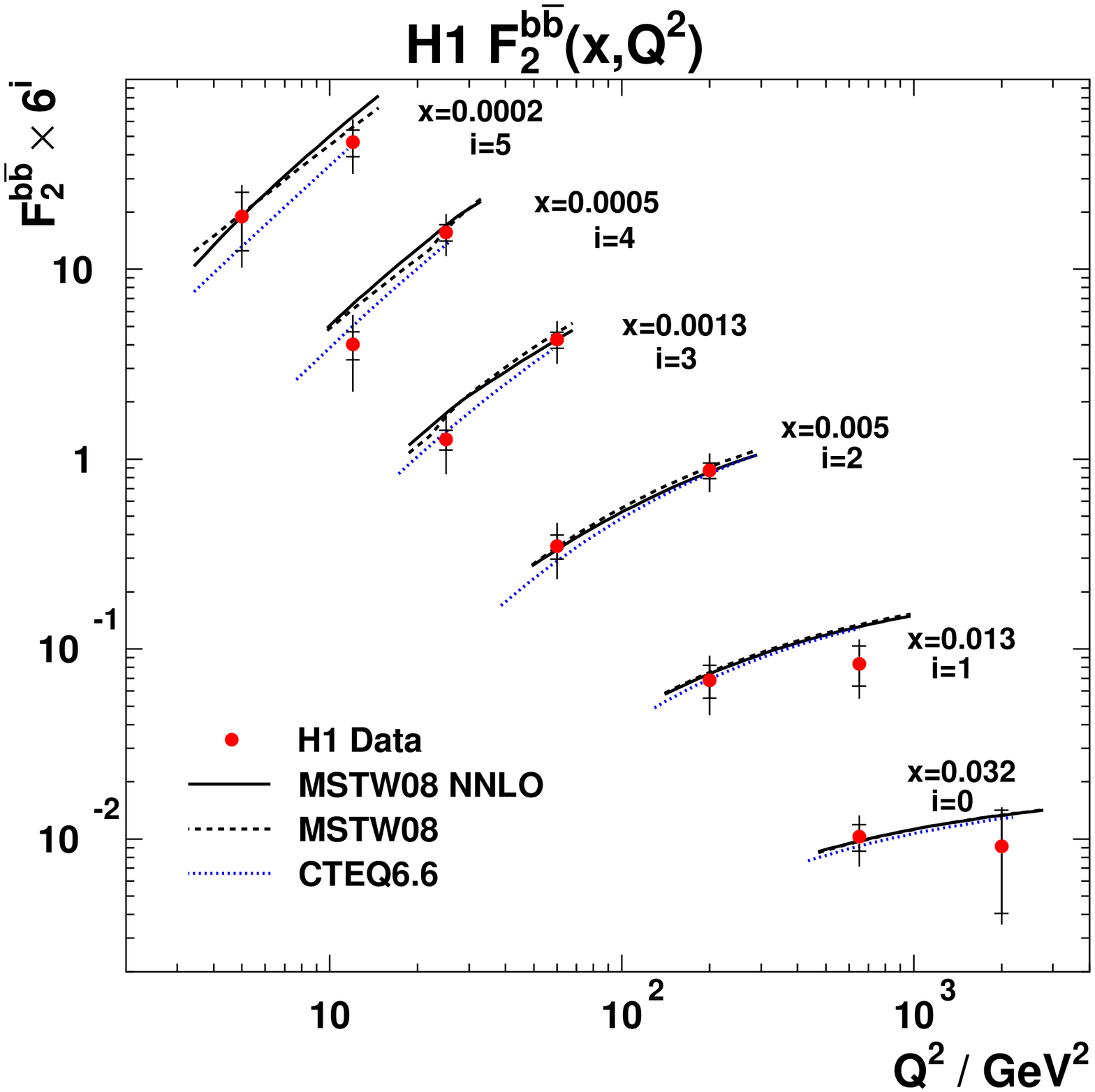} &
  \includegraphics[width=0.5\columnwidth]{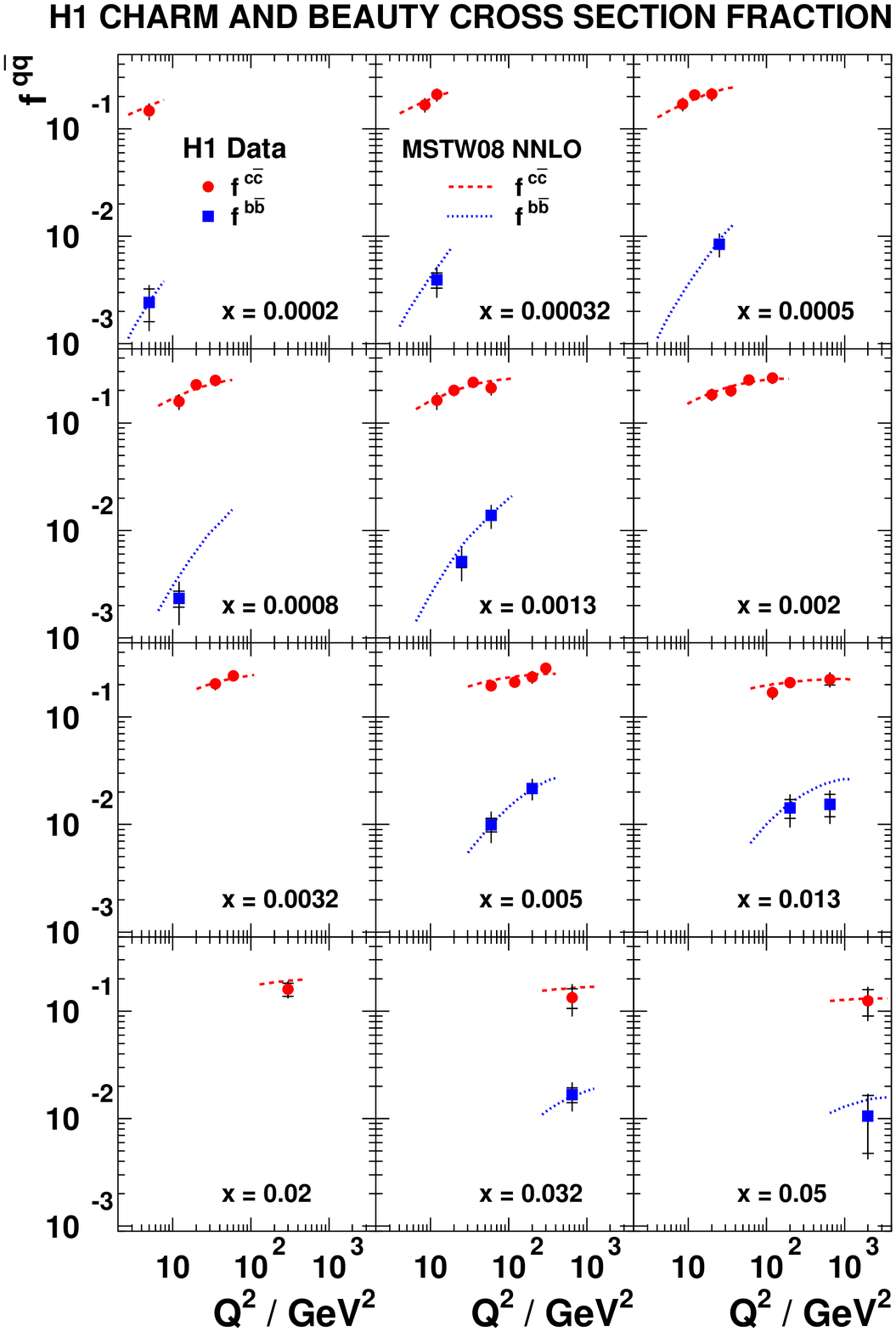}
 \end{tabular}
}
\caption{
\sl Left: $F_2^{b{\bar b}}$ measured by the H1 experiment using a lifetime technique. Right: the fractions 
of charm and beauty
in the proton, derived from the same analysis. From~\cite{H1F2BBNEW}.
}
\label{fig:F2bb}
\end{figure}
for example the $D^* \rightarrow D^0 \pi_{\rm{slow}} \rightarrow K \pi \pi$ decay
chain, or the $b \rightarrow \mu X$ decays, see 
e.g.~\cite{Chekanov:2008yd, Adloff:1998vb,Chekanov:2009kj,Aktas:2005zc}), 
and via semi-inclusive
measurements which exploit the long lifetime of the charmed and beauty hadrons,
using silicon vertex devices around the interaction points~\cite{H1F2BBNEW,Aktas:2005iw}.
Figure~\ref{fig:F2bb} (left) shows the $F_2^{b\bar{b}}$ measured by the 
H1 experiment~\cite{H1F2BBNEW}.
Just as for the inclusive $F_2$, it shows large scaling violations at low $x$.
In Fig.~\ref{fig:F2bb} (right), the charm fraction in the proton
is shown to be about $20\%$ independently of $Q^2$, while the beauty fraction increases rapidly
with $Q^2$, reaching $\sim 1\%$ at high $Q^2$. 
The precision of the measurements of $F_2^{c \bar{c}}$ and $F_2^{b \bar{b}}$ is about
$\sim 15 \%$ and $\sim 30 \%$, respectively. These measurements provide an important
test of the theoretical schemes within which observables involving heavy flavours are calculated
(see section~\ref{sec:fits_theochoices}).

\subsubsection{Dedicated measurements at very low $\boldsymbol{Q^2}$ }

Extending the $F_2$ measurements down to very low $Q^2$
requires dedicated techniques or detectors. 
The squared momentum transfer $Q^2$ can be written as $Q^2 = 2 E^0_e E_e ( 1 + \cos \theta_e)$,
where $E^0_e$ denotes the energy of the incoming lepton in the laboratory frame, $E_e$ that
of the scattered lepton, and $\theta_e$ is the angle of the scattered lepton with respect to the
direction of the incoming proton. Thus it can be seen by inspection
that to go down to low $Q^2$, one needs to access larger angles $\theta_e$,
or to lower the incoming electron energy $E^0_e$.
This can be achieved by:
\begin{itemize}
 \item using a dedicated apparatus, as the ZEUS Beam Pipe Tracker (BPT), which consisted of
       a silicon strip tracking detector and an electromagnetic calorimeter very close
       to the beam pipe in the backward electron direction;
 \item shifting the interaction vertex in the forward direction. Two short
       runs were taken with such a setting, where the nominal interaction point
       was shifted by $70$~cm;
 \item exploiting QED Compton events: when the lepton is scattered at a large
       angle $\theta_e$, it may still lead to an observable electron (i.e. within the
       detector acceptance) if it radiates a photon.
 \item using events with initial state photon radiation which can lower the
   incoming electron energy $E^0_e \rightarrow E^0_e -
   E_{\gamma}$ where $E_{\gamma}$ is the energy of the radiated
   photon.

\end{itemize}
All these methods have been exploited at HERA~\cite{LowQ2Measurements}.
In particular, it was observed that $F_2$ continues to rise at low $x$, even 
at the lowest $Q^2$, $Q^2 \sim 0.5$~GeV$^2$.
Note that these measurements are usually not
included in QCD analyses determining parton distribution functions, since 
they fail the lower cut
in $Q^2$ that is usually applied to DIS measurements, to ensure that they are not affected by non-perturbative
effects.

%======================================================
\subsubsection{Jet cross sections in DIS at HERA}
%======================================================

\begin{figure}[tb]
\centerline{
   \includegraphics[width=0.5\columnwidth]{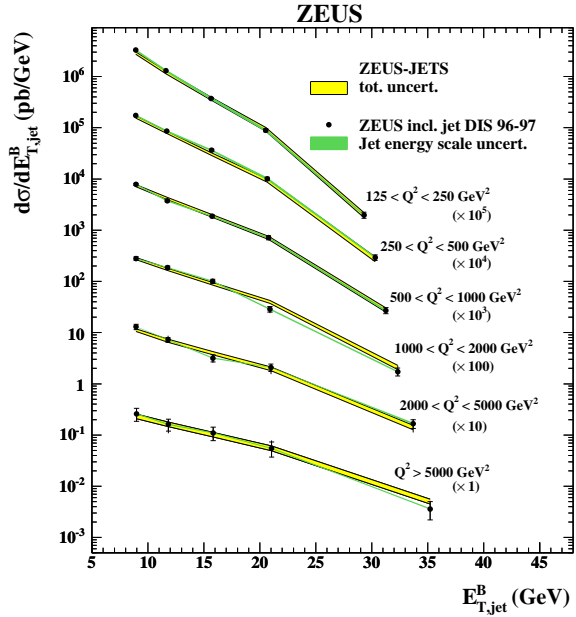}
}
\caption{ \sl
Differential cross sections of inclusive jet production in DIS measured
by the ZEUS experiment, together
with their systematic uncertainties. From~\cite{Chekanov:2005nn}.
}
\label{fig:Zeus_jets}
\end{figure}

The H1 and ZEUS experiments have measured inclusive jet cross sections
in the so-called ``Breit'' frame as a function of several variables,
for example differentially with respect to the jet energy and in several $Q^2$
bins~\cite{Chekanov:2002be,Chekanov:2006xr,:2007pb,:2009he,Chekanov:2005nn}.
The Breit frame~\cite{BreitFrame} is of particular interest for jet
measurements at HERA since it provides a maximal separation between the products
of the beam fragmentation and the hard jets. In this frame, the exchanged
virtual boson $V^*$ is purely space-like and is collinear with the incoming parton,
with $\vec{q} = (0, 0, -Q)$. For parton-model processes, $V^* q \rightarrow q$, the
virtual boson is absorbed by the struck quark, which is back-scattered with zero transverse
momentum with respect to the $V^*$ direction. On the other hand, for $\mathcal{O}(\alpha_s)$ processes
like QCD-Compton scattering ($V^* q \rightarrow qg$) and boson-gluon fusion ($V^* g \rightarrow q \bar{q}$),
the two final-state partons lead to jets with a non-vanishing transverse momentum.
Hence, the inclusive requirement of at least one jet in the Breit frame with
a high transverse momentum selects $O(\alpha_s)$ processes and suppresses parton-model processes.
Example measurements of inclusive jet production in DIS obtained by the ZEUS
experiment are shown in Fig.~\ref{fig:Zeus_jets}.
With small systematic uncertainties of typically $\sim 5 \%$, such data can bring
constraints on the gluon density in the medium $x$ range, $x = 0.01 - 0.1$.
However, when included in global QCD fits which also include other jet data, the
impact of these measurements is limited~\cite{ReviewThorne}.

Jet production has also been measured in the photoproduction regime of
$Q^2 \rightarrow 0$. However, these measurements
are usually not included in QCD analyses of the proton structure because of their potential
sensitivity to the photon parton densities.

%\clearpage

\subsection{Experiments with hadronic beams}
 \label{exp:nonDIS}

In interactions where no lepton is involved in the initial state,
the cross sections depend on products of parton distribution functions as shown
by Eq.~\ref{eq:factor}.
Hadro-production experiments, using either a fixed target or two colliding beams,
provide a wealth of measurements
that nicely complement those made in lepto-production.
In particular, they set specific constraints on some parton distribution functions
that are not directly accessed in DIS experiments.
The corresponding measurements, performed by fixed target experiments
and by the D0 and CDF experiments at the Tevatron collider, are described in
this section.

\subsubsection{Kinematics in hadro-production}

In $pp$, $pd$ or $p \bar{p}$ collisions, the production of a final state of invariant mass
$M$ involves two partons with Bjorken-$x$ values $x_1$ and $x_2$ 
related by
\begin{equation}
 M^2 = x_1 x_2 s
\end{equation}
where $s$ denotes the square of the energy in the centre of mass of the hadronic collision.
The minimum value of $x_{1,2}$ is thus $x_{min} = \tau$ with
\begin{equation}
 \tau \equiv M^2 / s
\label{eq:tau}
\end{equation}
In the rest frame of the two hadrons and neglecting the hadron masses,
 the rapidity $y$ of the final state $X$ is
\begin{equation}
y = \frac{1}{2} \ln \frac{E-p_z}{E+p_z} = \frac{1}{2} \ln \frac{x_1}{x_2}
\label{eq:DY_y}
\end{equation}
where the hadron that leads to the parton with momentum fraction $x_1$ defines the positive
direction along the beam(s) axis. Hence $x_1$ and $x_2$ can be written as
\begin{equation}
x_1 = \sqrt{\tau} e^y \qquad , \qquad x_2 = \sqrt{\tau} e^{-y}
\label{eq:DY_kinematics}
\end{equation}
In fixed target experiments, the positive direction is usually defined by
the direction of the incident beam, such that $x_1$ denotes the Bjorken-$x$ of
the parton in the beam hadron, and $x_2$ that of the parton in the target hadron.
In $p \bar{p}$ collisions, the positive direction can be defined by the proton beam,
in which case $x_1$ ($x_2$) denotes the Bjorken-$x$ of the parton in the proton (anti-proton).

%=================================================================================
\subsubsection{Drell-Yan di-muon production in fixed target experiments}
\label{sec:DY_E866}
%=================================================================================

The experiments E605, E772 and E866/NuSea have measured di-muon
production in Drell-Yan interactions of a proton off a fixed target.
They used an $800$~GeV proton beam extracted from the Fermilab
Tevatron accelerator that was transported to the east beamline of the
Meson experimental hall.
While changes were made to the spectrometer for E772 and E866/NuSea, the basic
design has remained the same since the spectrometer was first used for E605 in the
early 1980s. The core consists of three large dipole magnets that allow
the momentum of energetic muons
to be measured and deflect soft particles out of the acceptance.
Different targets have been employed: copper for E605, liquid deuterium for
E772 and E866, and liquid hydrogen for E866.
The centre-of-mass energy of the Drell-Yan process for these
experiments is $\sqrt{s} = 38$~GeV. A broad range of di-lepton invariant
mass $M$ could be covered, extending up to $M \sim 20$~GeV.

\paragraph{Differential cross sections}

The experiments published~\cite{Moreno:1990sf, McGaughey:1994dx}
double-differential cross sections in $M$ and either in the rapidity
of the di-lepton pair $y$, or in Feynman $x_F$, defined as
$x_F = 2 q_z / \sqrt{s}$ where $q$ denotes the four-vector of the
Drell-Yan pair in the hadronic centre-of-mass frame and $q_z$ its projection on the longitudinal axis.
At leading order, $x_F = x_1 - x_2$ 
and
the leading order differential cross sections can be written as:
\begin{eqnarray}
\frac{d^2 \sigma}{dM^2 dy} &= & \frac{4 \pi \alpha^2}{9 M^2 s} \sum_i e^2_i \left[ q_i(x_1) \bar{q}_i(x_2) + \bar{q}_i(x_1) q_i(x_2) \right ]  
\label{eq:dy_eq} \\
\frac{ d^2 \sigma}{dM^2 dx_F} &= & \frac{1}{x_1 + x_2} \frac{d\sigma}{dM^2 dy}   \qquad .
\end{eqnarray}
The experiments made measurements in the range $4.5 < M < 14$~GeV and $0.02 < x_F < 0.75$,
corresponding to $x_1 \sim 0.1 - 0.8$ and
$x_2 \sim 0.01 - 0.3$, the acceptance of the detector being larger for
$x_1 \gg x_2$. In this domain, the first term dominates in Eq.~\ref{eq:dy_eq},
and the measurements bring important information on the sea densities
$\bar{u}(x)$ and $\bar{d}(x)$, especially for $x$ larger than about $0.1$ where DIS
experiments poorly constrain the sea densities.

\paragraph{The ratio $\boldsymbol{pp/pd}$ from E866 }

E866/NuSea made measurements using both a deuterium and a hydrogen 
target~\cite{Hawker:1998ty,Towell:2001nh,Webb:2003ps}
from which ratios of the differential cross sections
$\sigma_{pp} / \sigma_{pd}$ could be extracted.
These measurements have brought an important insight
on the asymmetry $\bar{d} - \bar{u}$ at low $x$.
Indeed, the cross sections in the phase space where $x_1 \gg x_2$ 
can be written as:
\begin{eqnarray}
\sigma_{pp} & \propto & \frac{4}{9} u(x_1) \bar{u}(x_2) + \frac{1}{9} d(x_1) \bar{d}(x_2), \qquad   \label{eq:DY_pp} \\
 \sigma_{pn} & \propto & \frac{4}{9} u(x_1)\bar{d}(x_2) + \frac{1}{9} d(x_1)\bar{u}(x_2)
\end{eqnarray}
such that:
$$ \frac{\sigma_{pd}} { 2 \sigma_{pp} } \sim \frac{1}{2} 
   \frac{ 1 + \frac{1}{4} \frac{d(x_1)}{u(x_1)}} { 1 + \frac{1}{4} \frac{d(x_1)}{u(x_1)} 
   \frac{ \bar{d}(x_2)} {\bar{u}(x_2)}  } \left[ 1 + \frac{ \bar{d}(x_2)}{\bar{u}(x_2)}   \right] . $$
In the relevant domain of $x_1$, the ratio $d(x_1)/u(x_1)$ is quite well known, such that
the ratio $\sigma_{pd} / 2 \sigma_{pp}$ gives access to the ratio
$\bar{d} / \bar{u}$ at low and medium $x$, $x \sim 0.01 - 0.3$.

This idea was first used by the NA51 experiment at CERN~\cite{Baldit:1994jk}, which confirmed the indication,
previously obtained by the NMC experiment, that $\bar{d} \neq \bar{u}$ (see section~\ref{section:NMC}).
But the acceptance of the NA51 detector was limited, and their result for $\bar{d} / \bar{u}$ 
($\bar{d} / \bar{u} \sim 2$) could
be given for a single $x$ value only.

\begin{figure}[bt]
 \centerline{
   \begin{tabular}{cc}
   \includegraphics[width=0.5\columnwidth]{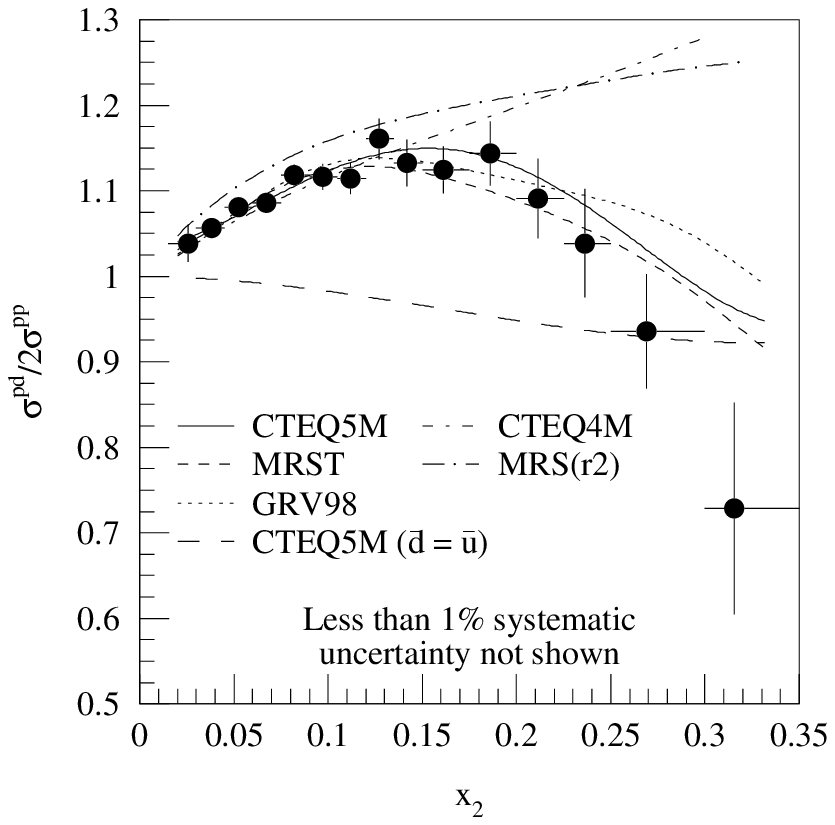} &
   \includegraphics[width=0.5\columnwidth]{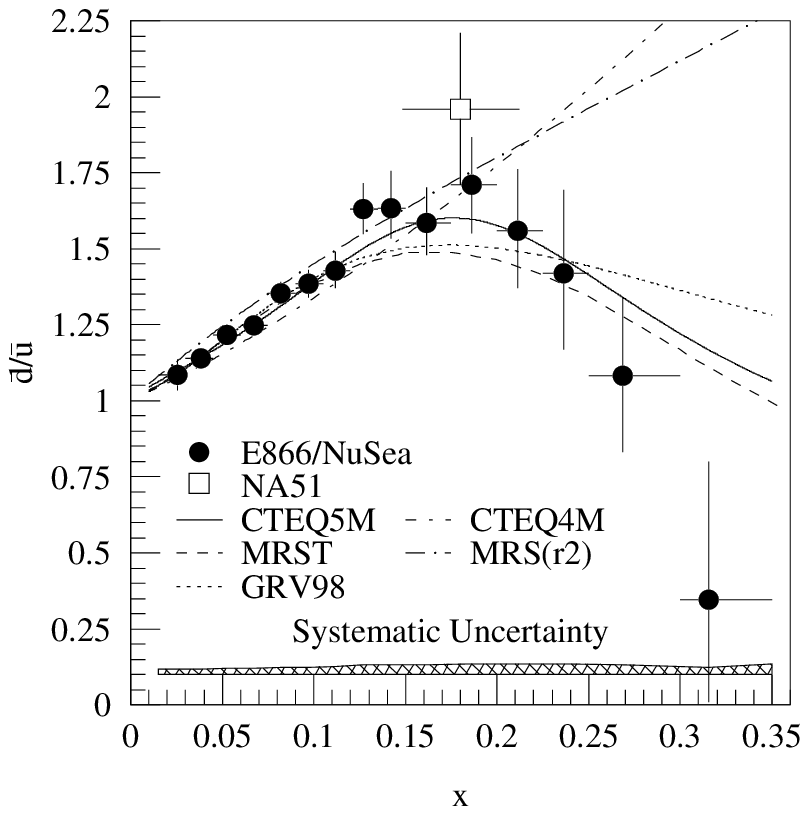}
   \end{tabular}
 }
 \caption{\sl 
    Left: The ratio $\sigma_{pd} / 2 \sigma_{pp}$ measured by the E866 experiment.
    Right: The ratio $\bar{d} / \bar{u}$ extracted from this measurement.
    The previous result from the NA51 experiment is also shown as the open square.
    From~\cite{Towell:2001nh}.
 }
 \label{fig:e866}
 \end{figure}

The E866 experiment was the first to measure the $x$-dependence of $\bar{d} / \bar{u}$.
Fig.~\ref{fig:e866} shows the obtained measurement~\cite{Towell:2001nh},
which extends down to $x_2 \sim 0.03$.
Note the spread of the theoretical predictions before
these data were included in the fits. The ratio $\bar{d} / \bar{u}$ as extracted by
E866 is shown in Fig.~\ref{fig:e866} (right), and clearly demonstrates that
$\bar{d} > \bar{u}$. The asymmetry between $\bar{d}$ and $\bar{u}$ is largest
for $x \sim 0.2$ and decreases with decreasing $x$; what happens as $x \rightarrow 1$ 
remains unclear. 

%

%======================================================
\subsubsection{The D0 and CDF experiments at Fermilab}
%======================================================
 \label{exp:tevatron}

The D0 and CDF experiments were located at the Tevatron collider at Fermilab,
which collided protons and anti-protons.
In a first phase of operation (``Run I'', from $1992$
to $1998$), the Tevatron
was operated at a centre-of-mass energy of $1.8$~TeV. 
The second phase, ``Run II'', started in $2001$ following significant upgrades
of the accelerator complex and of the experiments, with a centre-of-mass energy
of $1.96$~TeV. The data taking has stopped in $2011$.
 
The measurements of the D0 and CDF experiments provide several important constraints on proton structure:
\begin{itemize}
 \item measurements of lepton charge asymmetry from $W$ decays bring 
       constraints on the ratio $d/u$ at $x > 0.05$, and hence on the
       $d$ density, which is less well known than the $u$ density;
 \item measurements of the $Z$ rapidity distribution in $Z \rightarrow l^+ l^-$
       decays bring constraints on the quark densities at $x > 0.05$, which are complementary
       to those obtained from DIS measurements;
 \item the cross sections for inclusive jet production in several rapidity
       bins provide constraints on the gluon and the quark distributions
       for $0.01 < x < 0.5$. In particular, they set the strongest constraints
       on the gluon density at high $x$.
\end{itemize}

A detailed description of the D0 detector can be found in~\cite{Abazov:2005pn}.
The inner most part is a central tracking system surrounded
by a $2$~T superconducting solenoidal magnet. The two
components of the central tracking system, a silicon microstrip
tracker and a central fibre tracker, are used to
reconstruct interaction vertices and provide the measurement
of the momentum of charged particles in the pseudo-rapidity range
$\mid \eta \mid < 2$. The tracking
system and magnet are followed by the calorimetry
system that consists of 
electromagnetic and hadronic uranium-liquid argon sampling
calorimeters.
Outside of the D0 calorimeter
lies a muon system which consists of layers of drift tubes
and scintillation counters and a 1.8 T toroidal magnet.

The CDF II detector is described in detail in~\cite{Acosta:2004yw}.
The detector has a charged particle tracking
system that is immersed in a $1.4$~T solenoidal magnetic
field coaxial with the beam line, and provides coverage
in the pseudo-rapidity range $ \mid \eta \mid < 2$. Segmented sampling
calorimeters, arranged in a projective tower geometry, 
surround the tracking system and measure the energy
of interacting particles for $\mid \eta \mid < 3.6$.

%.................................................................
\paragraph{Lepton charge asymmetry from $\boldsymbol{W}$ decays}
%.................................................................

In $pp$ or $p \bar{p}$ collisions, the production of $W^+$ bosons 
proceeds mainly via $u \bar{d}$ interactions, or via $\bar{u} d$ for
 $W^-$ production.
At large boson rapidity $y_W$, the interaction involves one
parton with 
$x = \sqrt{ \tau } \exp {|y_W|}$ (see Eq.~\ref{eq:DY_kinematics}) where $\sqrt{ \tau } = M_W / \sqrt{s} \sim 0.04$.
In $p \bar{p}$ collisions at the Tevatron, this medium to high $x$ parton is most
likely to be a $u$ quark  picked up from the proton
in the case of $W^+$ production, or a
$\bar{u}$ anti-quark from the anti-proton in $W^-$ production; this follows from the
fact that $u(x) > d(x)$ at medium and high $x$.
Hence, $W^+$ bosons are preferably emitted in the direction of the
incoming proton and $W^-$ bosons in the anti-proton direction, leading
to an asymmetry between the rapidity distributions of $W^+$ and $W^-$
bosons.
This asymmetry can be written as:
\begin{eqnarray}
 A( y_W ) &=  & \frac { d \sigma(W^+) / d y_W - d \sigma(W^-) / d y_W } { d \sigma(W^+) / d y_W + d \sigma(W^-) / d y_W }   \\
             &\sim  & \frac { u(x_1) d(x_2) - d(x_1) u (x_2) } { u(x_1) d(x_2) + d(x_1) u (x_2) }   \\
             &=  & \frac { R(x_2) - R(x_1) } { R(x_2) + R(x_1) }   \sim  y_W \sqrt{ \tau } R'(\sqrt{\tau}) / R(\sqrt{\tau}) 
  \label{eq:asym_R}
\label{eq:wasym_tevatron}
\end{eqnarray}
where $R(x) = d(x) / u(x)$ and $R'$ denotes the derivative of $R$. 
It can be seen from Eq.~\ref{eq:asym_R} that the $W$ charge asymmetry is directly
sensitive to the $d / u$ ratio in the range $ x \sim 0.01 - 0.3$, and to its slope 
at $x \sim 0.04$.

\begin{figure}[tb]
\centerline{
 \begin{tabular}{cc}
   \includegraphics[width=0.5\columnwidth]{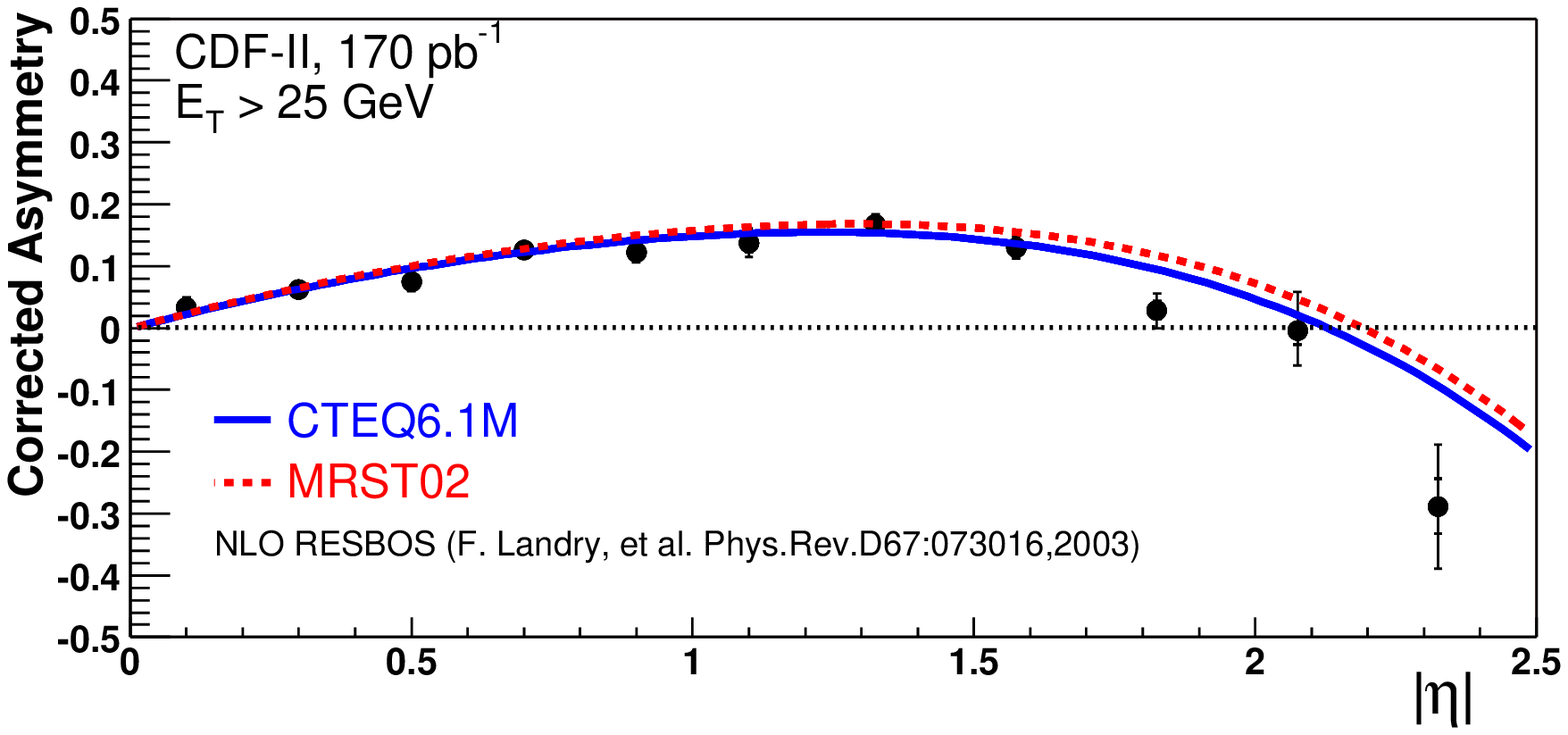} &
   \includegraphics[width=0.5\columnwidth]{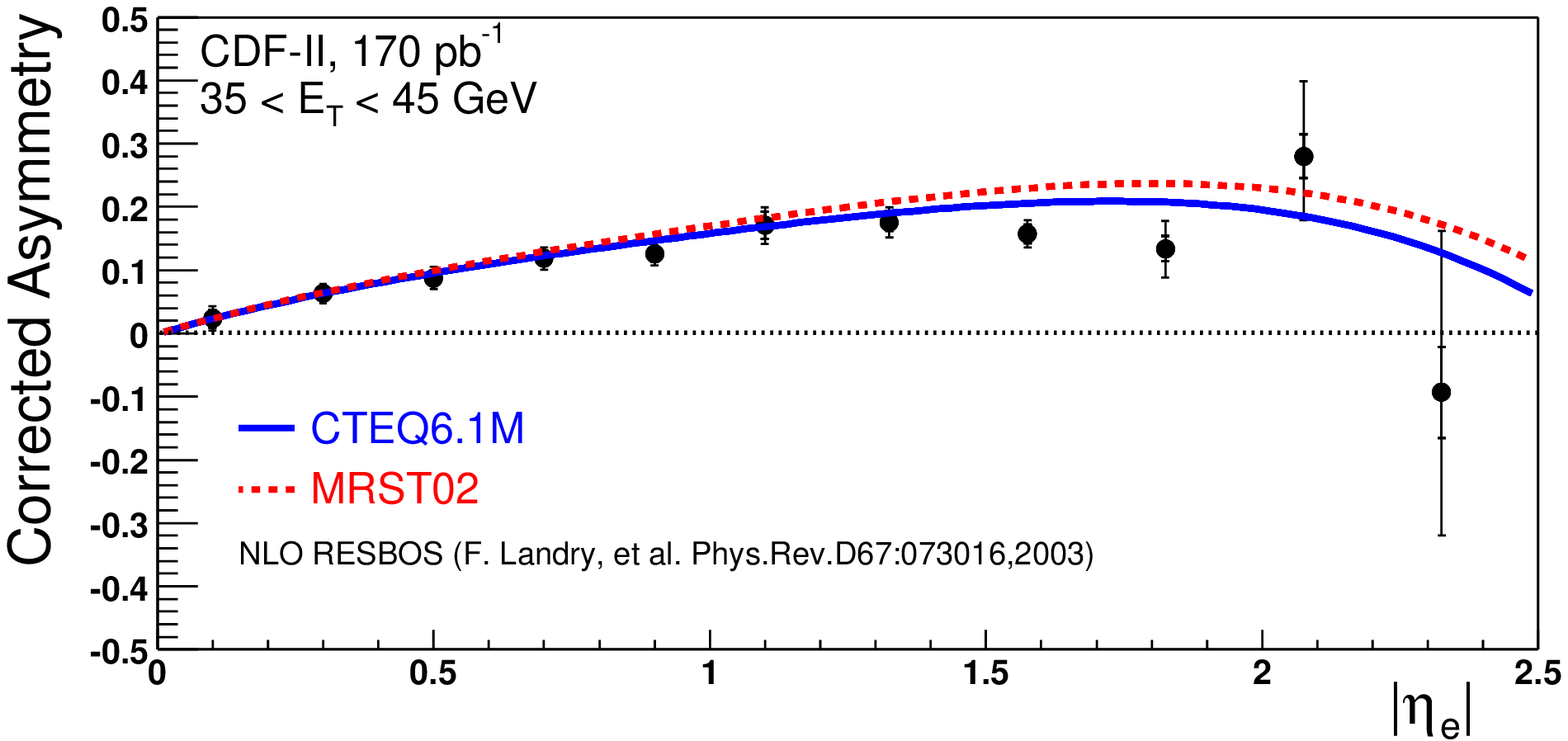}
 \end{tabular}
}
\caption{\sl 
 Lepton charge asymmetry from $W$ decays, as a function of the pseudo-rapidity 
 of the charged lepton. The asymmetry is shown for two ranges of the
 lepton transverse energy. From~\cite{Acosta:2005ud}.
}
 \label{fig:w_asym}
\end{figure}

This asymmetry remains, though diluted, when measuring the
experimentally observable\footnote{A measurement of $A(y_W)$ was actually performed in~\cite{Aaltonen:2009ta}.} rapidity of the charged lepton
coming from the $W$ decay~\cite{Abe:1998rv, Acosta:2005ud, Abazov:2007pm, Abazov:2008qv}.
Figure~\ref{fig:w_asym} shows example measurements from the CDF experiment, in
two bins of the transverse energy $E_T$ of the lepton. At low $E_T$, the
measured asymmetry is also sensitive to the anti-quark densities, via subleading interactions
involving an anti-quark coming from the proton and a quark from the anti-proton, which were
neglected in the approximate formula Eq.~\ref{eq:wasym_tevatron}.

%...............................................................................
\paragraph{Rapidity distribution in $\boldsymbol{Z \rightarrow l^+ l^-}$ events}
%...............................................................................

The large integrated luminosity delivered by the Tevatron allows the $Z$ rapidity
distribution to be precisely measured by the D0 and CDF experiments~\cite{
Abazov:2007jy,Aaltonen:2010zz}.
The $Z / \gamma^*$ rapidity distribution is measured in a di-lepton
mass range around the $Z$ boson mass, extending up to $ \mid y \mid \sim 3$. 
The measurements provide constraints on the quark densities at $Q^2 \sim M^2_Z$, over
a broad range in $x$. 
Neglecting the $Z / \gamma^*$ interference terms, which are small in the considered
mass range and well below the
experimental uncertainties, the differential cross section reads as:
\begin{eqnarray}
 \frac{d \sigma}{dy} &= 
 &\frac{ \pi G_F M^2_Z \sqrt{2} } {3s} \sum_{i} c_{i} \left[ q_i (x_1) \bar{q}_i (x_2) + \bar{q}_i(x_1) q(x_2) \right]  
 \end{eqnarray}
where $c_i = v_i^2 + a_i^2$ is the sum of the squares of the vector and axial couplings of 
the quarks to the $Z$ boson. Hence, the measured cross sections mainly probe the combination
\begin{eqnarray}
\sim 0.29 u(x_1)u(x_2) + 0.37 d(x_1) d(x_2) \quad {\rm{with}} \quad x_{1,2} = \frac{M_Z}{\sqrt{s}} e^{\pm y}
\label{eq:DY_ewcoupling}
\end{eqnarray}
complementary to the combination $ \sim 4 u \bar{u} + d \bar{d} $ probed
by $pp$ Drell-Yan production in fixed target experiments (see 
section~\ref{sec:DY_E866} and Eq.~\ref{eq:DY_pp}).
These D0 and CDF measurements 
bring interesting constraints on the $d$ distribution and, in the 
forward region, on the quark densities at high $x$.

%......................................
\paragraph{Inclusive jet cross sections}
%.....................................

The Tevatron measurement of the jet production
cross section with respect to the jet transverse momentum $p_T$, $d^2\sigma/d p_Tdy$,
in several bins of jet rapidity $y$, provide constraints on the quark and gluon
densities for $x$ larger than a few $10^{-2}$. 
For example, in the central rapidity region, the production of jets with
$p_T = 200$~GeV involves partons with $x \simeq 0.2$, and at least one of them
is a gluon in $\sim 70\%$ of interactions. Hence, these measurements provide
crucial constraints on the gluon density at high $x$.

\begin{figure}[tb]
\centerline{
  \begin{tabular}{c}
   \includegraphics[width=0.5\columnwidth]{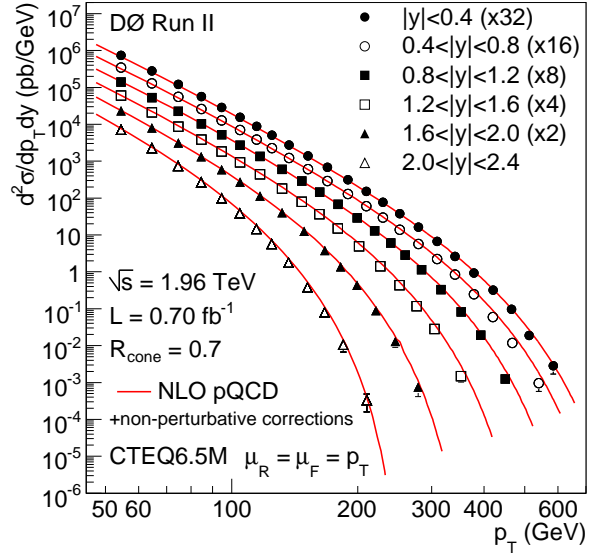}  
 \end{tabular}
}
\caption{
 \sl
 Inclusive jet cross section as a function of jet $p_T$, as measured
 by the D0 experiment.
 From~\cite{D0JETSRUNII}.
}
 \label{fig:d0_incjets}
\end{figure}

The jet measurements from Run I~\cite{Abe:1996wy,Abbott:2000ew}
preferred a rather high gluon density at high $x$,
in some tension with the other experimental measurements available at that time,
as discussed in section~\ref{sec:fits_gluon}. 
As will be seen in section~\ref{sec:qcdana}, this tension is much reduced with
the Run II measurements~\cite{CDFJETSRUNII, D0JETSRUNII,Aaltonen:2008eq,Abazov:2011vi}.
An example of these measurements, from the D0 collaboration, is shown in Fig.~\ref{fig:d0_incjets}.
These measurements are presented in six bins of jet rapidity extending out to $\mid y \mid = 2.4$.
The cross section extends over more than eight orders of magnitude from $p_T = 50$~GeV to
$p_T > 600$~GeV.
Compared to previous Run I results, the systematic uncertainties have been reduced by up to a factor of two,
to typically $\sim 10 - 15\%$.
This has been made possible by extensive studies of the jet response, which lead to
a relative uncertainty of the jet $p_T$ calibration of about $1 \%$ for jets measured
in the central calorimeter, for $p_T$ in $150-500$~GeV.

%======================================================
\subsubsection{Prompt photon production}
%======================================================
\label{exp:promptphotons}

In hadronic interactions, the production of prompt photons,
i.e. photons that do not arise from the decay of a hadron produced
in the interaction, is sensitive to the gluon density via the
QCD Compton process $q g \rightarrow \gamma q$.
However, measurements of inclusive prompt photon production 
performed at low energy ($\sqrt{s} = 20 -40$~GeV) by the fixed target E706
experiment~\cite{Alverson:1993da,Apanasevich:1997hm,Apanasevich:2004dr} could not easily
be included in QCD analyses of proton PDFs, as they were systematically
higher than theoretical predictions~\cite{Baer:1990ra,Aurenche:1992yc}. Consequently, once precise jet measurements
from the Tevatron experiments became available, they were used instead of
the prompt photon data to constrain the gluon density at medium and high $x$,
and the usage of prompt photon measurements in QCD fits was abandoned.

Since then, the compatibility of prompt photon measurements with pQCD predictions
has been discussed at length, and the current status is reviewed in~\cite{DaviddEnterria}.
With respect to older measurements,
measurements performed
in hadronic collisions, at higher $\sqrt{s}$, are less affected by non-perturbative
effects such as intrinsic $k_T$ broadening~\cite{Apanasevich:1997hm,Apanasevich:1998ki}. 
Moreover, the requirement that the photon
be isolated reduces the contribution of photons coming from
fragmentation processes. These fragmentation photons are less well
understood and are subject to large uncertainties.
The analysis performed in~\cite{DaviddEnterria} considered the measurements
of isolated prompt photon production carried out by\footnote{Recent measurements made by
the ATLAS and CMS experiments at the LHC were also studied in~\cite{DaviddEnterria} and
are described in section~\ref{sec:LHCphotons}.}:
\begin{itemize}
 \item the Tevatron experiments, at $\sqrt{s} = 0.63 - 1.96$~TeV; photons with a
  transverse energy between $7$ and $400$~GeV were measured, corresponding to a
  range in $x$ of $10^{-3}$ to $0.4$;
 \item the UA1 experiment at the CERN Sp$\bar{{\rm{p}}}$S, at $\sqrt{s} = 0.55 - 0.63$~TeV;
  photons of $12$ to $100$ GeV were measured, covering a range of $0.01$ to $0.5$ in $x$;
 \item the PHENIX experiment at the RHIC collider at the Brookhaven National Laboratory.
  RHIC is the only collider than can collide polarised protons, as well as several species
  of heavy ions such as gold or uranium. In~\cite{RHIC}, unpolarised $pp$ cross sections were
  reported at $\sqrt{s} = 200$~GeV, by averaging over the spin states of the proton 
  beams, for the production
  of photons of $3$ to $16$~GeV, corresponding to a range in $x$ of $0.03$ to $0.2$.
\end{itemize}
Generally a good agreement with NLO pQCD predictions has been found.
However, these data are not included in the QCD fits described in section~\ref{sec:GlobalFits}.

%\clearpage
%%%%%%%%%%%%%%%%%%%%%%%
\subsection{Summary}
%%%%%%%%%%%%%%%%%%%%%%%

\label{sec:exp_summary}

\begin{figure}[htb]
 \centerline{\includegraphics[width=0.5\columnwidth]{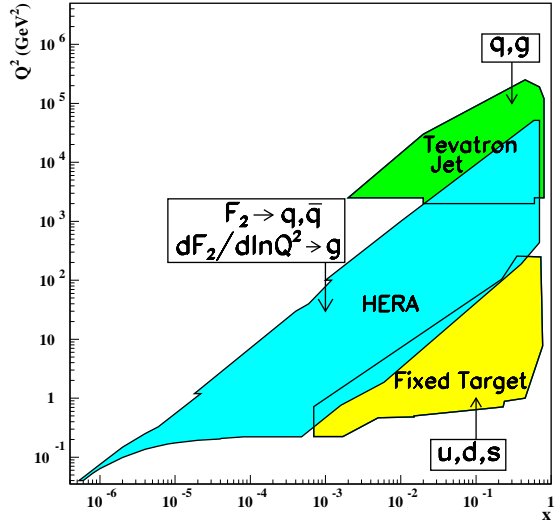}}
 \caption{\sl The kinematic plane in $(x, Q^2)$ accessed by the DIS
 and hadron collider experiments. From~\cite{pdg}.}
 \label{fig:KinPlane}
 \end{figure}

The complete range of measurements used in proton structure
determinations now span six orders of magnitude in both $x$ and $Q^2$
and is shown in Fig.~\ref{fig:KinPlane}. The region of high $x$ and
low $Q^2$ is covered by the fixed target data with charged lepton and
neutrino DIS experiments as well as proton beams on nuclear targets.
The large $\sqrt{s}$ of the HERA collider provides access to a wide
kinematic range.
Finally the range of the Tevatron $p\bar{p}$ collider data
operating at $\sqrt{s}=1.96$~TeV is shown providing access to $Q^2\sim
10^5$~GeV$^2$ through the inclusive jet measurements. A summary of the
main features of the experimental measurements is provided in
Tab.~\ref{dataset}.

The PDF flavour decomposition of the proton is achieved by combining
data from different types of experiment each of which brings its own
constraints and are summarised in Tab.~\ref{constraints}. The fixed target
charged lepton DIS data provide stringent constraints on the light
quark PDFs in the valence region of $x>0.1$ as well as medium $x
\simeq 0.01$. 
The combination of $F_2$ measurements in NC DIS off a proton and a 
deuterium target provides the primary $U$-type and $D$-type flavour separation.
For example the $F_2^p / F_2^d$ ratio of DIS inclusive measurements from NMC
constrains the ratio of $u/d$ in a rather wide $x$ domain. 
Measurements of Drell-Yan di-muon
production data are sensitive to the combination $q\bar{q}$ which at
high $x$ constrains the anti-quarks.
Neutrino induced DIS structure functions from CCFR, NuTeV and CHORUS
disentangle the contribution of sea quarks from that of valence quarks,
through the measurements of the CC structure functions $W_2$ and $xW_3$
at $x>0.01$. The di-muon production measurements allow the $s$ and
$\bar{s}$ components to be ascertained via the process $\nu
N\rightarrow\mu^+\mu^- +X$ (and the charge conjugate reaction)
mediated by $W+s$ and $W+\bar{s}$ fusion.

The fixed target data generally benefit from large event rates in the
high $x$ region but can be complicated by the use of nuclear targets.
The HERA DIS data are not affected by these issues and the NC
measurements place tight constraints on the low $x$ gluon distribution
as well as the sea quark PDFs at low $x$. The CC measurements have
moderate statistical precision but provide sufficient flavour
separation to allow PDF extractions using only HERA data. 

\begin{figure}[htb]
\centerline{\includegraphics[width=0.8\columnwidth]{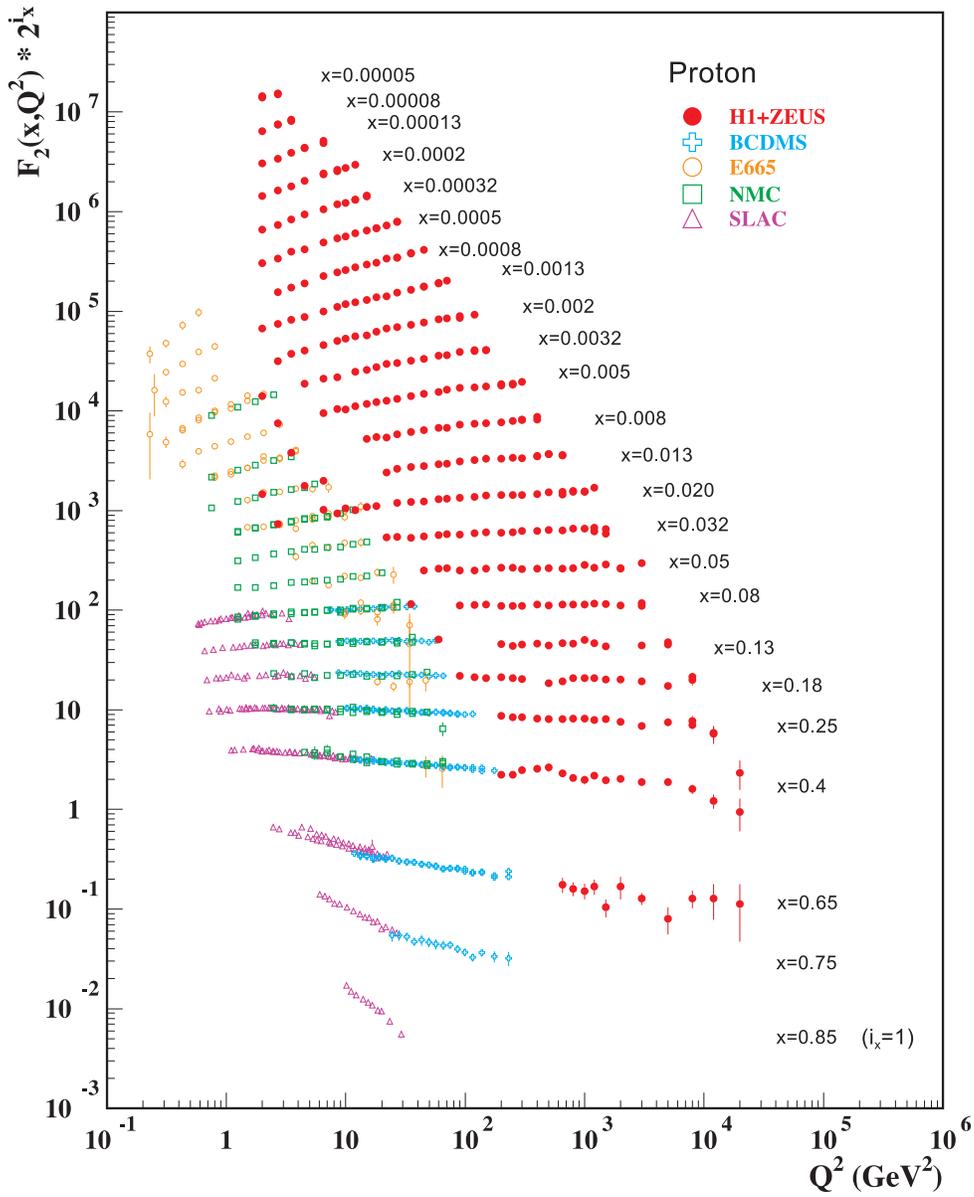}}
%%\centerline{\includegraphics[width=0.8\columnwidth]{f2p_q2x.ps}}
\caption{\sl The proton structure function $F_2(x, Q^2)$ measured in a wide
kinematic range by various DIS experiments. From~\cite{pdg}.
}
\label{fig:f2summary}
\end{figure}

\begin{figure}[htb]
\centerline{
\begin{tabular}{cc}
 \includegraphics[width=0.49\columnwidth]{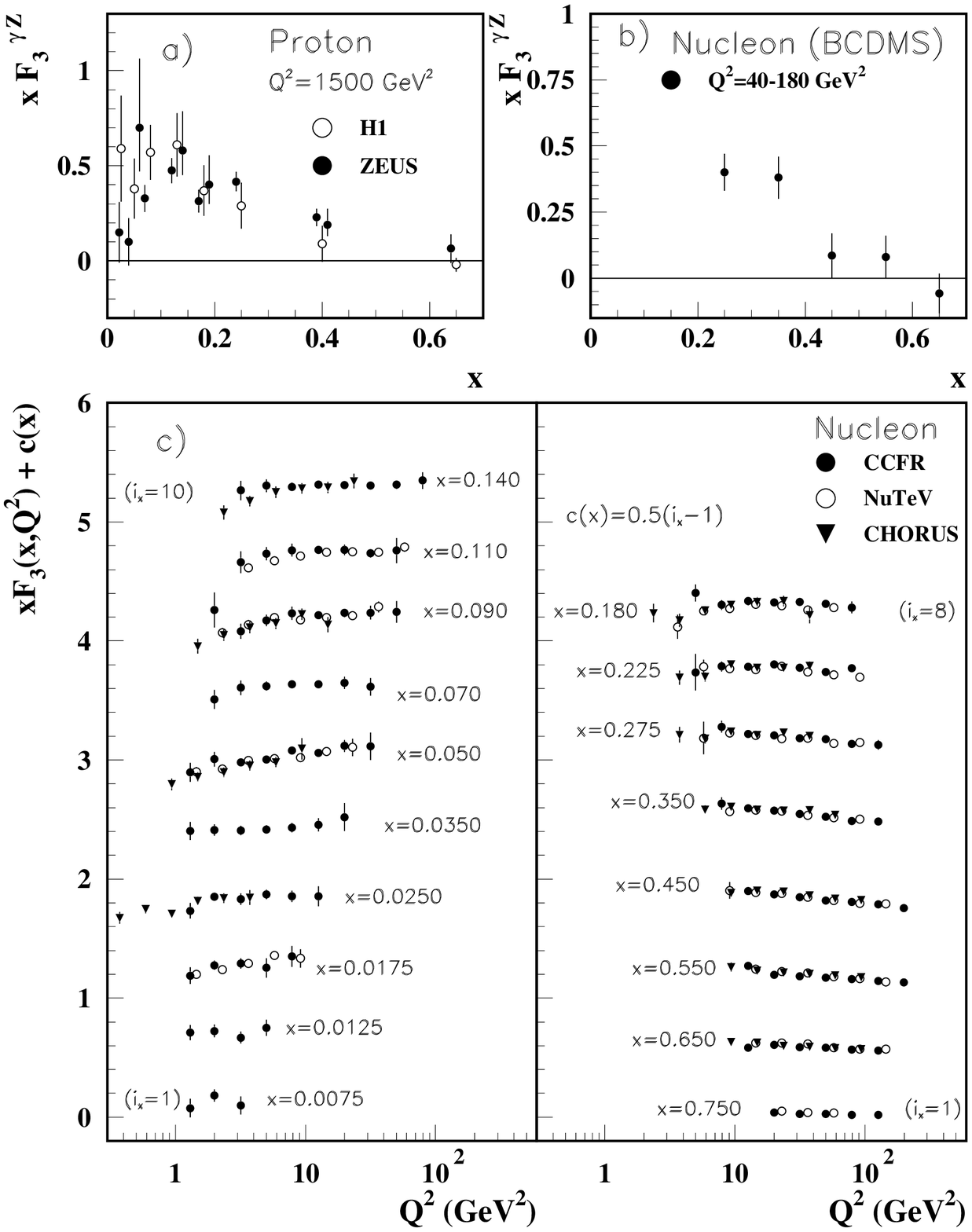}
 \includegraphics[width=0.49\columnwidth]{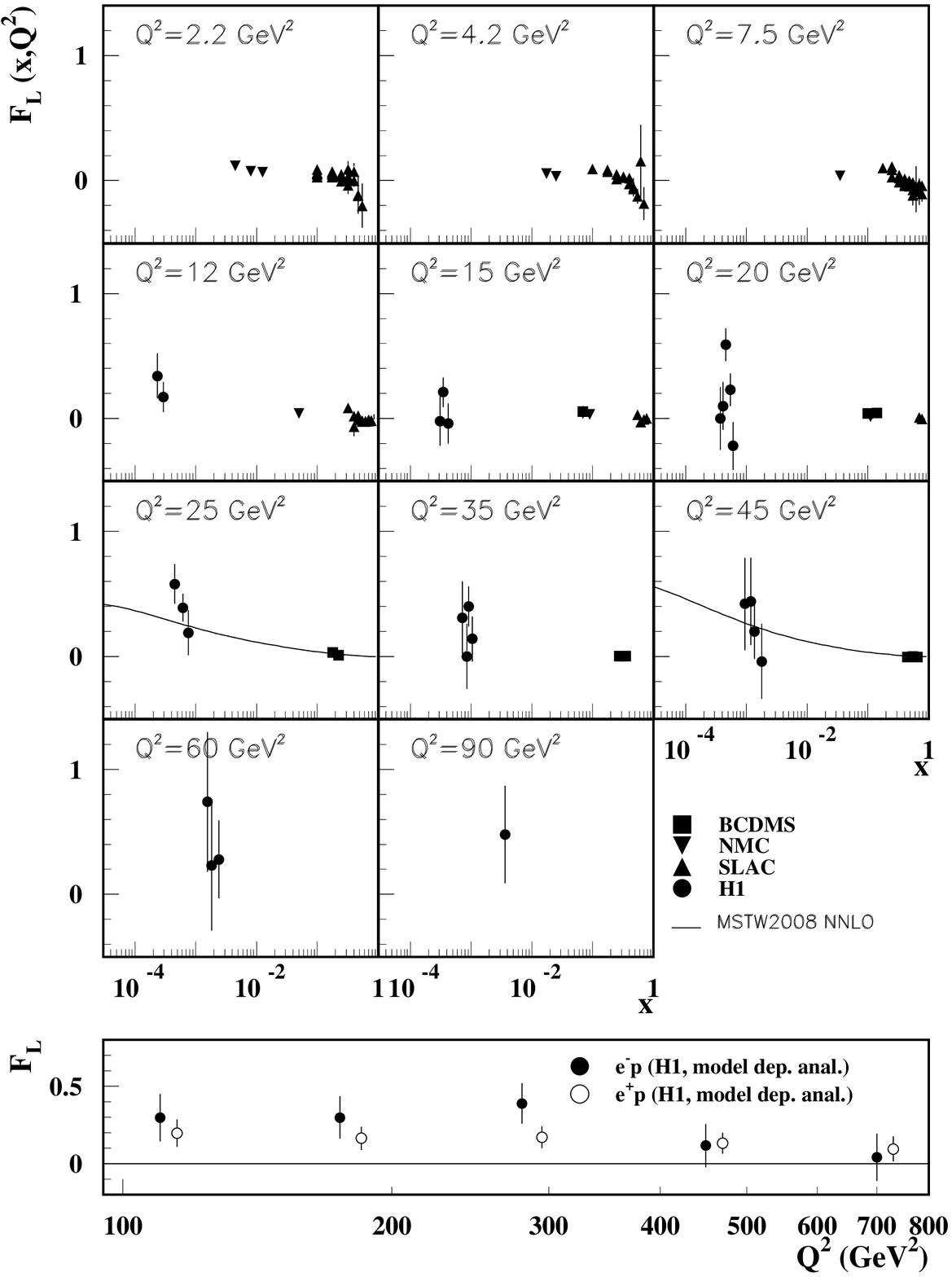}
 \end{tabular}
}       
\caption{\sl The proton structure functions $xF_3(x, Q^2)$ and $F_L(x, Q^2)$. From~\cite{pdg}.}
\label{fig:flsummary}
\end{figure}

The existing measurements of $F_2$
are summarised in Fig.~\ref{fig:f2summary};
it demonstrates the scaling violations which are most prominent at low $x$.
The charged lepton $xF_3$ data are consistent albeit within large
uncertainties as shown in Fig.~\ref{fig:flsummary}. The neutrino
induced DIS measurements also shown have much greater precision. 
Direct measurements of the $F_L$ structure
function are also shown in Fig.~\ref{fig:flsummary}. The precise fixed
target data at high $x$ are compared to the HERA measurements lying at
much lower $x$ with much larger uncertainties.
There is good overall consistency of the variety of measurements from
different experiments.
The compatibility
between the various experimental datasets is further
discussed in section~\ref{sec:fits_compatibility}.

Combinations of the experimental observables listed here allow flavour separated PDFs
of the proton to be extracted in pQCD analyses,
as discussed in chapter~\ref{sec:qcdana}.

\begin{table}[htb]
\scriptsize
\begin{center}
\begin{tabular}{|cc|c|l|l|r|r|r|c|}
\hline
\multicolumn{2}{|c|}{\multirow{2}{*}{Experiment} } & 
\multirow{2}{*}{Measurement} & \multirow{2}{*}{$x_{min}$} & \multirow{2}{*}{$x_{max}$} & $Q^2_{min}$ &$Q^2_{max}$ &
\multirow{2}{*}{ref.}            & Beam Energy   \\

                     &   &          &       &           &  (GeV$^2$)  & (GeV$^2$)  &
                    &   (GeV)     \\
 \hline
SLAC          & $ep$, $ed$    & $F_2^p$, $F_2^d$   & $0.06   $ &  $0.9$        &  $0.6$   & $30$
              & \cite{Whitlow:1991uw,Whitlow:1990dr}
              &  $4-20$ \\

SLAC          &  $ep$, $ed$    & $R^p$, $R^d$      & $0.1   $ &  $0.9$        &  $0.6$   & $20$
              & \cite{Whitlow:1990gk}
              &  $4-17$ \\

BCDMS         & $\mu p$, $\mu d$  & $F_2^p$, $F_2^d$  & $0.07   $ &  $0.8$        &  $7.5$   & $230$
              & \cite{Benvenuti:1989fm,Benvenuti:1989rh}
              &  $100$, $120$, $200$, $280$ \\

BCDMS         & $\mu p$, $\mu d$  & $R^p$, $R^d$      & $0.07   $ &  $0.7$        &  $15$    & $50$
              & \cite{Benvenuti:1989fm,Benvenuti:1989rh}
              & $100$, $120$, $200$, $280$\\

NMC           & $\mu p$, $\mu d$  & $F_2^p$, $F_2^d$ & $0.008   $ &  $0.5$        &  $0.8$   & $65$
              & \cite{Arneodo:1996qe} 
              & $90$, $120$, $200$, $280$ \\

NMC           & $\mu p$, $\mu d$  & $R^p$, $R^d$     & $0.0045  $ &  $0.11$      &  $1.4$   & $21$
              & \cite{Arneodo:1996qe}   
              &  $90$,$120$, $200$, $280$ \\

NMC           & $\mu p$, $\mu d$  & $F_2^p/F_2^d$    & $0.002   $ &  $0.7$        &  $0.2$   & $100$
              & \cite{Arneodo:1996kd}    
              &  $90$,$120$, $200$, $280$ \\

E665          & $\mu p$, $\mu d$  & $F_2^p$, $F_2^d$ & $0.0009  $ &  $0.4$        &  $0.2$   & $64$
              & \cite{Adams:1996gu}      &
              $470$ \\
\hline
\multicolumn{8}{|c|}{\hspace{2.5cm} HERA} 
& $\sqrt{s}$~(GeV)\\
\hline
H1/ZEUS       & $ep$  & $\tilde{\sigma}_{NC},\tilde{\sigma}_{CC}$ & $10^{-5}$ &  $0.65$  & $0.1$   & $30\,000$
              & \cite{:2009wt}      &
              $301,319$ \\
H1/ZEUS       & $ep$  & $F_L$ & $0.00003$ &  $0.01$  & $1.5$   & $120$
              & \cite{H1FL,Chekanov:2009na}      &
              $225,250$ \\
\hline
\multicolumn{8}{|c|}{\hspace{2.5cm}Neutrino Experiments} 
& Beam Energy~(GeV)\\
\hline
CCFR          & $\nu_{\mu}Fe/\bar{\nu}_{\mu}Fe$   & $F_2$, $xF^{\nu}_3$& $0.015  $ &  $0.65$        &  $1.2$   & $126$
              & \cite{Oltman:1992pq}      &
              $30-500$ \\
NuTeV          & $\nu_{\mu}Fe/\bar{\nu}_{\mu}Fe$  & $F_2$,$xF^{\nu}_3-xF^{\bar{\nu}}_3$ & $0.015$ &  $0.75$  & $1.2$   & $125$
              & \cite{Tzanov:2005kr,Yang:2000ju,unki}      &
              $30-360$ \\
CCFR\&NuTeV   & $\nu_{\mu}Fe/\bar{\nu}_{\mu}Fe$  & $\nu N/\bar{\nu}N\rightarrow\mu^+\mu^-X$ & $0.02$ &  $0.36$  & $1.2$   & $117$
              & \cite{Goncharov:2001qe}      &
              $30-500$ \\

CHORUS        & $\nu_{\mu}Pb/\bar{\nu}_{\mu}Pb$   & $F_2$,$xF_3$ & $0.02$ &  $0.65$  & $0.3$   & $82$
              & \cite{Onengut:2005kv}      &
              $30-360$ \\
\hline
\multicolumn{8}{|c|}{\hspace{2.5cm}Hadron Beam Experiments } 
& $\sqrt{s}$~(GeV)\\
\hline
E605          & $pCu$        & $\sigma_{pN}$ & $0.14$ &  $0.65$        &  $51$   & $286$
              & \cite{Moreno:1990sf}      & $800$ \\
E772          & $pd$         & $\sigma_{pN}$ & $0.026 $ &  $0.56$      &  $23$   & $218$
              & \cite{McGaughey:1994dx}      & $800$ \\
E866/NuSea    & $pp/pd$      & $\sigma_{pp}/\sigma_{pd}$ & $0.026 $ &  $0.56$        &  $21$   & $166$
              & \cite{Towell:2001nh}      & $800$ \\
CDF/D0        & $p\bar{p}$  & $\sigma_{W,Z}$ , $Y_Z$, $A(Y_W)$& $\sim0.003$ &  $\sim0.8$  & $M_W^2$   & $M_Z^2$
              & \cite{Abazov:2007jy,Aaltonen:2010zz,Aaltonen:2009ta}      &
              $1800,1960$ \\
CDF/D0        & $p\bar{p}$  & $p\bar{p}\rightarrow$ incl. jets & $\sim0.003$ &  $\sim0.8$  & $0.1$   & $\sim 10^5$
              & \cite{CDFJETSRUNII,D0JETSRUNII}      &
              $1800,1960$ \\

\hline
\end{tabular}
\end{center}
\caption[RESULT] {\sl Table of datasets generally used in
  current QCD fits. The kinematic range of each measurement in $x$ and
  $Q^2$ and the incident beam energy are also given. The normalisation
  uncertainties of the charged lepton scattering experiments are
  typically $2-3\%$.}
\label{dataset} 
\end{table}

\begin{table}
  \begin{center}
    \begin{tabular}{llll}
      \hline
      \hline
      Process & Subprocess & Partons & $x$ range \\ \hline
      $\ell^\pm\,\{p,n\}\to\ell^\pm\,X$ & $\gamma^*q\to q$ & $q,\bar{q},g$ & $x\gtrsim 0.01$ \\
      $\ell^\pm\,n/p\to\ell^\pm\,X$ & $\gamma^*\,d/u\to d/u$ & $d/u$ & $x\gtrsim 0.01$ \\
      $pp\to \mu^+\mu^-\,X$ & $u\bar{u},d\bar{d}\to\gamma^*$ & $\bar{q}$ & $0.015\lesssim x\lesssim 0.35$ \\
      $pn/pp\to \mu^+\mu^-\,X$ & $(u\bar{d})/(u\bar{u})\to \gamma^*$ & $\bar{d}/\bar{u}$ & $0.015\lesssim x\lesssim 0.35$ \\
      $\nu (\bar{\nu})\,N \to \mu^-(\mu^+)\,X$ & $W^*q\to q^\prime$ & $q,\bar{q}$ & $0.01 \lesssim x \lesssim 0.5$ \\
      $\nu\,N \to \mu^-\mu^+\,X$ & $W^*s\to c$ & $s$ & $0.01\lesssim x\lesssim 0.2$ \\
      $\bar{\nu}\,N \to \mu^+\mu^-\,X$ & $W^*\bar{s}\to\bar{c}$ & $\bar{s}$ & $0.01\lesssim x\lesssim 0.2$ \\\hline
      $e^\pm\,p \to e^\pm\,X$ & $\gamma^*q\to q$ & $g,q,\bar{q}$ & $0.0001\lesssim x\lesssim 0.1$ \\
      $e^+\,p \to \bar{\nu}\,X$ & $W^+\,\{d,s\}\to \{u,c\}$ & $d,s$ & $x\gtrsim 0.01$ \\
      $e^\pm p\to e^\pm\,c\bar{c}\,X$ & $\gamma^*c\to c$, $\gamma^* g\to c\bar{c}$ & $c$, $g$ & $0.0001\lesssim x\lesssim 0.01$ \\
      $e^\pm p\to\text{jet}+X$ & $\gamma^*g\to q\bar{q}$ & $g$ & $0.01\lesssim x\lesssim 0.1$ \\ \hline
      $p\bar{p}\to \text{jet}+X$ & $gg,qg,qq\to 2j$ & $g,q$ & $0.01\lesssim x\lesssim 0.5$ \\
      $p\bar{p}\to (W^\pm\to\ell^{\pm}\nu)\,X$ & $ud\to W,\bar{u}\bar{d}\to W$ & $u,d,\bar{u},\bar{d}$ & $x\gtrsim 0.05$ \\
      $p\bar{p}\to (Z\to\ell^+\ell^-)\,X$ & $uu,dd\to Z$ & $d$ & $x\gtrsim 0.05$
      \\
      \hline
      \hline    \end{tabular}  \end{center}
  \caption{\sl The main processes included in current global PDF analyses
    ordered in three groups: fixed target experiments, HERA and the
    Tevatron.  For each process an indication of their dominant
    partonic subprocesses, the primary partons which are probed and
    the approximate range of $x$ constrained by the data is
    given. From~\cite{MSTW08}.}
  \label{constraints} 
\end{table}

%\clearpage

%Chapter 3
\clearpage
\section{Determination of parton distribution functions}
\label{sec:qcdana}
%

%------------------------------------------------
\subsection{Introduction and generalities}
%------------------------------------------------

Parton distribution functions are determined from fits of perturbative QCD calculations,
based on the DGLAP evolution equations, to various sets of experimental data.
These fits are regularly updated to account for new experimental input and
theoretical developments.
Most fits are performed at NLO, although leading
order fits are still of interest, for example for Monte-Carlo simulations.
With the recent, full calculation of the DIS cross section at NNLO~\cite{nnlo},
first NNLO fits are becoming available (see section~\ref{sec:NNLO}).

In the following, we will discuss in particular the latest NLO fits performed by
the CTEQ/CT group (CTEQ6.6~\cite{CTEQ6.6} and CT10~\cite{CT10}), the MSTW group (MSTW08~\cite{MSTW08})
and the NNPDF collaboration (NNPDF2.0~\cite{NNPDF2.0} and NNPDF2.1~\cite{NNPDF2.1}),
which try to include
all relevant experimental data. Other groups, e.g. GJR~\cite{GJR} or ABKM/ABM~\cite{ABM11,ABKM, ABM},
also provide fits using a subset of data. The H1 and ZEUS
collaborations have also published QCD fits
based on their inclusive DIS data only (see e.g. HERAPDF1.0~\cite{:2009wt}). 
The main differences
between these various fits are described in the next sections.
These fits are publicly available via the LHAPDF interface~\cite{LHAPDF}, which also provides access
to older proton PDF fits as well as to photon and pion PDFs.

The general ansatz used in QCD fits is the parameterisation of parton distributions at a
so-called starting scale $Q^2_0$, using a flexible analytic form.
For example, one can choose
to parameterise  the gluon density $xg(x)$, the valence quark densities $xu_{v}(x) = x(u(x) - \bar{u}(x))$ 
and $xd_{v}(x) = x(d(x) - \bar{d}(x))$, the light sea
distribution 
defined as $xS(x) = x \left[ 2 ( \bar{u}(x) + \bar{d}(x)) + s(x) + \bar{s}(x) \right]$,
and $x \Delta(x) \equiv x (\bar{d}(x) - \bar{u}(x))$.
Most QCD analyses make use of a simple functional form, like:
\begin{equation}
 % x f_i(x, Q^2_0) = A_i x^{\alpha_i} (1-x)^{\beta_f} ( 1 + p_{0,f} x + p_{1,f} \sqrt{x} + ..) \qquad , 
 x f_i(x, Q^2_0) = A_i x^{\alpha_i} (1-x)^{\beta_i} P_i(x)
\label{eqn:pdf_functional}
\end{equation}
where $P_i(x)$ can be e.g. a polynomial function in $x$ or $\sqrt x$.
The parameterisation can also be based on interpolation 
polynomials or non-linear functions. The latter
approach is exploited by the NNPDF collaboration and is described in
section~\ref{sec:ParamChoice} in more detail.

The DGLAP evolution equations are used to obtain, from these parameterised densities at $Q^2_0$,
the parton densities $x f(x, Q^2)$ at any $Q^2$.
This allows the theoretical cross sections of the processes of interest (DIS, Drell-Yan di-lepton production, jet
production, ...) to be computed.
The parameters that define the PDFs at the starting scale (e.g. $\alpha_i$, $\beta_i$, .. in Eq.~\ref{eqn:pdf_functional}) 
can then be obtained
by fitting these theoretical predictions to the experimental measurements.

This is achieved by minimising a $\chi^2$ function. A usual choice for this function is:
\begin{equation}
         \chi^2 = \sum_{exp} \chi^2_{exp}
 \label{eq:chi2_a}
\end{equation}
where the individual contribution of each independent dataset is given by~\cite{Pascaud:1995qs, PDFUncertainties}:
\begin{equation}
  \chi^2_{exp} = \sum_i \frac{ (d_i - \sum_k \beta_{k,i} s_k - t_i)^2 }{ \alpha_i^2 } + \sum_k s^2_k  \qquad .
 \label{eq:chi2_b}
\end{equation}
In this equation,
\begin{itemize}
 \item $d_i$ denote the measurements and $t_i$ the corresponding theoretical predictions;
 \item the total uncorrelated uncertainty affecting the measurement $i$ is given by 
       % $\alpha_i^2 = \sigma^2_{i, stat} + \sigma^2_{i, uncorr.}$;
       $\alpha_i$ which sums in quadrature the statistical and the
       uncorrelated systematic uncertainties;
 \item $k$ labels the sources of correlated systematic uncertainties;
 \item  $\beta_{k,i}$ is the amount of
        change of $d_i$ when the source $k$ (for example an energy scale) is shifted by
        one standard deviation ($s_k = 1$); the values of $\beta_{k,i}$ are taken from the
        correlated systematic error tables published by the experiments;
 \item the second term in Eq.~\ref{eq:chi2_b} introduces a quadratic penalty $s_k^2$ when
       the data points are moved coherently by $\beta_{k,i} s_k$ and restricts large deviations from $s_k = 0$.
\end{itemize}

When the parameters $s_k$ in Eq.~\ref{eq:chi2_b} are fixed to zero, the fit is performed to
the raw data points published by the experiments, but the correlated systematic errors are ignored.
Instead, the $s_k$ can be free parameters of the fit and determined by the $\chi^2$ minimisation.
Technically, they are obtained analytically~\cite{Pascaud:1995qs, PDFUncertainties}: the $\chi^2$ is quadratic in
$s_k$, hence $\partial \chi^2 / \partial s_k = 0$ leads to a simple matrix
equation for the $s_k$. This means that the fit is not performed
to the raw data, but to the data shifted by the optimal setting for
the systematic error sources as determined by the fit.
In that case, at the $\chi^2$ minimum, Eq.~\ref{eq:chi2_b} is mathematically equivalent
to the standard $\chi^2$ expression involving the correlation matrix between
the measurements,
$$\chi^2 = \sum_{i,j} (d_i - t_i) (V^{-1})_{i,j} (d_j - t_j) $$
with $V_{ij} = \alpha_i^2 ( \delta_{ij} + \sum_{k}  \beta_{ki} \beta_{kj} / \alpha_i^2)$.
However, inverting the large matrix $V_{ij}$ makes this expression inconvenient, hence
Eq.~\ref{eq:chi2_b} is preferred. Moreover, it also facilitates the determination of
the fit uncertainties, as will be discussed in section~\ref{sec:systematic_errors}.

%------------------------------------------------
\subsection{Choices and assumptions}
%------------------------------------------------

\subsubsection{Experimental input}
\label{sec:fits_expinput}

QCD fits may not include all experimental measurements described in section~\ref{sec:expt},
despite the fact that they all provide sensitivity to some parton distribution functions.

Including as much experimental input as possible provides maximal constraints. 
This ``global fit" approach is followed by the CTEQ/CT, MSTW and NNPDF collaborations, which include
typically $3000$ experimental points in their latest analyses.
However, tensions
between different datasets may require the PDF uncertainties resulting from the fit to be enlarged
(see section~\ref{sec:tolerance}).
If these tensions were due to problems with one experimental dataset (e.g. wrong calibration or 
underestimated systematic uncertainties), the usual procedure, that is equivalent to inflating the
experimental errors of all measurements included in the fit, would result in an overestimation of
the PDF uncertainties. On the other hand, apparent inconsistencies between datasets may also arise
because the fit does not have enough flexibility, or because some underlying assumptions (see
section~\ref{sec:ParamChoice}) are not correct.
In that case, the enlarged PDF uncertainties would not necessarily
be over-conservative.

The fits performed by other groups include only a subset of all available data. 
For example, the latest fits performed by GJR and ABKM include most
measurements from deep inelastic
scattering experiments ($eN$, $\mu N$ and $\nu N$), Drell-Yan
measurements from fixed target 
experiments and jet production data from the Tevatron experiments, but Tevatron data on $W$ and $Z$ production
are not included.
The HERAPDF fits use only the measurements from the H1 and ZEUS experiments\footnote{Only the inclusive NC 
and CC measurements were included in HERAPDF1.0. More recent (preliminary) fits, HERAPDF1.5 
and HERAPDF1.7~\cite{MandyReview}, also include
exclusive HERA measurements, such as jet or charm production.}, which are known to be fully consistent with each other and
for which the systematic uncertainties are very well understood.
Despite the much reduced experimental input, 
with about $600$ points included in HERAPDF1.0,
good constraints can be obtained on most parton densities,
over a very large kinematic range, as will be shown in section~\ref{sec:HERAPDF1.0}.
Note however that additional assumptions are made to compensate for the lack of sensitivity of $ep$ measurements 
alone on the flavour separation.

Besides the choice of datasets, some data points within a given dataset can be excluded deliberately.
That is the case in particular for the DIS measurements at very low $x$, e.g. $x < 10^{-5}$, where DGLAP evolution
may break down.
The same holds for data points at very low $Q^2$ where $\alpha_s (Q^2)$ would become
too large to ensure a good convergence of the perturbative series. A typical cut
$Q^2_{min} \sim 2$~GeV$^2$ is usually chosen. Data points at very high $x$ and low $W^2$
are often removed as well, as higher twists corrections
proportional to $1 / Q^2$ are enhanced in this domain (see section~\ref{sec:highertwists}). 
A requirement that $W^2$ be above $\sim 10 - 15$~GeV$^2$ is usual.
An alternative approach was followed in~\cite{ABM11}, where a cubic spline function was chosen
to parameterise the function $H_i(x)$ that describes the higher twist  corrections to $F_2$ 
(see Eq.~\ref{eq:highertwists}),
and the parameters of this function were fitted together with the PDF parameters.
They found that the higher twist corrections improve considerably the description of the
data of the SLAC experiments, even in the region $W^2 > 12$~GeV$^2$.

\subsubsection{Theoretical choices}

\label{sec:fits_theochoices}

The most important theoretical ingredients that can lead to different fit results are the following.

\begin{itemize}
 \item For fits performed beyond LO, one needs to choose the renormalisation
       scheme, usually taken to be the $\overline{MS}$ scheme. One also needs to choose
       the factorisation and renormalisation scales, $\mu_F$ and $\mu_R$, which
       are used in the theoretical calculation. For DIS, a common choice is
       to take $\mu_F^2 = \mu_R^2 = Q^2$.
 \item There is no unique way to treat the heavy flavours (HF).
       In the zero-mass variable flavour number scheme (ZM-VFNS),
       the charm density, for example, is
       set to zero below $Q^2 \sim m^2_c$. Above this threshold, the charm is generated
       by gluon splitting and is treated as massless. The drawback of this approach is
       that it ignores charm mass effects in the threshold region.
       In contrast, in the fixed flavour number scheme (FFNS), there is no PDF for the
       charm and bottom, i.e. there are only three active flavours. For $W^2$ above
       the production threshold, the DIS production of charm proceeds via photon-gluon
       fusion, $\gamma g \rightarrow c \bar{c}$. The drawback of this treatment is that
       the calculations involve terms in $\ln (Q^2 / m^2_c)$ which become large
       at high $Q^2$ and would need to be resummed.
       The state-of-the-art approach, called general-mass variable flavour
       number scheme (GM-VFNS), can be seen as an interpolation between the ZM-VFNS and the FFNS
       (this is put on rigorous grounds in~\cite{Collins:1998rz}).
       Such a scheme is not easy to implement, especially at NNLO, and the various fit groups
       choose different prescriptions. The so-called ACOT scheme~\cite{Kramer:2000hn,Aivazis:1993pi} is 
       used by the CTEQ/CT group since CTEQ6.5~\cite{CTEQ65},
       NNPDF2.1 uses the FONLL scheme~\cite{Forte:2010ta, Cacciari:1998it} (previous versions of the NNPDF fit were performed
       in the FFNS scheme), and the Thorne-Roberts prescription~\cite{hep-ph/0601245, hep-ph/9709442} is used in MSTW08.
       These are reviewed in~\cite{Thorne:2008xf}.

 \begin{figure}[htb]
 \centerline{\includegraphics[width=0.4\columnwidth]{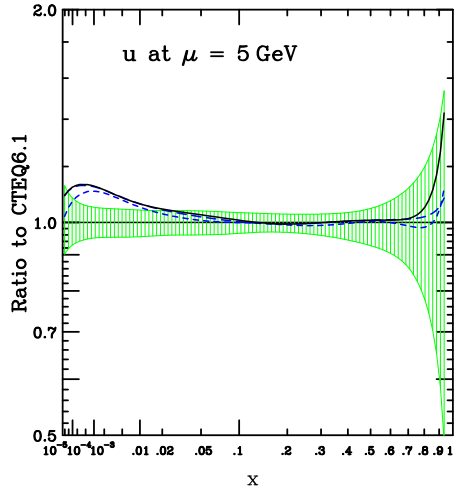}}
 \caption{\sl Ratio of the CTEQ6.5 $u$ density to that of CTEQ6.1, at a
 scale $Q = 5$~GeV. The contour shows the uncertainty at $90 \%$ confidence level
of the CTEQ6.1 density. From~\cite{CTEQ65}. }\label{Fig:cteq65_u}
 \end{figure}

	A different treatment of heavy flavours in QCD fits can lead to sizable differences.
An important update of the CTEQ6.5 fit compared to the previous release, CTEQ6.1, came from the treatment of heavy
quarks in a general mass variable flavour number scheme.
Using this scheme instead of the ZM-VFNS scheme lead to a considerable improvement
of the fit, the $\chi^2$ of the fit being reduced by $\sim 200$ units for $\sim 2700$ data points.
The resulting CTEQ6.5 PDFs mainly differ from the previous CTEQ6.1 fit by
larger $u$ and $d$ distributions in the region $x \sim 10^{-3}$, for a wide
range in $Q^2$, as illustrated in Fig.~\ref{Fig:cteq65_u}.
This resulted in a $\sim 8\%$ increase of the predicted $W$ and $Z$ cross sections
at the LHC, compared to previous CTEQ estimates, and brought the CTEQ-based prediction closer to that obtained
using the MSTW08 parton distributions. 
The uncertainties associated to the remaining freedom in defining a GM-VFNS at NLO or NNLO have been
studied in~\cite{Thorne:2012qh}. At NLO, they result in a $2 - 3 \%$ uncertainty on the $Z$ production
cross section at the LHC.

 \item The values of the heavy quark masses $m_c$ and $m_b$ differ
       between analyses, and in some cases are treated as free
       parameters of the fit. The values chosen also depend on the
       renormalisation scheme used to define them and to calculate
       heavy quark related observables. The on-shell scheme uses the
       pole mass, defined to coincide with the pole of the heavy quark
       propagator at each order of pQCD. This definition is chosen in
       most QCD analyses.  Instead, the $\overline{MS}$ scheme, used
       in the recent ABM11 analysis~\cite{ABM11}, introduces a running
       mass $m_q (Q^2)$. It was shown in~\cite{Alekhin:2010sv} that
       the perturbative stability of predictions for the heavy quark
       structure functions is better in terms of the $\overline{MS}$
       mass, thus leading to reduced theoretical uncertainties due to
       variations of the renormalisation and factorisation scales.  In
       most QCD analyses, the masses $m_c$ and $m_b$ are fixed in the
       fit, and fits obtained when the masses are varied from the
       chosen central values may be provided together with the central
       fit.  In the ABM11 analysis, $m_b(Q = m_b)$ and $m_c(Q = m_c)$
       are fitted together with the PDF parameters, with external
       constraints given by their world average values. The HERA
       experiments have also performed preliminary fits where the PDFs
       are fitted together with the pole mass of the charm
       quark~\cite{MandyReview}.
       
 \item The value of the strong coupling
       constant $\alpha_s (M_Z)$ is an important consideration in any
       QCD analysis, unless it is treated as a free parameter in the fit.
       Many experimental observables can be used to measure $\alpha_s (M_Z)$. 
       The world averages of $\alpha_s (M_Z)$ do not include the measurements of scaling violations 
       of the structure function $F_2$, as $\partial F_2 / \partial \ln (Q^2)$ is sensitive to the product of $\alpha_s$ times
       the gluon density. However, $\alpha_s (M_Z)$ can be determined together with the parton
       distribution functions in a combined QCD analysis. In order to disentangle the gluon density from
       the strong coupling constant, these QCD fits should include jet data in addition to structure
       function measurements~\cite{Wobisch:2000dk, Adloff:2000tq}. 
       In the central fit of the MSTW08 and ABM11 analyses, $\alpha_s(M_Z)$ is fitted together with the other parameters that define
       the PDFs at the starting scale.
       In contrast, other groups fix $\alpha_s(M_Z)$ and
       provide several sets of fits, corresponding to a range of fixed $\alpha_s$ values\footnote{To facilitate comparisons
       with results from other groups, MSTW08 and ABM11 also provide fits for a range of (fixed) $\alpha_s(M_Z)$ values.}.

 \item Different numerical methods can be used to calculate the theoretical cross sections that are needed in the fit. Indeed, while
       the NLO calculation of inclusive DIS cross sections can be done relatively fast, the exact calculation of jet
       cross sections, for example, using standard techniques requires huge CPU resources and, in practice, numerical approximations
       have to be used when such processes are included in a NLO QCD fit (where the calculation of cross sections has to be
       done for every iteration of the fit).
       For example, the FastNLO technique~\cite{FASTNLO} allows rapid calculations for a large number of jet cross sections, with a
       very high accuracy. The method rewrites the cross sections as a sum of products where the time-consuming
       step is factorised out, such that it needs to be done only once. 
       This approach is used by the MSTW and NNPDF groups to calculate the jet cross sections. 
       A similar technique is used in APPLgrid~\cite{APPLGRID}
       which covers a broader range of processes and in the FastKernel approach~\cite{NNPDF2.0} developed by the NNPDF collaboration.
       Alternatively, or for processes for which no fast NLO calculation is available, a ``K-factor" approximation
       can be used. For each bin of the experimental measurement, the factor $K = \sigma_{NLO} / \sigma_{LO}$ is
       first calculated for a given PDF. In the calculations performed in each iteration of the fit, only the leading
       order cross section  is calculated, and it is multiplied by this factor $K$ to account for the higher order corrections.
       Usually the procedure is repeated in which the $K$-factor is re-evaluated using the PDFs from the
       converged fit and another series of fit iterations is performed.
\end{itemize}

Moreover, the fits usually require basic consistency constraints to be satisfied. For example, a set of parameters
that would lead to negative cross sections, negative values of $F_2$ or of $F_L$, should not be
considered as a valid fit. On the other hand, a valid fit may result, for example, in a negative
gluon distribution at low $x$, since a parton distribution function is not a physical observable.

\subsubsection{Parameterisation choices and assumptions}
\label{sec:ParamChoice}

In any QCD analysis, a priori choices have to be made to define the starting conditions.
These are: the starting scale of the fit, the set of densities that are parameterised,
and the functional form chosen for the parameterisation. 
Some assumptions usually complement these choices.
The systematic uncertainty that should be associated with these choices is not easy to
assess and is usually not estimated explicitly.

Large values of the starting scale $Q^2_0$ are unpractical. Indeed,
       since DGLAP evolution has the effect of washing out the
       $x$-dependence of PDFs with increasing $Q^2$, parameterising the densities at a low scale offers
       more freedom in the fit.
       However, the scale $Q^2_0$ cannot be too low, since the DGLAP evolution should be valid down to that scale.
       Typical values for $Q^2_0$ range between $1$~GeV$^2$ and a few GeV$^2$.
       For fits that treat the heavy flavours in a GM-VFNS scheme, it is convenient to set the scale
       $Q^2_0$ below the charm threshold, $Q^2_0 < m^2_c$, such that\footnote{
       A non-perturbative ``intrinsic charm" component of the proton~\cite{Brodsky:1980pb}  would result in
       $c(x,Q^2_0) \neq 0$ even below the charm threshold. That possibility was explored for
       example in~\cite{Pumplin:2007wg}.
       } $c(x,Q^2_0) = 0$.
       The GJR group uses a lower starting scale, for example $Q^2_0 = 0.5$~GeV$^2$ in~\cite{GJR}. They assume
       that all PDFs have a valence form (i.e. the parameter $\alpha$ in Eq.~\ref{eqn:pdf_functional}
       is positive) at a low scale; the gluon and sea quarks tend to zero at low $x$, and the
       large values for these PDFs at $Q^2$ above a few GeV$^2$ are entirely generated by the
       DGLAP evolution\footnote{In the original idea~\cite{GRV1, GRV2}, the proton was only made of valence
       quarks at a very low scale, the gluons and sea quarks being dynamically generated by the
       DGLAP evolution as $Q^2$ increases. The experimental data did not support this assumption,
       which was then revised in order to also include the gluon and sea quarks at the starting scale,
       still keeping the assumption of a valence-like shape.}.

The set of parton densities that are parameterised at the starting scale of the fit, and that are the
input to the evolution equations, has to be chosen depending on the experimental measurements
that are included in the fit.
       One does not fit the eleven quark and gluon distributions since
       the data do not contain enough information to disentangle them all. Instead, well-defined combinations
       of PDFs are fitted. In addition to the gluon density, at least two quark
       distributions are needed
       (one singlet distribution that evolves as Eq.~\ref{eq:dglap_s}, one non-singlet that follows Eq.~\ref{eq:dglap_ns}).
       A QCD fit to H1 NC DIS cross sections only, using $xg(x)$ and two quark distributions as input
       to the DGLAP equations, was performed in~\cite{H1Rainer} and allowed the gluon distribution
       to be extracted. Additional measurements are necessary in order to also extract the quark densities, 
       and at least four quark distributions need to be parameterised.
       For example, the HERAPDF1.0 fit~\cite{:2009wt}, described in more detail in~\ref{sec:HERAPDF},
       parameterises $g(x)$, $\bar{U}(x)$, $\bar{D}(x)$ and the valence distributions $u_v(x)$ and $d_v(x)$, 
       at a starting scale just below the charm threshold.
       Since the included data have no sensitivity to constrain the strange density, $ s(x)$ is assumed to be proportional 
       to $\bar{d} (x)$ at $Q^2_0$ and $s(x) = \bar{s}(x)$ is assumed.
       The MSTW08 analysis also parameterises the gluon and valence densities, $g$, $u_v$ and $d_v$, together with the
       light sea density $S$
       and the $\bar{d} - \bar{u}$ asymmetry.
       Moreover, since it includes strange-sensitive measurements,
       the total strangeness density $s + \bar{s}$ and the strange asymmetry $\bar{s} - s$ are also fitted.

The functional form that is used to parameterise these input densities at the starting scale
differs depending on the analyses.
It should be flexible enough to allow for a good
fit. However, too much freedom in the parameterisation should be avoided, as this leads to
unstable fits and secondary minima.
A functional form like Eq.~\ref{eqn:pdf_functional}
is often chosen. The low $x$ behaviour is motivated by Regge phenomenology, which
suggests (see for example~\cite{CooperSarkar:1997jk})
that $x g(x)$ and $x \bar{q} (x)$ behave at low $x$ as $ (1/x) ^ { \alpha_P(0) - 1}$ with
$\alpha_P (0) \sim 1.08$, while the valence distributions depend instead on the ``Reggeon intercept"
$\alpha_R(0) \sim 0.5$, i.e. $x q_v(x) \sim (1/x) ^ { \alpha_R (0) - 1} \sim x^{0.5}$.
The high $x$ behaviour can be motivated by simple dimensional arguments~\cite{Brodsky1973}  based on
the energy dependence of the scattering cross section\footnote{These arguments predict
that, at high $x$,
$x q(x) \sim (1 - x)^{\beta}$ with $\beta = 2 n_s -1$, where $n_s$ is the number of ``spectator
quarks" that are attendant to the parton in the Fock expansion of the proton wave-function,
i.e. $n_s = 2$ for valence quarks ($qqq$), $n_s = 3$ for the gluon ($qqqg$), and $n_s = 4$ for anti-quarks ($qqq \, q\bar{q}$).
Note that these predictions are not well-defined in the context of pQCD, since they do not
provide the scale at which they should hold. }.
While these predictions were approximately borne out by early measurements, the
existing experimental data now show that the PDFs can not be described by a single power-law
at high $x$, hence the correction function $P_i(x)$ in Eq.~\ref{eqn:pdf_functional} is
necessary.

The HERAPDF or MSTW analyses take $P_i(x)$ to be of the form
$$ P_i(x) = 1 + a_i x + b_i \sqrt{x} + ... \qquad,$$
whereas in the fits performed by the CTEQ/CT collaboration, it is chosen as
$$P_i(x) = \exp ( a_i x + b_i \sqrt x + c_i x + d_i x^2) \qquad. $$
In recent global fits, additional flexibility is needed for the gluon distribution, in order to obtain a good fit
to all the data.
The MSTW08 analysis uses
\begin{equation}
xg(x) = A_g x^{\alpha_g} (1-x)^{\beta_g} ( 1 + a_g x + b_g \sqrt{x}) + A'_g x^{\alpha'_g} (1-x)^{\beta'_g}
 \label{eq:MSTW08_gluon}
\end{equation}
which allows the gluon density to become negative at low $x$, while, in the CT10 analysis,
extra freedom at low $x$ is given by 
\begin{equation}
 xg(x) = A_g x^{\alpha_g} (1-x)^{\beta_g} \exp ( a_g x + d_g x^2 - e_g / x^{k_g}  )\qquad.
 \label{eq:CT10_gluon}
\end{equation}

The number of free parameters is usually reduced by imposing the number sum rules,
Eq.~\ref{eq:countingrule} and Eq.~\ref{eq:countingrule2}, together with the
momentum sum rule, Eq.~\ref{eq:sumrule}, which helps fix the gluon normalisation 
and connects the low $x$ and high $x$ behaviour of the gluon density.
Additional assumptions are often made. For example, the CT10 analysis assumes the same
low $x$ power-law for the input distribution $x\bar{u}$ and $x\bar{d}$, i.e.
$\alpha_{\bar{u}} = \alpha_{\bar{d}}$ in Eq.~\ref{eqn:pdf_functional}, as well as
the equality of the normalisation parameters, $A_{\bar{u}} = A_{\bar{d}}$, such that
$\bar{d} - \bar{u} \rightarrow 0$ as $x \rightarrow 0$. A similar assumption is made
for the HERAPDF1.0 fit.
All in all, there are $10$ free PDF parameters in the HERAPDF1.0 analysis and $26$
free parameters in the CT10 fit.
The MSTW08 fit has $29$ free parameters, including the value of $\alpha_s(M_Z)$ 
which is fitted together with the parton densities\footnote{Three additional parameters associated with
nuclear corrections are also fitted in this analysis.}.

\begin{figure}[htb]
 \centerline{\includegraphics[width=0.65\columnwidth]{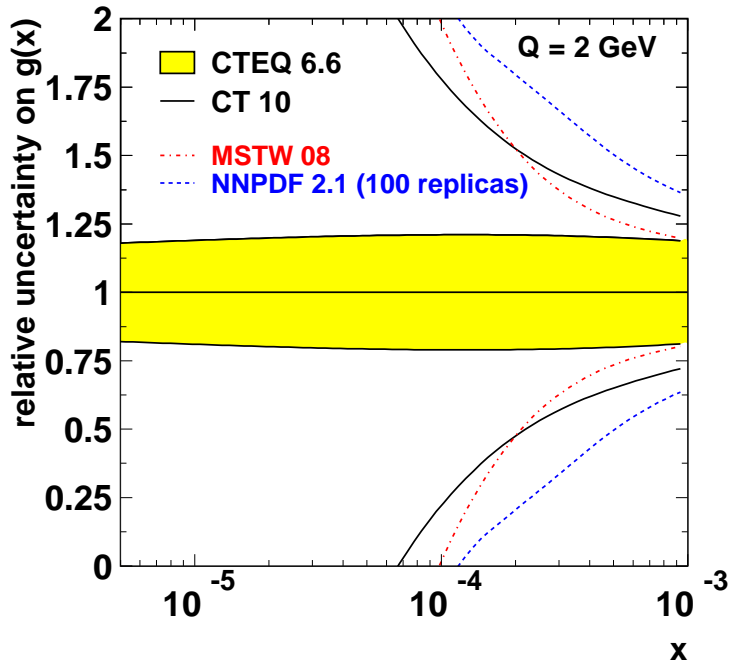}}
 \caption{\sl Relative uncertainty ($90 \%$ confidence level contours) on the gluon density at $Q^2 = 4$~GeV$^2$, as obtained from the
  CTEQ6.6, CT10, MSTW08 and NNPDF2.1 analyses. }
 \label{fig:gluon_lowx}
 \end{figure}

Although the functional form chosen to parameterise the densities at $Q^2_0$ is
rather flexible, a potential parameterisation bias does remain, that
is difficult to avoid.
Moreover, the uncertainties, obtained as explained in section~\ref{sec:systematic_errors},
can be considerably underestimated since they usually do not include any parameterisation uncertainty. 
An example of a parameterisation bias is illustrated in Fig.~\ref{fig:gluon_lowx}, which shows
the relative uncertainty on the gluon density at low $x$, at a scale $Q^2 = 4$~GeV$^2$.
The huge difference between the uncertainty obtained with CT10 and that resulting from the
previous fit CTEQ6.6 is due to the more flexible gluon parameterisation used in the former fit.
At low $x$, the CTEQ6.6 parameterisation is equivalent to a single power-law, $x g(x) \sim A x^{\alpha}$.
Neglecting the correlation between the $A$ and $\alpha$ parameters,  this parameterisation
prevents the relative uncertainty from growing faster than linearly with respect to $\ln x$,
since $\Delta (xg) / g = (\Delta \alpha) \ln x$. As there is no experimental data at $x$ below $10^{-5}$ and
$Q^2$ above the 
typical lower $Q^2$ cut used in the fits,
the small uncertainty band resulting from the CTEQ6.6
fit can only be artificial.
Indeed, using a very similar experimental input, but the more flexible parameterisation
Eq.~\ref{eq:CT10_gluon} for the gluon density, the uncertainty
increases dramatically as shown by the CT10 contour
in Fig.~\ref{fig:gluon_lowx}.
With this more flexible parameterisation, the CT uncertainty becomes comparable to that obtained
by  MSTW08, where extra freedom to the low $x$ gluon is provided by Eq.~\ref{eq:MSTW08_gluon}.

The parameterisation bias can be assessed,
to some extent, by varying the starting scale $Q^2_0$, and / or by making variations around the chosen functional form
at the given $Q^2_0$ and providing an additional parameterisation uncertainty, as pioneered
in~\cite{Aaron:2009kv} and also estimated in the HERAPDF analysis; however, the resulting uncertainties remain subjective.

Alternative choices 
for the densities $x f_i(x)$ can be based on Chebyshev polynomials, on any interpolation  
polynomials, or on non-linear functions. The latter
approach is exploited by the NNPDF collaboration,
which uses neural networks to parameterise the densities.
The formalism is described in~\cite{NNPDF2.0} and references therein. Neural networks are just another
functional form, that generalises parameterisations like $xf(x) = \sum_n \alpha_n P_n(x)$
based on interpolation polynomials $P_n(x)$. They allow non-linear dependencies of
the function on the fitted parameters $\alpha_n$.

The analysis presented in~\cite{NNPDF2.0} 
fits the gluon density together with the six densities for light quarks and
anti-quarks\footnote{The flexibility of the neural network allows any decomposition to be used,
as was checked explicitly in~\cite{NNPDF1.0}.}, $u, \bar{u}, d, \bar{d}, s, \bar{s}$. 
The neural networks chosen
to parameterise these densities have $37$ free parameters each. Hence, the resulting
parameterisation has a total of $7 \times 37 = 259$ free parameters, which is much larger than
the number of free parameters, ${\cal{O}}(25)$, which are fitted in QCD analyses
based on a standard functional form like Eq.~\ref{eqn:pdf_functional}.
The use of such a flexible parameterisation scheme 
considerably reduces any parameterisation bias.

\subsection{Treatment of experimental systematic uncertainties}

\subsubsection{The various methods}
\label{sec:systematic_errors}

A lot of work has been done over the past $ \sim 15$ years to assess
the uncertainties on parton densities extracted from
QCD fits~\cite{PDFUncertainties}.        
The task is not trivial since, as soon as many experimental data points are
included in the fit, a standard statistical approach does not appear to be
adequate, as will be discussed in~\ref{sec:tolerance}.

While most of the fits now minimise a $\chi^2$ function similar to Eq.~\ref{eq:chi2_a} and
Eq.~\ref{eq:chi2_b},
this $\chi^2$ definition can be used in different ways:
\begin{itemize}
        \item in the so-called ``offset method" : the parameters $s_k$ are set to
        zero in the central fit - i.e., the central fit is performed without
        taking into account the correlated systematic errors.
        Then, for each source of systematic error, $s_k$ is set to $\pm 1$ and
        the fit is redone. The uncertainty of a given quantity (e.g. a parton density)
        is calculated by adding in quadrature all differences to the quantity
        obtained in the central fit.
\item in the ``Hessian method" : the $s_k$ are not fixed, but are parameters of the fit.
        This means that the central fit is performed
        to the data shifted by the optimal setting for
        the systematic error sources as determined by the fit.
	The errors on the fitted PDF parameters $(p_{\alpha})$ are obtained from $\Delta \chi^2 = T^2$
        with $T=1$ or larger, see section~\ref{sec:tolerance}. The error on any given
   	quantity $F$ is then obtained from standard error propagation:
	\begin{equation}
	\sigma^2_F = \Delta \chi^2 \left( \sum_{\alpha,\beta} \frac{\partial F}{\partial p_{\alpha}}C_{\alpha,\beta} \frac{\partial F}{\partial p_{\beta}} \right) 
	\label{eqn:error_propagation}
	\end{equation}
	where the covariance matrix $ C = H^{-1}$ is the inverse of the Hessian matrix defined by
	$ H_{\alpha,\beta} = \frac{1}{2} \partial^2 \chi^2 / \partial p_{\alpha} \partial p_{\beta}$, evaluated at
	the $\chi^2$ minimum.
        The method developed in~\cite{PDFUncertainties}, which is now widely used by most groups,
        allows the Hessian matrix to be diagonalised without numerical instabilities; the error $\sigma_F^2$ 
        can then be calculated simply as
        \begin{equation}
         \sigma^2_F = \frac{1}{2} \sum_i \left( F ( S_i^+) - F (S_i^-) \right)^2
        \end{equation} 
        where the sum runs over the eigenvectors and $S_i^+$ and $S_i^-$ are PDF sets displaced 
        by $\Delta \chi^2$ along the $i^{th}$ eigenvector direction.
        For fits that use the Hessian method to determine the uncertainties, the PDF sets $S_i^+$ and $S_i^-$
        are stored in the public LHAPDF library, together with the set corresponding to the central fit.
\end{itemize}

The offset method gives fitted theoretical predictions which are as close as
possible to the ``raw" data points.
It does not use the full statistical power of the fit to correct the data for
the best estimate of the systematic shifts, since it distrusts that systematic
uncertainties are Gaussian distributed.
The offset method thus appears to be more conservative than the Hessian method.
It usually results in larger uncertainties than what is obtained from the Hessian
method, when the criterion $\Delta \chi^2 = 1$ is used to obtain the error bands\footnote{However,
when the systematic errors are smaller than the statistical errors, both methods give very
similar results.}.
With the Hessian method, model uncertainties (e.g. varying $\alpha_s$, $Q^2_0$, ...)
are often larger than the fit uncertainties. This is because each model choice
can result in different values of the systematic shifts, i.e. when changing the
model one does not fit the same data points.

Alternatively, a Monte-Carlo approach can be used (see for example~\cite{ForteReview}), which avoids the usage of 
the error matrix formalism of Eq.~\ref{eqn:error_propagation}. A set of $N_{rep}$ replicas
of the $n$ experimental measurements is built by sampling the probability distribution
defined by the data, such that the means, variances and covariances given by the
replicas are those of the experimental measurements. The fit is then performed separately
on each Monte-Carlo replica. The best fit is defined as the average over the
replicas, and uncertainties on physical quantities are obtained as standard variances.
This method can be used for any choice of the parameterisation of parton densities,
but it is mostly convenient when the parameterisation is more complex than
the standard functional form of Eq.~\ref{eqn:pdf_functional}, leading 
to a much larger number of free parameters. QCD fits using standard
parameterisations like Eq.~\ref{eqn:pdf_functional} usually make use of the Hessian matrix
method to propagate the systematic uncertainties - that is the case of the analyses
performed by the CTEQ and the MSTW groups - although the fit performed by the H1 collaboration
in~\cite{h1new} estimated the uncertainties with the Monte-Carlo approach. 
The NNPDF collaboration, which
uses a more flexible parameterisation,
always propagates the systematic uncertainties with the Monte-Carlo method.
For the NNPDF2.0 analysis presented in~\cite{NNPDF2.0}, an
ensemble of $N_{rep} = 1000$ replicas of the measurements was used. The result of the
$N_{rep}$ fits performed over these replicas (i.e. $N_{rep}$ sets of $259$ parameters each)
is stored in the LHAPDF package, and can be used to predict mean values and uncertainties
on physical observables.

\subsubsection{The tolerance parameter $\boldsymbol{\Delta \chi^2 = T^2}$ in global fits with ``Hessian" uncertainties }
\label{sec:tolerance} 

Ideally, the error bands corresponding to $68 \%$ (one standard deviation) confidence level (CL) should
be obtained from the well-known criterion $\Delta \chi^2 = 1$, or $\Delta \chi^2 = 2.71$
for the $90 \%$ (two standard deviation) contours.
This would be appropriate when fitting consistent datasets to a well defined theory,
with systematic uncertainties being Gaussian distributed.
However, in practice, these conditions are not necessarily fulfilled.
For example, when fitting data from various experiments, it can happen that some
datasets are marginally compatible with the others, possibly because some
systematic uncertainty has been underestimated. Such datasets should
not be dropped from the fit unless there is a clear experimental evidence that
the measurement is incorrect.
Instead, the level of inconsistency between the datasets should be reflected in
the uncertainties of the fit. 
This can be done by considering the sets of PDF parameters as alternative hypotheses,
and by allowing all fits for which a desired level of consistency is
obtained for all datasets. If a dataset consists of $N$ experimental points,
its partial $\chi^2$ should be about $N \pm \sqrt{2N}$.
Practically, a tolerance parameter $T$ is chosen such that the criterion
$\Delta \chi^2 = T^2$ ensures that each dataset is described within the desired
confidence level. An example procedure to obtain the numerical value of $T$ can
be found in ~\cite{Pumplin:2002vw}.
For example, the $90 \%$ CL contours of CTEQ6.6 correspond to
$T = 10$ ($T \sim 6$ for $68 \%$ CL), while the MRST fits~\cite{Martin:2007bv} used $T = \sqrt{50} \sim 7$.
The MSTW08 analysis uses a ``dynamical tolerance"~\cite{MSTW08} where $T$ can be different for
the various eigenvectors of the Hessian matrix, with values  ranging between $T \sim 1$ and $T \sim 6.5$
for the $68 \%$ CL contours.

While this approach is well motivated and based on how far the parameters can be varied
while still giving an acceptable description of all the datasets, one should keep
in mind that setting
$\Delta \chi^2 = 100$ or $50$ corresponds to an increase of the errors of
all experiments by a factor of typically $5-6$, including those for which the measurements are very well controlled.

\subsection{QCD fits to DIS data}

\label{sec:HERAPDF}

As seen in section~\ref{hera}, the HERA experiments performed 
high precision measurements of NC and CC DIS in a large kinematic domain,
both in $e^+p$ and $e^-p$ collisions.
In particular:
\begin{itemize}
\item the precise measurement of the scaling violations of the structure function $F_2$ in NC DIS, i.e.
  its logarithmic dependence on the four-momentum transfer squared $Q^2$,
  gives access to the gluon density at low and medium $x$;
\item the precise measurement of $F_2$ in NC DIS at low and medium $x$
   sets strong constraints on the combination 
   $ 4 ( U + \bar{U} ) + ( D + \bar{D})$ where $U$ and $D$ denote the combined up-type and down-type 
   quark densities, $U = u + c$ and $D  =d + s + b$;
\item the CC DIS measurements provide two constraints:
   one on a linear combination of $D$ and $\bar{U}$ ($e^+ p$ data), and one on a 
   linear combination of $U$ and $\bar{D}$ ($e^- p$ data);
\item the measurement of $xF_3$ obtained from the difference of $e^+p$ and $e^-p$
  NC DIS cross sections provides a constraint on a linear combination of $U - \bar{U}$ 
  and $D - \bar{D}$.
\end{itemize}
As a result, good constraints can be obtained on the gluon density as well
as on $U$, $D$, $\bar{U}$ and $\bar{D}$ from HERA data alone.
The separation between $U$ and $D$, which in fits to HERA data 
is provided by the CC measurements,
can be further improved
by including e.g. DIS measurements
on a deuterium target.
In fits based only on inclusive DIS measurements,
separating further $U$, $D$, $\bar{U}$ and $\bar{D}$ along the individual
quark flavours mostly relies on assumptions.

\subsubsection{Fit of the combined HERA-I inclusive datasets}
\label{sec:HERAPDF1.0}

The HERAPDF1.0 parton densities~\cite{:2009wt} were extracted using only the
averaged H1 and ZEUS DIS measurements presented in section~\ref{sec:HERAaverage}.
The averaged HERA dataset corresponds to $741$ cross section measurements: $528$ ($145$) measurements 
of $e^+ p$ ($e^- p$) NC DIS and  $34$ measurements of both $e^+ p$ and $e^- p$ CC DIS.
Following a cut $Q^2 > Q^2_{min}$ with $Q^2_{min} = 3.5$~GeV$^2$, imposed to remain in the kinematic domain where pQCD is
reliable, $592$ data points are included in the QCD fit of~\cite{:2009wt}.
The fit is performed at NLO within the
general mass variable flavour number scheme of~\cite{hep-ph/9709442, hep-ph/0601245}.
The starting scale is chosen to be slightly below the charm threshold,
$Q^2_0 = 1.9$~GeV$^2$.

The initial parton distributions\footnote{Since $c(x) =0$ and $b(x) = 0$ at the
chosen starting scale, $U(x) = u(x)$ and $D(x) = d(x) + s(x)$ at $Q^2_0$.}
$xf = xg, xu_v, x d_v, x \bar{U}$ and $x \bar{D}$
are parameterised at $Q^2_0$ using the generic form:
\begin{equation}
     xf(x) = A x^B (1-x)^C (1 + \epsilon \sqrt{x} + D x + E x^2)
 \label{eqn:generic_HERAPDF}
\end{equation}

At low $x$ the assumptions $\bar{d} / \bar{u} \rightarrow 1$ as $x \rightarrow 0$
and $B_{u_v} = B_{d_v}$ are made, which together with the number and momentum sum rules,
removes $6$ free parameters of the fit.
The strange quark distribution, $x \bar{s} = f_s x \bar{D}$,  is expressed as
a $x-$independent fraction, $f_s = 0.31$, of the down-type sea at the starting scale.

A $9$-parameter fit is first performed by setting to zero all
$\epsilon$, $D$ and $E$ parameters in Eq.~\ref{eqn:generic_HERAPDF}.
These parameters are then introduced one by one, the best $10$-parameter
fit having $E_{u_v} \neq 0$. 
As a result\footnote{
Setting $E_{u_v} \neq 0$ and introducing an
eleventh parameter in the fit does not reduce the $\chi^2$ significantly,
with the exception of the fit having both $E_{u_v} \neq 0$ and $D_g \neq 0$.
However, the latter leads to very low valence quark distributions at high $x$,
with in particular $d_v (x) < \bar{d}(x)$. As this would dramatically fail to
describe e.g. $\nu d$ fixed target measurements of the structure
function $xF_3$~\cite{WA25}, this solution is discarded for the ``central fit".
However it is included in the parameterisation uncertainty discussed below.
},
the central fit of HERAPDF1.0 is a
$10$-parameter fit corresponding to the following parameterisation:
\begin{eqnarray}
xg(x) &=   & A_g x^{B_g} (1-x)^{C_g}  \label{eqn:herapdf_gluon}  \\
xu_v(x) &=  & A_{u_v} x^{B_{u_v}}  (1-x)^{C_{u_v}} \left[ 1 + E_{u_v} x^2 \right] ,  \label{eqn:herapdf_uv} \\
xd_v(x) &=  & A_{d_v} x^{B_{d_v}}  (1-x)^{C_{d_v}} ,  \label{eqn:herapdf_dv}  \\
x\bar{U}(x) &=  & A_{\bar{U}} x^{B_{\bar{U}}} (1-x)^{C_{\bar{U}}} ,  \label{eqn:herapdf_barU}   \\
x\bar{D}(x) &= & A_{\bar{D}} x^{B_{\bar{D}}} (1-x)^{C_{\bar{D}}} .   \label{eqn:herapdf_barD}
\end{eqnarray}

\begin{figure}[tbh]
\begin{tabular}{cc}
\includegraphics[width=0.5\columnwidth]{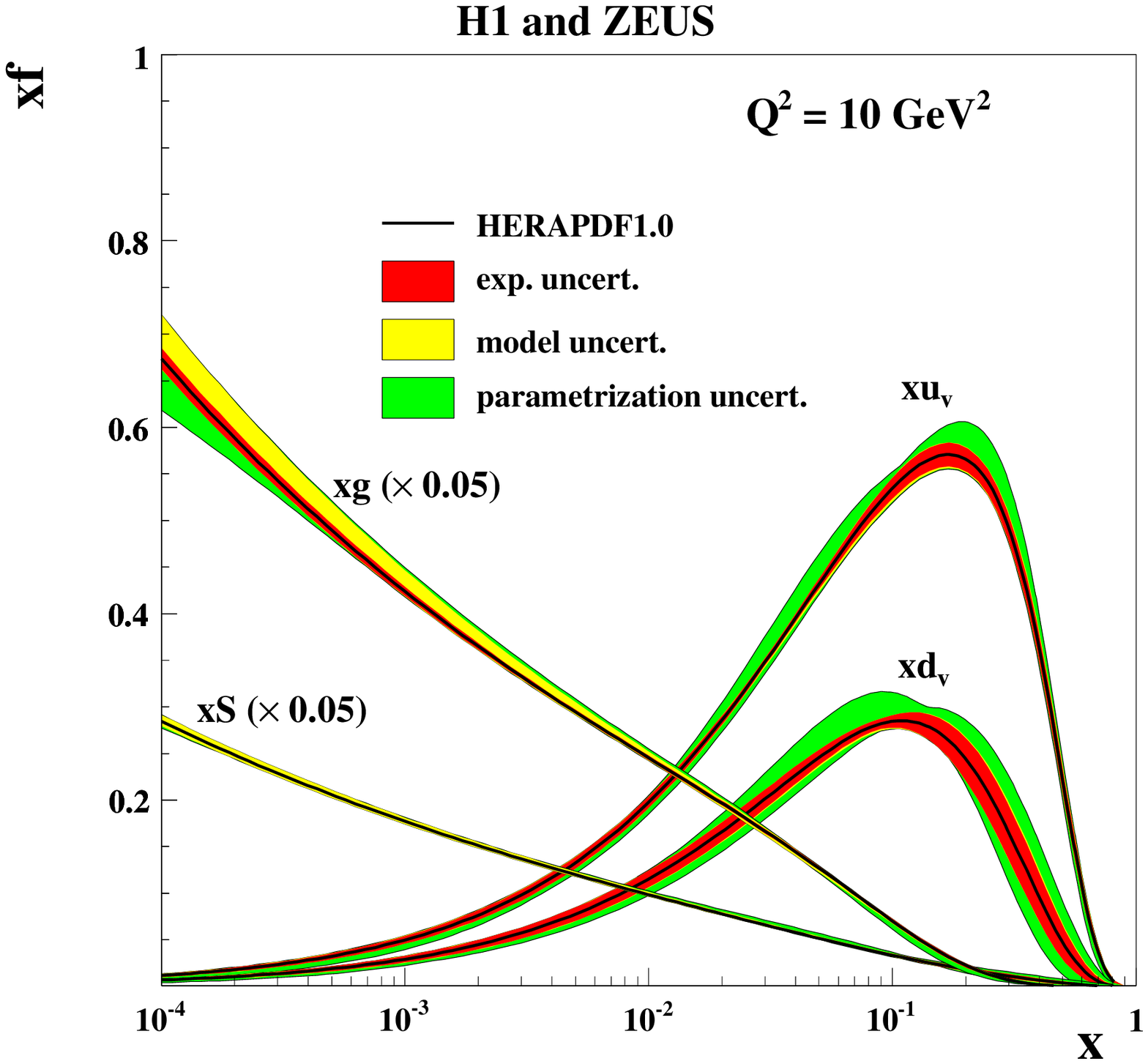} &
\includegraphics[width=0.5\columnwidth]{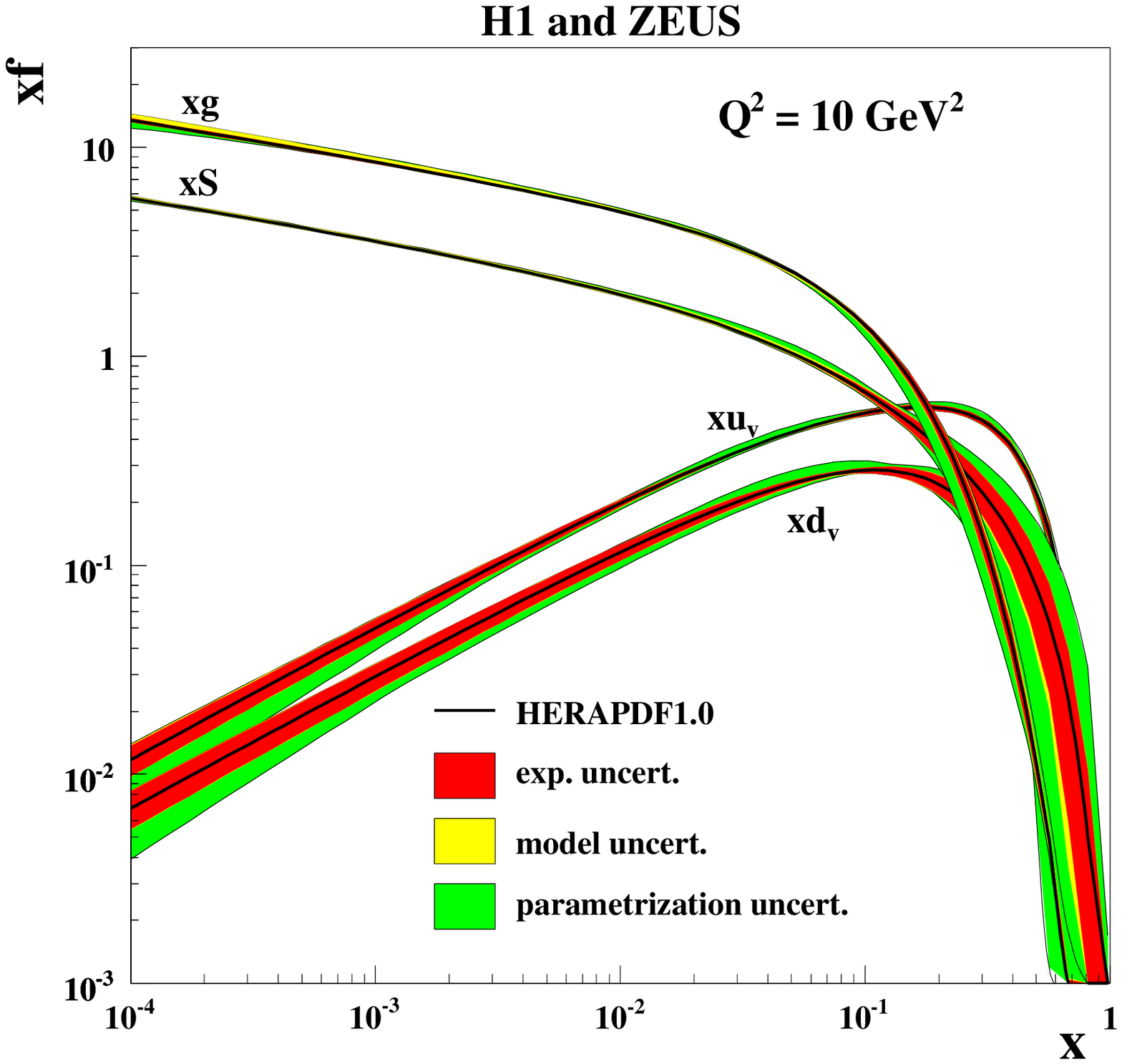}
\end{tabular}
\caption{\sl
Distributions of valence quarks densities $x u_v$ and $x d_v$, of
the gluon density $xg$ and of the density of sea quarks $xS$ obtained from
the fit to the HERA-I dataset, at $Q^2 = 10$~GeV$^2$, together with one standard deviation
uncertainties.
From~\cite{:2009wt}.
}
\label{Fig:hera_fits}
\end{figure}

The fit has a $\chi^2$ of $574$ for $582$ degrees of freedom.
Example PDFs at the scale $Q^2 = 10$~GeV$^2$ are shown in Fig.~\ref{Fig:hera_fits},
together with their uncertainties obtained as described below.

The experimental uncertainty of the HERAPDF1.0 PDFs 
(shown as the red band in Fig.~\ref{Fig:hera_fits})
is determined using the Hessian method
described in section~\ref{sec:systematic_errors}, taking into account the $110$ sources of
systematic errors of the individual measurements together with their correlations.
The tolerance criterion $\Delta \chi^2 = 1$ is used to determine the
$1 \sigma$ error bands.  Three additional errors are included which
account for different treatment of the systematic uncertainties in the
averaging procedure of the H1 and ZEUS measurements. These are the
largest uncertainties and are included using the more conservative
offset method\footnote{For the other $110$ sources of systematic
  uncertainties, the offset method yields very similar results as the
  Hessian method, since these uncertainties are smaller than the
  statistical uncertainties.}.
Very good constraints are obtained on $U$,
$D$, $\bar{U}$ and $\bar{D}$, which are the combinations the measurements are
directly sensitive to, in a large range of $x$ extending down to $10^{-4}$ and up
to ${\cal{O}}(0.1)$. As expected, the best constraints are obtained on $U$, with an
uncertainty that remains below $10 \%$ for $x \lsim 0.5$.
The gluon PDF is also well constrained up to $x \sim 0.2$, and 
down to\footnote{Large uncertainties are actually obtained at low $x$ for the starting scale $Q^2 = 1.9$~GeV$^2$,
but are quickly
washed out by the DGLAP evolution.} $x \sim 10^{-4}$.

The model uncertainty, shown as the yellow band in Fig.~\ref{Fig:hera_fits}, is obtained
by varying the input values of $f_s$, $m_c$, $m_b$ and $Q^2_{min}$ and repeating the fit.
The largest effect comes from the variation of $f_s$ and of the heavy quark masses, which
affects considerably the strange and charm densities.

Following~\cite{Aaron:2009kv}, an assessment of the additional uncertainty that
is introduced by the parameterisation choice is made in this analysis. 
This is particularly relevant since the number of parameters in the central
parameterisation of this analysis (Eq.~\ref{eqn:herapdf_uv}-\ref{eqn:herapdf_barD})
is rather small ($10$).  
The variation of the starting scale $Q^2_0$ is included in this 
parameterisation uncertainty. Indeed, when the DGLAP equations are used to
perform a backward evolution of the gluon distribution obtained from the central fit,
from $Q^2_0 = 1.9$~GeV$^2$
down to $1.5$~GeV$^2$, the resulting function cannot be fitted by the simple form
of Eq.~\ref{eqn:herapdf_gluon}. Consequently, repeating the fit with a lower starting scale
$Q^2_0 = 1.5$~GeV$^2$ results in large differences compared to the central fit if the same
parameterisation is kept, because Eq.~\ref{eqn:herapdf_gluon} is
not flexible enough to describe the gluon distribution at low scales.
Hence, when the fit is repeated with $Q^2_0 = 1.5$~GeV$^2$, additional freedom is
given for the gluon distribution by subtracting a term $A'_{g} x^{B'_g} (1-x)^{C'_g}$
to Eq.~\ref{eqn:herapdf_gluon}, as first suggested in~\cite{MSTW08}, where $C'_g$ is fixed
to a large value which ensures that this additional term does not contribute at high $x$.
Moreover, an additional fit is performed using the parameterisation of the central fit
but relaxing 
the assumption $B_{u_v} = B_{d_v}$.
Alternative
$11$-parameter fits with $E_{u_v} \neq 0$ are also considered, including those which lead
to good fit quality but peculiar behaviour at large $x$. 
An envelope is constructed, representing the maximal deviation, at each $x$ value,
between the central fit and the fits obtained using these parameterisation variations.
This envelope defines the parameterisation uncertainty of HERAPDF1.0
and is shown separately in Fig.~\ref{Fig:hera_fits} as the green band. 
The gluon and sea PDFs at low $x$ are mostly affected by the variation of $Q^2_0$
while at large $x$ (and over the whole $x$ range for the valence distributions), adding 
an eleventh parameter in the fit dominates the parameterisation
uncertainty. 
The total PDF uncertainty is obtained by adding in quadrature the experimental, model and
parameterisation uncertainties.
In particular, the parameterisation uncertainty increases considerably the uncertainty
of the gluon PDF at $x \sim 0.1$ and beyond.

\subsubsection{Impact of jet data on the fits to HERA data}

As seen in Fig.\ref{Fig:hera_fits}, the precision on the gluon density
is limited at medium and high $x$ when only inclusive HERA data are used.
Adding HERA data from jet production in DIS and in photoproduction was
shown to lead to better constraints~\cite{ZEUS-JETS}.
Although HERA jet data do not bring strong constraints on the gluon density at high $x$
due to the limited statistics (better constraints at high $x$ are brought by Tevatron jet data),
they can be useful
for medium $x$, since HERA jet cross sections have small systematic uncertainties
(typically $5 \%$, a factor of at least $2$ smaller than the systematic uncertainties of 
jet cross sections measured by the Tevatron experiments, see section~\ref{exp:tevatron}).
For the fits performed using only HERA DIS inclusive data~\cite{ZEUS-JETS},
the uncertainty of the gluon density was reduced by
a factor of $\sim 2$ in the mid-$x$ region, $x = 0.01 - 0.4$, when measurements
of inclusive jet production at HERA (see Fig.~\ref{fig:Zeus_jets}) were included.
It is also interesting
to note that both fits, with and without the jet data, lead to the same shape for
the gluon density, indicating that there is no tension between the HERA jet and 
inclusive DIS data.
Similar conclusions were reached in preliminary fits using the full statistics
of HERA data~\cite{MandyReview}.

\subsubsection{Fits performed using preliminary combinations of HERA data}

Following HERAPDF1.0, several QCD fits have been performed to preliminary
combinations of HERA data. 
As for HERAPDF1.0, the extraction of the HERAPDF1.5 PDFs relies on inclusive DIS data only,
but a preliminary combination of H1 and ZEUS measurements from HERA-I and
HERA-II was used instead of the published HERA-I combined dataset.
This fit was also performed with additional freedom given to the gluon and the
$u_v$ parameterisation in Eq.~\ref{eqn:herapdf_gluon} and~\ref{eqn:herapdf_uv},
leading to $14$ free parameters instead of $10$ (HERAPDF1.5f). This extended parameterisation
was also used for the extraction of NNLO PDFs, HERAPDF1.5 NNLO. These fits, although unpublished, 
are available in the LHAPDF interface.
H1 and ZEUS jet data were added in order to extract the HERAPDF1.6 (NLO) PDFs.
The most recent preliminary fit, HERAPDF1.7, also includes the $F_2^{c \bar{c}}$ measurements
and the data taken in $2007$ with a lower proton beam energy.
Further details can be found in~\cite{MandyReview, HERAWebPage}.

\subsubsection{Impact of fixed target DIS data}
\label{sec:impactofFixedTarget}

\begin{figure}[htb]
\begin{tabular}{cc}
\includegraphics[width=0.5\columnwidth]{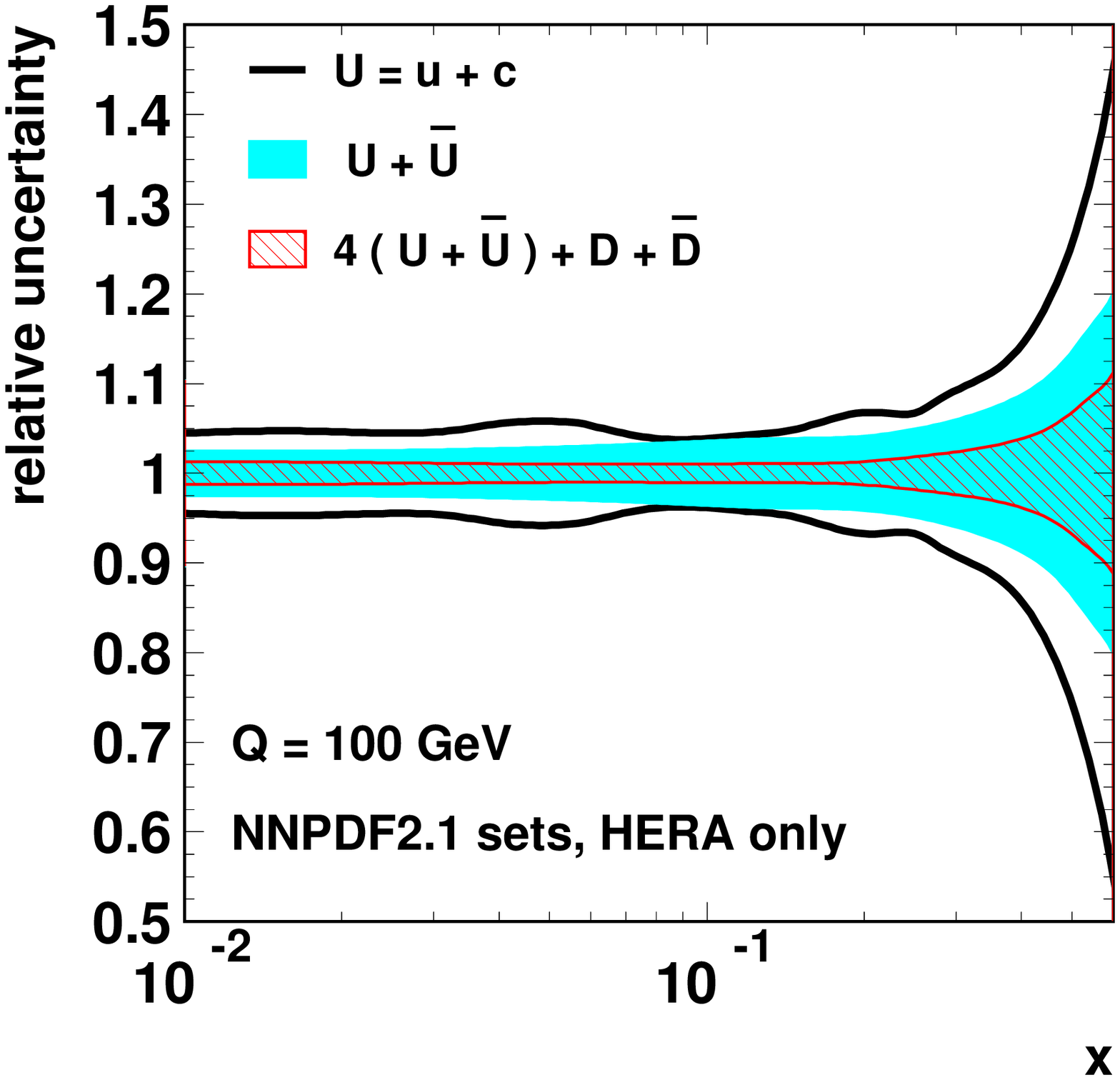} &
\includegraphics[width=0.5\columnwidth]{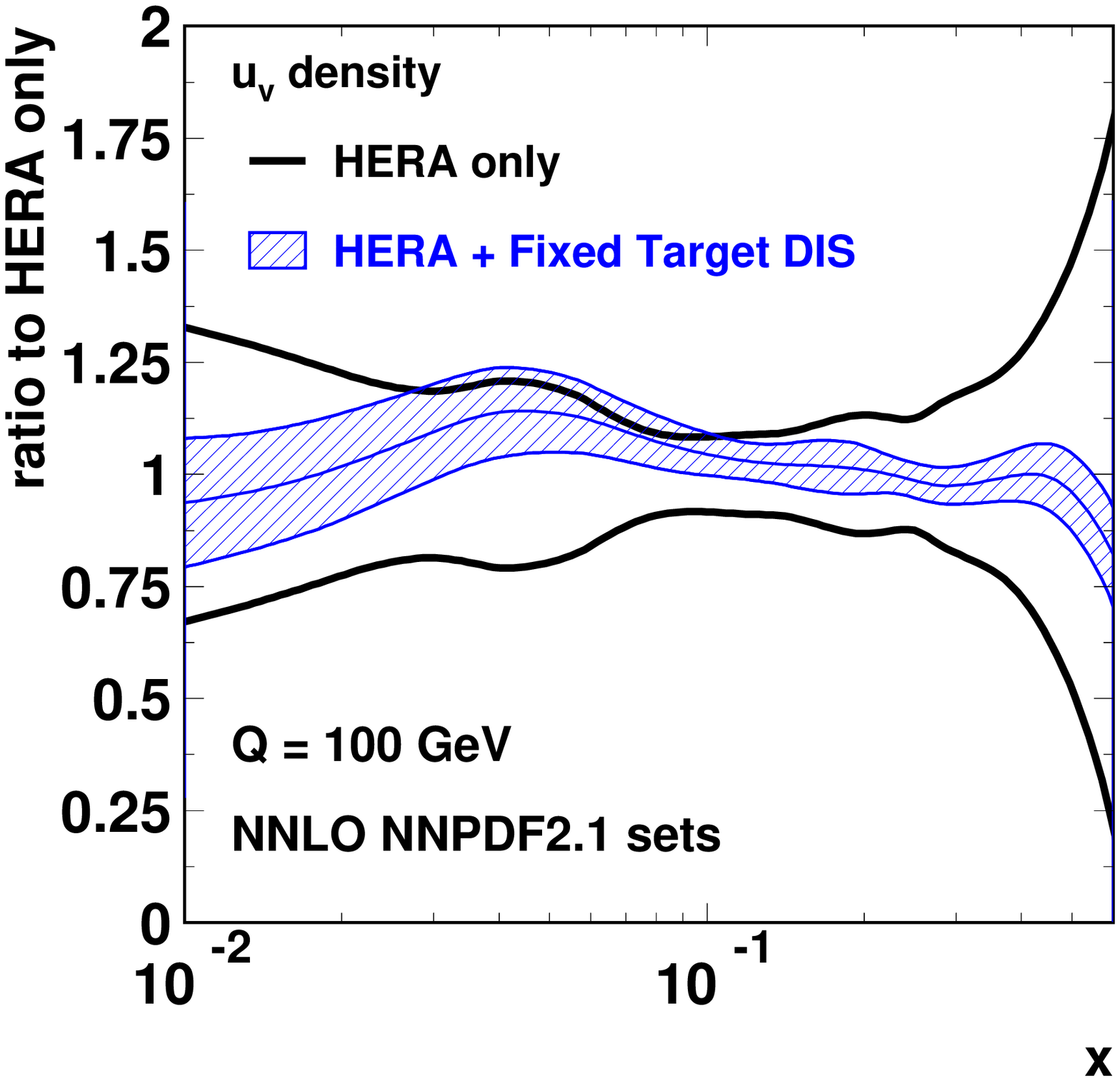} \\
\includegraphics[width=0.5\columnwidth]{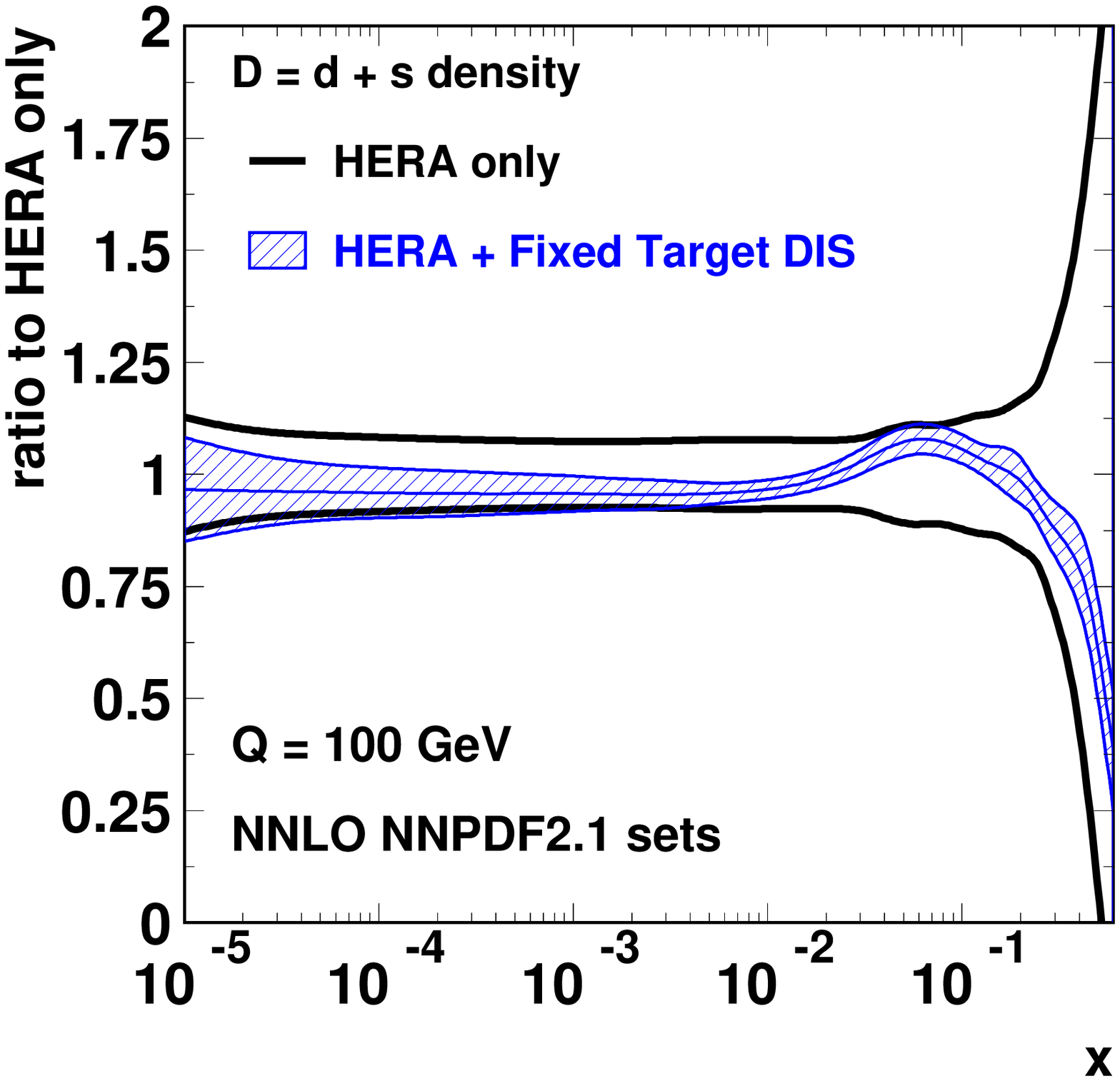} &
\includegraphics[width=0.5\columnwidth]{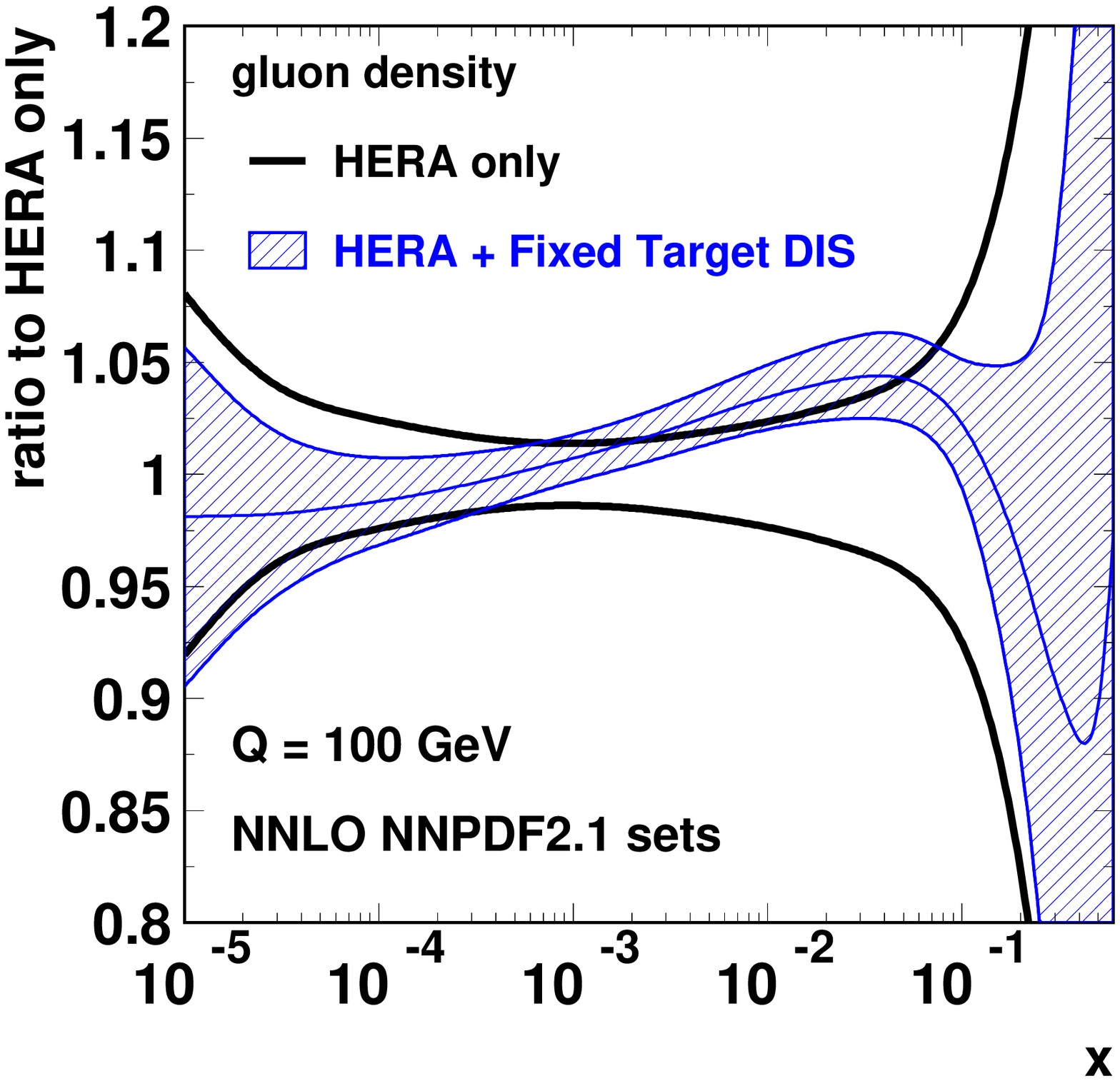}
\end{tabular}
\caption{\sl
 Top left: relative uncertainties on example quark densities obtained from
 the NNPDF fit to HERA data alone. Top right and bottom: comparison of the
 results of the fits with and without the fixed target data for the $u_v$,
 $D$ and gluon distributions.
 The contours correspond to one standard deviation uncertainties.
}
\label{Fig:fixedtarget}
\end{figure}

In fits based on HERA data alone, the flavour separation, as well as
the separation between quarks and anti-quarks (i.e. between valence and
sea quarks)  is provided by the high $Q^2$ measurements of CC cross sections
and of $x F_3$ in NC interactions. 
This separation, in particular at high $x$, can be further improved by
adding measurements of fixed target DIS experiments to the fitted data.
In particular, the $F_2^d$ measurements
of BCDMS and NMC mostly set a constraint on $4d + u$ which
nicely complements that on $4(U + \bar{U}) + (D + \bar{D})$ set by lepton-proton
measurements of $F_2^p$. Moreover, the measurements made in neutrino DIS provide direct access to
the distribution of valence quarks, assuming that nuclear corrections are
under control.

In~\cite{ZEUS2003}, the improved determination of quark distributions brought by
the addition of fixed target measurements in a fit to HERA data was studied 
within the fitting framework of the ZEUS experiment.
Here, the complementarity between HERA data and fixed target DIS measurements is illustrated 
by using the series of fits 
performed by the NNPDF collaboration, identical to
the NNPDF2.1 analysis but using subsets of experimental data. These fits are described
in~\cite{NNPDF2.1} and have been released in the LHAPDF package.
In particular, a fit has been performed using only the HERA data (the combined HERA-I NC and
CC datasets from H1 and ZEUS, 
inclusive $e^- p$ HERA-II measurements from ZEUS~\cite{Chekanov:2009gm,Chekanov:2008aa},
as well as $F_L$ and $F_2^{c \bar{c}}$ measurements). 
With respect to HERAPDF1.0
the fitting method used here largely avoids any parameterisation bias.
A similar fit has been performed by also including data from the
fixed target DIS experiments described in section~\ref{sec:FixedTargetDIS}.

Figure~\ref{Fig:fixedtarget} (top left) shows relative uncertainties
on example quark densities resulting from the fit to HERA data only.
While the combination $4(U + \bar{U}) + (D + \bar{D})$ that is directly probed 
by the $F_2^p$ measurements is well constrained over the full $x$ range,
the uncertainty increases at high $x$ when one tries to separate up-like from
down-like distributions, and quarks from anti-quarks.
The much improved separation of valence and sea densities provided
by the fixed target DIS data is illustrated in Fig.~\ref{Fig:fixedtarget} (top
right) with the example of the $u_v$ distribution.
The distributions of down-like quarks obtained from the two fits are
compared in Fig.~\ref{Fig:fixedtarget} (bottom left). Again a much better
determination at medium and high $x$ is achieved when the fixed target data are 
included in the fit, resulting in a reduced high $x$ distribution. 
In contrast, the constraints at low $x$ are largely coming
from HERA data. Figure~\ref{Fig:fixedtarget} (bottom right) shows that the
gluon determination is not improved significantly by the addition of fixed target 
data in the fit.

The next section will show how the determination of quark densities at high $x$,
in particular the separation between quarks and anti-quarks, can be further
improved by including Drell-Yan measurements in the fits (section~\ref{sec:DrellYanAndValenceSea}).
It should also be noted that this determination will benefit from the
stronger constraints brought by 
the full $e^-p$ and $e^+p$ HERA-II measurements at high $Q^2$, which show a much
better precision than the HERA-I measurements for example on $x F_3$, and from the final HERA combination.
The specific impact of the $e^-p$ and $e^+p$ HERA-II data from H1 was studied in~\cite{h1new} within
an analysis framework similar to that used for HERAPDF1.0 
and found indeed to be significant.

%-------------------------------
\subsection{Global QCD fits}
%-------------------------------

\label{sec:GlobalFits}

Although HERA data alone can determine the distributions of all 
partons, albeit with a limited precision at high $x$,
the determination of parton densities in
the proton is considerably improved by including additional datasets. The following datasets
are routinely included in ``global" pQCD analyses of the proton structure.

 \begin{figure}[bht]
 \centerline{\includegraphics[width=0.7\columnwidth]{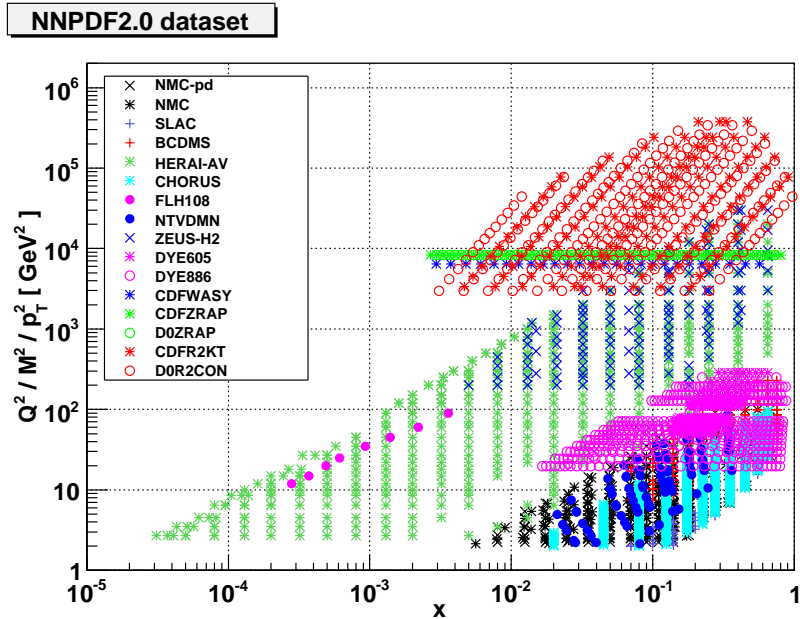}}
 \caption{\sl Experimental data which enter in the NNPDF global analysis (from~\cite{NNPDF2.0}).}
 \label{fig:globalFit_kin}
 \end{figure}

As shown in section~\ref{sec:impactofFixedTarget}, the inclusive NC DIS 
measurements from fixed target experiments using 
a deuterium target and the CC DIS measurements from the $\nu N$ experiments
improve the flavour separation and allow better disentanglement between the quarks and
the anti-quarks (i.e. the sea and the valence distributions).
The Drell-Yan measurements $pN \rightarrow \mu \mu$ mostly constrain the sea
densities. In particular, they set important constraints on the anti-quark densities
at medium and high $x$, which are not well known from DIS data alone.
Comparing the Drell-Yan cross sections measured in $pp$ and $pd$ provides
important constraints on the ratio $\bar{d} / \bar{u}$ at medium $x$.

The exclusive production of muon pairs in neutrino-nucleon scattering,
$\nu_{\mu} s \rightarrow \mu c \rightarrow \mu \mu X$, is the only pre-LHC process
that sets direct constraints on the strange density\footnote{Measurements of
multiplicities of strange hadrons, performed by the HERMES
experiment~\cite{HermesStrange},
should also constrain the strange PDF, once the experimental observables are corrected
for fragmentation effects. However, they have not yet been included  
in any global QCD analysis.}.

The Tevatron measurements of inclusive jet cross sections set the strongest
constraints on the gluon density at high $x$.
The measurements of $W$ and $Z$ production at the Tevatron mostly constrain
the $u$ and $d$ densities in the valence domain. Since the $u$ density is already
well constrained by the DIS experiments, they improve our knowledge of the $d$
density and of the ratio $d / u$ at medium $x$.

The addition of non-HERA data in the QCD analyses typically leads to ${\cal{O}} ( 3000)$
data points to be included in the fit. 
Fig.~\ref{fig:globalFit_kin} shows how 
the experimental data included in the NNPDF2.0 and NNPDF2.1 analyses are distributed in the $(x, Q^2)$ plane.
These two fits include $2841$ points from DIS experiments (with $743$ HERA points), 
$318$ points of Drell-Yan production
in fixed target experiments, $186$ points of jet production at the Tevatron, and $70$ points
of vector boson production by D0 and CDF.

In this section we mostly discuss results from the MSTW08, CT10 and NNPDF2.1 NLO analyses.
They are based on a similar experimental input and use a GM-VFNS for the
treatment of heavy flavours.
The MSTW08 analysis parameterises $g$, the valence quark densities $u_v$ and $d_v$, 
the light sea $S = 2(\bar u + \bar d) + s + \bar s$, the asymmetry $\Delta = \bar d - \bar u$, the
total strangeness $s^+ = s + \bar s$ and the strange asymmetry $s^- = s - \bar{s}$ at a starting scale $Q^2_0 = 1$~GeV$^2$.
The CT10 analysis parameterises $g$, $u_v$, $d_v$, $\bar u$, $\bar d$, $s$ at $Q^2_0 = 1.69$~GeV$^2$
and assumes $s = \bar s$. 
The NNPDF2.1 analysis parameterises the gluon density and the six quark and anti-quark light
flavours at a scale $Q^2_0 = 2$~GeV$^2$.
In the MSTW08 analysis, $\alpha_s (M_Z)$ is fitted together with the PDF parameters, while
in the CT10 and NNPDF2.1 fits it is set to a constant value.
\begin{figure}[htb]
\begin{tabular}{cc}
\includegraphics[width=0.45\columnwidth]{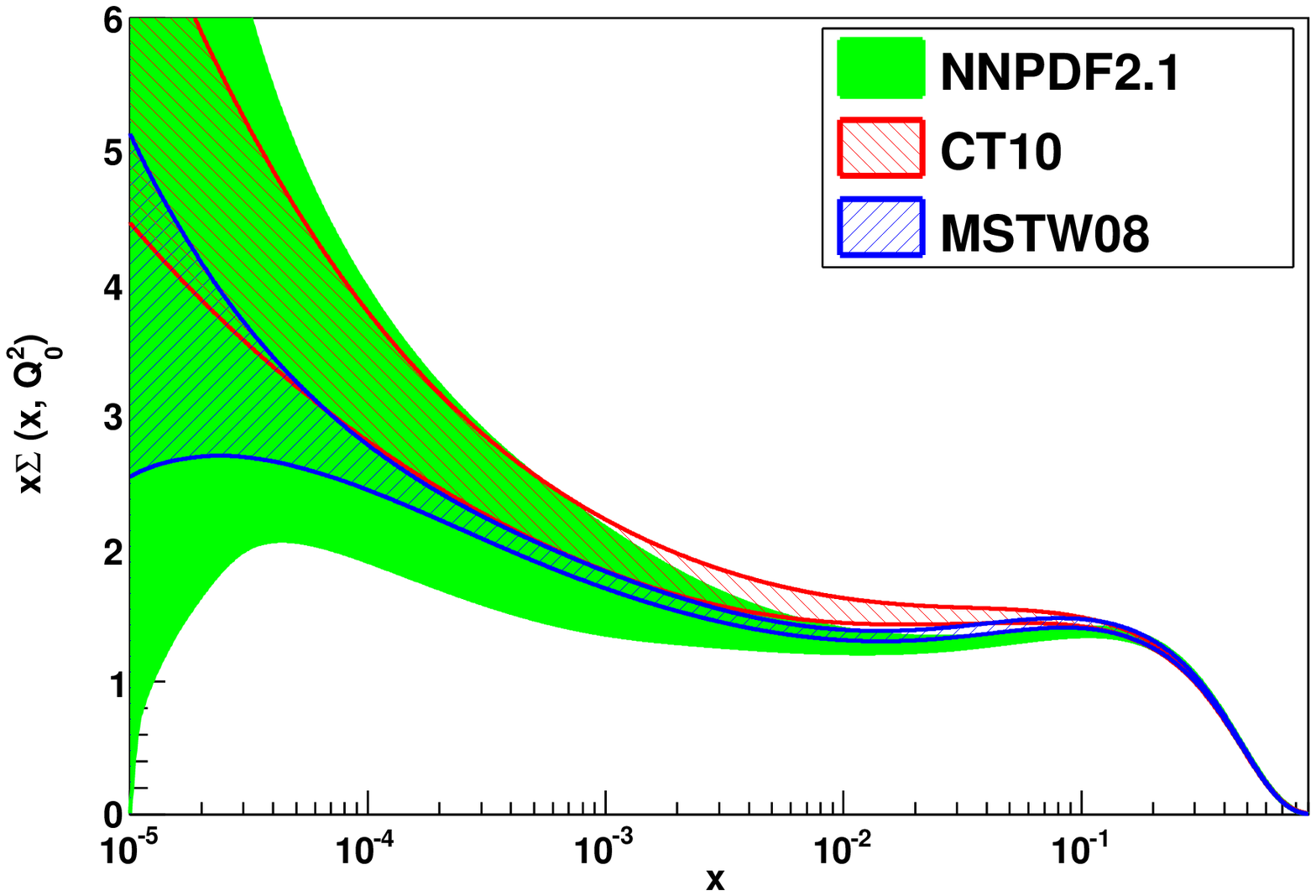} &
\includegraphics[width=0.45\columnwidth]{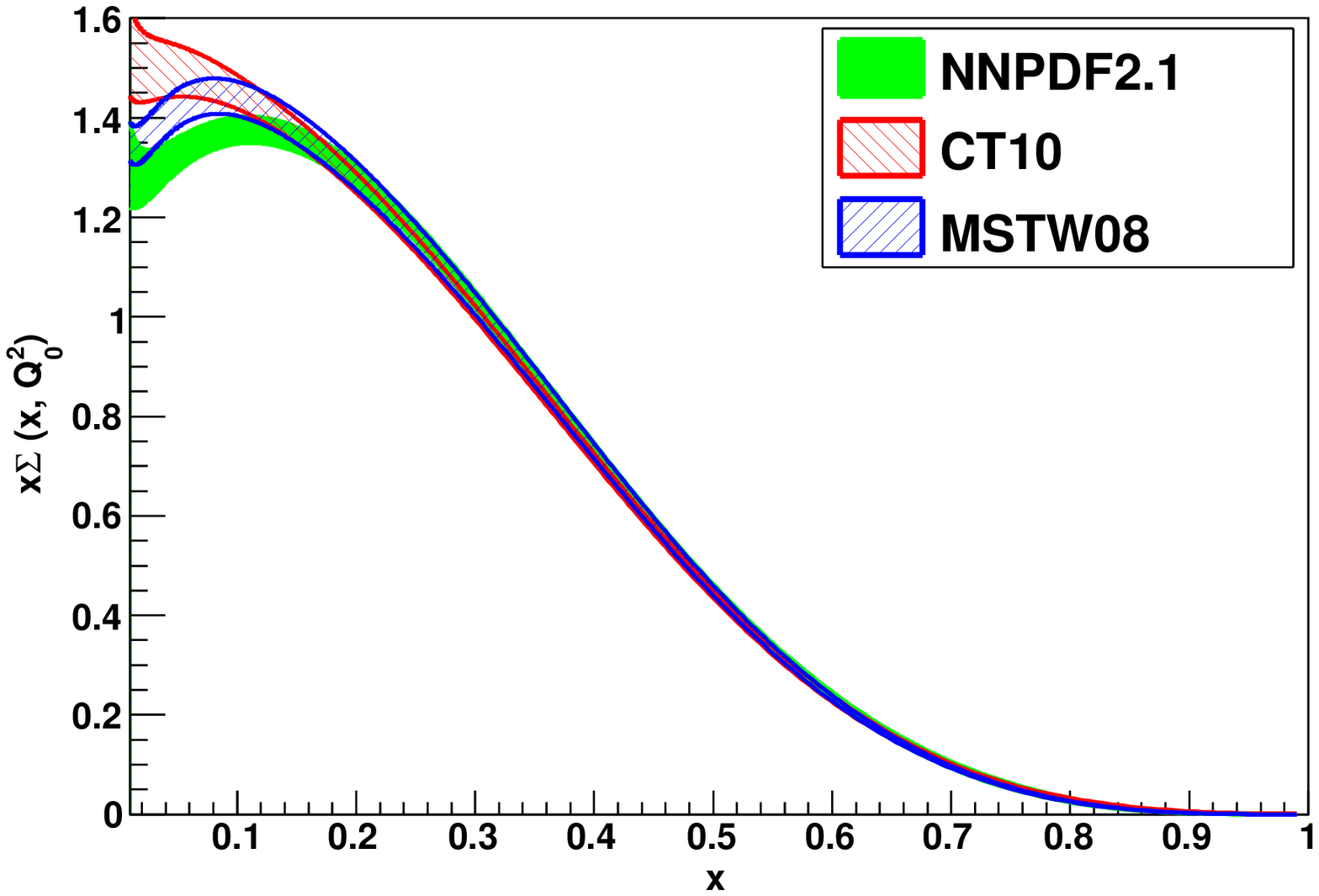} \\
\includegraphics[width=0.45\columnwidth]{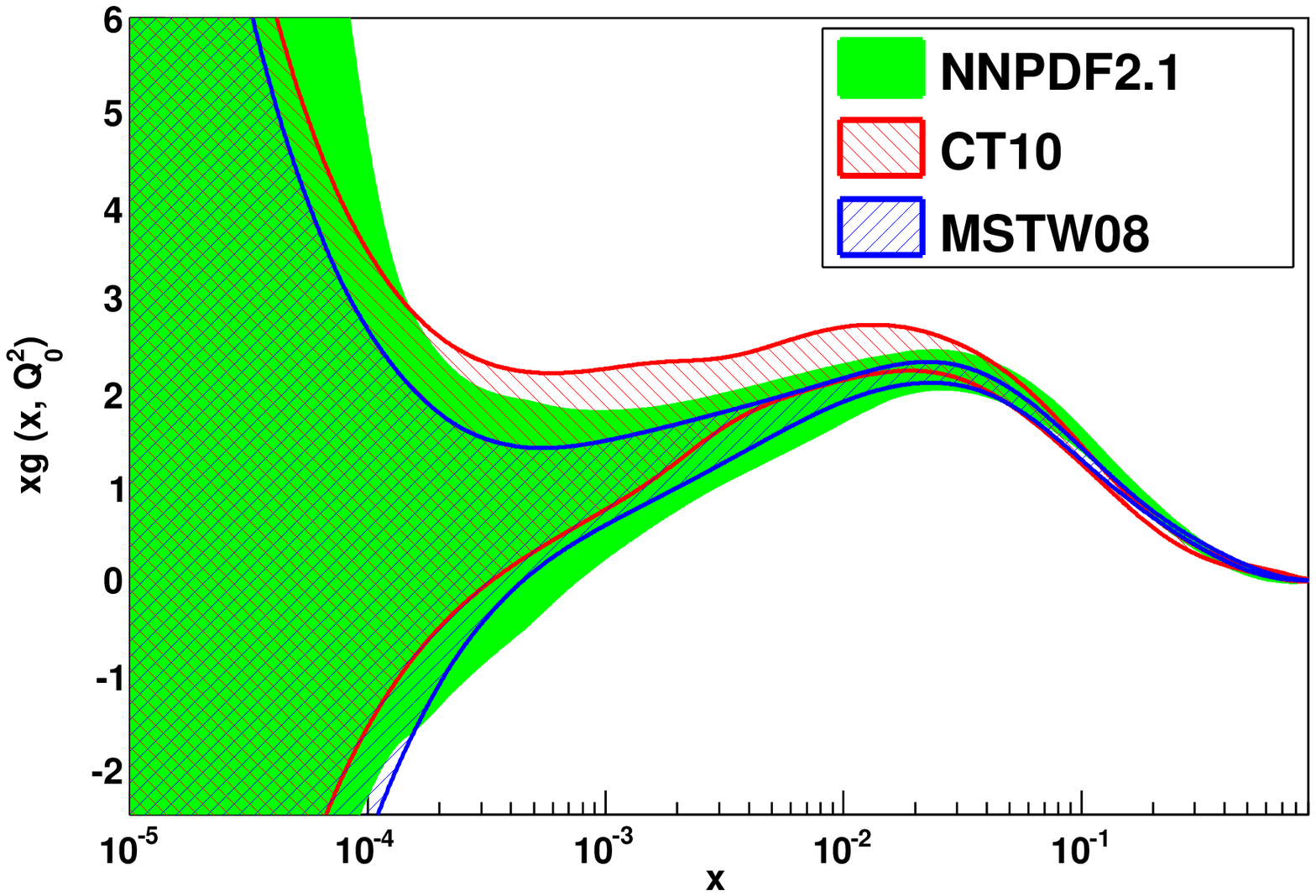} &
\includegraphics[width=0.45\columnwidth]{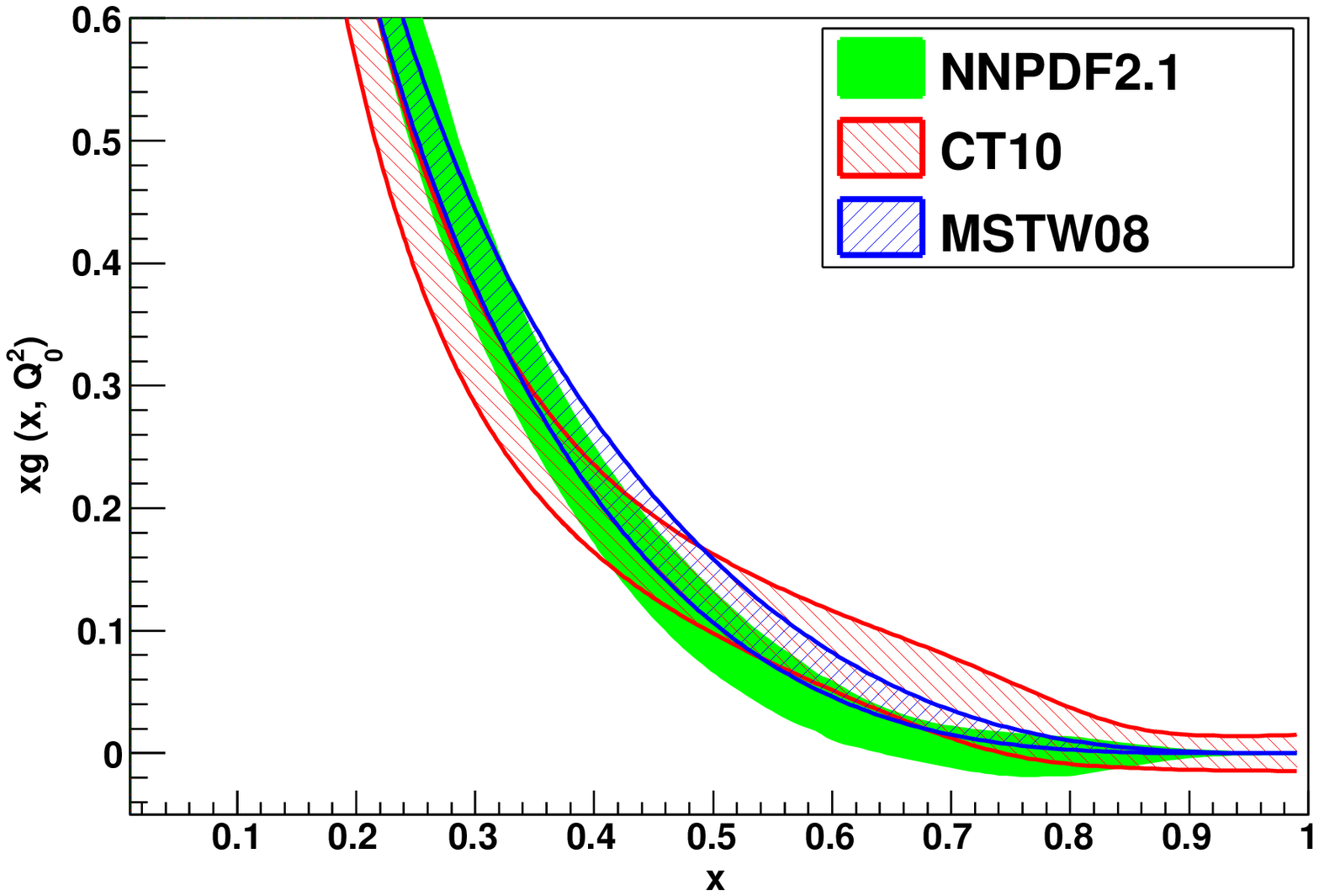} \\
\includegraphics[width=0.45\columnwidth]{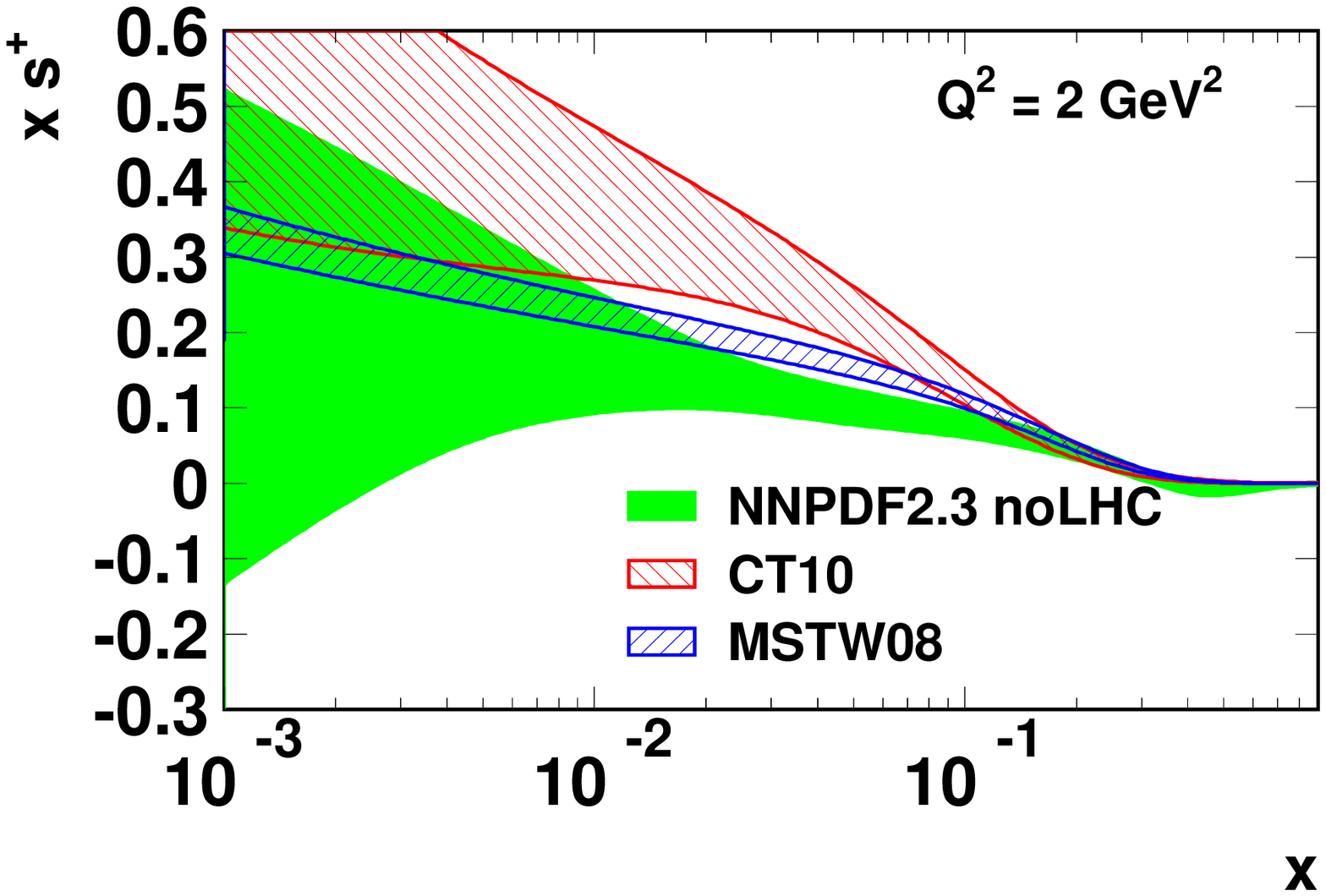} &
\includegraphics[width=0.45\columnwidth]{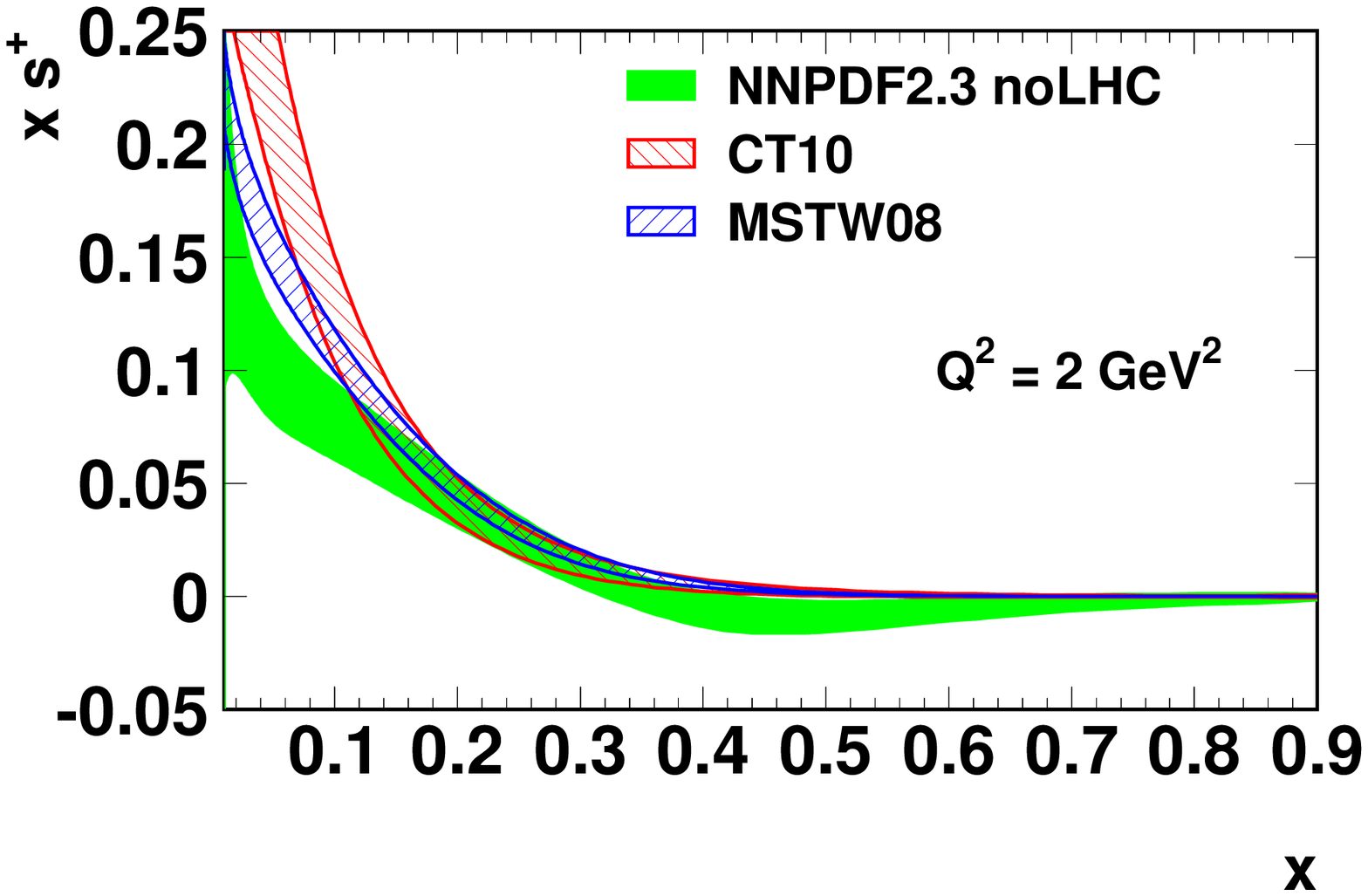} \\
\includegraphics[width=0.45\columnwidth]{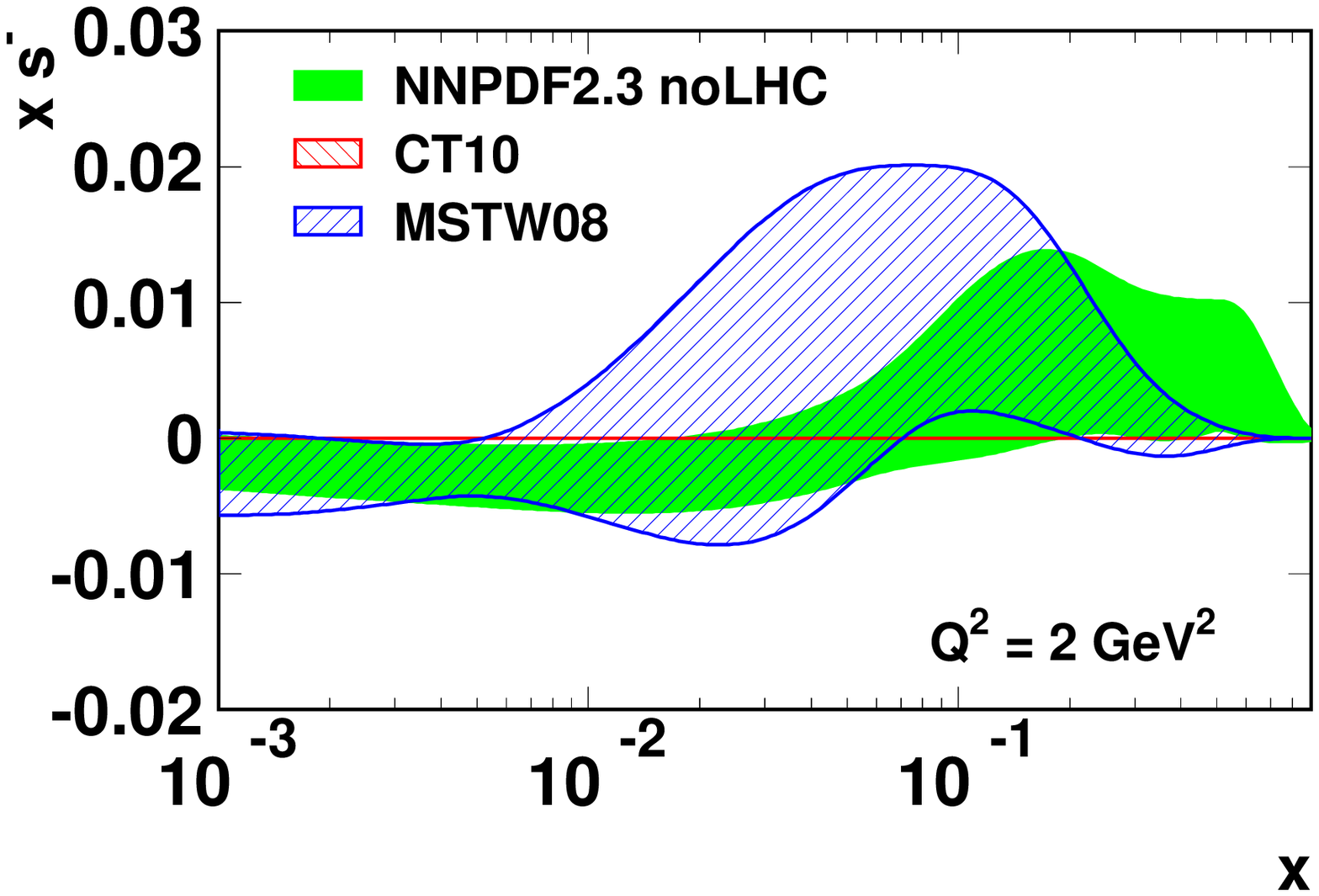} &
\includegraphics[width=0.45\columnwidth]{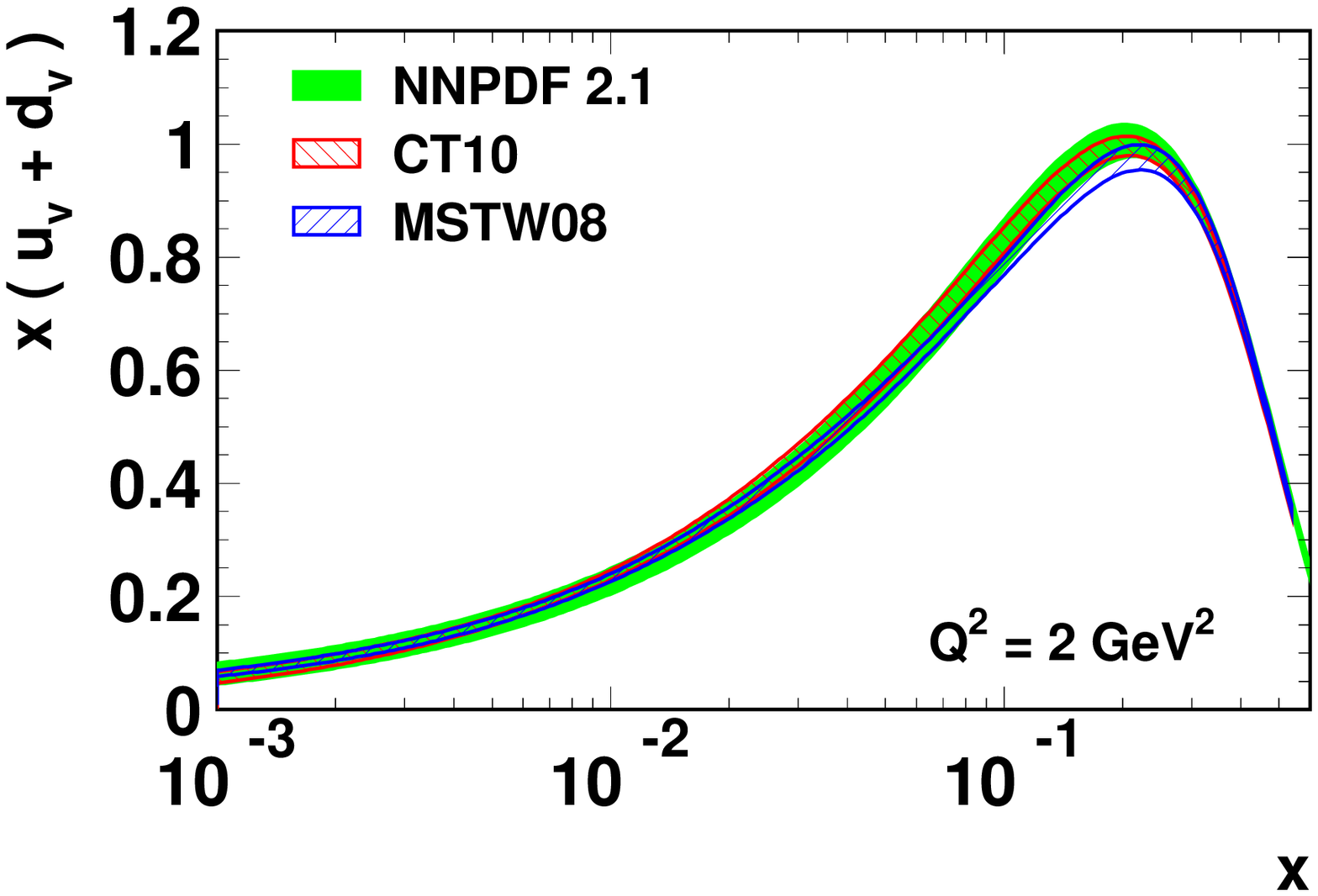}
\end{tabular}
\caption{\sl Comparison of recent global fits at $Q^2 = 2$~GeV$^2$, for the singlet
density $\sum = \sum ( q + \bar{q})$, the gluon density (from~\cite{NNPDF2.1}), the total strangeness
$s^+ = s + \bar{s}$, the strange asymmetry $s^- = s - \bar{s}$ and the total valence
distribution.
Contours at $68 \%$ confidence level are shown.
}
\label{fig:globalFits_Comparison}
\end{figure}

Figure~\ref{fig:globalFits_Comparison} compares a few example PDFs
at $Q^2 = 2$~GeV$^2$, as extracted from MSTW08, CT10 
and NNPDF2.1\footnote{The NNPDF2.1 analysis was affected by an error in the
calculation of di-muon production in neutrino DIS scattering, which had a significant
effect on the strange distribution. This error has been corrected
in~\cite{NNPDF2.3} and in the fit labelled NNPDF2.3-noLHC which is shown in Fig.~\ref{fig:globalFits_Comparison}
for the $s^+$ and $s^-$ distributions, which is based on the same experimental input as NNPDF2.1.}
The agreement for the gluon density at low and medium $x$ is rather good. In particular,
the error band of CT10 is much larger than the one previously predicted from CTEQ6.6,
which used a less flexible parameterisation for the gluon density (see Fig.~\ref{fig:gluon_lowx}), and agrees with
that obtained with MSTW08 and NNPDF2.1. At high $x$ however, the gluon densities
predicted by the three fits show sizable differences. 
Moreover, the three fits lead to very different predictions for the 
strange densities $s^+$ and $s^-$,
although they all use the same strange-sensitive datasets.
The NNPDF fit makes no assumption on the shape of the total strangeness density 
$s^+ = s + \bar{s}$, in contrast to the MSTW08 and CT10 fits (see section~\ref{sec:StrangeSea}). 
This results in a larger
error band, which impacts the uncertainty of other flavour PDFs at low $x$, especially that
of the down quark, since the main
constraint on low $x$ quarks comes from the HERA measurement of $F_2$, which probes 
charge weighted sums of quark PDFs.
Some differences can also be seen in the valence distribution, in particular for $x \sim 0.1$.
Since the error band of the NNPDF2.1 fit is not much larger than that of the other fits, it is unlikely
that this difference comes solely from a parameterisation bias; it could be due to, for example,
differences in the treatment of nuclear corrections to neutrino DIS data, which set
important constraints on the valence densities~\cite{ReviewThorne}.
The next paragraphs show a more detailed comparison of these three fits, together with
the specific impact of the non-DIS datasets.

\subsubsection{Tevatron data on $\boldsymbol{W}$ and $\boldsymbol{Z}$ production and the $\boldsymbol{d/u}$ ratio }
\label{sec:Wasym_Tevatron}

As shown in section~\ref{exp:tevatron}, the shape of the rapidity distribution of $W$ and
$Z$ bosons at the Tevatron provides interesting constraints on the $u$ and $d$ densities
at $x \gsim 0.01$. 
The $W$ charge asymmetry constrains the ratio $d / u$ and its slope. This ratio is otherwise
mostly constrained by the ratio $F_2^p / F_2^d$ measured by the NMC experiment, and by
the deuterium measurements of BCDMS. In practice, the Tevatron constraints on $d / u$ are
mostly constraints on the down density, since the density of up quarks is much better known
in that range.

\begin{figure}[htb]
 \centerline{\includegraphics[width=0.7\columnwidth]{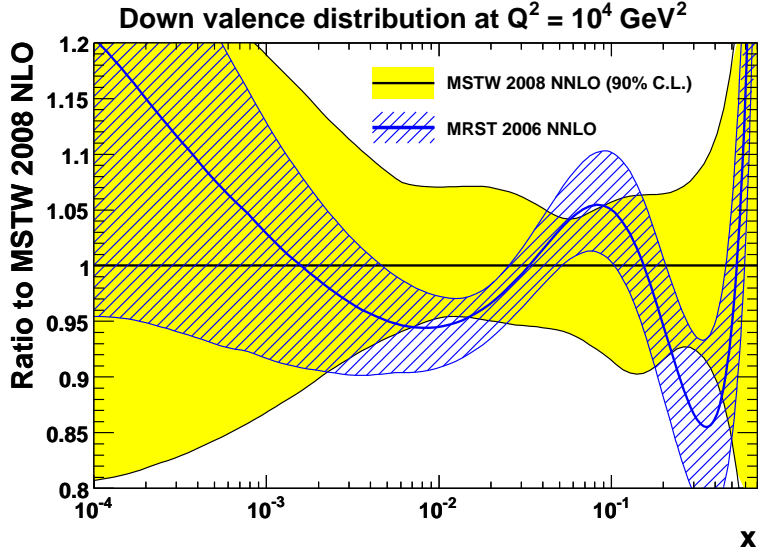}}
 \caption{\sl Comparison of $d_{v}$ as extracted from the MSTW analysis, when taking into
   account the CDF~\cite{Acosta:2005ud} and D0~\cite{Abazov:2007pm} Run II $W$ asymmetry data (MSTW08), or when using only the CDF Run I 
   measurements (MRST 2006). From~\cite{MSTW08}.  }
 \label{fig:mstw08_dvalence}
 \end{figure}

CDF Run I data on the $W$ asymmetry in the electron channel~\cite{Abe:1998rv} have been included for long in the
QCD analyses. They are now complemented by more precise Run II data~\cite{Aaltonen:2009ta,Acosta:2005ud,Abazov:2007pm, Abazov:2008qv}, 
some of them~\cite{Acosta:2005ud, Abazov:2008qv} being also available in several bins of the lepton transverse energy.
Fig.~\ref{fig:mstw08_dvalence} shows the effect of the Run II measurements from CDF~\cite{Acosta:2005ud} and D0~\cite{Abazov:2007pm},
corresponding to a luminosity of $170$~pb$^{-1}$ and $300$~pb$^{-1}$ respectively,  on the extracted
down valence distribution. The larger uncertainty obtained in the MSTW08 fit is due to
additional freedom introduced in the $d_{v}$ parameterisation, compared to  the previous
fit MRST2006~\cite{Martin:2007bv}. A change in the shape is clearly visible, with a significant increase
in $d_{v}$ for $ x \sim 0.3$, which is compensated (because of the number sum rule)
by a decrease for lower $x$ values. 

\begin{figure}[htb]
\begin{tabular}{cc}
\includegraphics[width=0.45\columnwidth]{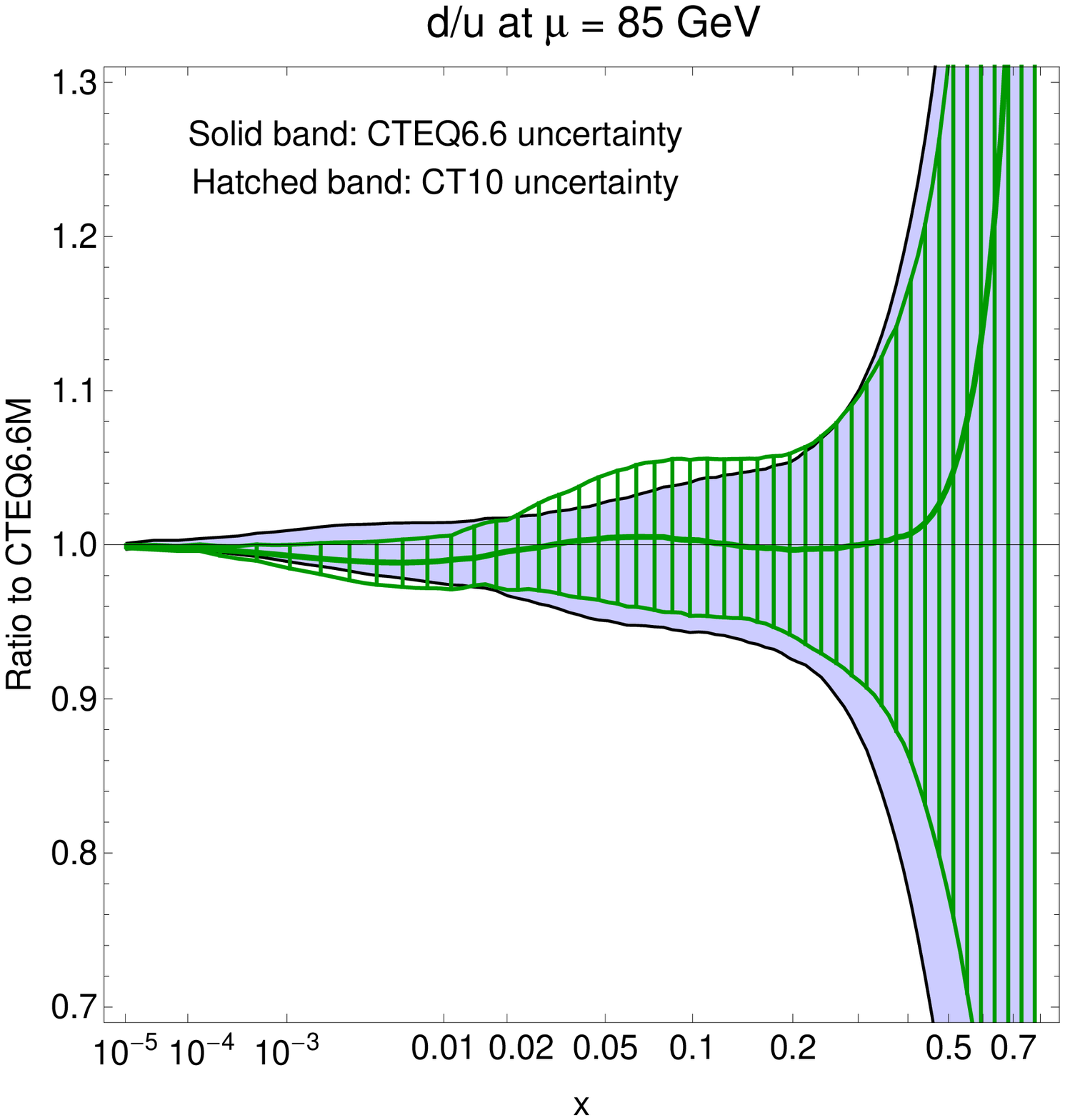} &
\includegraphics[width=0.45\columnwidth]{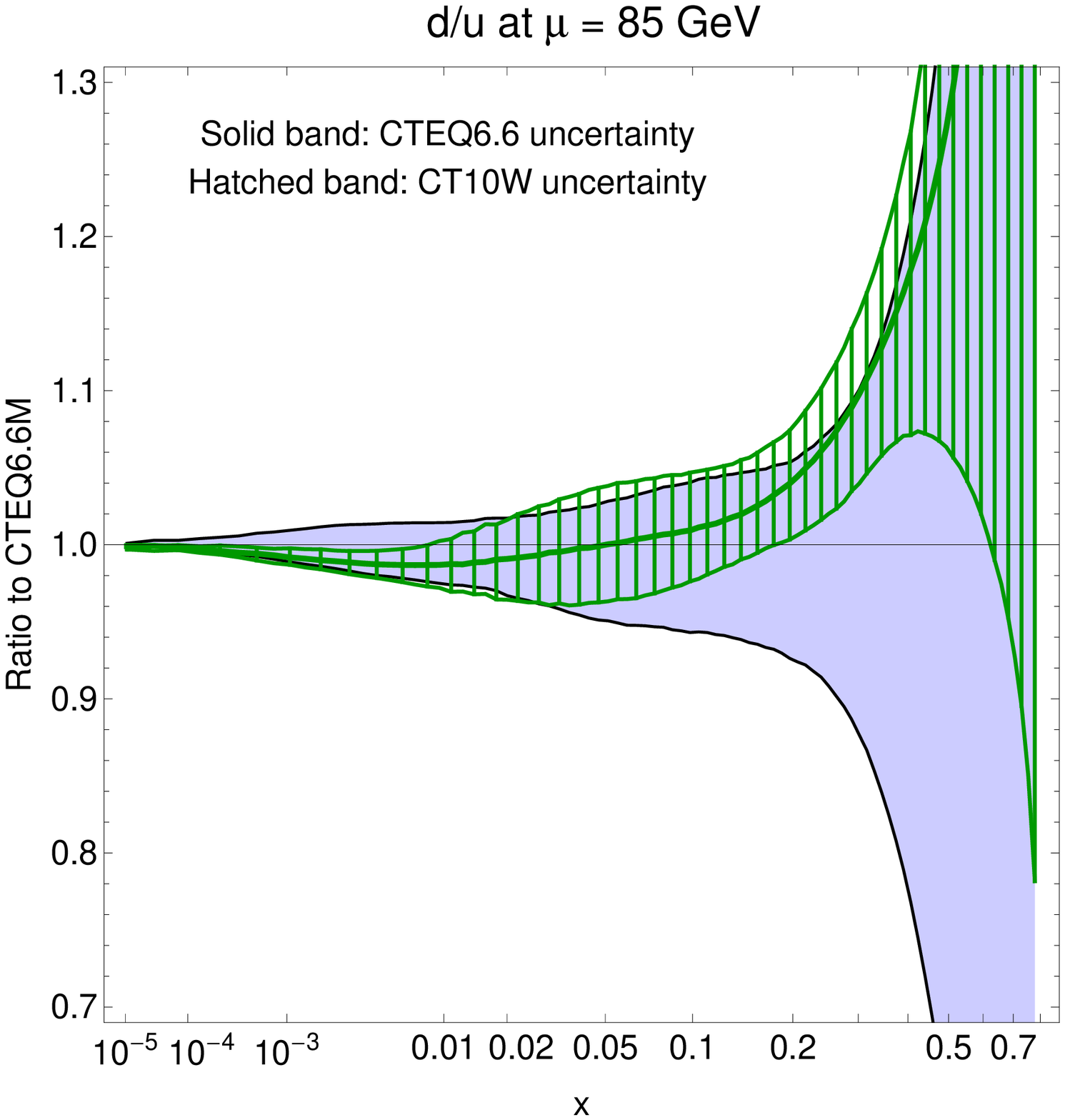}
\end{tabular}
 \caption{\sl The $d/u$ ratio obtained from the latest CT10 fit without including the Run II D0 measurements
  on the $W$ charge asymmetry (left), and with these measurements included (right). The ratio is
  normalised to that derived from the previous CTEQ fit (CTEQ6.6). The uncertainty bands correspond to
  two standard deviations. From~\cite{CT10}. }
 \label{fig:ct10_doveru}
 \end{figure}

It is interesting to note that the inclusion of the measurements of $W$
charge asymmetry made by D0 in several $E_T$ bins using $750$~pb$^{-1}$ of Run II data~\cite{Abazov:2008qv} 
is problematic, as consistently observed by all analyses.
These data show some incompatibility with the DIS structure function data, in particular
the NMC measurement of $F_2^p / F_2^d$ and the BCDMS $F_2^d$ measurement, and they
also show some tension within themselves.
The MSTW08 analysis decided not to include the D0~\cite{Abazov:2008qv} and CDF~\cite{Aaltonen:2009ta} 
high luminosity Run II measurements in their fit,
pending further investigation.
The D0 Run II data of~\cite{Abazov:2007pm, Abazov:2008qv} are also excluded from the NNPDF analyses and
from the main CT10 fit. However the CTEQ group also
provides a fit with these data included (CT10W), obtained by artificially increasing the weight
of these datasets in the global fit.
Figure~\ref{fig:ct10_doveru} compares the $d / u$ ratios obtained from these two fits. The ratio obtained from
the CT10W fit has a markedly different slope at $x > 0.01$, and a much reduced
uncertainty as compared to CT10.  While the outcome of the CT10W fit has to be used with care
until the compatibility with other data is better understood, this shows the potentially large
implications that precise $W$ asymmetry data from the Tevatron can have on the $d / u$ ratio,
and hence on the down density at large $x$. 

\subsubsection{The asymmetry of the light sea}

\begin{figure}[htb]
 \begin{tabular}{cc} 
  \includegraphics[width=0.5\columnwidth]{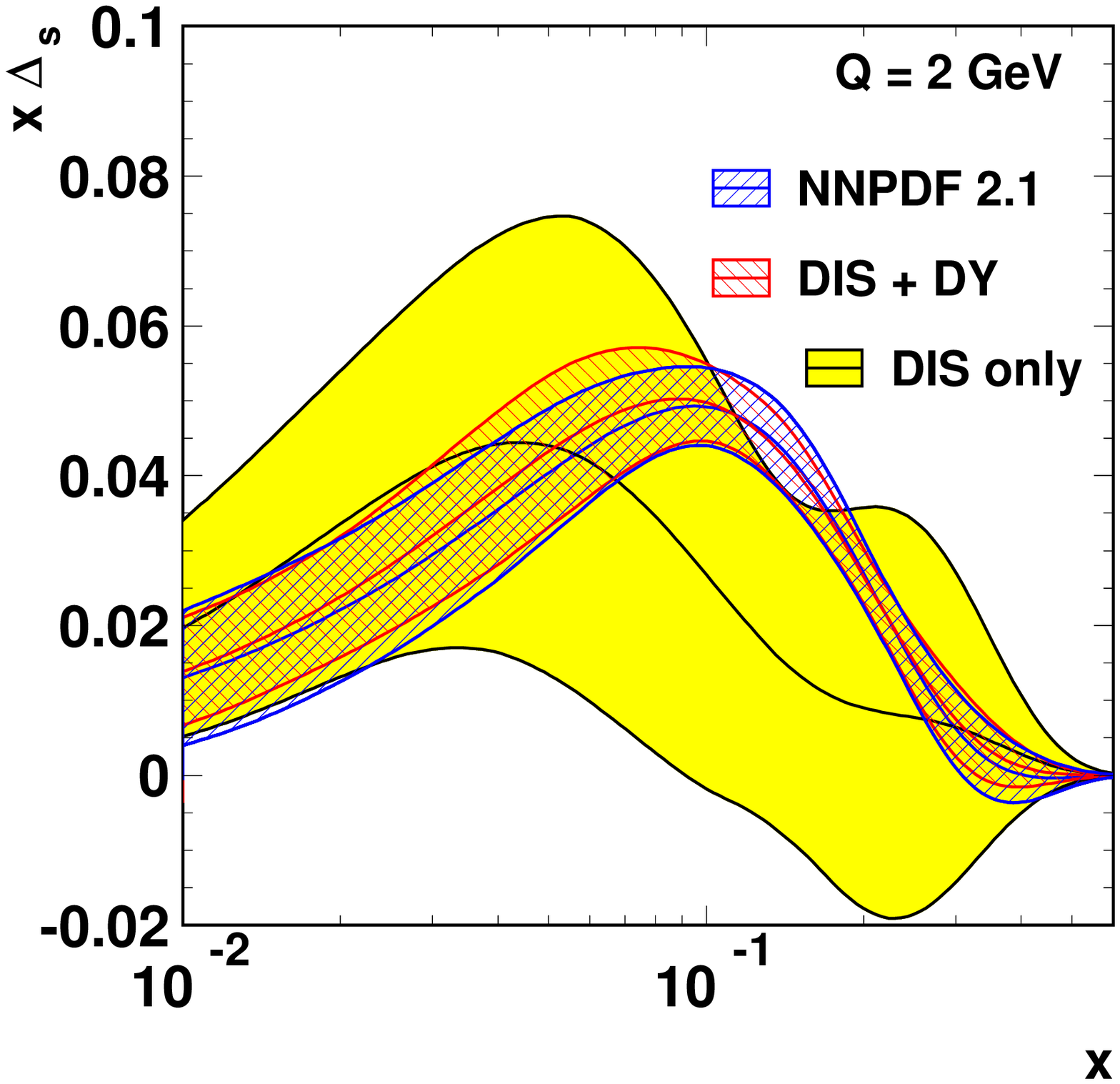} &
  \includegraphics[width=0.5\columnwidth]{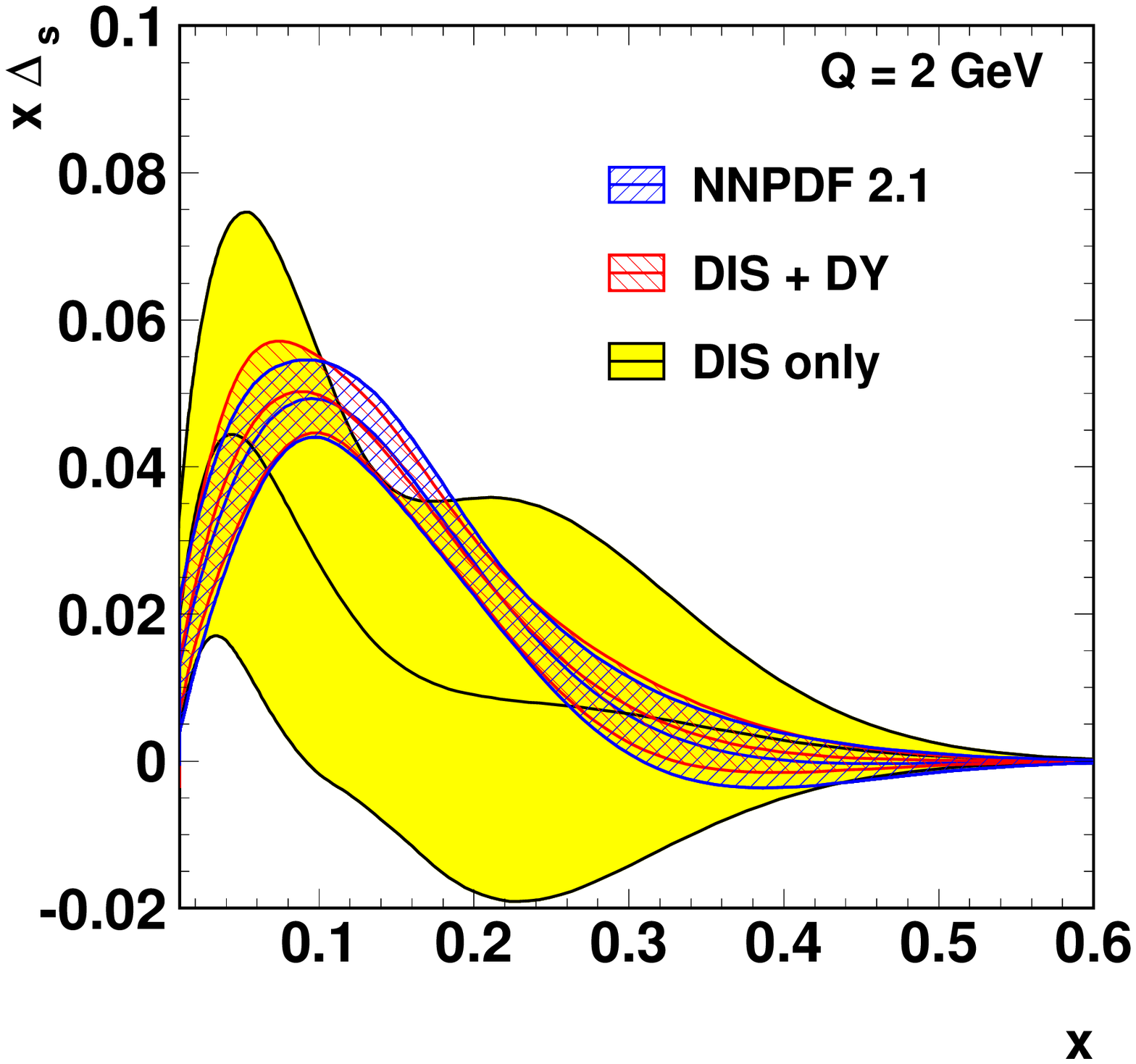}
 \end{tabular}
 \caption{\sl The asymmetry of the light sea $\Delta_s(x) = \bar{d}(x) - \bar{u}(x)$ and its 
  one standard deviation uncertainty at 
 $Q = 2$~GeV as obtained from the NNPDF analysis, when only DIS data are included in the fit
 (yellow contour), when Drell-Yan data are included in addition (red hashed contour),
 and from the reference NNPDF2.1 fit (blue hashed contour).
 % From~\cite{Forte:2010dt}.
 }
 \label{fig:impact_DY_fits}
 \end{figure}

The combination of constraints from muon-proton and muon-deuteron DIS, from HERA data, and from
neutrino DIS data, is not enough to determine the
light sea asymmetry $\bar{d} - \bar{u}$, which is very loosely constrained by 
a fit which includes DIS data only. The inclusion of Drell-Yan data 
(mostly proton and deuteron fixed target data) dramatically improves
this determination, as shown in Fig.~\ref{fig:impact_DY_fits}.
Including Tevatron data in addition does not further reduce the uncertainty in a
significant manner.
The figure was obtained from a series of fits performed by the NNPDF collaboration,
identical to the NNPDF2.1 analysis but using subsets of experimental data, which 
have been released 
in the LHAPDF package.

\subsubsection{Quarks and anti-quarks at high $x$}

\label{sec:DrellYanAndValenceSea}

\begin{figure}[htb]
  \begin{tabular}{cc}
 \includegraphics[width=0.5\columnwidth]{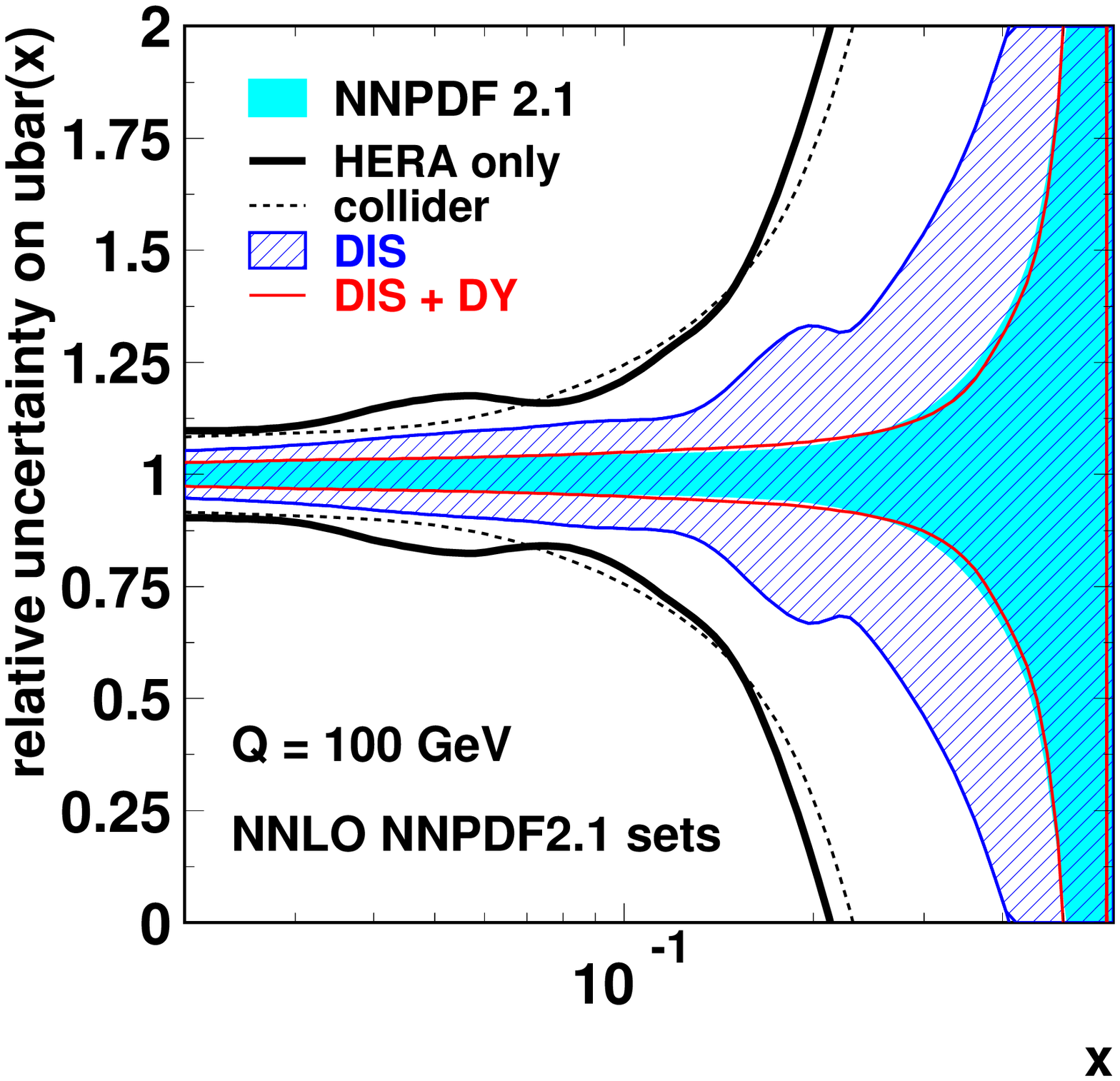} &
 \includegraphics[width=0.5\columnwidth]{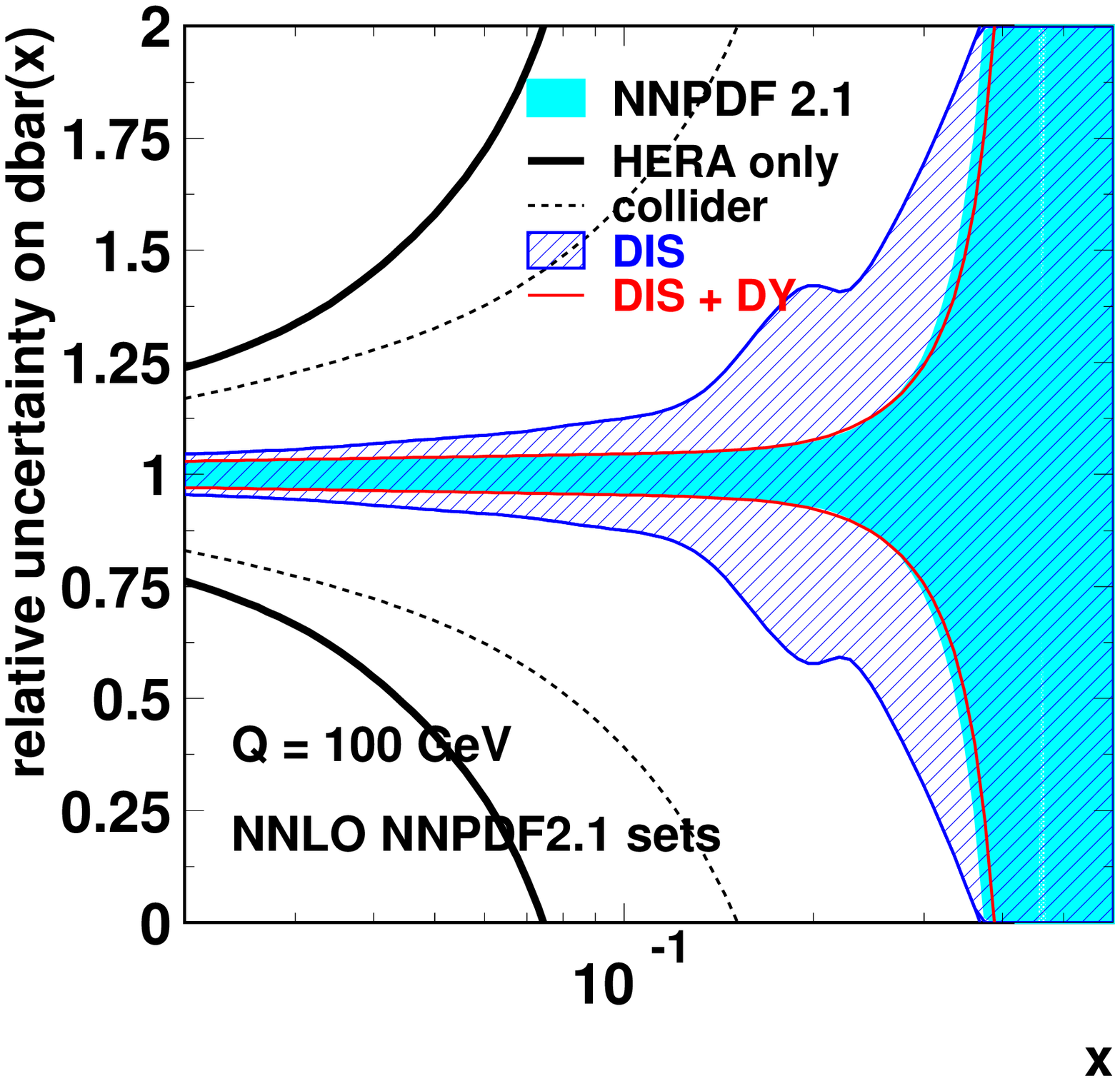} 
  \end{tabular}
 \caption{\sl One standard deviation uncertainties on $\bar{u}(x)$ (left) and on $\bar{d}(x)$ (right) at $Q^2 = 10^4$~GeV$^2$
  for $x$ between $0.02$ and $0.6$,
  as obtained from the NNPDF2.1 global fit (filled area), and from the same fit but applied to a subset
  of experimental data.}
 \label{fig:highx_antiquarks} 
 \end{figure}

The Drell-Yan measurements from fixed target experiments are also extremely useful to constrain the
quark and anti-quark densities at high $x$.
This is illustrated in Fig.~\ref{fig:highx_antiquarks}, which uses again the NNPDF2.1 sets 
released in the LHAPDF interface.
The HERA data alone provide very little constraints on the anti-quark densities at $x \sim {\cal{O}}(0.1)$.
Fixed target neutrino DIS experiments, which measured both $\nu N$ and $\bar{\nu} N$ cross sections,
provide a separation between the valence and the sea quarks at high $x$. As a result, a fit to
the full DIS data reduces the uncertainty on the anti-quark densities at high $x$.
However, the resulting uncertainty on $\bar{d}(x)$ remains large, e.g. $\sim 40 \%$ at $x \sim 0.2$.
With the addition of the fixed target Drell-Yan data, this uncertainty is reduced down
to $\sim 10 \%$.
The other datasets included in NNPDF2.1 do not reduce further the uncertainties.

Figure~\ref{fig:highx_antiquarks} also shows the uncertainties obtained in a fit 
using only data from the collider experiments (H1 and ZEUS, D0 and CDF). 
Although the Tevatron data help constrain the $\bar{d}(x)$
distribution at high $x$,
their impact is not as large as that of the Drell-Yan data, and their impact
on the uncertainty of $\bar{u}(x)$ at high $x$ is marginal.
The measurement
of high mass di-lepton production at the LHC will obviously bring further constraints on
anti-quarks at high $x$, assuming that effects of physics beyond the Standard Model do
not distort the mass spectrum\footnote{Several new phenomena may lead to an enhancement or to a
reduction of the production of a high mass di-lepton pair at the LHC as e.g. $qqll$ contact interactions,
quark substructure, or ``towers" of Kaluza-Klein gravitons in models with large extra spatial dimensions.}.

\subsubsection{Tevatron jet data and the gluon distribution}
\label{sec:fits_gluon}

The inclusive jet production at the Tevatron experiments is very sensitive to
the gluon density at high $x$. Early CDF Run I measurements reported in $1995$~\cite{Abe:1996wy}
indicated an excess of high $p_T$ jets compared to NLO QCD predictions based on the
PDFs available at that time. The possibility that this excess could be an indication of quark
compositeness created quite some excitement, until it was shown~\cite{CTEQ4} that these jet data
could be satisfactorily accommodated in a global fit, which resulted in a larger gluon density
at high $x$. While this showed the important role of jet constraints on the gluon density,
it also triggered intensive efforts in order to provide PDFs with associated uncertainties,
which lead to the state-of-the-art presented above.

The inclusive jet production cross sections measured by D0 and CDF are included
in the global QCD analyses since CTEQ4~\cite{CTEQ4}, MRST2001~\cite{MRST2001}
and NNPDF2.0.
Since CT10 and MSTW08, the Run II measurements are used in place of the Run I results.
Indeed, these new datasets have a much higher statistics and smaller systematic uncertainties,
and the experiments have provided the full correlation matrix of systematic errors.
These Run II measurements prefer a smaller high $x$ gluon distribution than
the Run I data.

\begin{figure}[htb]
 \begin{tabular}{cc}
 \includegraphics[width=0.5\columnwidth]{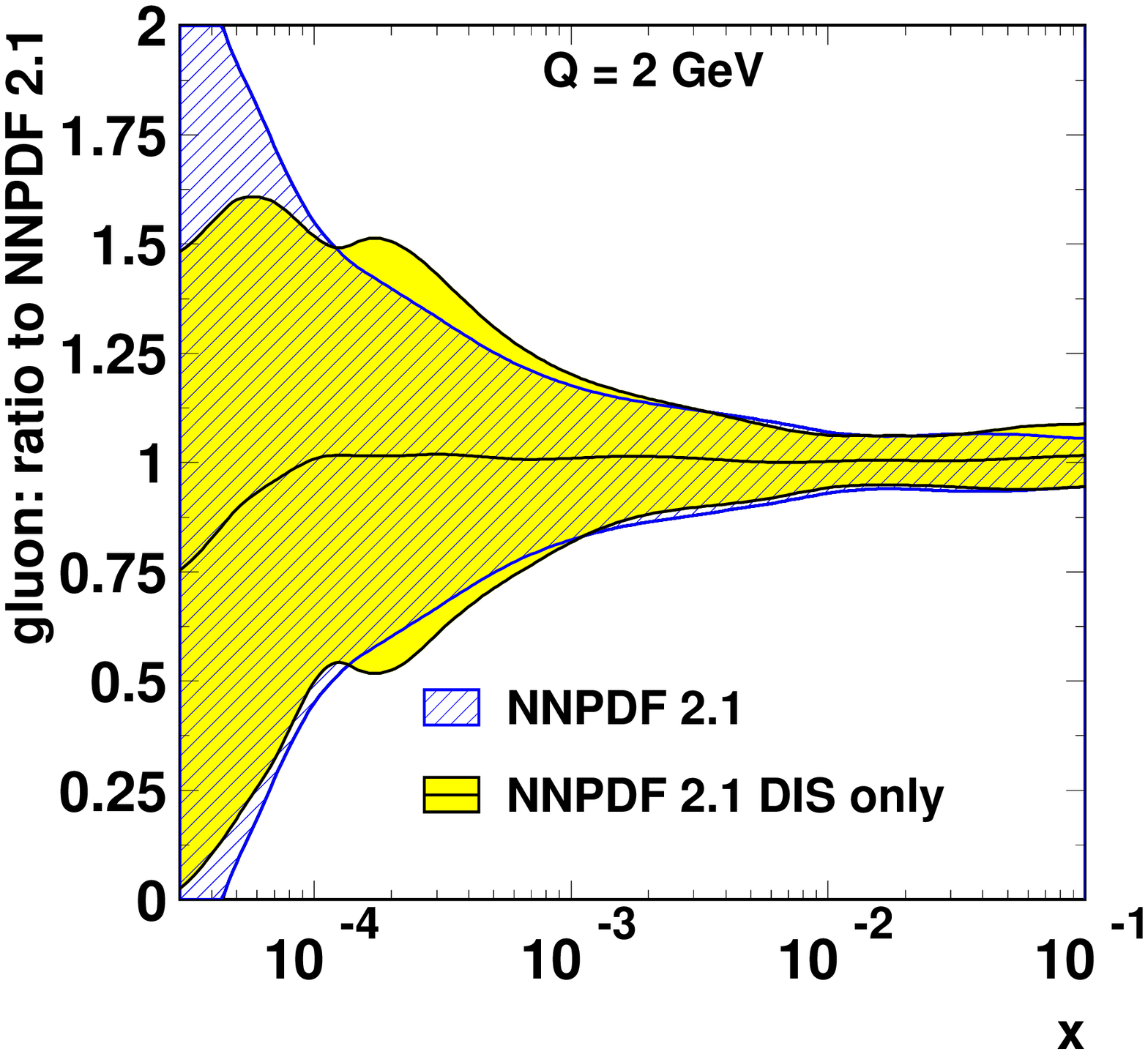} &
 \includegraphics[width=0.5\columnwidth]{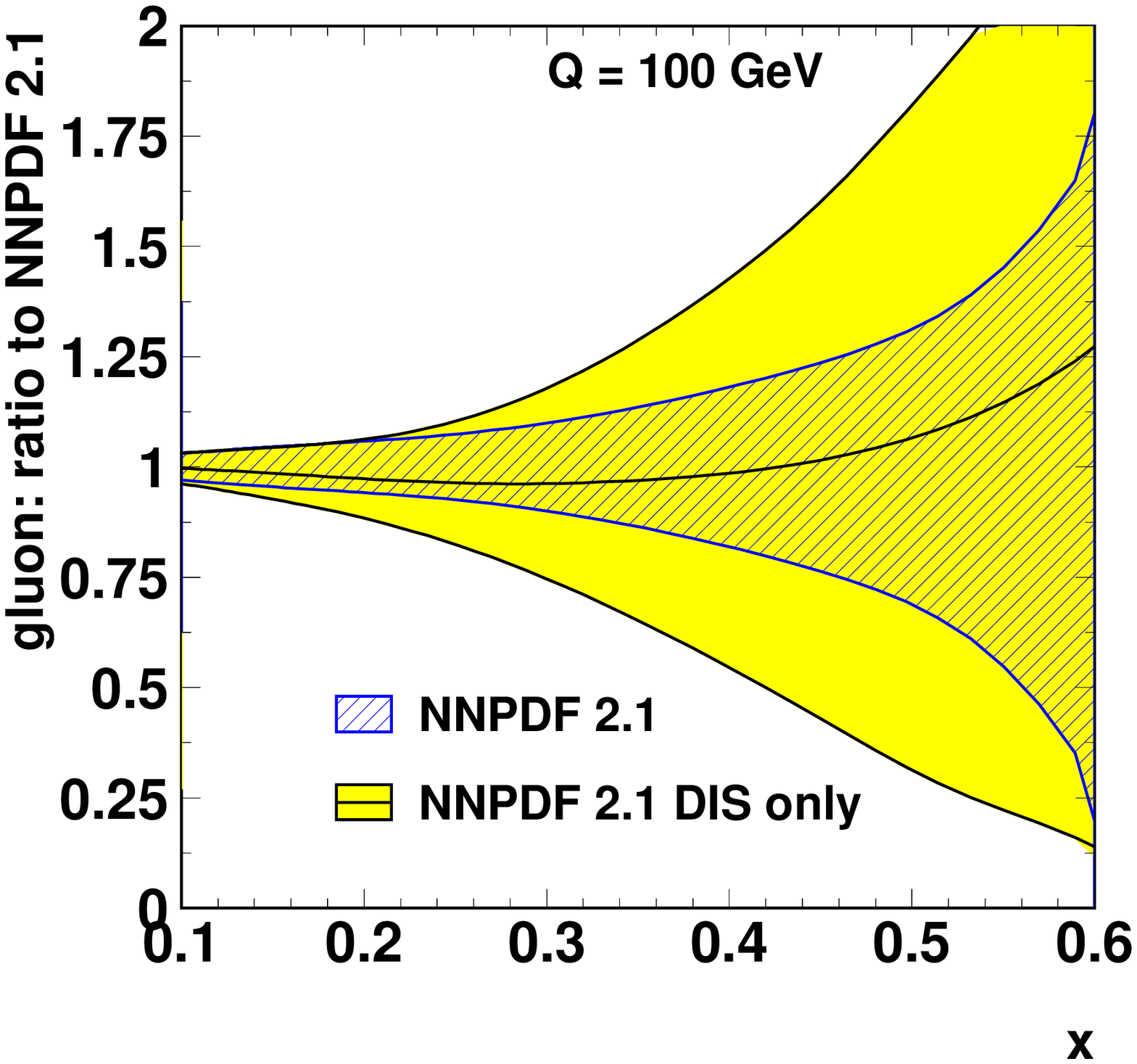} 
 \end{tabular}
  \caption{\sl 
  Comparison of the gluon density obtained from NNPDF2.1 and from a similar fit 
  restricted to the DIS data, at low and medium $x$ and $Q = 2$~GeV (left), and at 
  high $x$ and $Q = 100$~GeV (right). The ratio to the NNPDF2.1 density is shown, 
  together with $68 \%$ confidence level contours. 
 }
 \label{fig:impact_jets_fits}
 \end{figure}

Figure~\ref{fig:impact_jets_fits} shows the impact of Tevatron jet data on the determination
of the gluon density.
A fit similar to that of NNPDF2.1 has been performed, using DIS data only, and it  is compared
to the standard fit of NNPDF2.1.
At low $x$, both fits lead to a very similar gluon density, with the same relative uncertainty,
meaning that most of the constraints on the gluon density are
coming from DIS data (mainly HERA). 
However, at medium and high $x$, the non-DIS datasets (mainly the jet measurements from the
Tevatron experiments) provide a significantly improved uncertainty on the gluon density.

\subsubsection{The strange sea}
\label{sec:StrangeSea}

As seen in section~\ref{sec:CCFR-NuTev}, the exclusive production of di-muon
events in neutrino DIS experiments, measured by the NuTeV and CCFR experiments,
sets constraints on the density of strange quarks
for $ x \gsim 10^{-2}$ and $ x \lsim 0.3 - 0.4$, via the subprocess $W s \rightarrow c$.
The recent inclusion of these data in global fits
allows the strange content of the nucleon
to be studied in more details~\cite{MSTW08, Lai}. Previous fits assumed that
$s + \bar{s}$ was a constant fraction of the non-strange sea $\bar{u} + \bar{d}$
at the starting scale,
$$ s + \bar s = \kappa_s ( \bar u + \bar d)$$
 with $\kappa_s \sim 0.4-0.5$ reflecting the
suppressed probability to produce $s \bar{s}$ pairs compared to $u \bar{u}$ or $d \bar d$ pairs.
Recent fits however, give more freedom to the strange density
in particular at high $x$. This additional freedom leads indeed to an improved $\chi^2$.

Both the CTEQ6.6 and the MSTW08 analyses assume that, at low $x$, the 
strange density follows the same power-law as the light sea density,
i.e. $s + \bar s \propto x^\alpha$ with $\alpha$ set to the low-$x$ power of the light sea
(MSTW08) or to that of the $\bar{u}$ and $\bar{d}$ densities (CTEQ6.6).
They parameterise the strange density $s + \bar{s}$ as
$$ x s + x \bar{s} = A x^\alpha (1-x)^{\beta} P(x)$$
and fit the normalisation $A$ and the high-$x$ power $\beta$. Parameters defining  the polynomial
function $P(x)$ are also fitted in the CTEQ6.6 analysis, while in MSTW08 they are fixed to be the
same as those defining the $g(x)$ of the total sea.
The CTEQ6.6 analysis still assumes $s = \bar{s}$, while the MSTW08 analysis parameterises
the strange asymmetry as
$x s - x \bar{s} = A x^\alpha (1-x)^\beta (1 - x/x_0) $
where $x_0$ is given by the number sum rule of zero strangeness, and $\alpha$ is fixed to $0.2$
as the data do not constrain $A$ and $\alpha$ independently.

\begin{figure}[htb]
 \begin{tabular}{cc}
 \includegraphics[width=0.5\columnwidth]{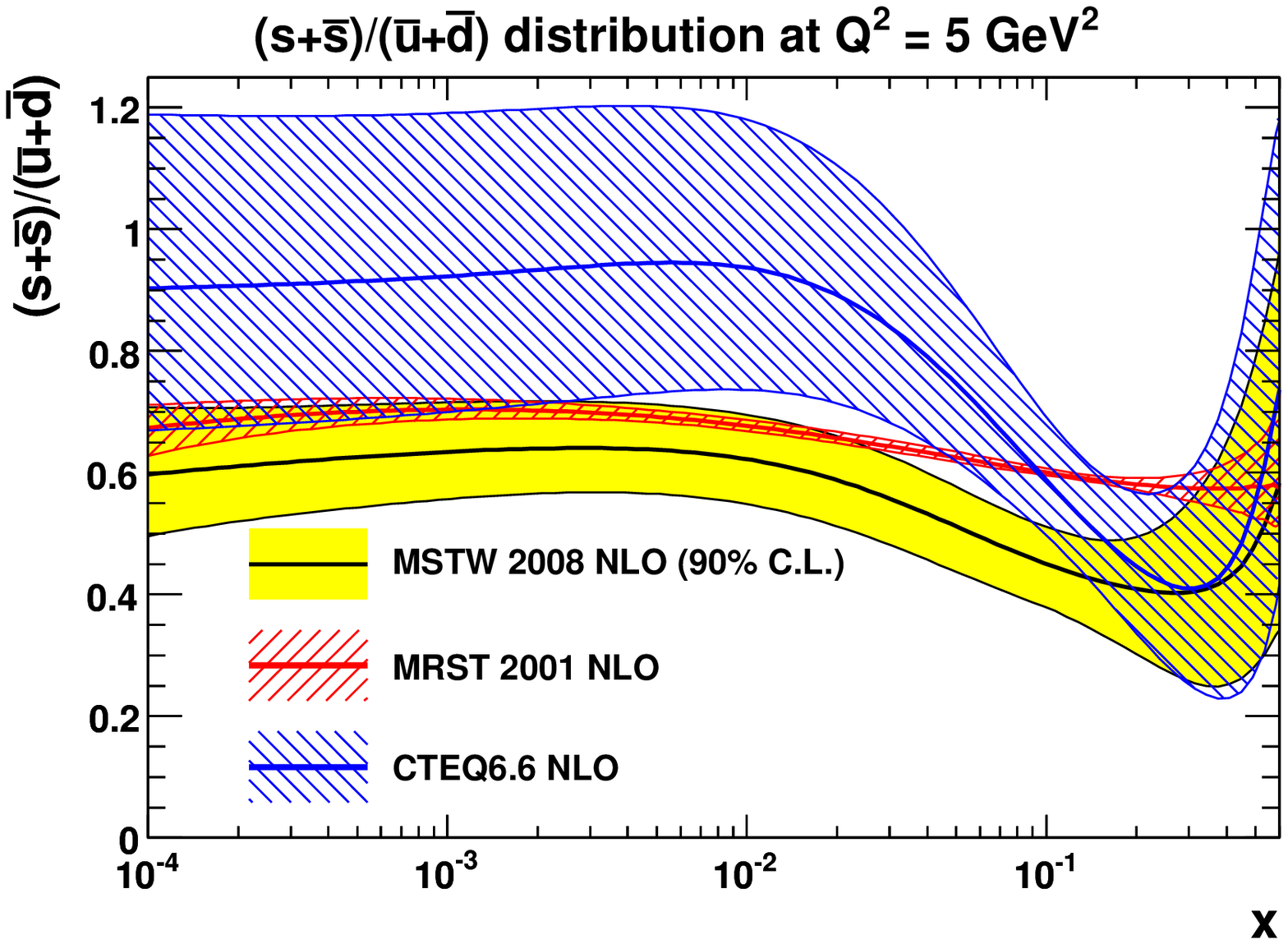} &
 \includegraphics[width=0.5\columnwidth]{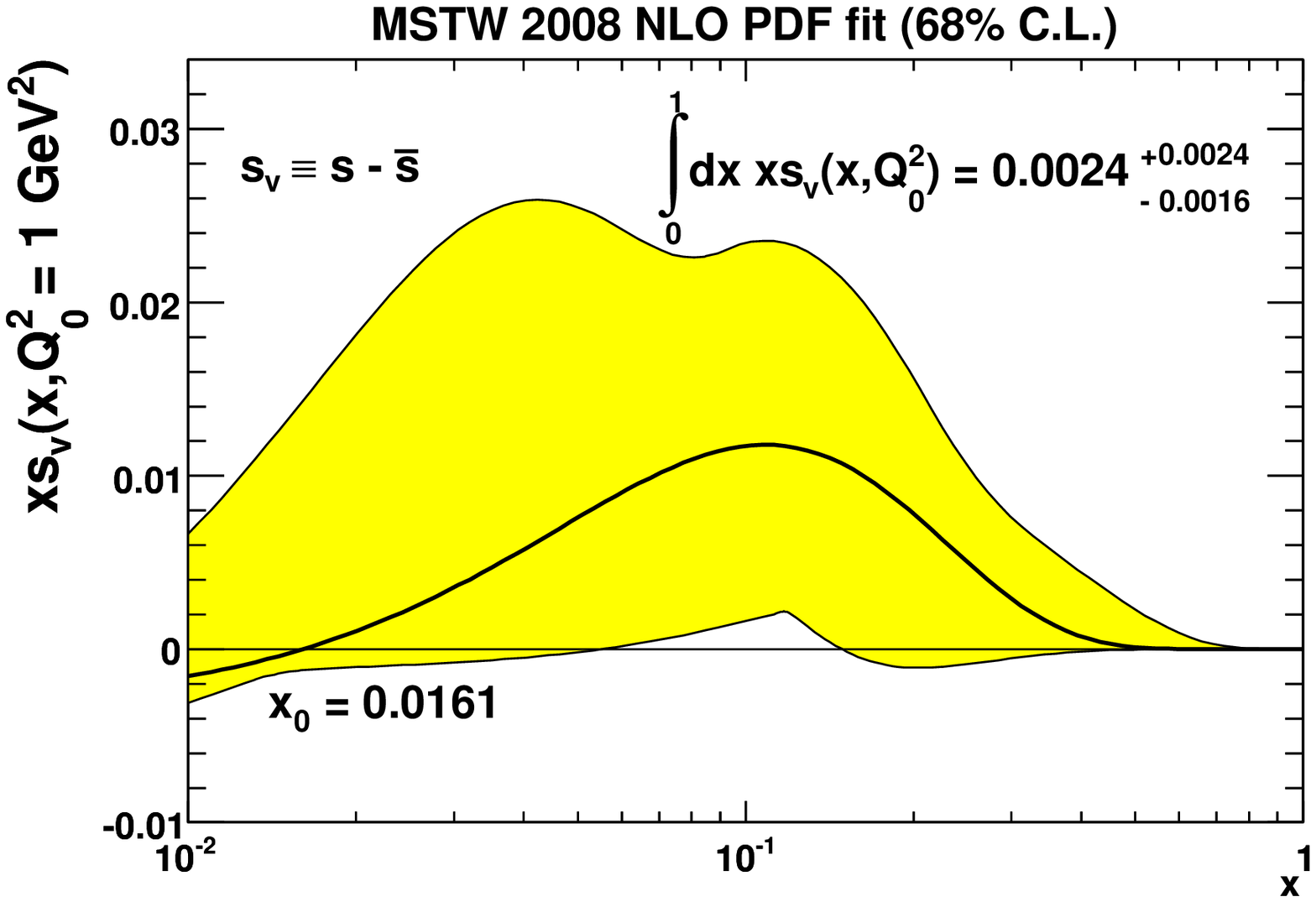}
 \end{tabular}
 \caption{\sl Left: The ratio of the input $s + \bar{s}$ distribution to that of
 the non-strange sea $\bar{u} + \bar{d}$ at $Q^2 = 5$~GeV$^2$, as obtained
 in the MSTW08, MRST2001 and CTEQ6.6 fits, with $2 \sigma$ uncertainty contours. Right: The strange asymmetry $xs - x \bar s$
  at $Q^2_0 = 1$~GeV$^2$, together with the $1 \sigma$ uncertainty. From~\cite{MSTW08}. }
 \label{Fig:splus}
 \end{figure}

Figure~\ref{Fig:splus} (left) shows the strange density obtained in the two fits, at $Q^2 = 5$~GeV$^2$.
In the MSTW08 analysis, $s + \bar{s}$ is smaller than $(\bar{u} + \bar{d})/2$, 
especially\footnote{The strange quark mass could explain this additional suppression
at high $x$, as this corresponds to low $W^2$, i.e. close to the production threshold.}
at large $x$.
The strange density from the CTEQ6.6 fit is considerably larger,
even in the range $10^{-2} < x < 10^{-1}$ which is directly constrained by the 
data. The larger uncertainty obtained in the CTEQ6.6 analysis is probably due to the
more flexible parameterisation. The uncertainty from the NNPDF2.1 fit, where
any parameterisation bias is largely removed, was seen to be even larger (see Fig.~\ref{fig:globalFits_Comparison}).
Note that the small uncertainty obtained in MRST2001
is due to the assumption $s + \bar s = \kappa_s ( \bar u + \bar d)$ that was made
in that fit.

The strange asymmetry $s - \bar s$ is actually very loosely constrained, as shown
in Fig.~\ref{Fig:splus} (right).
The existing data seem to indicate a positive value for the momentum asymmetry 
$\int dx \,x (s - \bar s) $ of the strange sea. This asymmetry has important consequences
for the $\sin^2 \theta_W$ anomaly reported by the NuTeV collaboration~\cite{Zeller:2001hh}.
From the asymmetry between $\sigma ( \nu_{\mu} N \rightarrow \nu_{\mu} X)$ and
$\sigma ( \bar{\nu}_{\mu} N \rightarrow \bar{\nu}_{\mu} X)$, NuTeV extracted a value
for $\sin^2 \theta_W$ that is $3 \sigma$ above the global average. Half that discrepancy 
can be explained by isospin violations~\cite{MRST2001}, and a positive value for 
$\int dx \,x (s - \bar s) $ would further reduce this NuTeV anomaly.

\subsubsection{Compatibility between the datasets}
\label{sec:fits_compatibility}

Except for some datasets of electroweak boson production at the Tevatron discussed
in section~\ref{sec:Wasym_Tevatron}, 
the global QCD analyses find in general a very good consistency of all
datasets with each other and with NLO QCD.

Some amount of tension is however observed between the $F_2$ data of
fixed target $\mu p$ experiments and the rest of the data, although not
consistently in all analyses.
In the NNPDF and CTEQ analyses, the $\chi^2$ of the NMC $F_2$ data is a bit large.
Since this was already the case in the early NNPDF analyses where
a parameterisation of the structure function $F_2$ was constructed without using pQCD,
this could reflect the fact that the data within this dataset show point-by-point fluctuations
which are larger than what is allowed by their declared uncertainty~\cite{NNPDF2.0}.
Some tension is also seen in the MSTW08 analysis between the BCDMS $F_2$ data and
the rest of the data, with the BCDMS $\mu p$ data tending to prefer a higher gluon
at high $x$ in order to accommodate the observed $Q^2$ dependence.
A similar observation was made in~\cite{Pumplin:2009sc} within the framework of the CTEQ analysis.
As the degree of compatibility between the BCDMS data and the rest of the data becomes better
when a higher $Q^2_{min}$ cut is applied, this may be an indication of 
non perturbative effects
in these $\mu p$ data at low $Q^2$, and / or of deviations from NLO DGLAP in the HERA 
measurements at very low $x$.

Some inconsistencies are also observed with the neutrino DIS data.
Discrepancies between the NuTeV and the older CCFR structure function
measurements at high $x$ are now understood by both groups, and the
NuTeV dataset is believed to be more reliable (see section~\ref{sec:CCFR-NuTev}). 
However, the CHORUS
measurements (obtained with a lead rather than an iron target) also
disagree with the NuTeV data at high $x$. As a result, the MSTW08
analysis includes the NuTeV and CHORUS data (which replace the CCFR measurements)
only for $x < 0.5$.
These NuTeV and CHORUS data were analysed together with the latest Drell-Yan
measurements from E866 in~\cite{Owens}, in a global fit similar to those
performed by the CTEQ collaboration. This fit yields a $d/u$ ratio which flattens
out significantly at high $x$. A tension is observed at high $x$:
the NuTeV data pull the valence distributions upward (which pulls
against the BCDMS and NMC data), while the E866 measurements prefer lower
valence distributions at high $x$. This tension is actually amplified
by the nuclear corrections applied to the NuTeV data.  \\

%------------------------------------------------
\subsection{Fits and calculations at NNLO}
%------------------------------------------------
\label{sec:NNLO}

\subsubsection{Status of NNLO fits}

Over the past years, an increasing number of QCD calculations have become
available at NNLO with the goal of reducing
the scale uncertainties on the resulting predictions compared to
their NLO counterparts. 
Consequently, parton densities are now extracted at NNLO by several global analyses.
NNLO PDFs have been published for the MSTW08, NNPDF2.1 (in~\cite{NNPDF_NNLO}) and NNPDF2.3, ABKM09~\cite{ABKM} and ABM11~\cite{ABM11} fits.
Preliminary NNLO PDFs have also been extracted within the HERAPDF framework (HERAPDF1.5)
and the CT group plans to release soon a NNLO set~\cite{CTEQNNLO}.

The NNLO analyses of ABKM09 and ABM11 include DIS data and Drell-Yan measurements from
fixed target experiments. The MSTW08 and NNPDF analyses include additional datasets
as they use the same data
as in their NLO fits. However approximations have to be made in order to include 
the Tevatron jet data, since the full NNLO corrections to jet cross sections
are not available yet. Both groups use the approximate NNLO calculation obtained
from threshold resummation~\cite{Kidonakis:2000gi} and implemented in the FastNLO package.
ABM11 also used this approach in fits made to check the impact that the Tevatron jet
data would have on their analysis~\cite{ABM11}, but their central fit sticks to the datasets
for which the theoretical calculation is exact.
This approximation is however believed to be robust as the threshold correction should be the only source
of large NNLO corrections~\cite{MSTW08}.

\begin{figure}[htb]
 \begin{tabular}{cc}
 \includegraphics[width=0.5\columnwidth]{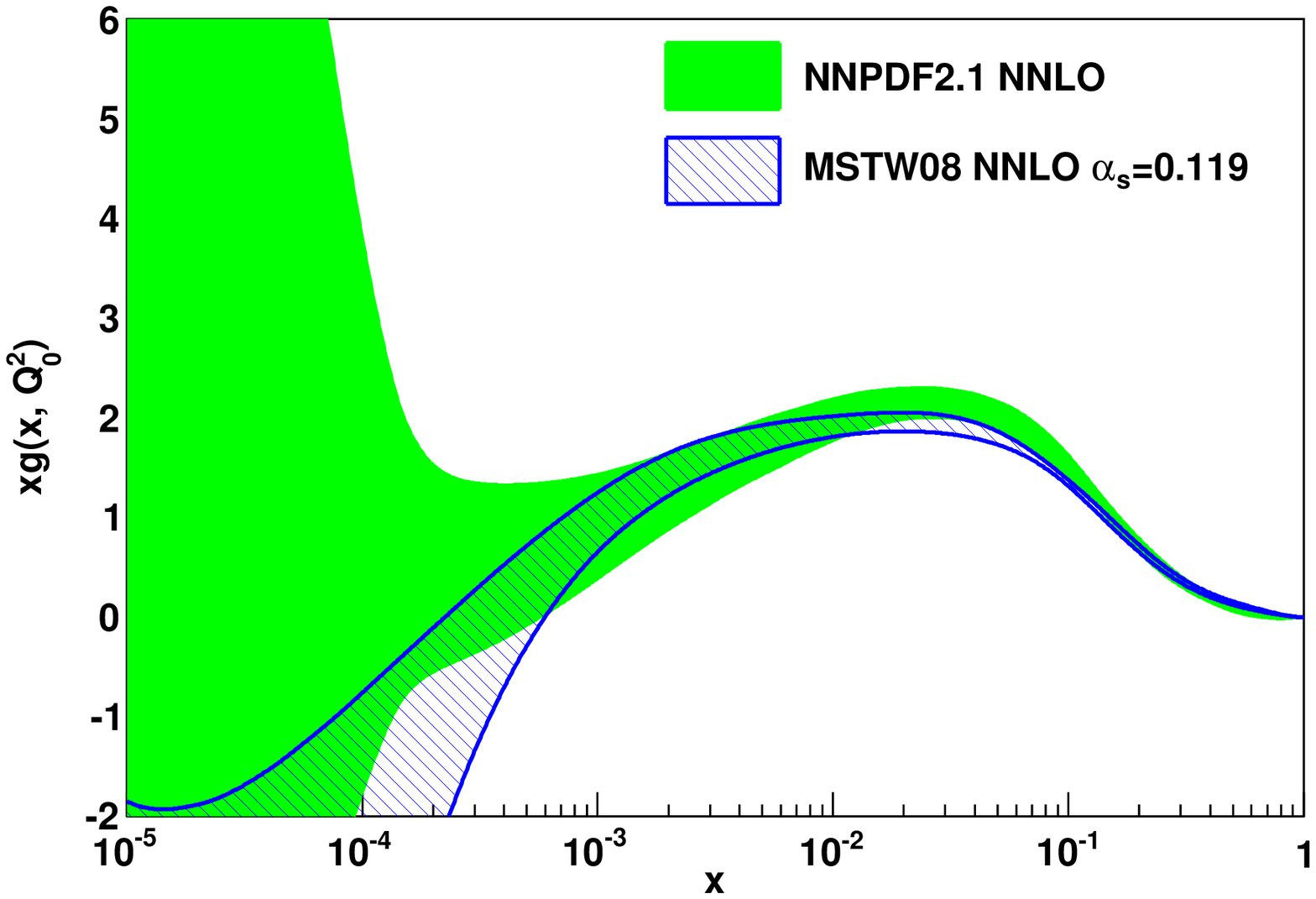} &
 \includegraphics[width=0.5\columnwidth]{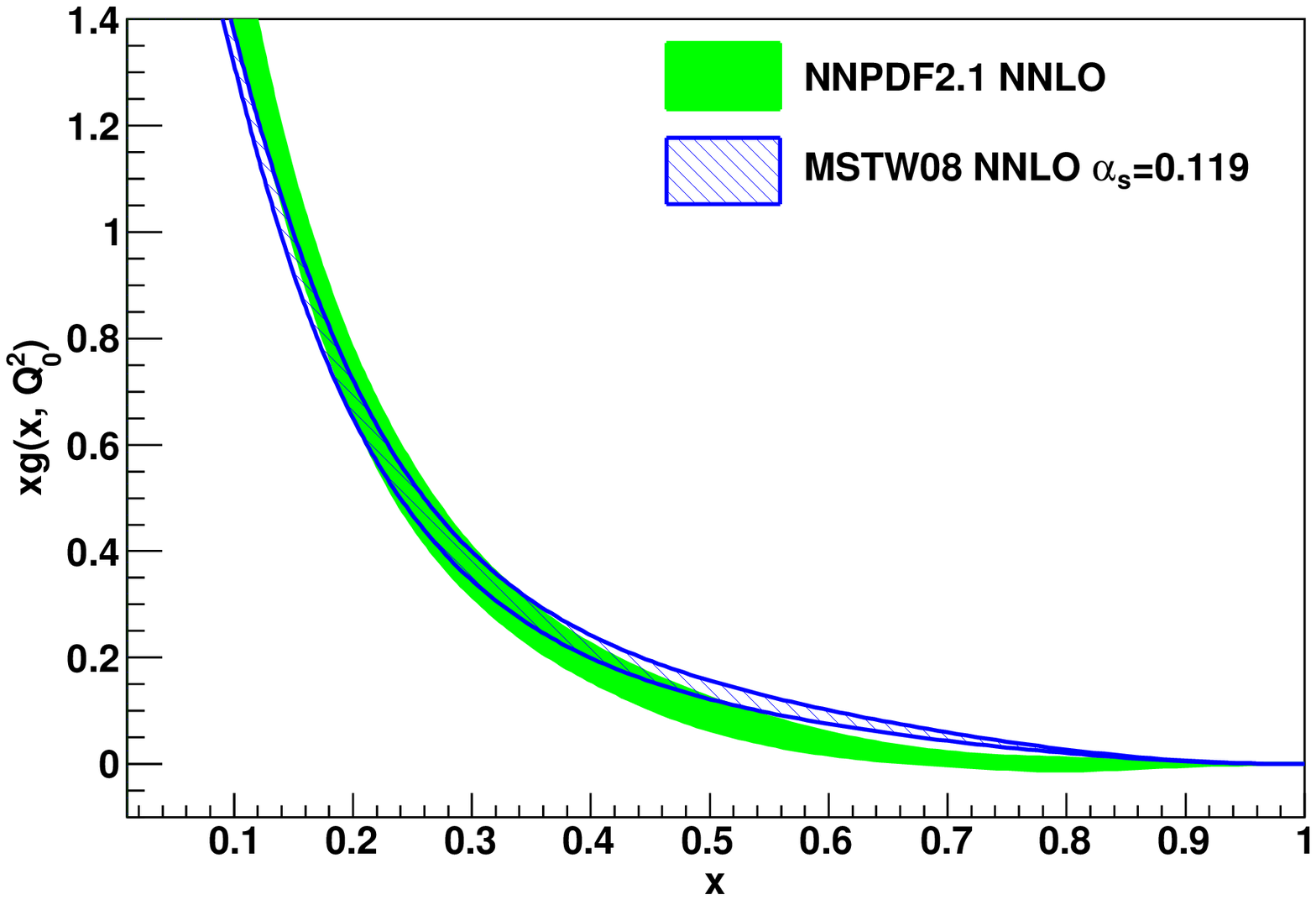}   \\
 \includegraphics[width=0.5\columnwidth]{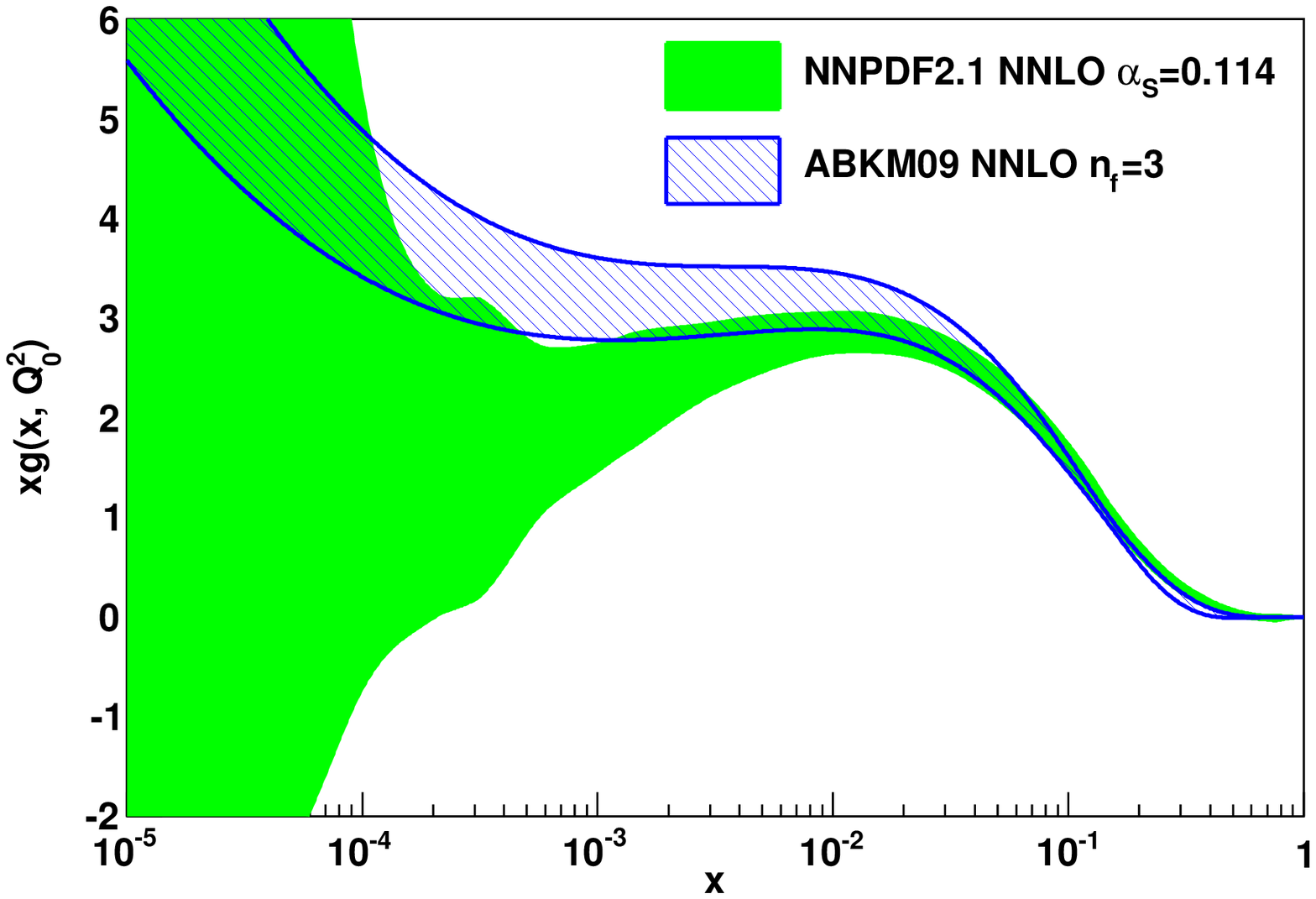} &
 \includegraphics[width=0.5\columnwidth]{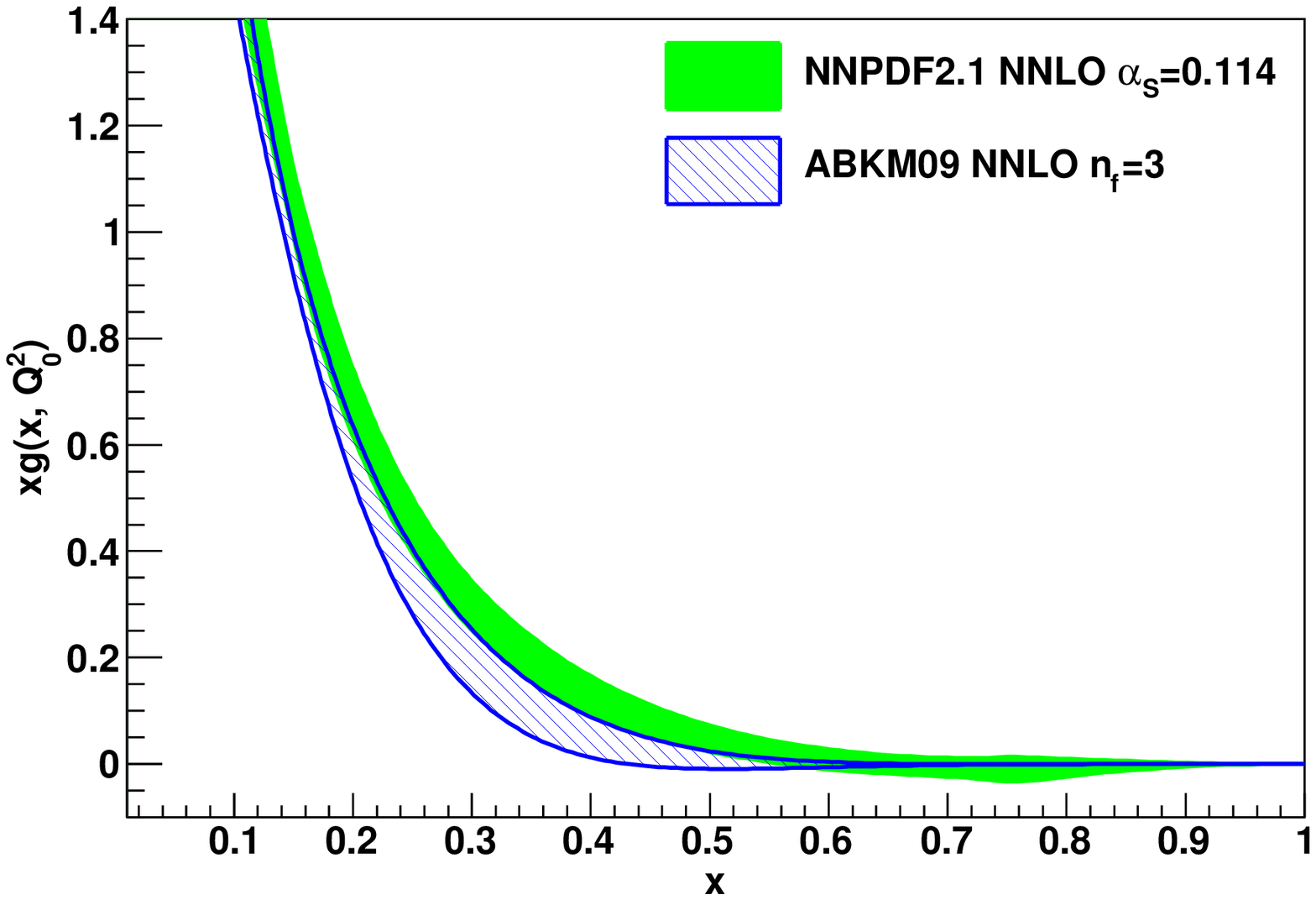}
 \end{tabular}
 \caption{\sl Comparison of the gluon density at $Q^2 = 2$~GeV$^2$ obtained in the NNPDF2.1,
 MSTW08 and ABKM09 NNLO analyses. The NNPDF2.1 density is shown for $\alpha_s = 0.119$ ($\alpha_s = 0.114$)
 in the top (bottom) row, such that it can be compared with the overlaid MSTW08 (ABKM09) density.
 Note that the ABKM uncertainties also include the uncertainty on $\alpha_s$ while for NNPDF and
 MSTW they are pure PDF uncertainties. From~\cite{NNPDF_NNLO}.}
 \label{fig:nnpdf_nnlo}
 \end{figure}

Figure~\ref{fig:nnpdf_nnlo} compares the gluon densities at $Q^2 = 2$~GeV$^2$, as obtained
by the NNPDF2.1, MSTW08 and ABKM09 analyses. 
At that low scale, the number of flavours is three in the GM-VFNS analyses of MSTW08 and NNPDF2.1,
hence the $n_f = 3$ set of the FFNS analysis of ABKM09 is used for the comparison.
Sizable differences can be observed in Fig.\ref{fig:nnpdf_nnlo}. 
The gluon distribution of ABKM09 is markedly different
from that of MSTW08, which at low scales becomes negative at low $x$. 
Part of the differences seen at low $x$ can be due to the fact that MSTW08 and ABKM09 use the individual
H1 and ZEUS data while NNPDF2.1 uses the more precise combined dataset. 
Indeed, the NNLO gluon density obtained from the ABM11 fit, which uses the combined HERA-I dataset,
is lower than that of ABKM09 and in better agreement with that of NNPDF2.1. The lower ABKM09
gluon distribution at high $x$ may come from the fact that the central NNLO fit of ABKM09
does not include the Tevatron jet data.
For other densities, the agreement between the central values is in general better,
although the uncertainty bands are different, being larger for NNPDF2.1.
Examples of NNLO predictions and of their PDF uncertainties for benchmark
processes at the LHC will be shown in chapter~\ref{sec:lhc}.

\subsubsection{The convergence of perturbative series and low ${\boldsymbol{x}}$ effects}
\label{sec:lowx}
 \begin{figure}[tbh]
 \centerline{\includegraphics[width=0.65\columnwidth]{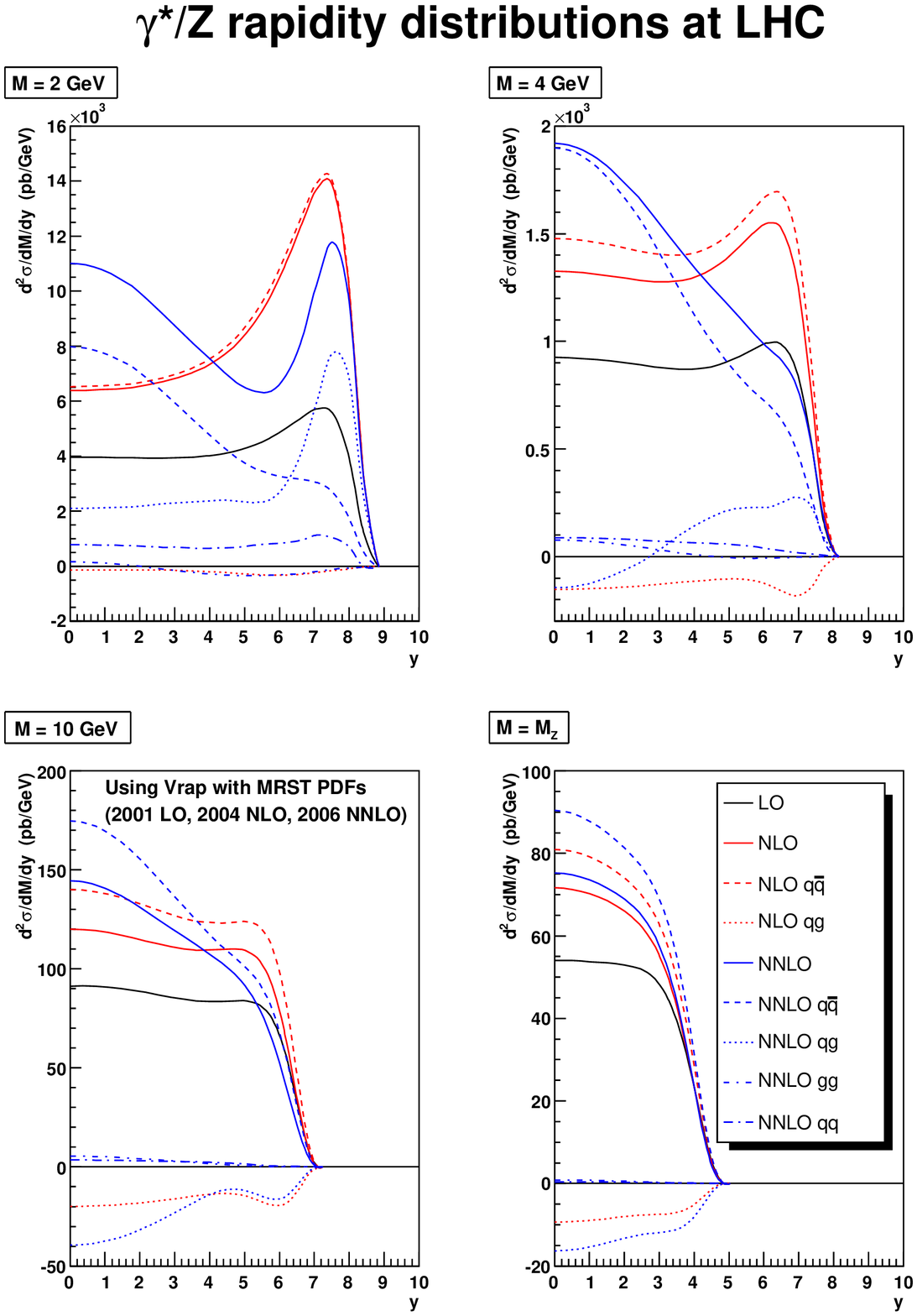}}
 \caption{\sl Drell-Yan cross section at the LHC in several mass bins, at LO, NLO and NNLO. From~\cite{Thorne:2008am}.
 }
 \label{fig:dy_thorne}
\end{figure}

For processes that do not involve low $x$ partons, calculating the cross sections to LO, NLO and NNLO
shows a reasonable convergence of the perturbative series.
For example, the NNLO cross sections for $W$ and $Z$ production at the LHC, obtained 
from the NNLO MSTW08 PDFs, is only $3-4 \%$
higher than the NLO cross section obtained from the NLO MSTW08 PDFs.
However, for processes involving low $x$ partons, convergence may not be reached
at NNLO. This is illustrated in Fig.~\ref{fig:dy_thorne} which shows the Drell-Yan cross sections
at LO, NLO and NNLO, in four mass bins. 
For di-lepton masses smaller than a few  $10$s of GeV, the NNLO and NLO predictions
are largely different, even in the central region; the difference being larger
for smaller masses.
This may indicate that, in part of the kinematic range where the LHC experiments will
make measurements (for example, LHCb should measure Drell-Yan at low masses in the
rapidity range $2 < y < 5$, see chapter~\ref{sec:lhc}),
a resummation of terms in $\ln(1/x)$ may be needed.

\begin{figure}[htb]
 \centerline{\includegraphics[width=0.7\columnwidth]{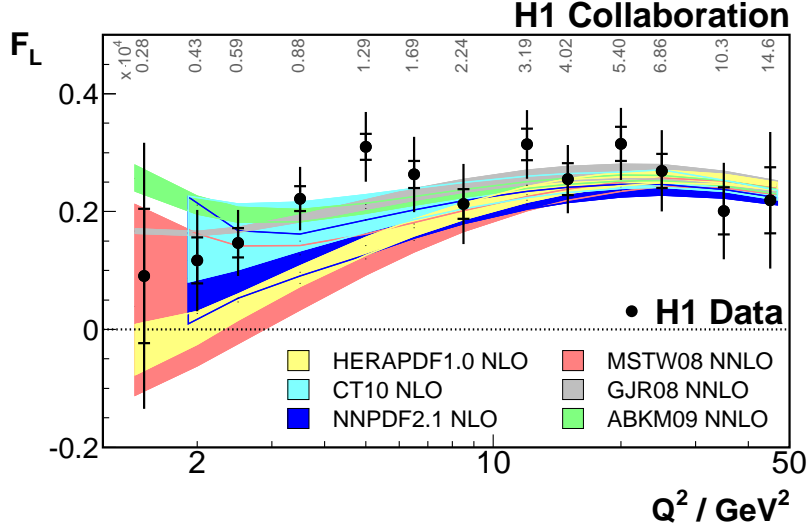}}
 \caption{\sl The longitudinal structure function $F_L$ measured by the H1 experiment, and compared with
  various NLO and NNLO predictions. From~\cite{H1FL} . }
 \label{fig:H1_FL}
 \end{figure}

The measurement of the longitudinal structure function $F_L$ at HERA~\cite{H1FL,Chekanov:2009na} provides
another test-bench for low $x$ effects that are not accounted for in the NNLO DGLAP equations.
A resummed calculation was shown to best describe the data~\cite{White:2006yh, ReviewThorne}.
However, as shown by Fig.~\ref{fig:H1_FL}, the fixed order DGLAP predictions are, in general,
in reasonable agreement with the measurement, within the rather large uncertainties.

The need for $\ln(1/x)$ resummations was also investigated
by studying exclusive final states, such as forward pions or forward jets at HERA.
The measurements were compared to fixed order DGLAP predictions, and to predictions
based on the BFKL equation~\cite{Kuraev:1977fs, Balitsky:1978ic}, which involve un-integrated parton densities\footnote{The
un-integrated gluon density needed for these predictions is calculated from the
usual gluon density via $\int^{Q^2} dk_T^2 / k_T^2 xf(x, k_T^2) = x g(x, Q^2)$. }.
No conclusive evidence for effects of BFKL dynamics was observed.

Besides resummations, the evolution of PDFs at very low $x$ is expected to be
affected by saturation effects, due to parton recombinations. 
Saturation would lead to a taming of the rise of $F_2$ at low $x$. Such an effect has
been looked for in HERA data,
by investigating the slopes of $F_2$~\cite{H1FL}, but was not observed in the $x$ and $Q^2$
range of the measurements.
A possible hint for saturation in HERA data may come from the energy dependence of diffractive
interactions, which was seen to be the same as that of the total cross section~\cite{Chekanov:2002bz}.
These aspects can be addressed within dipole models (see~\cite{Motyka:2008jk, Gelis:2010nm} and
references therein); a deeper discussion goes beyond the scope of this review.

%%%%%%%%%%%%%%%%%%%%%%%%%%%%%%%%%%%%%%%%%%%%%%%%%%%%%%%
\section{PDF Constraints from the LHC}
%%%%%%%%%%%%%%%%%%%%%%%%%%%%%%%%%%%%%%%%%%%%%%%%%%%%%%%
\label{sec:lhc}

The LHC $pp$ collision physics programme is driven by the search for
new physics and the 
understanding of electroweak symmetry breaking.
Precise theoretical
predictions of background processes are needed for a discovery,
whereas accurate predictions of new phenomena are needed for the
interpretation of exotic physics signals or for verification of the
Higgs boson properties. This programme is now well under way with
about $5$~fb$^{-1}$ of luminosity delivered to the ATLAS and CMS
experiments in 2011, and more than $20$~fb$^{-1}$ at
$\sqrt{s}=8$~TeV expected by 
the first long shutdown of the LHC in 2013.
As discussed in
section~\ref{exp:nonDIS} measurements from the Tevatron $p\bar{p}$
collider provide important PDF constraints beyond those obtained from
DIS data. Similarly it is expected that measurements from the LHC
experiments will also improve our knowledge of proton structure.

\subsection{The LHC experiments}

\begin{figure}[htb]
\begin{center}
\includegraphics[width=0.5\columnwidth]{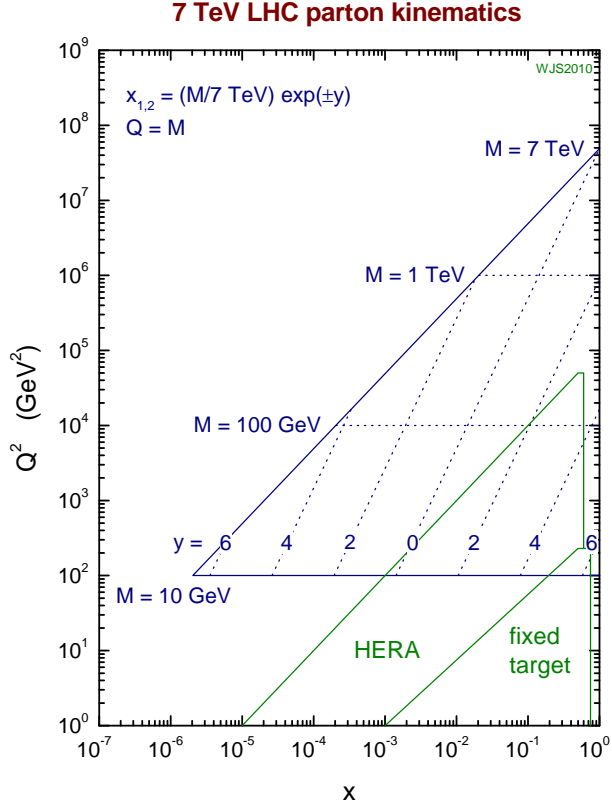}
\end{center}
 \caption{\sl Kinematic phase space accessible at the LHC with $pp$
 collisions at $\sqrt{s}$=$7$~TeV. 
 From~\cite{StirlingsPlots}, adapted from~\cite{Campbell:2006wx}. }
\label{fig:lhckine}
\end{figure}

The kinematic region opened up to the ATLAS, CMS and LHCb
experiments\footnote{The LHC experiment ALICE 
whose main goal is the study of heavy ion physics
will not be discussed in this
article.}  in the initial phase of LHC operation at $\sqrt{s}=7$~TeV
is shown in Fig.~\ref{fig:lhckine}. The lowest $Q$ is set by available
trigger thresholds and the lowest $x$ is determined by detector
rapidity ($y$) acceptance. The ATLAS and CMS experiments are largely
limited to $|y|<2.5$. For $W/Z$ production from partons with momentum
fractions $x_1$ and $x_2$ (here, $x_1 < x_2$ by convention) $M_{W,Z}^2=sx_1 x_2$ and the boson
rapidity is given by Eq.~\ref{eq:DY_y}.
This restricts the $x$ range at $\sqrt{s}=7$~TeV to approximately
$10^{-3} < x< 10^{-1}$. In contrast the LHCb experiment with more
forward instrumentation is able to access the region $2<y<5$ which
corresponds to $10^{-4} < x_1 < 10^{-3}$ and $0.1< x_2 < 1$ for the
same $Q=M_{W,Z}$. The overall reach in $x$ will be
extended by a further factor of two with $\sqrt{s}\simeq 14$~TeV operation
expected by 2015. 

\subsubsection{The ATLAS and CMS detectors}
The ATLAS~\cite{atlasdetector} and CMS~\cite{cmsdetector} detectors
are designed as multi-purpose experiments to exploit the full physics
potential at the LHC. They are segmented into a central barrel part
and two endcap regions. The innermost part of the detectors consist of
precision silicon pixel and strip tracking detectors close to the
nominal interaction points providing charged particle momentum
reconstruction over the region $|\eta|<2.5$. For ATLAS the silicon
trackers are supplemented by a surrounding straw-tube transition
radiation tracker for $|\eta|<2.0$ to enhance electron
identification. Both detectors have a large solenoid field axial with
the LHC beamline. The $2$~T field in the case of ATLAS encloses the
tracking and the electromagnetic calorimeter, whereas the trackers,
and both electromagnetic and hadronic calorimeters are immersed in the
$3.8$~T field of CMS.

The ATLAS electromagnetic and hadronic calorimeters extend to
$|\eta|<3.2$ and use a combination of liquid argon and tiled
scintillator technologies as the active media for energy sampling. A
very forward calorimeter provides additional coverage for
$3.2<|\eta|<4.9$. The CMS electromagnetic and hadronic calorimeters
use lead-tungstate crystals and scintillating plates for energy
sampling respectively, in the region $|\eta|<3.0$,
and are supplemented by an additional sampling Cherenkov calorimeter
in the forward region covering $3.0<|\eta|<5.0$.

Muons are measured in detectors located outside of the magnet
solenoids. The CMS design places the detectors inside a steel return
yoke for the solenoid which provides a bending field for the muons and
covers $|\eta|<2.4$. The ATLAS muon spectrometer uses three large
superconducting toroid magnet systems and is able to measure muons
over the range $|\eta|<2.7$.

\subsubsection{The LHCb detector}

The LHCb detector~\cite{lhcbdetector} is primarily designed to study properties of
$B$-meson decays at the LHC, including $CP$ violation and observation
of rare $B$ decays. It also has a programme of QCD and EW physics
measurements which are of relevance to this article. The detector is
a single arm forward spectrometer covering the region $2<\eta<5$. A
precision silicon strip vertex detector is located close to the
interaction region. Further strip silicon tracking devices are located
on either side of a dipole magnet, supplemented by straw-tube tracking
chambers. A ring imaging Cherenkov detector is used to help identify
charged hadrons. Electromagnetic and hadronic calorimeters located
downstream of the magnet distinguish electrons, photons and
hadrons. Muons are detected in multi-wire proportional chambers
furthest from the interaction region.

\subsection{Benchmark cross section predictions}
\label{lhc-benchmark}
\begin{figure}[htb]
\begin{center}
\includegraphics[width=0.49\columnwidth]{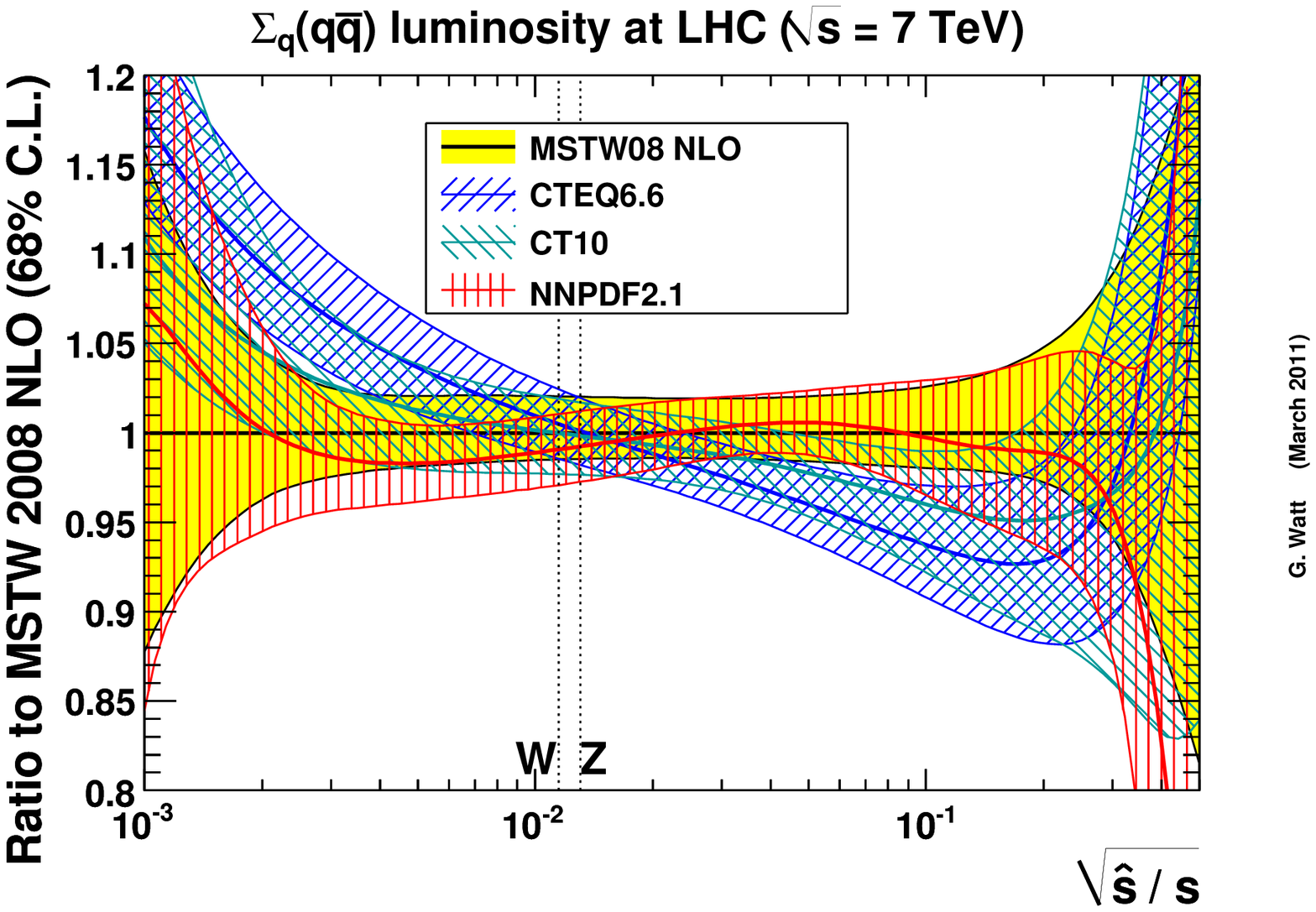}
\includegraphics[width=0.49\columnwidth]{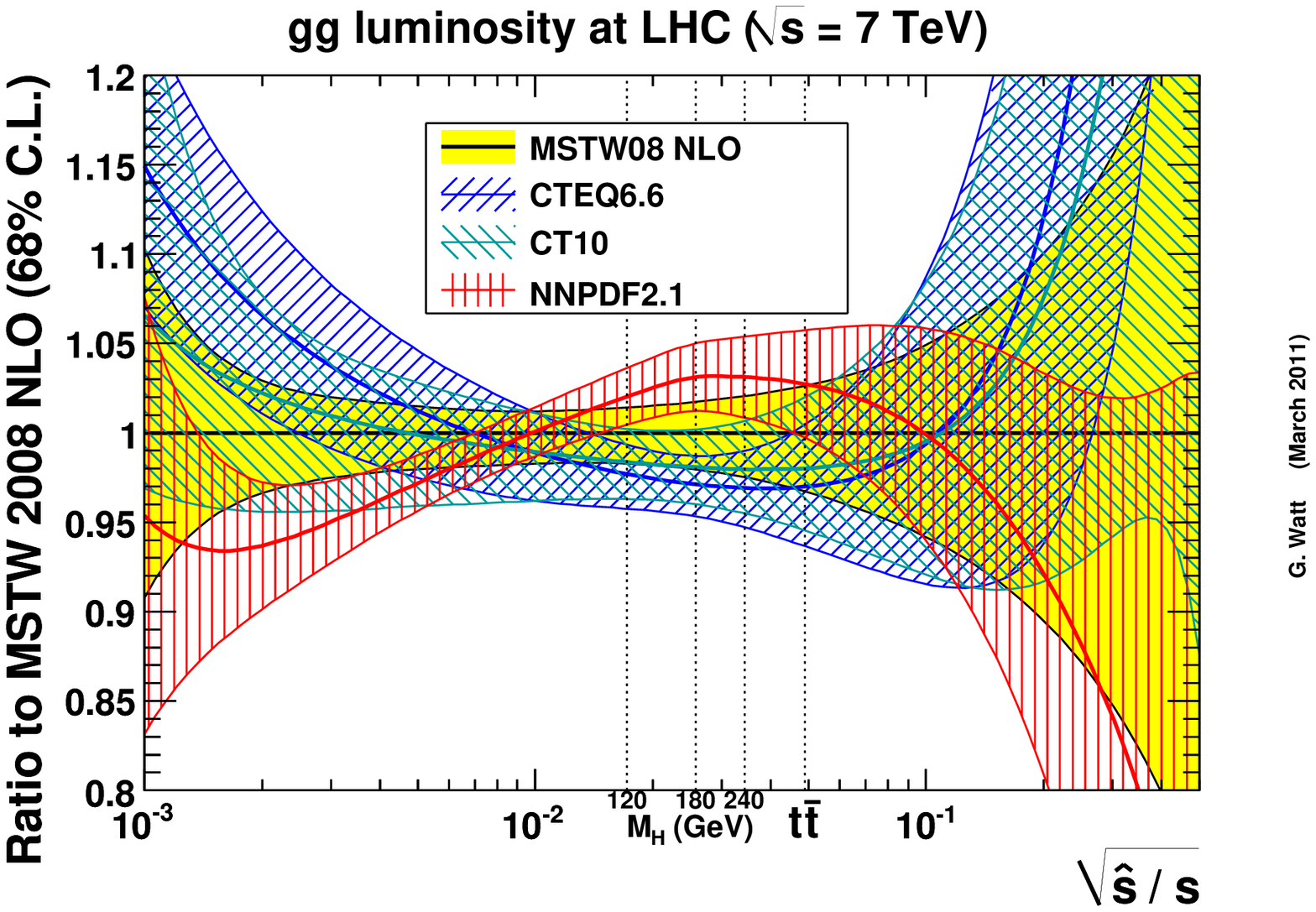}
\end{center}
\caption{\sl  Parton luminosities for the LHC at 7 TeV for the
combination $\sum_q q+\bar{q}$ (left) and $gg$
(right). From~\cite{Watt:2011kp}.  }
\label{fig:partonlumi}
\end{figure}

Parton luminosities are a convenient means of estimating the PDF
contributions to, and the $\sqrt{s}$ dependence of hadronic cross
sections for given combinations of partons~\cite{Ellis:1991qj}. The
parton luminosity for the combination $\sum_q q+\bar{q}$ is
relevant, for example, for $Z^0$ production, whereas the combination $gg$ is of
importance for Higgs production at the LHC. Using
Eq.~\ref{eq:tau}, $\tau=x_1\cdot
x_2=\hat{s}/s$ where $\hat{s}$ is the partonic centre-of-mass energy, the
differential luminosities
$\frac{\partial{\mathcal{L}}}{\partial\hat{s}}$ are defined as:
\begin{eqnarray}
\frac{\partial{\mathcal{L}_{\sum q+\bar{q}} }}{\partial\hat{s}}
&=& \frac{1}{s}\int^1_{\tau}\frac{dx}{x} \sum_{q} \left(f_q(x,\hat{s})f_{\bar{q}}(\tau/x,\hat{s})
+ [q \leftrightarrow \bar{q}] \right) \nonumber \\
\frac{\partial{\mathcal{L}_{gg} }}{\partial\hat{s}}
&=& \frac{1}{s}\int^1_{\tau}\frac{dx}{x}  f_g(x,\hat{s})
f_g(\tau/x,\hat{s}) \,\,.
\end{eqnarray}
The ratio of several NLO parton luminosities to MSTW08 are compared in
Fig.~\ref{fig:partonlumi} where very good agreement between all sets
is attained for the $W,Z$ resonance region, but diverge rapidly at
higher or lower fractional partonic centre-of-mass energy. The level of
agreement is similar for the $gg$ combination which shows a large
spread of predictions which are in some cases outside the uncertainty
bands of some of the predictions. This has led to a debate on the
best way to estimate PDF uncertainties for cross section predictions
incorporating the spread between PDF sets, and is discussed below.

A series of benchmark cross sections have been calculated at NLO and
NNLO~\cite{Watt:2011kp} in order to review the consistency of the most current PDF
sets available (MSTW08,
CTEQ6.6, CT10, NNPDF2.1,
HERAPDF1.0, ABKM09 and
GJR08). The chosen processes are $W$, $Z$ and $t\bar{t}$
production cross sections, as well as Higgs production with masses of
$M_H=120, 180, 240$ GeV. The cross sections are determined for fixed
values of $\alpha_s$.

\begin{figure}[htb]
\begin{center}
\includegraphics[width=0.49\columnwidth]{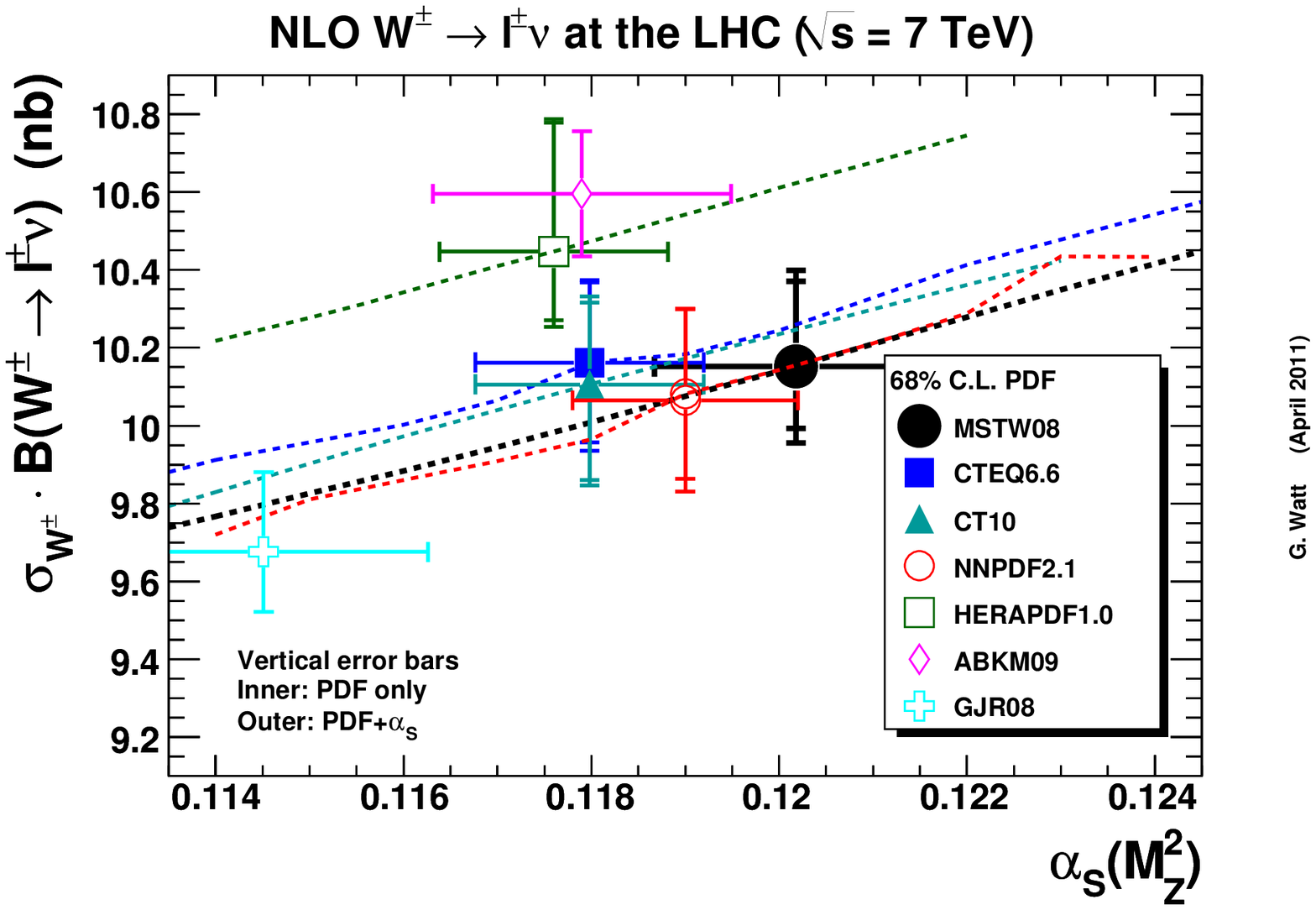}
\includegraphics[width=0.49\columnwidth]{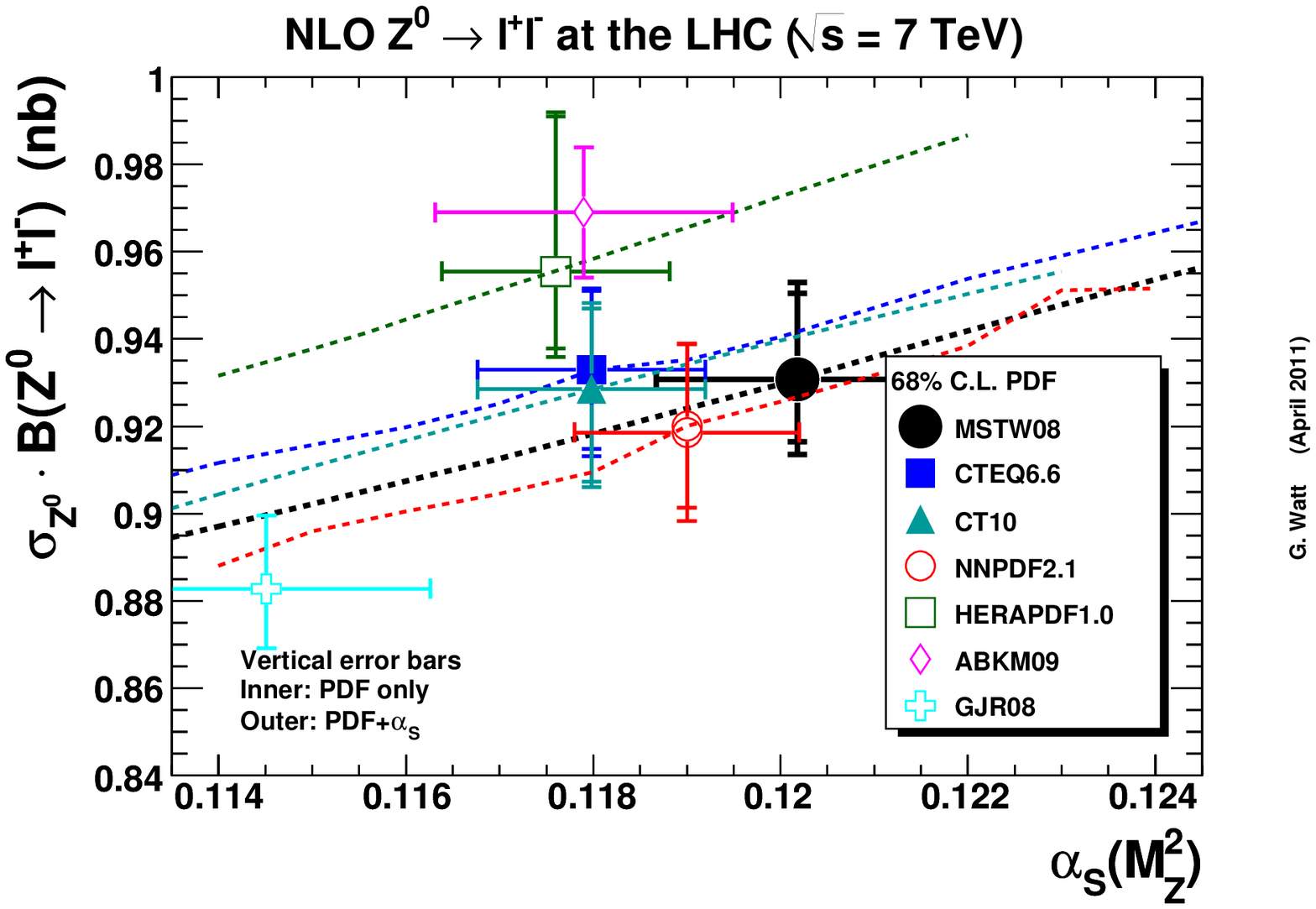}
\includegraphics[width=0.49\columnwidth]{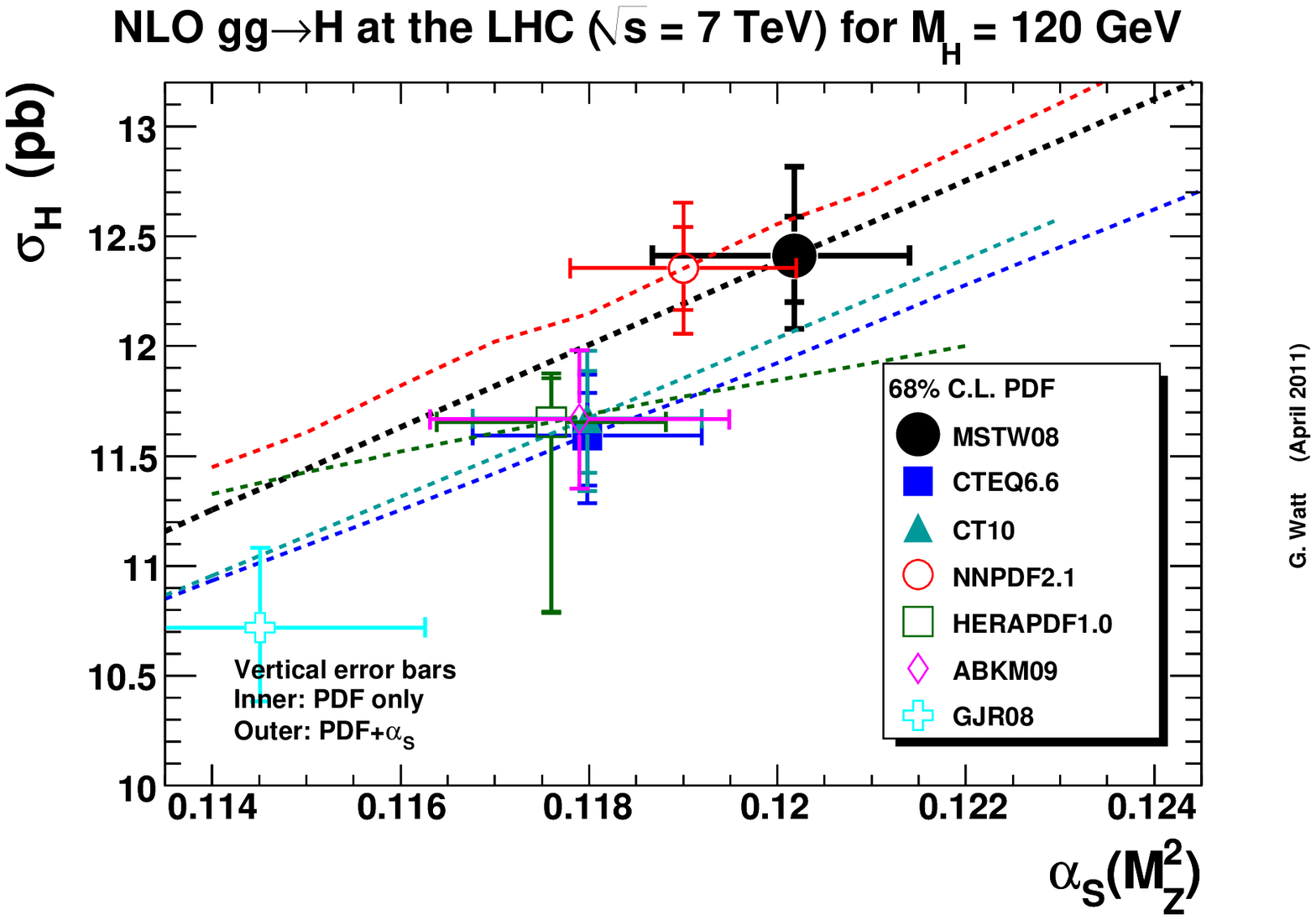}
\includegraphics[width=0.49\columnwidth]{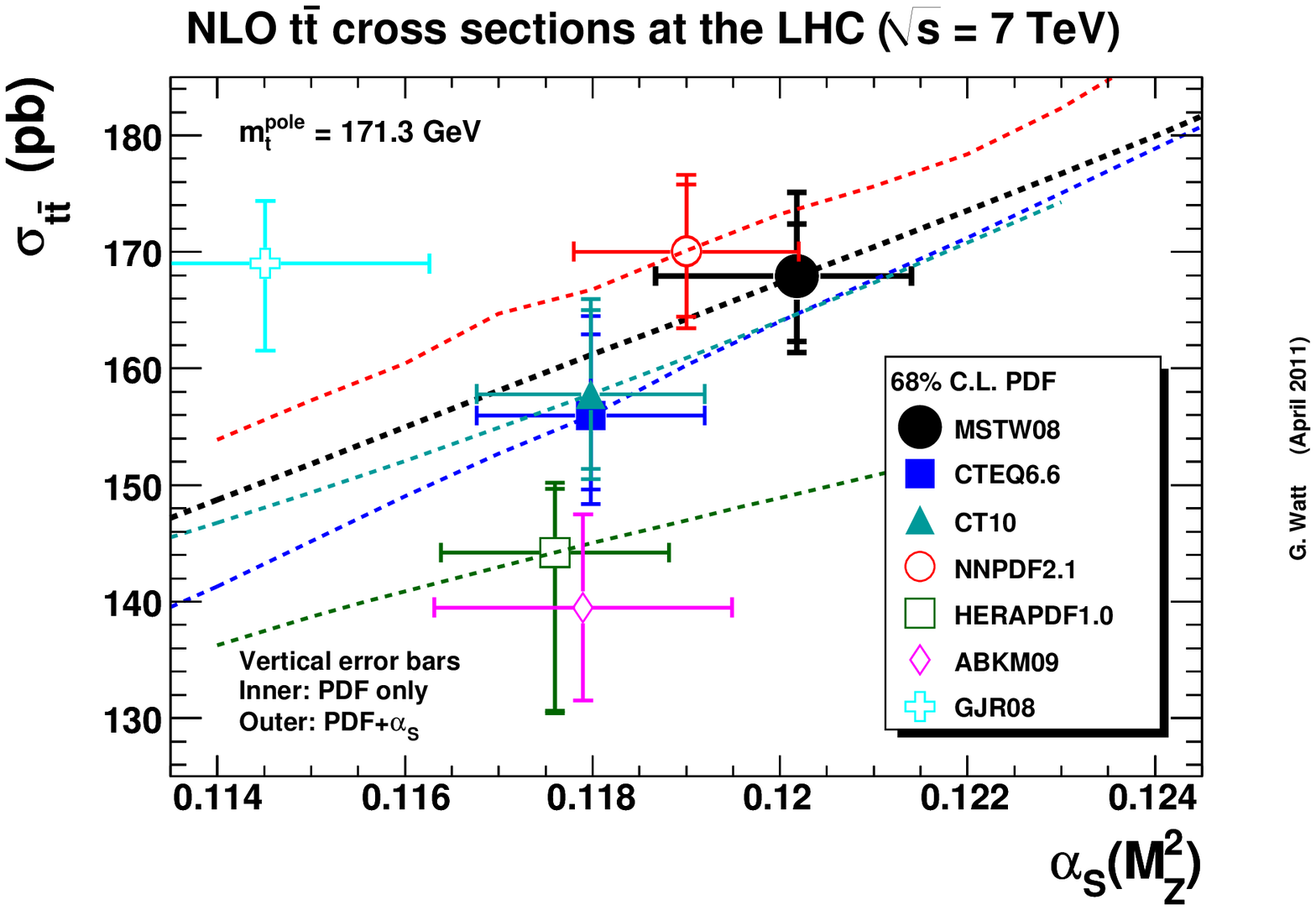}
\end{center}
 \caption{\sl Predicted NLO cross sections for the LHC at
 $\sqrt{s}=7$~TeV for $W^++W^-$ production (upper left), $Z^0$
 production (upper right),
 $gg\rightarrow H$ for $M_H=120$~GeV (lower left) and $t\bar{t}$
 production (lower right) with $68\%$ confidence level uncertainties
 as a function of $\alpha_s$. The inner vertical error bar corresponds
 to the PDF uncertainty and the outer error bar includes the
 $\alpha_S$ uncertainty. The horizontal error bar shows the
 $\alpha_S(M_Z^2)$ range considered for the uncertainty.
From~\cite{Watt:2011kp}.  }
\label{fig:wznnlo}
\end{figure}

An example of the NLO benchmark predictions is shown in
Fig.~\ref{fig:wznnlo}. Each point is plotted at the value of
$\alpha_s$ used in the central fit of the analysis. The dashed curves show the $\alpha_s$
variation using alternative PDFs from each group.  The precision of
the absolute cross section predictions at $68\%$ CL is broadly
similar for each PDF set, i.e. $\sim 2-3\%$ for $W^++W^-$ production
and $\sim 2\%$ for $Z$ production. It is however apparent that these
uncertainties do not fully cover the spread of predictions $\sim
6\%$. As expected the ratio of the $W^++W^-$ and $Z$ cross
sections (not shown) are obtained with greater precision and show a
smaller spread.  This is due to the fact that the numerator and
denominator of the ratio are both largely sensitive to the PDF
combination $u+d$, and that the $\alpha_s$ dependence almost cancels.

In addition to these variations the theoretical uncertainty should
also take into consideration the effects of neglected
higher-orders. These are usually estimated by varying the $\mu_F$ and
$\mu_R$ scales within a factor of two of the default choice usually
taken to be $\mu_F=\mu_R=M_{Z,W,H}$. The influence of the scale
uncertainty depends on the cross section under study, and can be
$~3\%$ for $Z$ production at NLO but is dramatically reduced to
$0.6\%$ at NNLO~\cite{Watt:2011kp}. 

Production of $t{\bar t}$ pairs at the LHC for $\sqrt{s}=7$~TeV is
dominated by $gg$ initial states which account for $80\%$ of the
production cross section and at threshold probes $x\sim
2m_t/\sqrt{s}=5\times10^{-2}$~\cite{Watt:2011kp}. By contrast $W$ and
$Z$ resonant production is dominated by $q{\bar q}$ pairs probing
$x\sim 2\times10^{-2}$. Thus $W,Z$ cross sections are anti-correlated
with $t{\bar t}$ production since an enhanced gluon distribution at
higher $x$ would lead to a reduced quark distribution at lower $x$
through the sum rules~\cite{Campbell:2006wx}. The predictions for
$t{\bar t}$ production are as yet only approximately known at
NNLO. The predictions with different PDF sets~\cite{Watt:2011kp} are
calculated at NLO and NNLO and show considerable range of about
$\pm10$\% which is larger than the uncertainties estimated from a
single PDF set at $68\%$~CL. The $90\%$~CL uncertainty bands give a
better reflection of the variation in the predictions.

\begin{figure}[htb]
\begin{center}
\includegraphics[width=0.49\columnwidth]{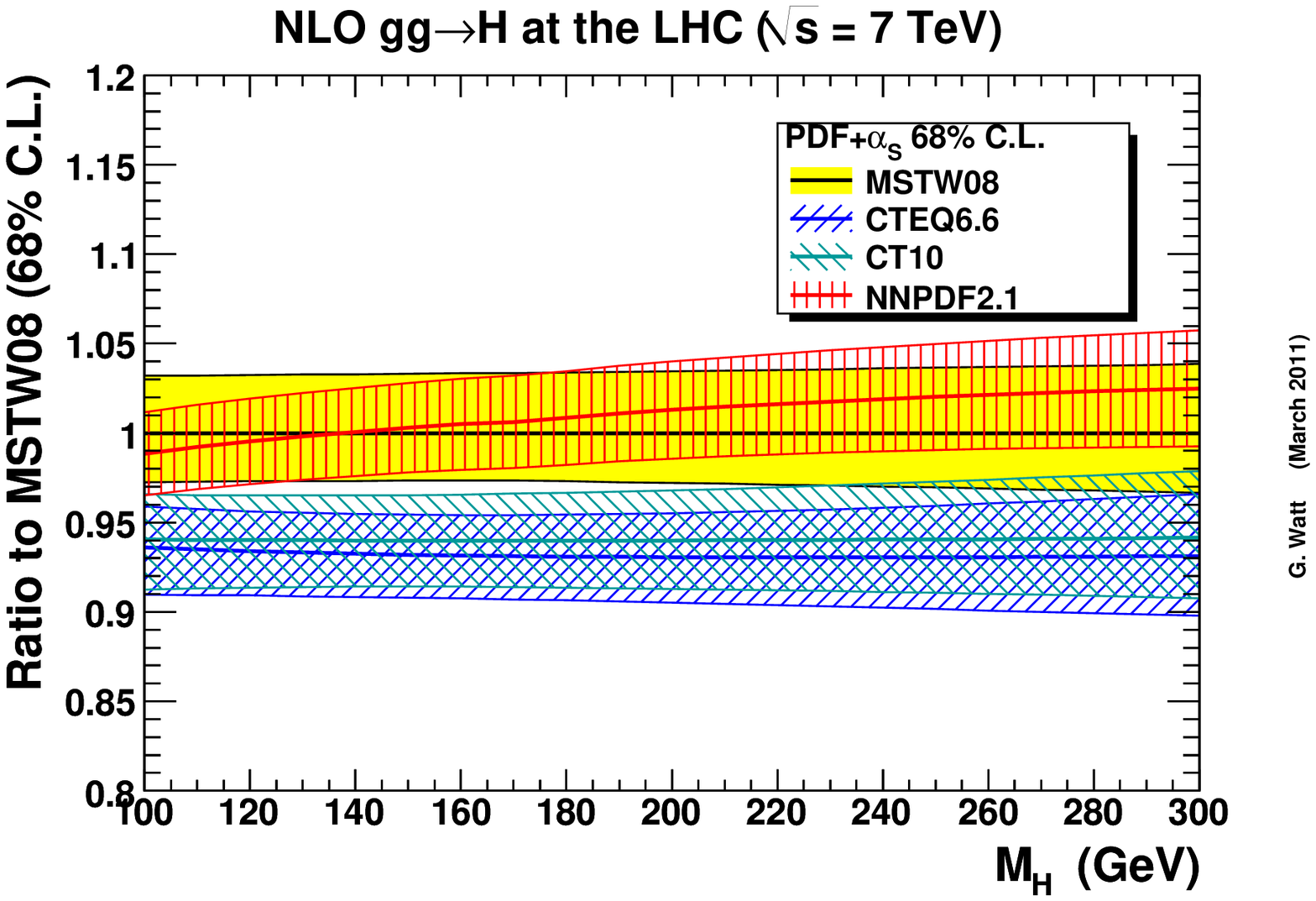}
\includegraphics[width=0.49\columnwidth]{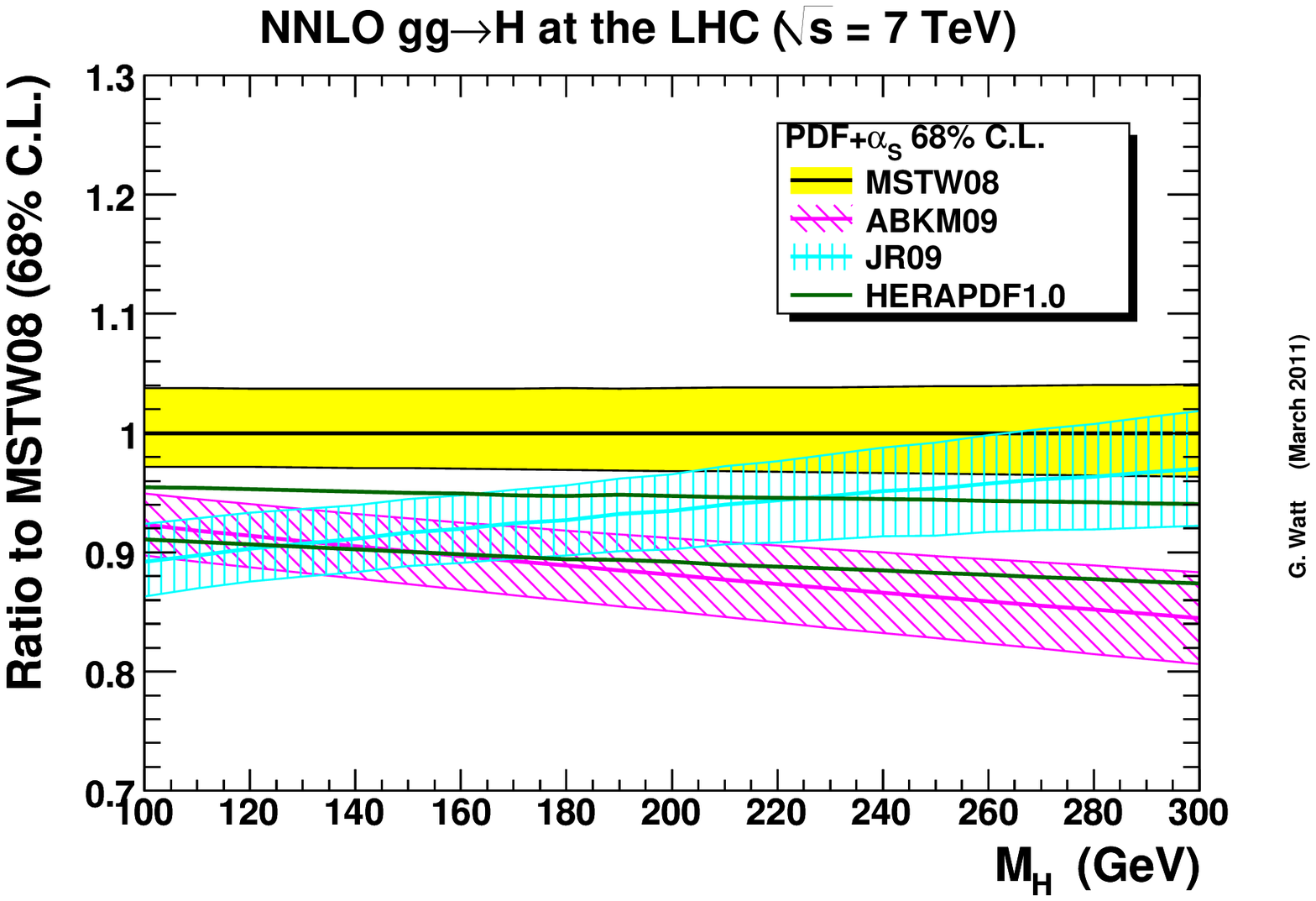}
\end{center}
\caption{\sl Predicted cross sections at NLO (left) and NNLO (right) for Higgs production with
 $68\%$ confidence level uncertainties as a function of $M_H$. The
 ratio to MSTW08 prediction is shown. From~\cite{Watt:2011kp}.  }
\label{fig:lhc-higgs}
\end{figure}

In Fig.~\ref{fig:lhc-higgs} a comparison of production cross sections
for the SM Higgs boson is shown (as a ratio to the MSTW08 prediction)
for a range of $M_H$. The NLO predictions each have an uncertainty of
$\pm3\%$ (at $68\%$ CL and including the uncertainty on $\alpha_s$)
although the spread between different PDF sets can be as large as
$10\%$. At NNLO the uncertainty bands are marginally larger, and the
spread of predictions is considerably larger than at NLO.  However,
scale uncertainty is not included in the error bands shown, and is
reduced by a factor of two to about $9\%$ at
NNLO~\cite{Moch:2005ky,Dittmaier:2011ti}.

These studies have been discussed in detail within the PDF4LHC working
group~\cite{PDF4LHC}. The group has made a recommendation on how to
determine NLO and NNLO PDF uncertainties for cross section predictions
which takes into account the spread between the PDF
groups~\cite{Botje:2011sn}. At NLO the prescription is based on the
MSTW08, CTEQ6.6 and NNPDF2.0 PDF sets which are commonly used by the
LHC experiments (although now CTEQ6.6 is superceded by CT10, and
NNPDF2.0 by NNPDF2.3). The recommendation is to calculate the envelope
of the three group's PDF $+\alpha_s$ uncertainty, and the mid-point
taken as the central value. At NNLO the recommendation is based on the
MSTW08 PDFs where the uncertainty of this set is increased by a scale
factor obtained from the ratio of the NLO PDF4LHC uncertainty band to
the MSTW08 NLO error band. This factor is found to be $\sim 2$ for the
$gg\rightarrow H$ process at the LHC.

It is argued~\cite{Watt:2011kp} that the prescription given above may
be overly complex to apply to all processes, for example in a process
where the theoretical uncertainty is dominated by scale variations. In
some cases it may be easier and statistically more correct to evaluate
the uncertainties according to the prescription of one PDF group using
the $90\%$ CL uncertainty.  For example the NNLO uncertainties
evaluated using the MSTW08 PDFs (and their prescription) including
scale uncertainties, $\alpha_s$ variations and the choice of $b$ and
$c$ quark masses are found to be $^{+2.9}_{-2.4}\%$ and
$^{+5.0}_{-4.7}\%$ for $Z$ production production at $68$\% and
$90$\%~CL respectively.

\subsection{First LHC measurements}

\subsubsection{Electroweak measurements}
The initial measurements of the $W$ and $Z$ production total and
differential cross sections have now been published by
ATLAS~\cite{Aad:2011dm},
CMS~\cite{CMS:2011aa,Chatrchyan:2011wt,Chatrchyan:2011cm} and LHCb~\cite{lhcb-wz}. The $Z$
production cross section is sensitive to the dominant combinations
$u{\bar u}+d{\bar d}+s{\bar s}$, whereas $W^+$ probes $u{\bar
d}+c{\bar s}$ and $W^-$ probes $d{\bar u}+s{\bar c}$. Thus the
flavour structure of the proton is accessible via measurements of
$W^+$ and $W^-$ production, or through the $W$ lepton charge asymmetry
$A(\eta)$:
\begin{eqnarray}
A(\eta) = \frac{{\rm d}\sigma/{\rm d}\eta (W^+\rightarrow l^+\nu) - {\rm d}\sigma/{\rm d}\eta (W^-\rightarrow l^-{\bar \nu}) }
               {{\rm d}\sigma/{\rm d}\eta (W^+\rightarrow l^+\nu) + {\rm d}\sigma/{\rm d}\eta (W^-\rightarrow l^-{\bar \nu}) }
\end{eqnarray}
which have recently been
published~\cite{Aad:2011dm,lhcb-wz,Aad:2011yn,Chatrchyan:2011jz,Chatrchyan:2012xt}.  The most
precise measurements of the asymmetry in $p{\bar p}$ collisions from
D0 show some tension with the CDF
measurements and to some extent with other DIS
data (see~\ref{sec:Wasym_Tevatron}). At the LHC the spread of
predictions for this observable can be as much as a factor of two larger than
the $90\%$ CL uncertainty from MSTW08\cite{Watt:2011kp}. Fits to the
di-muon production data in $\nu$ and $\bar{\nu}$ induced DIS prefer an
enhanced $s$ compared to $\bar{s}$ contribution (see
section~\ref{sec:StrangeSea}), although the significance of this
finding is weak.  Since the contribution of $s/{\bar s}$ to $Z$ and
$W$ production is large at the LHC (up to $20$\% and $27$\%
respectively at NLO~\cite{Nadolsky:2008cx}) new LHC data could help
resolve the issue and set interesting constraints in the strange
sector. First studies were carried out in~\cite{Aad:2012sb} and pursued in~\cite{NNPDF2.3}.

\begin{figure}[htb]
\begin{center}
\includegraphics[width=0.49\columnwidth]{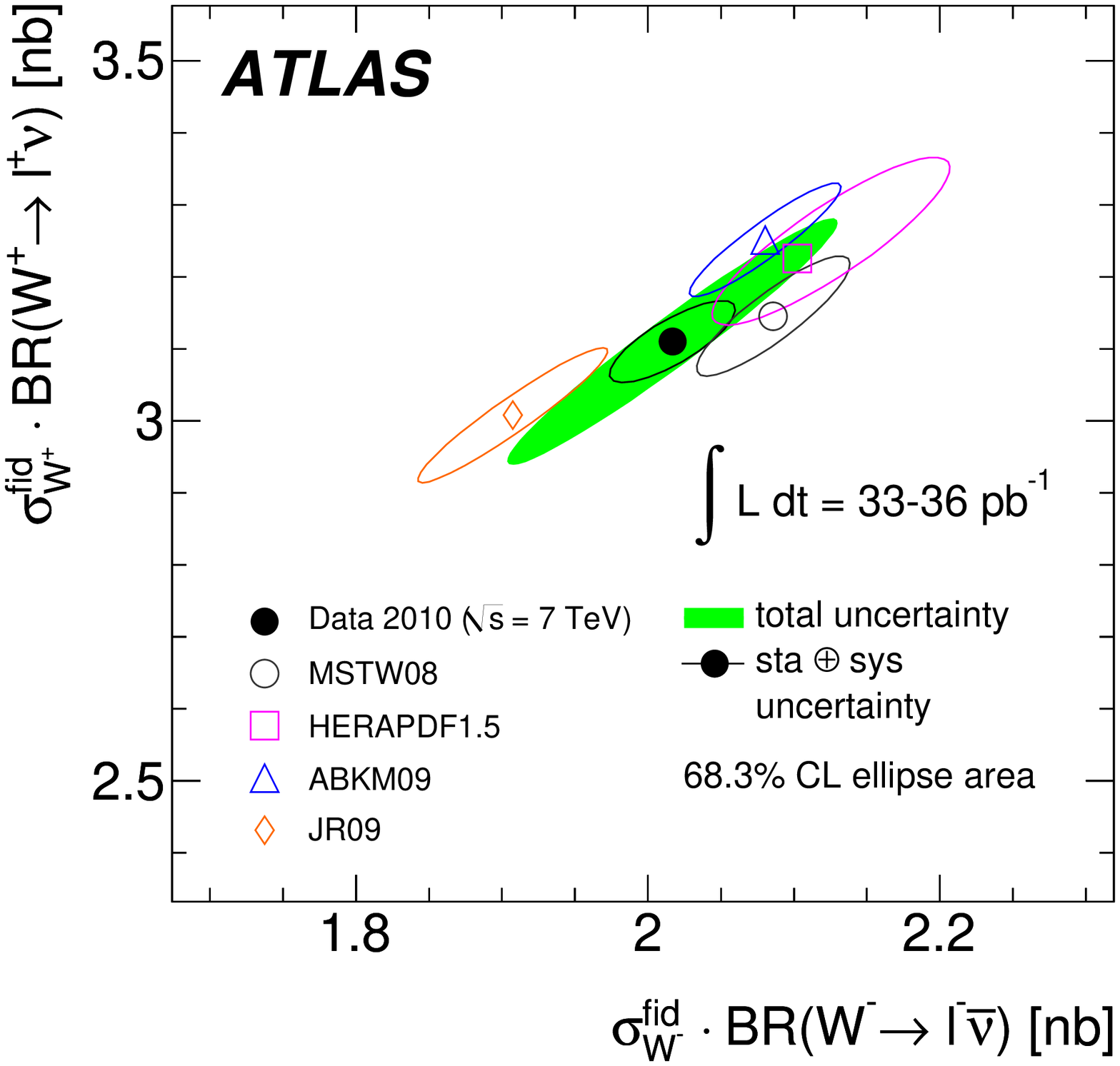}
\includegraphics[width=0.49\columnwidth]{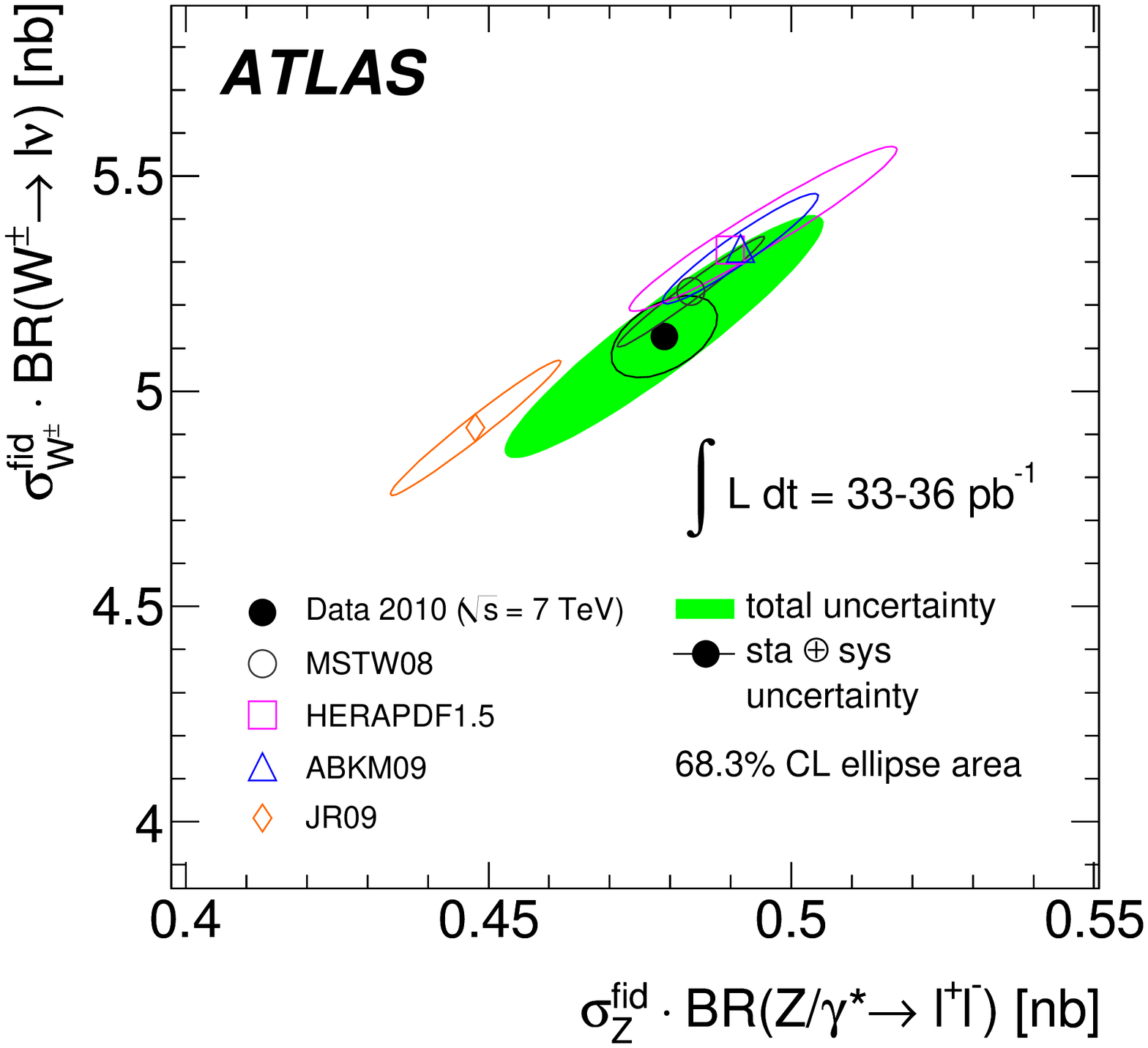}
\end{center}
\caption{\sl  The measured and predicted production cross sections times branching
ratio for $W^+$ vs. $W^-$ production (left), and for $W$ vs. $Z$
production (right). From~\cite{Aad:2011dm}.  }
\label{fig:atlas-wz}
\end{figure}

\paragraph{$\boldsymbol{W}$ and $\boldsymbol{Z}$ cross sections}
First measurements of the $W$ and $Z$ production cross sections in $e$
and $\mu$ decay channels at $\sqrt{s}=7$ TeV are available
~\cite{Aad:2011dm,CMS:2011aa,lhcb-wz} using $\sim 35$~pb$^{-1}$ integrated
luminosity recorded in 2010. The measurements are systematically
limited and both experiments have a precision of $\sim 1\%$ (excluding a 
$3-4\%$ luminosity uncertainty). Fig.~\ref{fig:atlas-wz} shows the
correlation of the $W^+$ to $W^-$ production cross sections, and the
$W^++W^-$ to $Z$ production from ATLAS. NNLO predictions compare
favourably with the measurements within their quoted uncertainties.

\paragraph{Differential Drell-Yan measurements}
The virtual $\gamma^*$ cross section below the $Z$ resonance provides
complementary information to that obtained at the $Z$ peak.  At low
$\gamma^*$ invariant mass the electromagnetic couplings dominate which
suppress the $d$-type contributions whereas the axial and vector EW
couplings to the $u$ and $d$ quarks, $v^2+a^2$, are of similar size
(see Eq.~\ref{eq:DY_ewcoupling}). Thus measurements at the $Z$
resonance peak and of the low mass continuum are sensitive to
different combinations of $u$-type and $d$-type quarks.

Since the virtual boson rapidity is related to the ratio of the quark
and anti-quark parton momentum fractions, an interesting measurement
is the $y$ spectrum in $Z/\gamma^*$ interactions. At large $y$ the
longitudinally boosted boson arises from increasingly asymmetric
momentum fractions of the $q,{\bar q}$ pair which provides
simultaneous access to the high $x$ and low $x$ kinematic regions.
Measurements of the low mass Drell-Yan cross section reach the region
of very low $x\sim~10^{-4}$ for ATLAS and CMS, and $10^{-5}$ for LHCb.
At very low mass however, fixed order calculations are not yet stable
(see~\ref{sec:lowx}). The PDF uncertainty for $M\sim15$~GeV is
estimated to be $3\%$ at NLO but the scale uncertainty can lead to
variations of as much as $30\%$ on the cross section (taking
$\mu_F=\mu_R={\hat s}$) which rapidly diminishes with increasing $M$,
and could limit the use of the
lowest $M$ data in PDF fits. At NNLO the scale uncertainty remains
sizeable at about $4\%$ but can be reduced by
choosing the scale appropriately such that the higher order
contributions are minimised~\cite{deOliveira:2012ji}.

\begin{figure}[htb]
\begin{center}
\includegraphics[width=0.44\columnwidth]{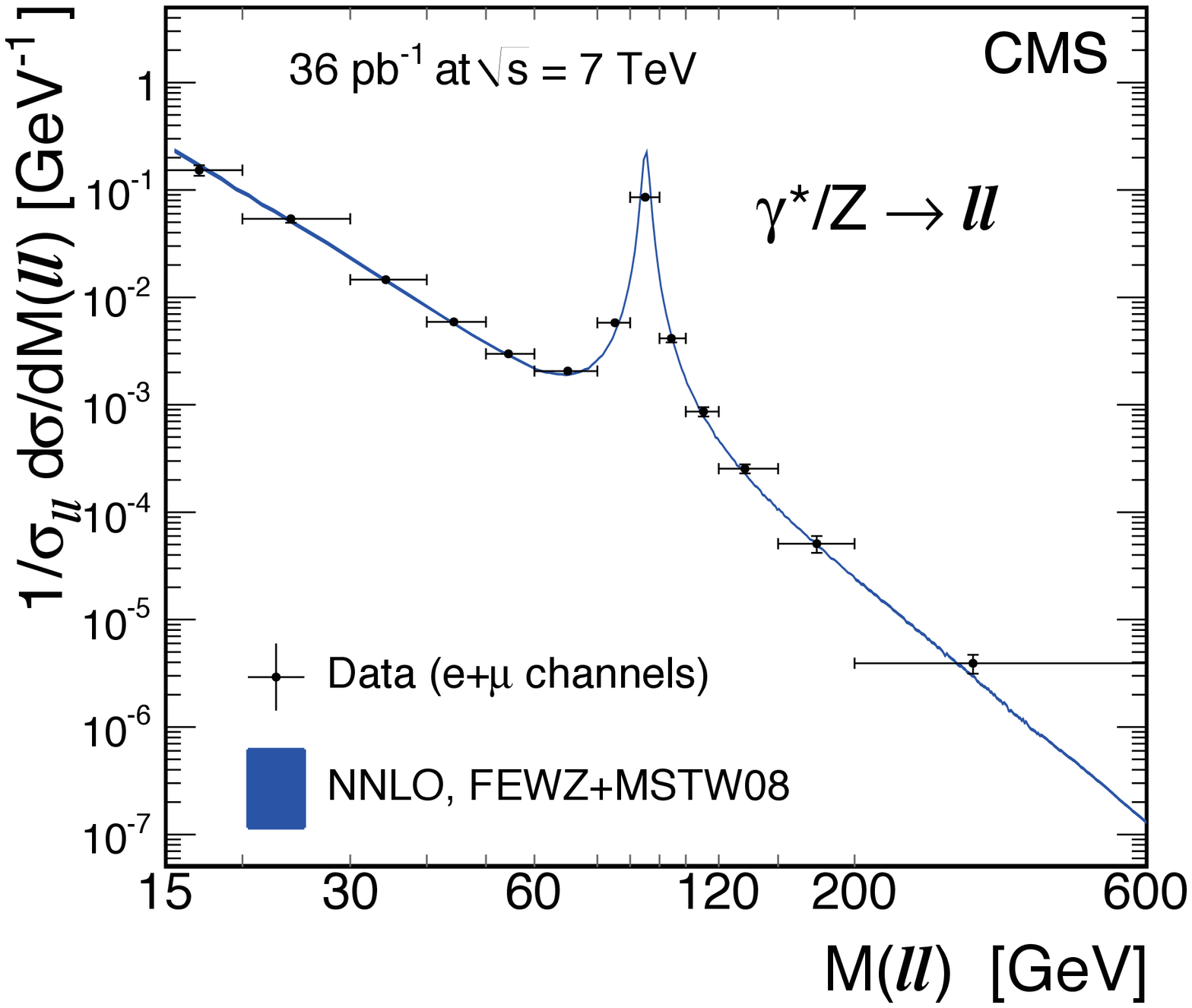}
\includegraphics[width=0.55\columnwidth]{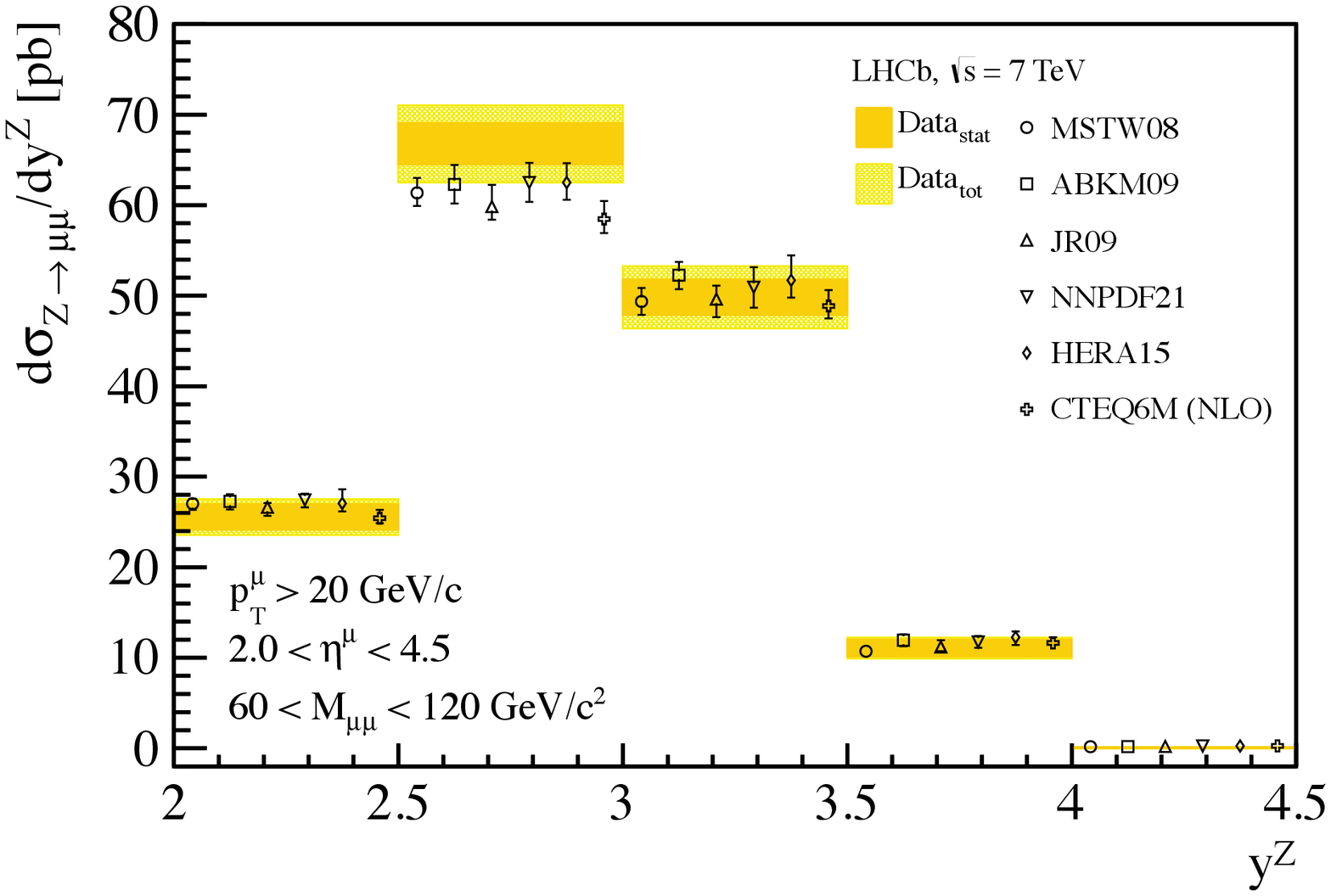}
\end{center}
\caption{\sl Left: 
The normalised differential Drell-Yan cross section vs. invariant mass
 of the virtual boson. Right: the differential rapidity
 spectrum of $Z^0$
 production. From~\cite{Chatrchyan:2011cm,lhcb-wz}.}
\label{fig:cms-zdiff}
\end{figure}

The first measurements of the differential invariant mass spectrum
from CMS~\cite{Chatrchyan:2011cm} are shown in
Fig.~\ref{fig:cms-zdiff} (left) spanning the range $15<M<600$~GeV compared to
NNLO predictions. Preliminary measurements from LHCb
down to $M=6$~GeV are also released~\cite{lhcb-dy}. Both measurements
are based on the 2010 datasets and have a moderate precision of
$9\%$ at $M=15-20$~GeV which is expected to improve.

Differential spectra for $W$ and $Z$ production have been published by
ATLAS~\cite{Aad:2011dm}, CMS~\cite{Chatrchyan:2011wt} and
LHCb~\cite{lhcb-wz} and the LHCb measurements are shown in
Fig.~\ref{fig:cms-zdiff} (right) compared to NNLO predictions. The data,
which in this case are based on the statistically limited 2010 data sample,
are not yet of sufficient precision to significantly constrain the PDFs 
although some deviation between theory and measurement is observed for
$2.5<y<3.0$. It will be interesting to see how this develops with the
new measurements with higher statistical precision and smaller systematic
uncertainties.

\begin{figure}[htb]
\begin{center}
\includegraphics[width=0.43\columnwidth]{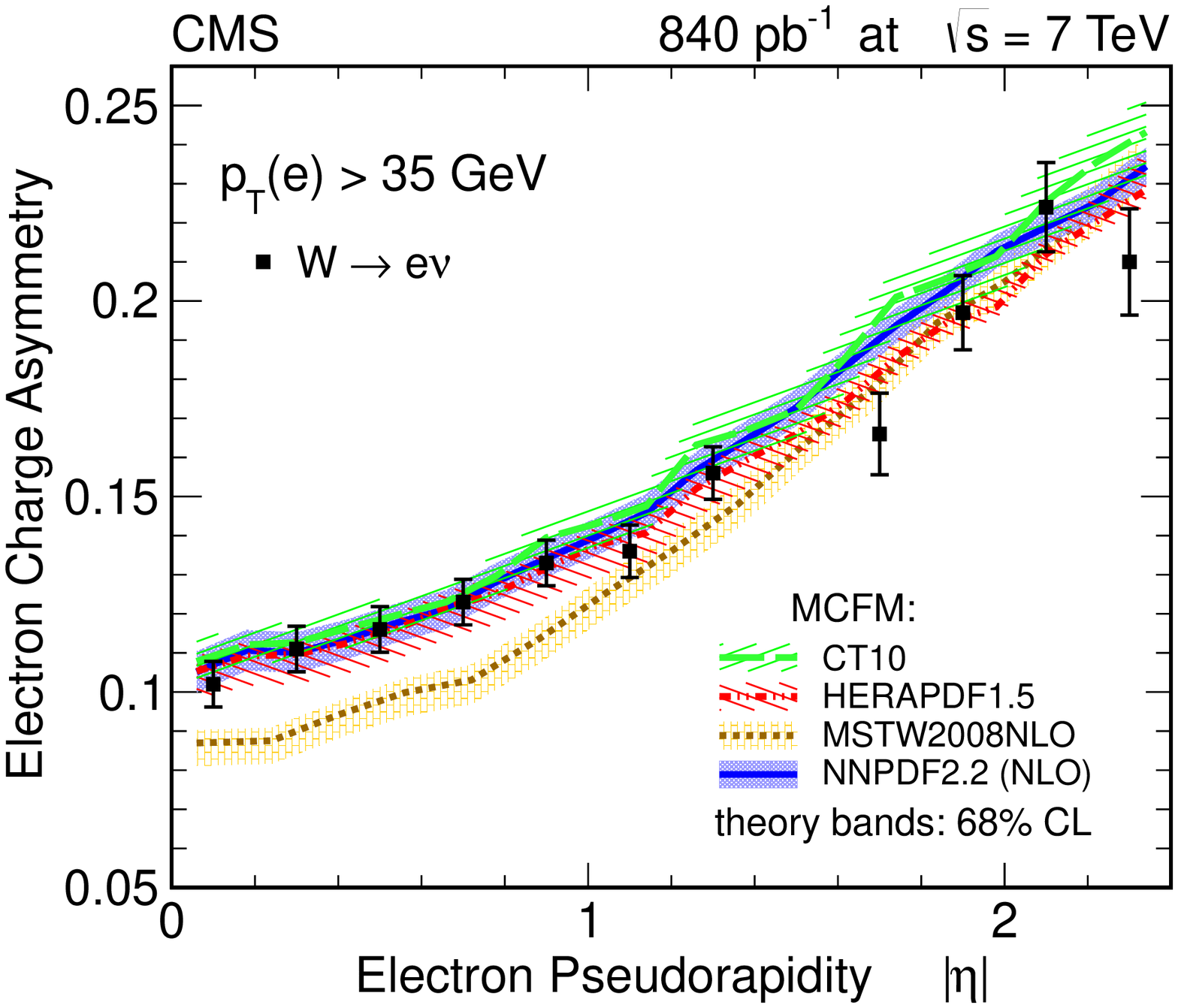}
\includegraphics[width=0.49\columnwidth]{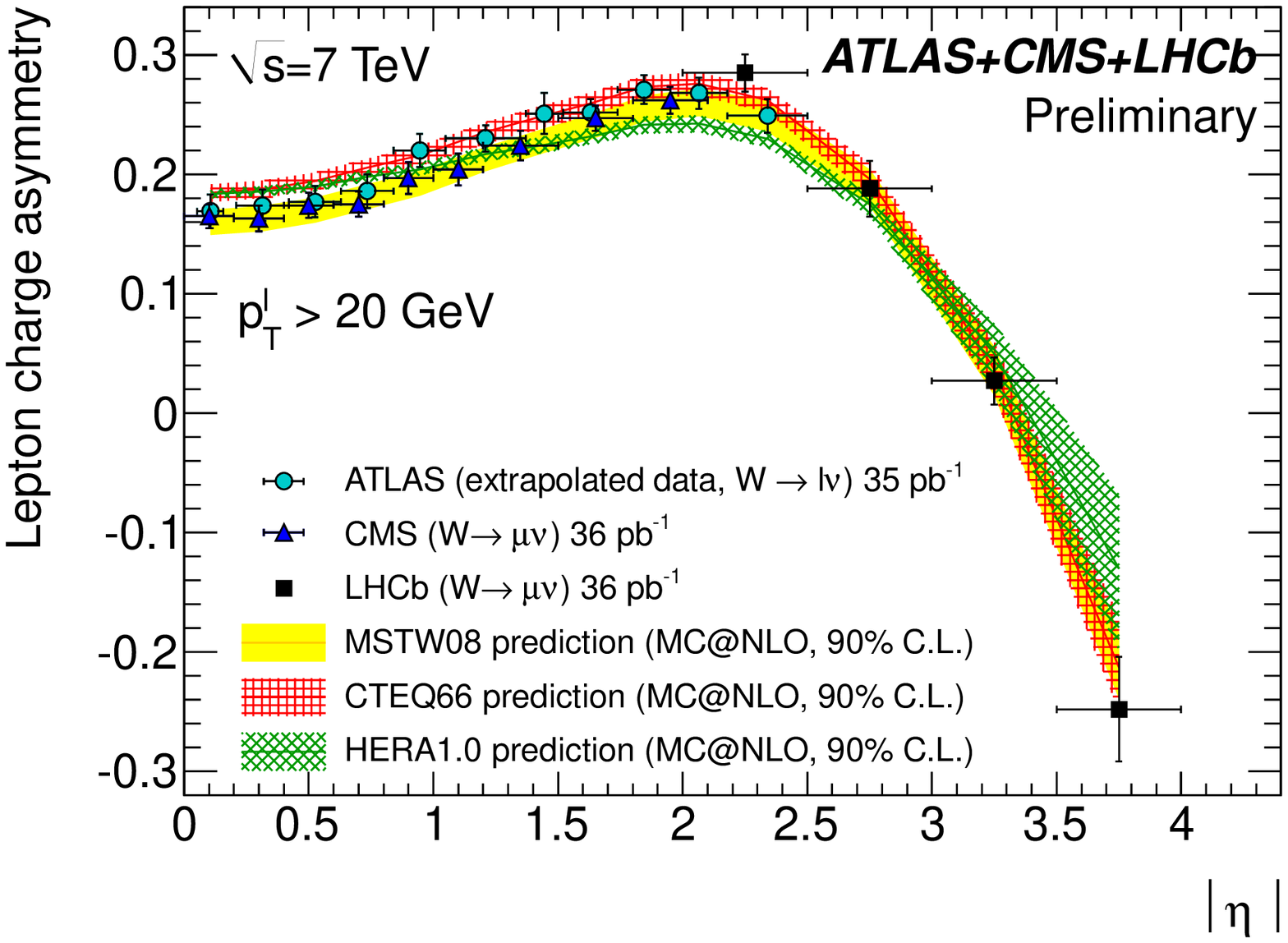}
\end{center}
\caption{\sl  Charged lepton asymmetry in $W$ decays at the LHC. Left:
CMS electron measurement for $p_T>35$~GeV with $840$~pb$^{-1}$
from~\cite{Chatrchyan:2012xt}. Right: combined electron and muon
channel measurements from ATLAS, CMS and LHCb for $p_T>20$~GeV with
$35-36$~pb$^{-1}$ from~\cite{lhc-wasym}.  }
\label{fig:atlas-wasym}
\end{figure}

\paragraph{$\boldsymbol{W}$ charged lepton asymmetry}
Measurements from ATLAS, CMS and LHCb of the $W$ charged lepton ($e+\mu$)
asymmetry are presented
in~\cite{Aad:2011dm,Aad:2011yn,Chatrchyan:2011jz,Chatrchyan:2012xt,lhcb-wz}. Fig.~\ref{fig:atlas-wasym}
shows the asymmetry as determined by all three experiments 
compared to fixed order NLO and NNLO predictions from several PDF
groups. The CMS electron channel measurement is shown in
Fig.~\ref{fig:atlas-wasym} (left) using $840$~pb$^{-1}$ of integrated
luminosity with a lepton $p_T$ cut of $35$~GeV. The data are in good
agreement with most NLO predictions, but at low lepton
rapidity $\eta_l$ the MSTW08 prediction undershoots the data.  The
differences between PDF sets are more pronounced in the predicted
$W^+$ and $W^-$ rapidity spectra leading the authors of~\cite{Aad:2011dm}
to argue that the individual spectra are more sensitive than the
lepton charge asymmetry. Better agreement between predictions and
measurements may be obtained when comparing to resummed calculations
at next-to-next-to-leading log order since these calculations give a
better description of the the $W$ $p_T$ spectrum as pointed out
in~\cite{Chatrchyan:2011jz}.

The charged lepton asymmetry measurements from ATLAS, CMS and LHCb are
shown together in Fig.~\ref{fig:atlas-wasym} (right) based on a
smaller data set of $35-36$~pb$^{-1}$ and less restrictive phase space
with the lepton $p_T$ required to be above $20$~GeV. The NLO
predictions of CTEQ6.6, HERAPDF1.0 and MSTW08 are in reasonable
agreement with the data shown, although here, better agreement with
MSTW08 is observed albeit within larger experimental uncertainties. Of
particular interest is the region accessed by the LHCb measurement for
$\eta_l>2.5$ where the predictions are in agreement with each other
and the data but with relatively large uncertainties. Thus current and
future measurements are expected to have a visible impact in reducing
the PDF uncertainties and improving the consistency between PDF sets
at large and small $\eta_l$.

\subsubsection{Inclusive jet cross sections}
\begin{figure}[htb]
\begin{center}
\includegraphics[width=0.49\textwidth]{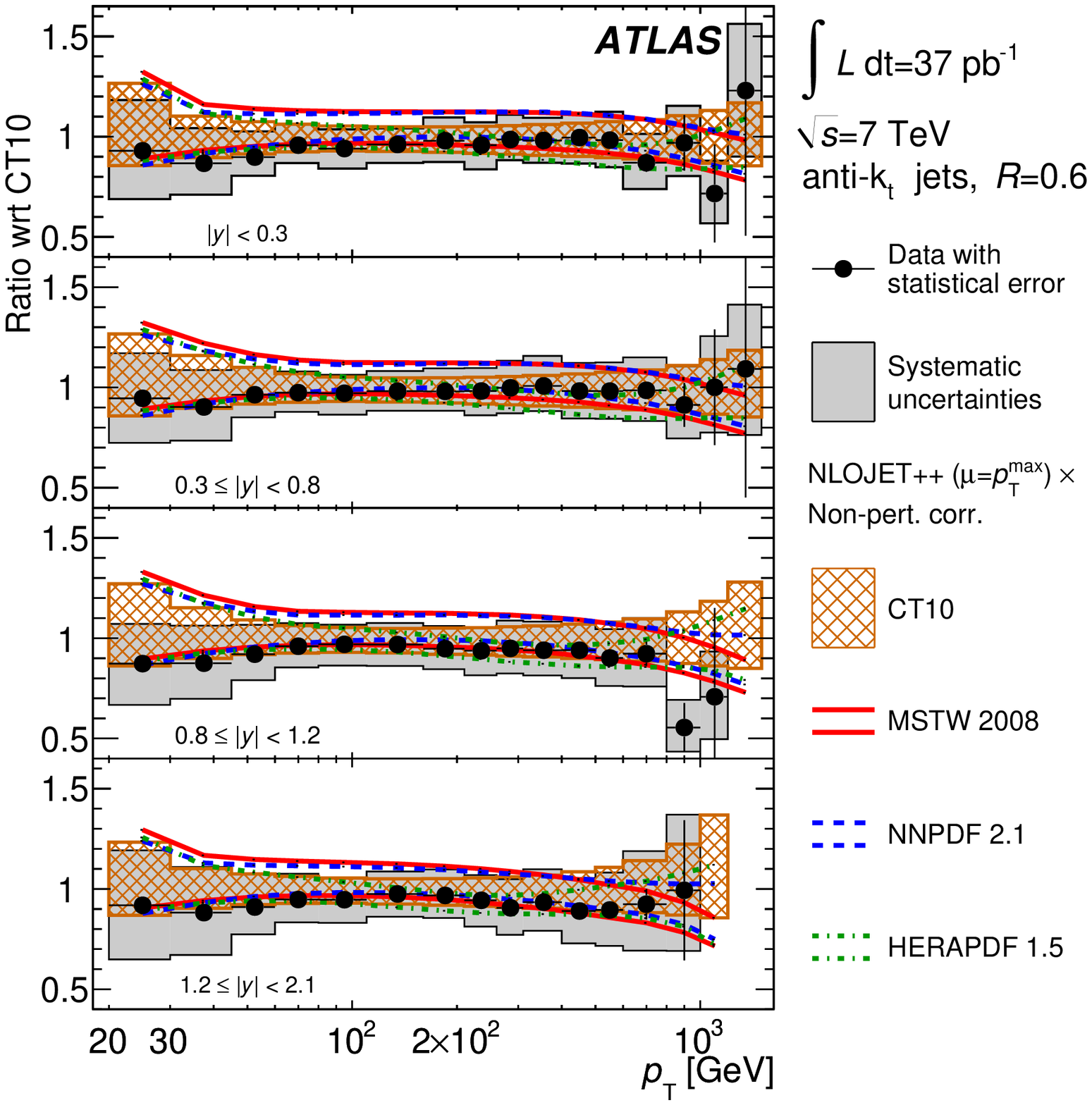}
\includegraphics[width=0.49\textwidth]{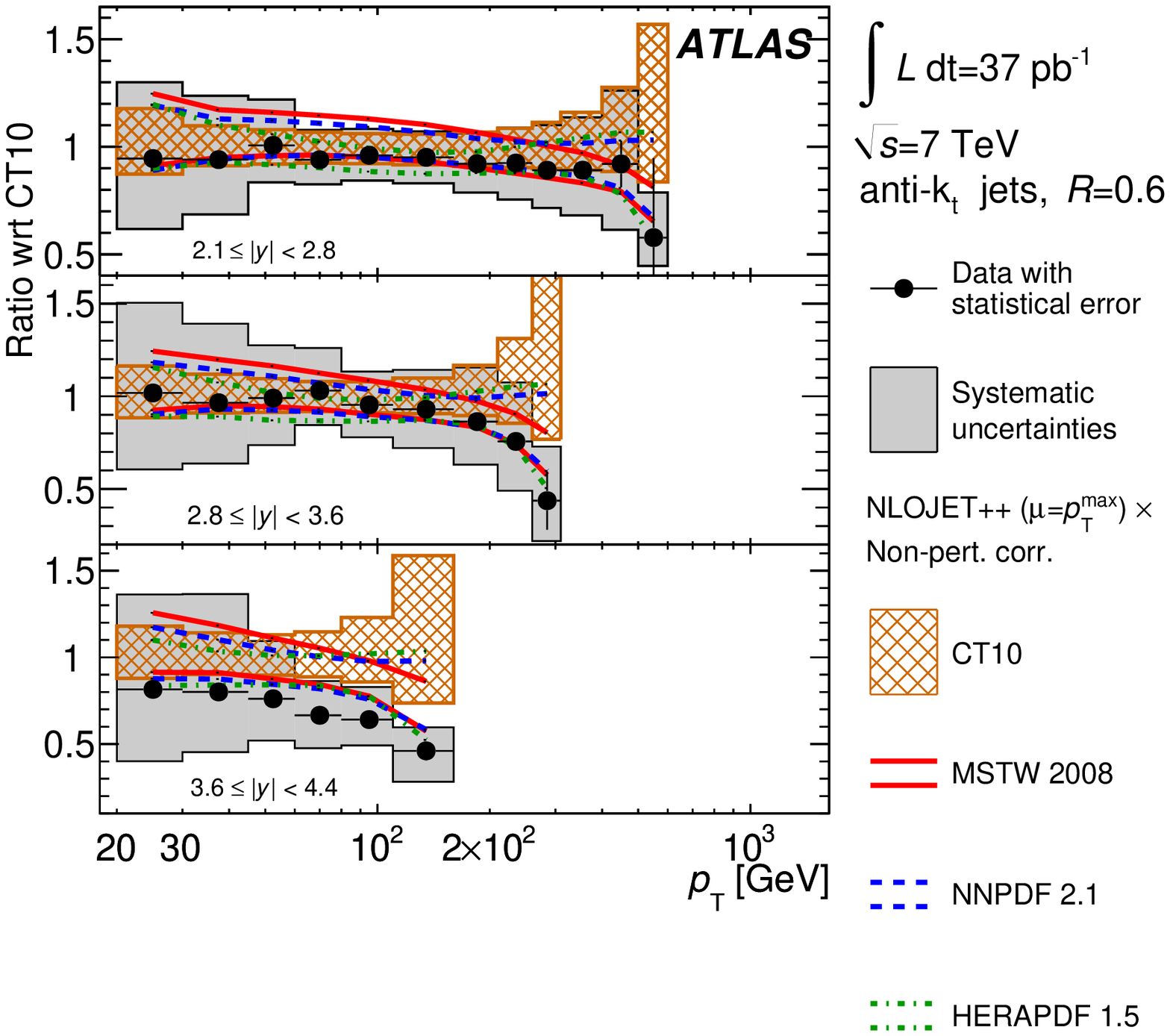}
\end{center}
\caption{\sl Ratio of the measured inclusive jet cross section
$d^2\sigma/dydp_T$ to the theoretical prediction using CT10
PDFs. The predictions from MSTW08, NNPDF2.1 and HERAPDF1.5 are also
shown. From~\cite{Aad:2011fc}.  }
\label{fig:atlas-incjets}
\end{figure}

Differential inclusive jet cross sections $d^2\sigma/dydp_T$ at
$\sqrt{s}=7$~TeV are available from ATLAS~\cite{Aad:2011fc} and
CMS~\cite{CMS:2011ab} and an example of the data can be seen in
Fig.~\ref{fig:atlas-incjets}. Even with the modest luminosity of
$\sim35$~pb$^{-1}$ the measurements extend to jet transverse momenta
of about $1.5$~TeV. The wide $\eta$ range of the ATLAS and CMS
calorimeters compared to the Tevatron experiments allows the
jet cross sections to be measured up to high rapidities of $4.4$. The
measurements are sensitive to partonic momentum fractions $x$ of
$\sim10^{-5}<x<0.9$, however the precision is limited by the
knowledge of the detector calibration. Jet cross sections exhibit a very
sharply falling jet $p_T$ spectrum (see for example
Fig.~\ref{fig:d0_incjets}), therefore small changes in the jet
energy scale lead to large correlated shifts in the cross
sections. Currently this leads to measurement uncertainties of about
$10-60\%$ dominated by a scale uncertainty of $3-4\%$ in the central
detector regions for moderate jet $p_T$ and rising to $\sim12\%$ at the
highest $y$.

\subsubsection{The NNPDF2.3 PDFs}

Global fits that include the early LHC measurements described above have been
performed by the NNPDF collaboration. 

In~\cite{Ball:2011gg}, the NNPDF2.1 NLO PDFs
were updated using a reweighting technique to include the first $W$ charged lepton asymmetry
measurements from ATLAS and CMS, leading to the NNPDF2.2 set.

\begin{figure}[thb]
\begin{tabular}{cc}
\includegraphics[width=0.45\columnwidth]{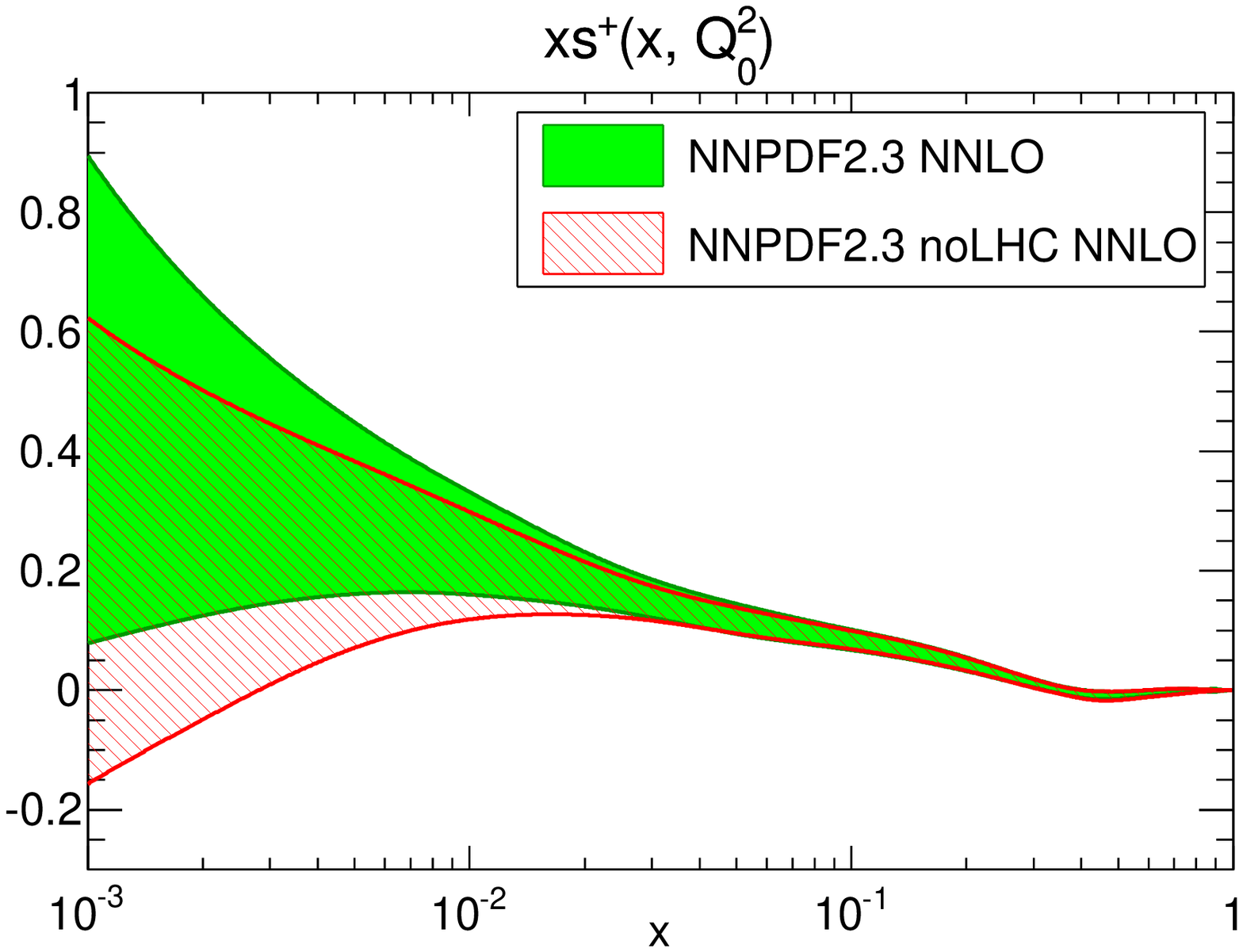} &
\includegraphics[width=0.45\columnwidth]{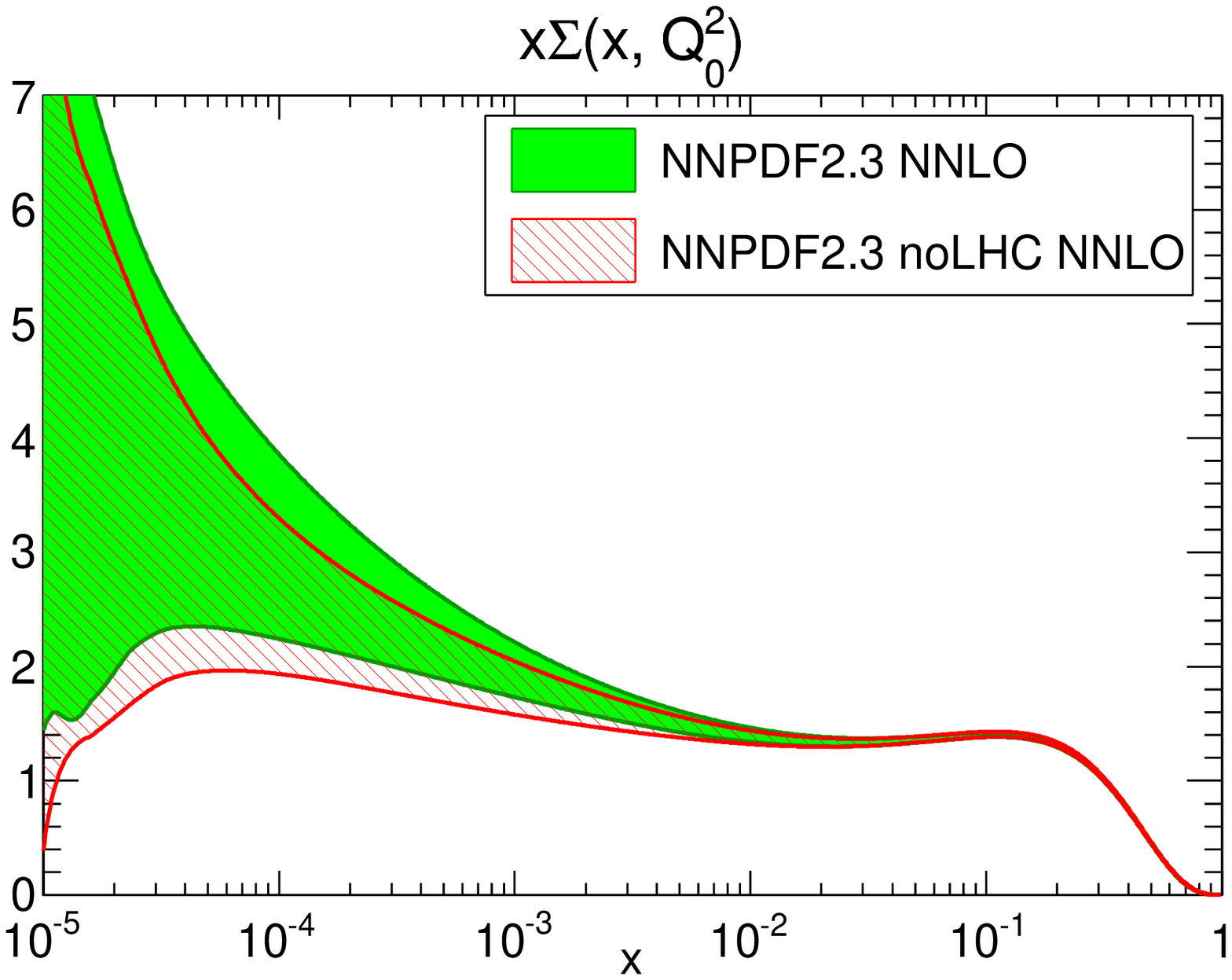}
\end{tabular}
\caption{\sl Comparison of the strangeness and singlet distributions obtained from NNPDF2.3,
which includes the LHC data, and from the same fit but restricted to the non-LHC measurements.
The contours correspond to uncertainties at $68 \%$ confidence level.
From~\cite{NNPDF2.3}.
}
\label{fig:nnpdf23}
\end{figure}

In~\cite{NNPDF2.3} new fits were performed which include, in addition
to the non-LHC data used in NNPDF2.1, the published measurements
from the LHC experiments for which the covariance matrix of the correlated systematic uncertainties
has been provided: the $W$ and $Z$ lepton rapidity distributions measured
by ATLAS~\cite{Aad:2011dm} and LHCb~\cite{lhcb-wz} using
the 2010 data, the $W$ electron asymmetry measured by CMS in the 2011 dataset~\cite{Chatrchyan:2012xt},
and the ATLAS inclusive jet cross sections measured in the 2010 data~\cite{Aad:2011fc}.
The corresponding NNPDF2.3 PDFs have been determined both at NLO and
at NNLO for a wide range of values of $\alpha_s$, in the same GM-VFNS scheme
as used for NNPDF2.1\footnote{PDFs obtained in the FFNS with $n_f = 4$ or $n_f = 5$
active flavours are also provided.}, and are available in the LHAPDF interface.
For the determination of NNLO PDFs, the NNLO predictions for electroweak boson
production at the LHC have been obtained via K-factors. For inclusive jet
production at the LHC, the NLO matrix elements have been used (together
with NNLO PDFs and $\alpha_s$) instead of the approximation usually made to
calculate NNLO jet production at the Tevatron (see section~\ref{sec:NNLO}), because the threshold
approximation is expected to be worse at the LHC energies.

The resulting NNPDF2.3 distributions provide a good description of all datasets included in the fit.
The comparison of these PDFs with those obtained from the same fit performed to
the non-LHC data only (the NNPDF2.3-noLHC set already mentioned at the beginning of~\ref{sec:GlobalFits})
allows to gauge the specific impact of the LHC data.
This impact is so far moderate but already visible~\cite{NNPDF2.3}:
the uncertainty on the gluon
distribution at high $x$ is somewhat reduced thanks to the LHC jet data; the electroweak
boson production data help improve the flavour decomposition;
and the strangeness fraction of the light sea
is pushed towards slightly higher values, although with a marginal statistical
significance.
As an example, Fig.~\ref{fig:nnpdf23} compares the strange
and singlet distributions obtained from NNPDF2.3 and NNPDF2.3-noLHC.

\paragraph{PDF fits based only on collider data}
PDFs derived from a fit restricted to data from
collider experiments were also extracted in~\cite{NNPDF2.3}. 
The motivation of this approach lies in the fact that the resulting PDFs are, by construction, independent of any nuclear
or higher twist corrections that may affect some fixed target measurements,
and could explain some of the tensions
reported in~\ref{sec:fits_compatibility}.
Restricting the fit to the collider measurements reduces by a factor of $\sim 3$ the number of fitted data points.
The resulting PDFs show no significant differences with those from NNPDF2.3, 
which indicates that any tension between collider and fixed target data can only be
moderate. However, some distributions resulting from this fit show very large
uncertainties. For example, the anti-quark PDFs at high $x$ are very poorly constrained
in such a fit, as shown\footnote{The LHC data, not included in the collider fit illustrated in Fig.~\ref{fig:highx_antiquarks},
do not reduce these uncertainties yet.} by the dashed curves in Fig.~\ref{fig:highx_antiquarks} of section~\ref{sec:DrellYanAndValenceSea}.

\subsubsection{Top production}
The production of $t\bar{t}$ pairs is dominated by $gg$ fusion at the
LHC, and at $\sqrt{s}=14$~TeV this subprocess contributes $90\%$ of the
total cross section. Therefore this provides an interesting probe of
the high $x$ gluon particularly at large $t\bar{t}$ invariant mass
$>1$~TeV. However, care should be taken in interpreting these cross sections
which are also used to constrain many models of new physics coupling
to the top quark. 

Latest measurements of the total production cross section based on $7$
and $8$~TeV centre-of-mass energy data from ATLAS and CMS have been
reported in a variety of decay modes including single and di-lepton
$W$ decays as well as purely hadronic modes, and measurements using
$b$ tagged jets, for
example~\cite{Chatrchyan:2011yy,CMS:ttbar8-1lep,CMS:ttbar8-2lep,ATLAS:2012aa,ATLAS:ttbar-1lep}
and references therein. The latest combination of measurements are
presented in~\cite{ATLAS:ttbar-comb,CMS:ttbar-comb} and an
experimental precision of $\sim5\%$ is now achieved (excluding the
luminosity uncertainty). Recent approximate NNLO
predictions~\cite{Kidonakis:2010dk} and NLO predictions with
next-to-next-to-leading log (NNLL)
corrections~\cite{Ahrens:2010zv} both have a similar
accuracy of about $\pm4\%$ for the scale variation uncertainty and
$\pm5\%$ for the PDF uncertainty evaluated using only the MSTW08 NNLO set at $90\%$
CL. This estimated PDF uncertainty is smaller than the spread of
different predictions as discussed in~\ref{lhc-benchmark}, and the
measurements are expected to constrain the differences between the PDF sets.

\begin{figure*}[htbp]
\begin{center}
\includegraphics[scale=0.37]{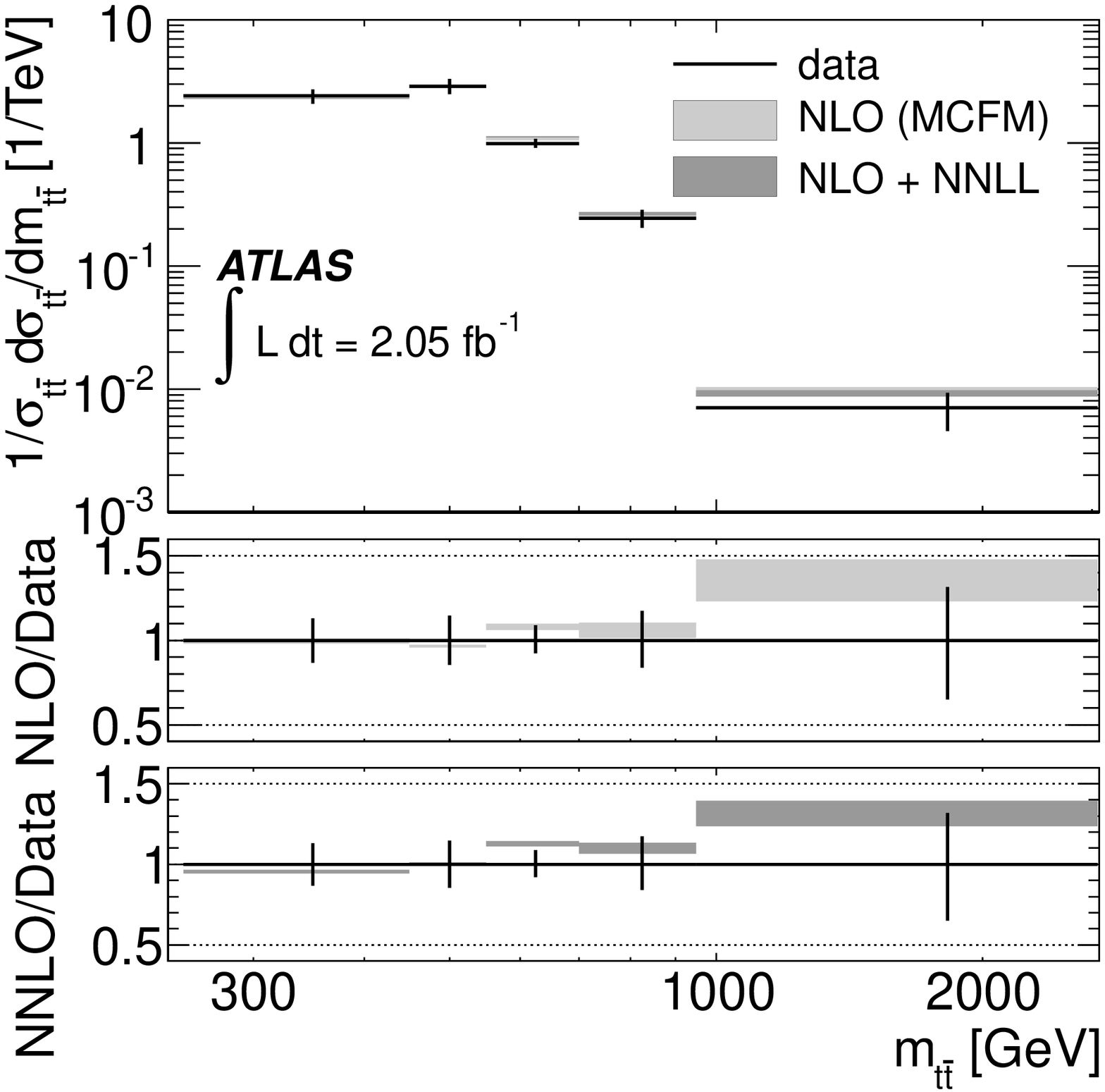}
\includegraphics[scale=0.37]{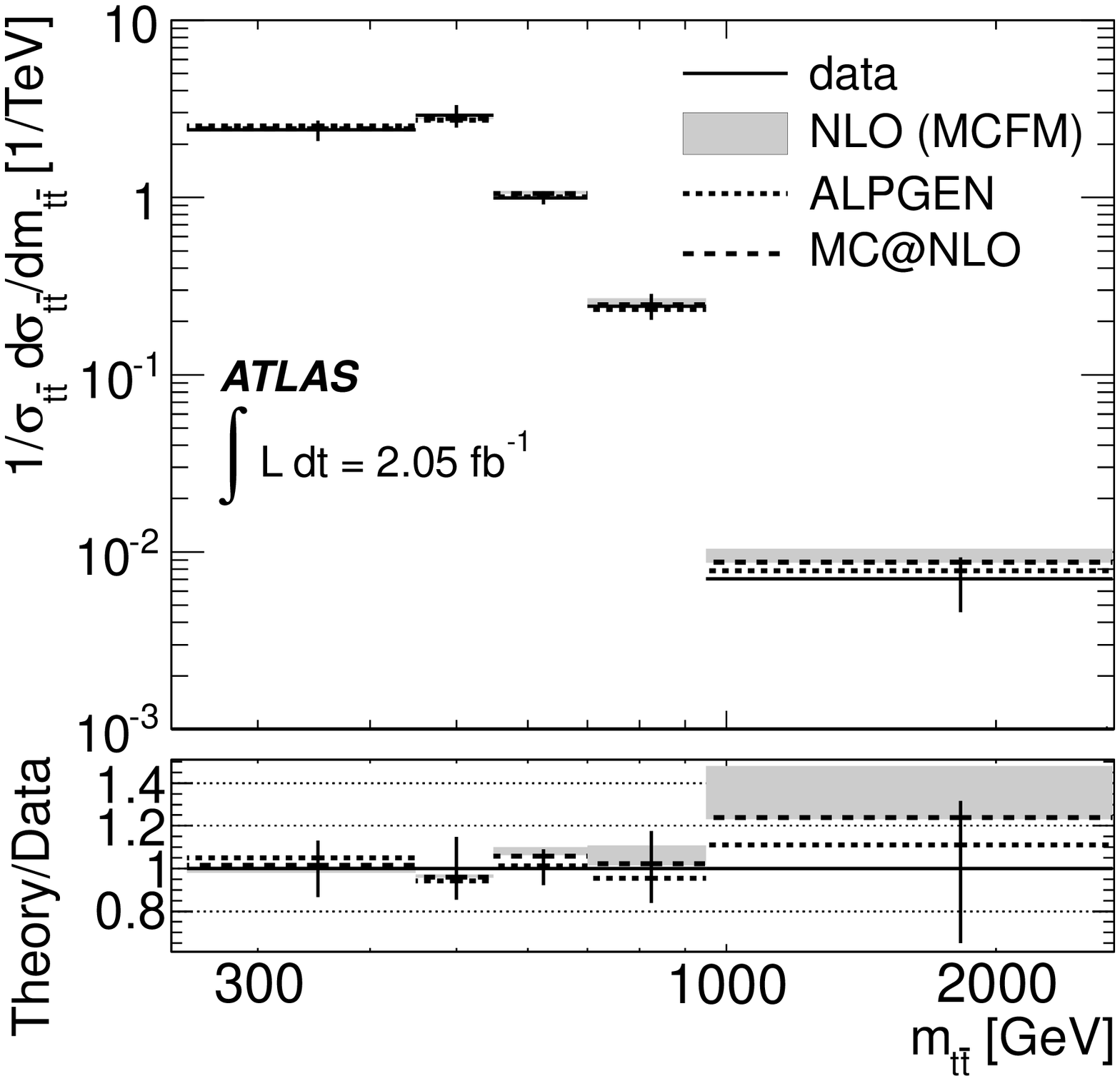}
\end{center}
\caption{\sl Normalised $t\bar{t}$ differential cross section at
$\sqrt{s}=7$~TeV compared to four NLO and NLO+NNLL predictions. From~\cite{:2012hg}.}
\label{fig:ttbar}
\end{figure*}

A first measurement of the normalised differential $t\bar{t}$ cross section is
now available performed with a $2$~fb$^{-1}$ data sample at
$\sqrt{s}=7$~TeV~\cite{:2012hg} in the single lepton ($e$+$\mu$)
channel. The data are shown in Fig.~\ref{fig:ttbar} and compared to NLO
and NLO+NNLL predictions. A precision of $10-20\%$ is achieved which is limited by
uncertainties related to the jet energy scale and resolution.

\subsubsection{Prompt photon production}
\label{sec:LHCphotons}

The potential sensitivity of measurements of isolated photon hadro-production 
on the gluon density has been mentioned in section~\ref{exp:promptphotons},
in the context of pre-LHC experiments. In $pp$ collisions at the LHC, the
relative contribution of the QCD Compton process $q g \rightarrow \gamma q$ 
to prompt photon production is enhanced compared to what happens in
$p \bar{p}$ collisions at the Tevatron, where $q \bar{q}$ annihilations
$q \bar{q} \rightarrow \gamma g$ also play an important role.
Moreover, in the large kinematic domain where the measurement can be performed
at the LHC,
the gluon density is involved in a broad range of Bjorken-$x$,
from ${\cal{O}}(10^{-3})$ at rapidities of $|\eta|\simeq2$ and low transverse energy
to ${\cal{O}}(0.1)$ at central pseudo-rapidities and high $E_T$~\cite{DaviddEnterria}.
Hence, the impact of LHC prompt photon measurements on the gluon PDF is
expected to be significant.

\begin{figure}[bht]
 \begin{tabular}{cc}
 {\includegraphics[width=0.45\columnwidth]{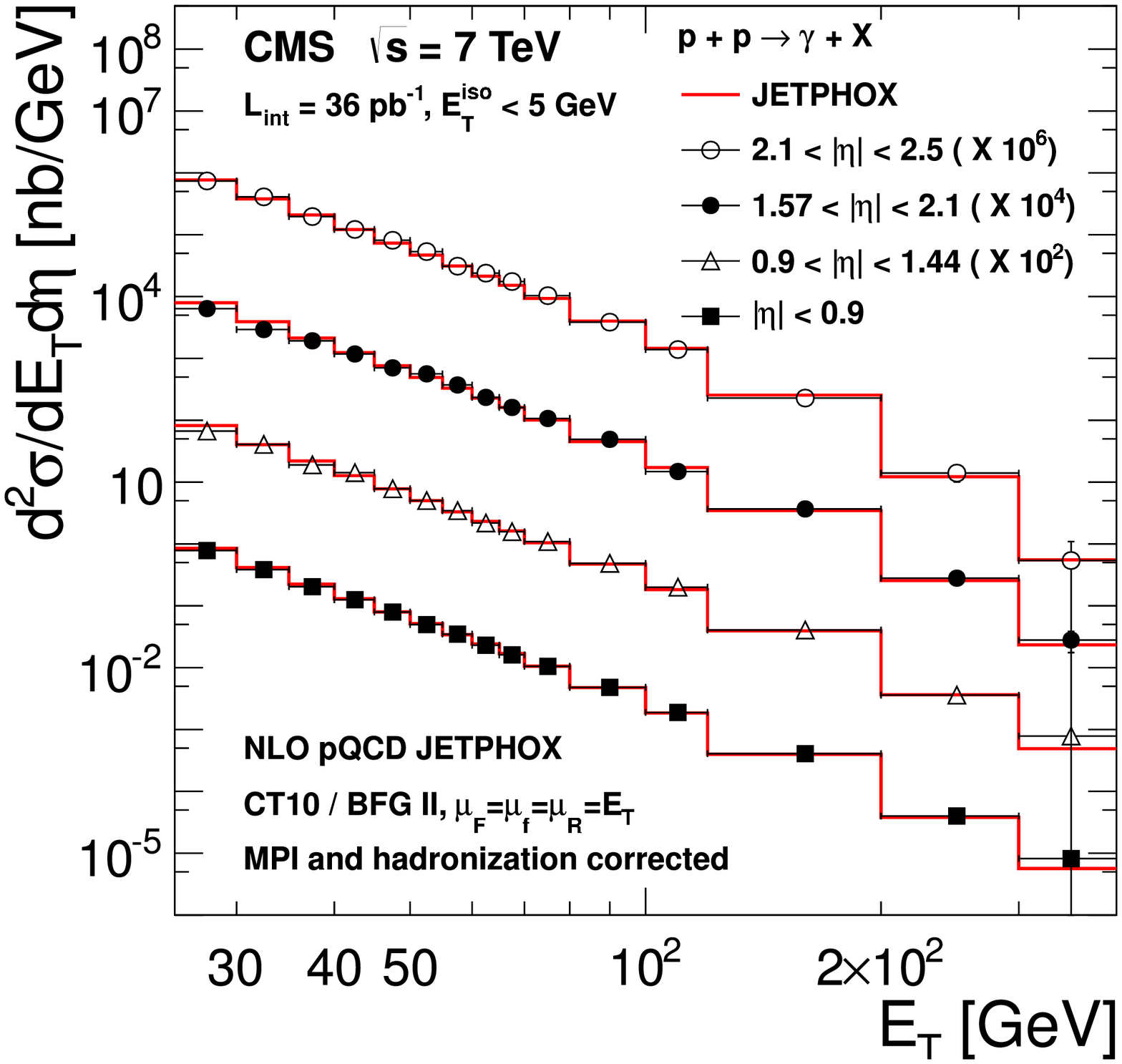}} & 
 {\includegraphics[width=0.45\columnwidth]{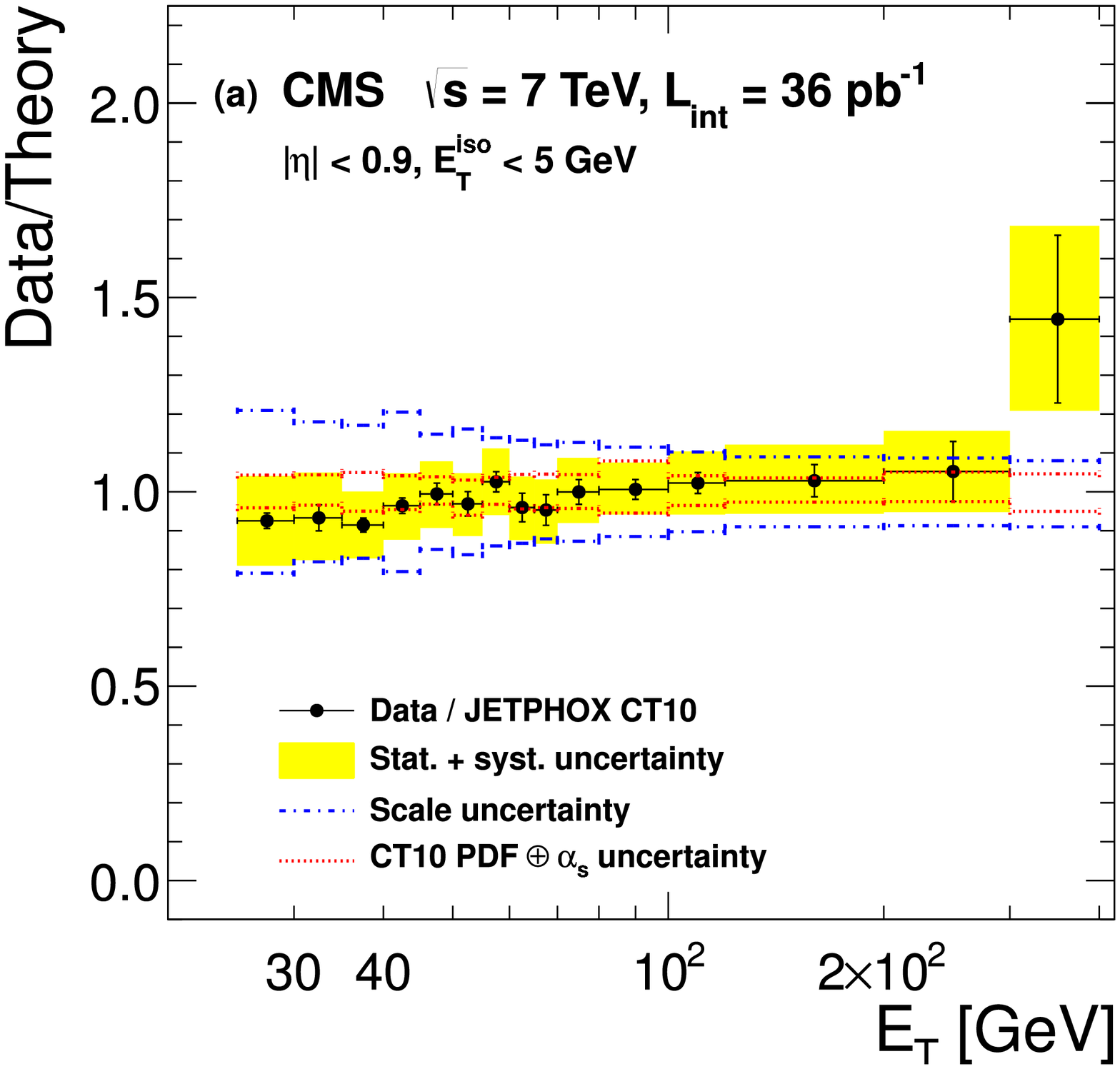}}
 \end{tabular}
 \caption{\sl Left: The isolated prompt photon cross section measured in 
  four pseudo-rapidity bins as a function of the photon transverse energy, 
  together with the NLO QCD prediction. 
  Right: Ratio of the measurement to the NLO prediction for the most central 
  bin; the vertical error bars show the statistical uncertainties, while the
  shaded areas show the total errors (not including a $4 \%$ normalisation uncertainty).
  From~\cite{Chatrchyan:2011ue}. }
 \label{fig:cmsphotons}
 \end{figure}

First measurements of isolated prompt photon production have been published by the
ATLAS~\cite{Aad:2011tw} and CMS~\cite{Chatrchyan:2011ue} experiments using $pp$ data\footnote{
Measurements have also been made at $\sqrt{s} = 2.76$~TeV,
in $pp$ and in Pb-Pb collisions~\cite{Chatrchyan:2012vq}.}
taken at 
$\sqrt{s} = 7$~TeV corresponding to an integrated luminosity of $\sim 35$~pb$^{-1}$.
For example, the CMS measurement, made in four pseudo-rapidity regions and in
the transverse energy range $25 < E_{T, \gamma} < 400$~GeV, 
is shown in Fig.~\ref{fig:cmsphotons}.
It is consistent with the NLO prediction from pQCD obtained from the JETPHOX 
program~\cite{Catani:2002ny,Aurenche:2006vj} using the CT10 PDFs.

In~\cite{DaviddEnterria}, the impact of these ATLAS and CMS data on the PDFs has
been quantified using the NNPDF reweighting technique mentioned previously~\cite{Ball:2011gg}.
Including these data in a fit similar to NNPDF2.1 leads to a significant reduction of the
uncertainty on the gluon PDF, of up to $20 \%$, most pronounced for $x \sim 0.01$. 
Moreover, the fit does not change significantly the
central value of the gluon density. This indicates that the constraints that these data
set on the gluon PDF at high $x$ have no tension with the constraints obtained from
the Tevatron jet data.

\subsubsection{Cross section ratios}
The centre-of-mass energy of the LHC is being increased in a step-wise
way with runs taken at $\sqrt{s}=7$~TeV and $8$~TeV and after the long
shutdown in 2013-2014, the machine is expected to operate at
$\sim14$~TeV. This gives rise to the possibility of measuring cross
section ratios at different $\sqrt{s}$ as well as double ratios of
hard process cross sections. The advantage of these ratios is that
experimentally many systematic uncertainties could cancel in the
measurements. Cancellation in the theoretical uncertainties on the
predictions are also expected~\cite{Mangano:2012mh} leading to very
precise predictions and measurements. These could offer interesting
new constraints on PDFs and enhanced sensitivity to new
physics. Taking the ratio between the $7$ and $8$~TeV data the $W$ and
$Z$ production cross section ratios are predicted to an accuracy of
$\sim0.2\%$ including PDF, $\alpha_s$, and scale uncertainties at
NNLO. However, for high mass $t{\bar t}$ production the predicted
uncertainty on the ratio is estimated to be $1\%$, and jet production
ratios for jets with $p_T>1$~TeV are claimed to be known to $\sim2\%$,
rising to $\sim6\%$ for jet $p_T>2$~TeV. Both of these processes probe
the very high $x$ PDFs and could therefore be used to constrain this
region. The ratios between $8$ and $14$~TeV data offer even larger
potential gains. It remains to be seen how well the experimental
uncertainties on the measured ratios will cancel but this is an
interesting proposal warranting further more detailed investigations.

%%%%%%%%%%%%%%%%%%%%%%%%%%%%%%%%%%%%%%%%%%%%%%%%%%%%%%%
\section{Conclusions and Outlook}
%%%%%%%%%%%%%%%%%%%%%%%%%%%%%%%%%%%%%%%%%%%%%%%%%%%%%%%
\label{sec:outlook}

A lot of progress has been made over the past $\sim 20$ years 
in the understanding of proton structure and in the determination of the
parton distribution functions. 
On the experimental side, the HERA collider has opened up the kinematic domain
of low $x$; high $p_T$ jet production at the Tevatron has shed light
on the gluon density at high $x$; the measurements from 
fixed target experiments have been finalised. On the theory side, the extraction
of PDFs has become more and more involved. While the first QCD fits
were made at leading order only, to a small number of $F_2$ data points, using simple
parameterisations with a few parameters, current QCD fits are now available up to NNLO;
they make use of about $3000$ data points, covering all processes that are sensitive
to proton PDFs; the parameterisations have typically
$25-30$ parameters (ten times more for the NNPDF fits); and uncertainties are
now delivered together with the central fits. The crucial need for obtaining error
bands for the PDFs has also lead the experimental collaborations to publish
their full correlated systematic uncertainties. This much improved knowledge
of proton structure comes together with lots of progress in QCD phenomenology
and theory: thanks to new calculation techniques, higher order calculations are 
now available for a wealth of processes;
several resummed calculations also exist; the development of new subtraction
techniques has allowed NLO calculations to be combined with the parton shower approach 
used in Monte-Carlos; new jet algorithms have been defined, that allow to better
compare experimental measurements with theoretical calculations. 
As a result, the theoretical predictions for the processes that are, or will be, observed
at the LHC, are much more robust than what they were one decade ago, at the start-up
of the Run II of the Tevatron.

Although most proton PDFs are now determined to a good precision, some open
issues remain. For example, the strange content of the proton is still very poorly known,
all PDFs are affected by large uncertainties at high $x$, and what happens at very low $x$
remains largely unknown.
The data that are being collected by the LHC experiments will further improve our
knowledge of proton structure - although the highest mass domain, which may be
affected by physics processes not accounted for in the Standard Model, may not
be best suited to constrain the PDFs at highest $x$.

In addition, other aspects of proton structure, which were not addressed in
this review, are far from being understood. The proton spin is one of those.
Since  the surprising finding by the EMC Collaboration that very little of the proton spin
is carried by the spins of quarks and anti-quarks, this issue has been tackled by 
several experiments, as (to mention only the most recent ones) the COMPASS experiment at CERN,
HERMES at DESY, CLAS at JLab, and the STAR and PHENIX experiments at RHIC.
One of the focus was the measurement of the contribution to the proton spin that is carried by gluons.
There is currently no experimental evidence that this contribution may be important, however the
uncertainties are large.
Another issue regards the transverse structure: the standard PDFs probe the longitudinal momentum
of the partons in a fast moving hadron, all information about the transverse structure is
integrated over.
This additional information is encoded within the Generalised Parton Distributions (GPDs),
which unify the concept of PDFs and that of hadronic form factors~\cite{GPDS}. The GPDs, which can be accessed via
exclusive processes as Deep Virtual Compton Scattering $ l p \rightarrow l p \gamma$, are
poorly constrained so far.

The study of proton structure has a continuing programme over the next
decade with, besides the LHC experiments, several new experiments and facilities in the planning,
construction or starting phase. Some of these are designed to focus on the high
$x$ region at low and moderate $Q^2$ whereas others are designed to
open a wider kinematic region than is currently accessible. 
They are briefly described below.

\paragraph{The Minerva experiment (E938) at Fermilab~\cite{MINERVA}}
 The Minerva detector is operated on the NuMI neutrino beamline. Its main goal
 is to perform precision measurements of neutrino scattering off several targets in the
 low energy regime, $E_{\nu} \sim 1 - 20$~GeV, which are needed by experiments
 studying neutrino oscillations. Following a low-energy run which ended in
 May $2012$, data taken starting from $2013$ with a higher energy beam will allow
 CC DIS to be further studied. The measurements should shed further light on
  the $d/u$ ratio at high $x$.

\paragraph{The Drell-Yan experiment E906/Seaquest~\cite{seaquest}}
  This Fermilab experiment continues the series of fixed target $pp$
  and $pd$ Drell-Yan measurements from E605, E772 and E866. Seaquest
  will operate with a $120$~GeV proton beam delivering an
  instantaneous luminosity of $10^{35}$~cm$^{-2}$s$^{-1}$, some $50$
  times the luminosity of E866. This will allow measurements of the
  $\bar{d}/\bar{u}$ ratio to be made with a factor $10$ improvement in
  precision in the region of $0.25<x<0.45$. The experiment will
  commence physics runs in 2013.

\paragraph{The COMPASS experiment at CERN} has a programme extending until $2016$ (see~\cite{Mallot:2012zc}).
In particular it will perform further measurements of DVCS in $2015-2016$ (this requires
major rearrangements of the spectrometer and the installation of a recoil detector).
Measuring the dependence in $t$, the momentum transferred at the proton vertex, will give access
to the nucleon transverse size.
Combined with the HERA data and the future JLab data (see below), a comprehensive picture of the evolution
of the nucleon's transverse size with $x_{Bjorken}$ will be achieved. 
Information on GPDs will also be obtained.

\paragraph{The upgrade of the accelerator complex at JLab\cite{JLAB}}
  The Jefferson laboratory hosts the CEBAF dual linac electron accelerator
  currently operating at $6$~GeV delivering beam to three experimental
  halls. The machine is being upgraded to operate at $12$~GeV and
  instantaneous luminosities of $10^{35}-10^{39}$~cm$^{-2}$s$^{-1}$
  which is necessary to explore the region of high $x\sim0.7$ and
  $Q^2 \lesssim 8$~GeV$^2$. Each hall houses one or more experiments
  and a fourth hall is under construction. The experiments cover a
  variety of measurements of nuclear and proton DIS including
  precision measurements of $F_L$ and $F_2$ at high $x$, measurements
  of $F_2$ neutron which will constrain the $d/u$ quark ratio at high
  $x$, DVCS measurements, as well as polarised scattering. The staged upgrade programme
  is underway and expected to commence physics operation in $2015$.

\paragraph{The Electron Ion Collider~\cite{EIC}}
  At the horizon of $\sim 2020$, a future EIC could be realised 
  as an upgrade to the existing RHIC ion collider (eRHIC). A $5-30$~GeV
  polarised electron beam would collide with polarised ion beams
  reaching a maximum of $325$~GeV for protons, and $130$~GeV/A for
  heavier nuclei. 
  Another option (MEIC / ELIC) would be to use the polarised electron beam of JLab
  and add a new ring for polarised protons or ions.
  The EIC would be the first lepton-proton collider with a polarised
  proton beam.
  It would shed further light on the proton's spin problem. 
  The contribution of gluons to the proton's spin will be measured precisely by an EIC.
  If this contribution turns out to be small, as indicated by the current experimental data,
  it would mean that  a large part of the proton
  spin is due to orbital angular momentum.
  The measurement of DVCS and of other exclusive processes
  as $J / \psi$ production, off transversely polarised  protons, will bring information on
  GPDs at low $x$. Together with the information on GPDs obtained, at higher $x$, by other experiments,
  it may then be possible
  to have a direct access to the parton angular momentum via Ji's angular momentum sum rule~\cite{Ji:1996ek} (this requires
  the GPDs to be reconstructed in a large kinematic domain).
  Moreover, the
  EIC physics programme also includes measurements of unpolarised proton and
  deuteron scattering at low $x$, measurements of $F_L$, and
  semi-inclusive DIS measurements sensitive to $s/\bar{s}$ content of
  the proton.

\paragraph{The LHeC project~\cite{LHeC}}
 This is a novel proposal to build
  a ring or linac electron machine to collide with an LHC proton/ion
  beam using interaction point IP2 in the LHC tunnel. The LHC and LHeC
  could run simultaneously with operation commencing in 2023 or later, after the
  long shutdown in preparation for LHC high luminosity running. The
  electron ring operating at $60$~GeV and $\sqrt{s}=1.3$~TeV could
  offer a luminosity of $10^{33}$~cm$^{-2}$s$^{-1}$, a factor $20$
  higher than HERA. A linac option could achieve higher $\sqrt{s}$ but
  the luminosity at these higher energies would be smaller.
  The physics programme would cover the
  measurement of precision NC and CC structure functions with a
  $20$-fold increase in kinematic reach for $Q^2$ and $1/x$ compared
  to HERA, improved accuracy in the determination of $\alpha_s$, and 
  the understanding of saturation and of non-linear dynamics.

The long term future of DIS experiments is not yet clear with the last
two projects described above still being discussed within the
appropriate committees. Nevertheless, our knowledge of proton
structure and QCD is expected to improve significantly within the next
decade, alongside more general developments across the field of
particle physics. Such developments will come in particular from the LHC
experiments which, at the time of writing, have just announced the
discovery of a new particle in their searches for the Standard Model
Higgs boson.

\section*{Acknowledgments}
The authors would like to thank colleagues from the PDF4LHC and HERAPDF working
groups where many useful discussions have taken place, and to Hubert
Spiesberger for helping our understanding of radiative corrections.
We would also like to thank Marc Cano Bret, Robert Hickling and Hayk Pirumov
for preparing several figures and for their help in preparing the
manuscript for publication.
The work of E.R. is supported by the STFC grant number STH001042-2.

\end{document}